\documentstyle[csp-cm,zed-cm,alg,12pt,side,times,psfig,html,acronym,fancyhead,lgrind,amstex,setspace]{myreport}

\ifx \htmlstyloaded\relax  \else\let\htmlstyloaded\relax\fi

\newcommand{\htmladdnormallink}[2]{#1}

\newcommand{\htmladdimg}[1]{}

\newcommand{\externallabels}[2]{}

\newcommand{\externalref}[1]{}

\makeatletter
\def\makeinnocent#1{\catcode`#1=12 }
\def\csarg#1#2{\expandafter#1\csname#2\endcsname}

\def\ThrowAwayComment#1{\begingroup
    \def\CurrentComment{#1}%
    \let\do\makeinnocent \dospecials
    \makeinnocent\^^L
    \endlinechar`\^^M \catcode`\^^M=12 \xComment}
{\catcode`\^^M=12 \endlinechar=-1 %
 \gdef\xComment#1^^M{\def\test{#1}
      \csarg\ifx{PlainEnd\CurrentComment Test}\test
          \let\html@next\endgroup
      \else \csarg\ifx{LaLaEnd\CurrentComment Test}\test
            \edef\html@next{\endgroup\noexpand\end{\CurrentComment}}
      \else \let\html@next\xComment
      \fi \fi \html@next}
}
\makeatother

\def\includecomment
 #1{\expandafter\def\csname#1\endcsname{}%
    \expandafter\def\csname end#1\endcsname{}}
\def\excludecomment
 #1{\expandafter\def\csname#1\endcsname{\ThrowAwayComment{#1}}%
    {\escapechar=-1\relax
     \csarg\xdef{PlainEnd#1Test}{\string\\end#1}%
     \csarg\xdef{LaLaEnd#1Test}{\string\\end\string\{#1\string\}}%
    }}

\excludecomment{comment}

\excludecomment{rawhtml}

\excludecomment{htmlonly}
\newcommand{\html}[1]{}

\newcommand{\hyperref}[4]{#2\ref{#4}#3}

\newcommand{\htmlimage}[1]{}

\newcommand{\htmladdtonavigation}[1]{}

\newtheorem{hyp}{Proposition}
\newcommand{\li}{\mbox{\ \underline{li}\ }}
\newcommand{\co}{\mbox{\ \underline{co}\ }}
\newcommand{\slice}{\mbox{\ \underline{sl}\ }}
\newcommand{\petripre}[1]{\mbox{pre-}#1}
\newcommand{\petripost}[1]{\mbox{post-}#1}
\newcommand{\slowSLPterm}{\tau_{task} - (\tau_{task} + \tau_{rb}) E[X]
- (\Delta_{vm} - {1 \over \lambda_{vm}}) E[Y]}
\newcommand{\slowterm}{\tau_{task} - (\tau_{task} + \tau_{rb}) E[X]
- (\Delta_{vm}S_{parallel} - {1 \over \lambda_{vm}}) E[Y]}

\newcommand{\hess}[1]{#1(x^*)}

\newcommand{\aparms}{$\lambda_{vm} = 0.03$ virtual messages per millisecond, 
$\Delta_{vm} = 30.0$ milliseconds, $\tau_{task} = 7.0$ milliseconds, 
$\tau_{rb} = 1.0$ milliseconds, $S_{parallel} = 1.5$ and $C_r=100$ where 
$C_r$ is the speedup gained from reading the cache over 
computing the result. These parameters were chosen to accommodate the 
\acl{ES} topology task which is currently the most time consuming \acl{RDRN} 
operation and precludes the \acl{ES} nodes from becoming mobile without 
\acl{VNC}}

\newcommand{\oparms}{$\lambda_{vm} = 0.1$ virtual messages per millisecond,
$\Delta_{vm} = 45.0$ milliseconds, $\tau_{task} = 10.0$ milliseconds,
$\tau_{rb} = 1.0$ milliseconds}
\newcounter{treecount}
\newcounter{branchcount}
\newsavebox{\parentbox}
\newsavebox{\treebox}
\newsavebox{\treeboxone}
\newsavebox{\treeboxtwo}
\newsavebox{\treeboxthree}
\newsavebox{\treeboxfour}
\newsavebox{\treeboxfive}
\newsavebox{\treeboxsix}
\newsavebox{\treeboxseven}
\newsavebox{\treeboxeight}
\newsavebox{\treeboxnine}
\newsavebox{\treeboxten}
\newsavebox{\treeboxeleven}
\newsavebox{\treeboxtwelve}
\newsavebox{\treeboxthirteen}
\newsavebox{\treeboxfourteen}
\newsavebox{\treeboxfifteen}
\newsavebox{\treeboxsixteen}
\newsavebox{\treeboxseventeen}
\newsavebox{\treeboxeighteen}
\newsavebox{\treeboxnineteen}
\newsavebox{\treeboxtwenty}
\newlength{\treeoffsetone}
\newlength{\treeoffsettwo}
\newlength{\treeoffsetthree}
\newlength{\treeoffsetfour}
\newlength{\treeoffsetfive}
\newlength{\treeoffsetsix}
\newlength{\treeoffsetseven}
\newlength{\treeoffseteight}
\newlength{\treeoffsetnine}
\newlength{\treeoffsetten}
\newlength{\treeoffseteleven}
\newlength{\treeoffsettwelve}
\newlength{\treeoffsetthirteen}
\newlength{\treeoffsetfourteen}
\newlength{\treeoffsetfifteen}
\newlength{\treeoffsetsixteen}
\newlength{\treeoffsetseventeen}
\newlength{\treeoffseteighteen}
\newlength{\treeoffsetnineteen}
\newlength{\treeoffsettwenty}

\newlength{\treeshiftone}
\newlength{\treeshifttwo}
\newlength{\treeshiftthree}
\newlength{\treeshiftfour}
\newlength{\treeshiftfive}
\newlength{\treeshiftsix}
\newlength{\treeshiftseven}
\newlength{\treeshifteight}
\newlength{\treeshiftnine}
\newlength{\treeshiftten}
\newlength{\treeshifteleven}
\newlength{\treeshifttwelve}
\newlength{\treeshiftthirteen}
\newlength{\treeshiftfourteen}
\newlength{\treeshiftfifteen}
\newlength{\treeshiftsixteen}
\newlength{\treeshiftseventeen}
\newlength{\treeshifteighteen}
\newlength{\treeshiftnineteen}
\newlength{\treeshifttwenty}
\newlength{\treewidthone}
\newlength{\treewidthtwo}
\newlength{\treewidththree}
\newlength{\treewidthfour}
\newlength{\treewidthfive}
\newlength{\treewidthsix}
\newlength{\treewidthseven}
\newlength{\treewidtheight}
\newlength{\treewidthnine}
\newlength{\treewidthten}
\newlength{\treewidtheleven}
\newlength{\treewidthtwelve}
\newlength{\treewidththirteen}
\newlength{\treewidthfourteen}
\newlength{\treewidthfifteen}
\newlength{\treewidthsixteen}
\newlength{\treewidthseventeen}
\newlength{\treewidtheighteen}
\newlength{\treewidthnineteen}
\newlength{\treewidthtwenty}
\newlength{\daughteroffsetone}
\newlength{\daughteroffsettwo}
\newlength{\daughteroffsetthree}
\newlength{\daughteroffsetfour}
\newlength{\branchwidthone}
\newlength{\branchwidthtwo}
\newlength{\branchwidththree}
\newlength{\branchwidthfour}
\newlength{\parentoffset}
\newlength{\treeoffset}
\newlength{\daughteroffset}
\newlength{\branchwidth}
\newlength{\parentwidth}
\newlength{\treewidth}
\newcommand{\ontop}[1]{\begin{tabular}{c}#1\end{tabular}}
\newcommand{\poptree}{%
\ifnum\value{treecount}=0\typeout{QobiTeX warning---Tree stack underflow}\fi%
\addtocounter{treecount}{-1}%
\setlength{\treeoffsettwo}{\treeoffsetthree}%
\setlength{\treeoffsetthree}{\treeoffsetfour}%
\setlength{\treeoffsetfour}{\treeoffsetfive}%
\setlength{\treeoffsetfive}{\treeoffsetsix}%
\setlength{\treeoffsetsix}{\treeoffsetseven}%
\setlength{\treeoffsetseven}{\treeoffseteight}%
\setlength{\treeoffseteight}{\treeoffsetnine}%
\setlength{\treeoffsetnine}{\treeoffsetten}%
\setlength{\treeoffsetten}{\treeoffseteleven}%
\setlength{\treeoffseteleven}{\treeoffsettwelve}%
\setlength{\treeoffsettwelve}{\treeoffsetthirteen}%
\setlength{\treeoffsetthirteen}{\treeoffsetfourteen}%
\setlength{\treeoffsetfourteen}{\treeoffsetfifteen}%
\setlength{\treeoffsetfifteen}{\treeoffsetsixteen}%
\setlength{\treeoffsetsixteen}{\treeoffsetseventeen}%
\setlength{\treeoffsetseventeen}{\treeoffseteighteen}%
\setlength{\treeoffseteighteen}{\treeoffsetnineteen}%
\setlength{\treeoffsetnineteen}{\treeoffsettwenty}%
\setlength{\treeshifttwo}{\treeshiftthree}%
\setlength{\treeshiftthree}{\treeshiftfour}%
\setlength{\treeshiftfour}{\treeshiftfive}%
\setlength{\treeshiftfive}{\treeshiftsix}%
\setlength{\treeshiftsix}{\treeshiftseven}%
\setlength{\treeshiftseven}{\treeshifteight}%
\setlength{\treeshifteight}{\treeshiftnine}%
\setlength{\treeshiftnine}{\treeshiftten}%
\setlength{\treeshiftten}{\treeshifteleven}%
\setlength{\treeshifteleven}{\treeshifttwelve}%
\setlength{\treeshifttwelve}{\treeshiftthirteen}%
\setlength{\treeshiftthirteen}{\treeshiftfourteen}%
\setlength{\treeshiftfourteen}{\treeshiftfifteen}%
\setlength{\treeshiftfifteen}{\treeshiftsixteen}%
\setlength{\treeshiftsixteen}{\treeshiftseventeen}%
\setlength{\treeshiftseventeen}{\treeshifteighteen}%
\setlength{\treeshifteighteen}{\treeshiftnineteen}%
\setlength{\treeshiftnineteen}{\treeshifttwenty}%
\setlength{\treewidthtwo}{\treewidththree}%
\setlength{\treewidththree}{\treewidthfour}%
\setlength{\treewidthfour}{\treewidthfive}%
\setlength{\treewidthfive}{\treewidthsix}%
\setlength{\treewidthsix}{\treewidthseven}%
\setlength{\treewidthseven}{\treewidtheight}%
\setlength{\treewidtheight}{\treewidthnine}%
\setlength{\treewidthnine}{\treewidthten}%
\setlength{\treewidthten}{\treewidtheleven}%
\setlength{\treewidtheleven}{\treewidthtwelve}%
\setlength{\treewidthtwelve}{\treewidththirteen}%
\setlength{\treewidththirteen}{\treewidthfourteen}%
\setlength{\treewidthfourteen}{\treewidthfifteen}%
\setlength{\treewidthfifteen}{\treewidthsixteen}%
\setlength{\treewidthsixteen}{\treewidthseventeen}%
\setlength{\treewidthseventeen}{\treewidtheighteen}%
\setlength{\treewidtheighteen}{\treewidthnineteen}%
\setlength{\treewidthnineteen}{\treewidthtwenty}%
\sbox{\treeboxtwo}{\usebox{\treeboxthree}}%
\sbox{\treeboxthree}{\usebox{\treeboxfour}}%
\sbox{\treeboxfour}{\usebox{\treeboxfive}}%
\sbox{\treeboxfive}{\usebox{\treeboxsix}}%
\sbox{\treeboxsix}{\usebox{\treeboxseven}}%
\sbox{\treeboxseven}{\usebox{\treeboxeight}}%
\sbox{\treeboxeight}{\usebox{\treeboxnine}}%
\sbox{\treeboxnine}{\usebox{\treeboxten}}%
\sbox{\treeboxten}{\usebox{\treeboxeleven}}%
\sbox{\treeboxeleven}{\usebox{\treeboxtwelve}}%
\sbox{\treeboxtwelve}{\usebox{\treeboxthirteen}}%
\sbox{\treeboxthirteen}{\usebox{\treeboxfourteen}}%
\sbox{\treeboxfourteen}{\usebox{\treeboxfifteen}}%
\sbox{\treeboxfifteen}{\usebox{\treeboxsixteen}}%
\sbox{\treeboxsixteen}{\usebox{\treeboxseventeen}}%
\sbox{\treeboxseventeen}{\usebox{\treeboxeighteen}}%
\sbox{\treeboxeighteen}{\usebox{\treeboxnineteen}}%
\sbox{\treeboxnineteen}{\usebox{\treeboxtwenty}}}
\newcommand{\leaf}[1]{%
\ifnum\value{treecount}=20\typeout{QobiTeX warning---Tree stack overflow}\fi%
\addtocounter{treecount}{1}%
\sbox{\treeboxtwenty}{\usebox{\treeboxnineteen}}%
\sbox{\treeboxnineteen}{\usebox{\treeboxeighteen}}%
\sbox{\treeboxeighteen}{\usebox{\treeboxseventeen}}%
\sbox{\treeboxseventeen}{\usebox{\treeboxsixteen}}%
\sbox{\treeboxsixteen}{\usebox{\treeboxfifteen}}%
\sbox{\treeboxfifteen}{\usebox{\treeboxfourteen}}%
\sbox{\treeboxfourteen}{\usebox{\treeboxthirteen}}%
\sbox{\treeboxthirteen}{\usebox{\treeboxtwelve}}%
\sbox{\treeboxtwelve}{\usebox{\treeboxeleven}}%
\sbox{\treeboxeleven}{\usebox{\treeboxten}}%
\sbox{\treeboxten}{\usebox{\treeboxnine}}%
\sbox{\treeboxnine}{\usebox{\treeboxeight}}%
\sbox{\treeboxeight}{\usebox{\treeboxseven}}%
\sbox{\treeboxseven}{\usebox{\treeboxsix}}%
\sbox{\treeboxsix}{\usebox{\treeboxfive}}%
\sbox{\treeboxfive}{\usebox{\treeboxfour}}%
\sbox{\treeboxfour}{\usebox{\treeboxthree}}%
\sbox{\treeboxthree}{\usebox{\treeboxtwo}}%
\sbox{\treeboxtwo}{\usebox{\treeboxone}}%
\sbox{\treeboxone}{\ontop{#1}}%
\sbox{\treeboxone}{\raisebox{-\ht\treeboxone}{\usebox{\treeboxone}}}%
\setlength{\treeoffsettwenty}{\treeoffsetnineteen}%
\setlength{\treeoffsetnineteen}{\treeoffseteighteen}%
\setlength{\treeoffseteighteen}{\treeoffsetseventeen}%
\setlength{\treeoffsetseventeen}{\treeoffsetsixteen}%
\setlength{\treeoffsetsixteen}{\treeoffsetfifteen}%
\setlength{\treeoffsetfifteen}{\treeoffsetfourteen}%
\setlength{\treeoffsetfourteen}{\treeoffsetthirteen}%
\setlength{\treeoffsetthirteen}{\treeoffsettwelve}%
\setlength{\treeoffsettwelve}{\treeoffseteleven}%
\setlength{\treeoffseteleven}{\treeoffsetten}%
\setlength{\treeoffsetten}{\treeoffsetnine}%
\setlength{\treeoffsetnine}{\treeoffseteight}%
\setlength{\treeoffseteight}{\treeoffsetseven}%
\setlength{\treeoffsetseven}{\treeoffsetsix}%
\setlength{\treeoffsetsix}{\treeoffsetfive}%
\setlength{\treeoffsetfive}{\treeoffsetfour}%
\setlength{\treeoffsetfour}{\treeoffsetthree}%
\setlength{\treeoffsetthree}{\treeoffsettwo}%
\setlength{\treeoffsettwo}{\treeoffsetone}%
\setlength{\treeoffsetone}{0.5\wd\treeboxone}%
\setlength{\treeshifttwenty}{\treeshiftnineteen}%
\setlength{\treeshiftnineteen}{\treeshifteighteen}%
\setlength{\treeshifteighteen}{\treeshiftseventeen}%
\setlength{\treeshiftseventeen}{\treeshiftsixteen}%
\setlength{\treeshiftsixteen}{\treeshiftfifteen}%
\setlength{\treeshiftfifteen}{\treeshiftfourteen}%
\setlength{\treeshiftfourteen}{\treeshiftthirteen}%
\setlength{\treeshiftthirteen}{\treeshifttwelve}%
\setlength{\treeshifttwelve}{\treeshifteleven}%
\setlength{\treeshifteleven}{\treeshiftten}%
\setlength{\treeshiftten}{\treeshiftnine}%
\setlength{\treeshiftnine}{\treeshifteight}%
\setlength{\treeshifteight}{\treeshiftseven}%
\setlength{\treeshiftseven}{\treeshiftsix}%
\setlength{\treeshiftsix}{\treeshiftfive}%
\setlength{\treeshiftfive}{\treeshiftfour}%
\setlength{\treeshiftfour}{\treeshiftthree}%
\setlength{\treeshiftthree}{\treeshifttwo}%
\setlength{\treeshifttwo}{\treeshiftone}%
\setlength{\treeshiftone}{0pt}%
\setlength{\treewidthtwenty}{\treewidthnineteen}%
\setlength{\treewidthnineteen}{\treewidtheighteen}%
\setlength{\treewidtheighteen}{\treewidthseventeen}%
\setlength{\treewidthseventeen}{\treewidthsixteen}%
\setlength{\treewidthsixteen}{\treewidthfifteen}%
\setlength{\treewidthfifteen}{\treewidthfourteen}%
\setlength{\treewidthfourteen}{\treewidththirteen}%
\setlength{\treewidththirteen}{\treewidthtwelve}%
\setlength{\treewidthtwelve}{\treewidtheleven}%
\setlength{\treewidtheleven}{\treewidthten}%
\setlength{\treewidthten}{\treewidthnine}%
\setlength{\treewidthnine}{\treewidtheight}%
\setlength{\treewidtheight}{\treewidthseven}%
\setlength{\treewidthseven}{\treewidthsix}%
\setlength{\treewidthsix}{\treewidthfive}%
\setlength{\treewidthfive}{\treewidthfour}%
\setlength{\treewidthfour}{\treewidththree}%
\setlength{\treewidththree}{\treewidthtwo}%
\setlength{\treewidthtwo}{\treewidthone}%
\setlength{\treewidthone}{\wd\treeboxone}}
\newcommand{\branch}[2]{%
\setcounter{branchcount}{#1}%
\ifnum\value{branchcount}=1\sbox{\parentbox}{\ontop{#2}}%
\setlength{\parentoffset}{\treeoffsetone}%
\addtolength{\parentoffset}{-0.5\wd\parentbox}%
\setlength{\daughteroffset}{0in}%
\ifdim\parentoffset<0in%
\setlength{\daughteroffset}{-\parentoffset}%
\setlength{\parentoffset}{0in}\fi%
\setlength{\parentwidth}{\parentoffset}%
\addtolength{\parentwidth}{\wd\parentbox}%
\setlength{\treeoffset}{\daughteroffset}%
\addtolength{\treeoffset}{\treeoffsetone}%
\setlength{\treewidth}{\wd\treeboxone}%
\addtolength{\treewidth}{\daughteroffset}%
\ifdim\treewidth<\parentwidth\setlength{\treewidth}{\parentwidth}\fi%
\sbox{\treebox}{\begin{minipage}{\treewidth}%
\begin{flushleft}%
\hspace*{\parentoffset}\usebox{\parentbox}\\
{\setlength{\unitlength}{2ex}%
\hspace*{\treeoffset}\begin{picture}(0,1)%
\put(0,0){\line(0,1){1}}%
\end{picture}}\\
\vspace{-\baselineskip}
\hspace*{\daughteroffset}%
\raisebox{-\ht\treeboxone}{\usebox{\treeboxone}}%
\end{flushleft}%
\end{minipage}}%
\setlength{\treeoffsetone}{\parentoffset}%
\addtolength{\treeoffsetone}{0.5\wd\parentbox}%
\setlength{\treeshiftone}{0pt}%
\setlength{\treewidthone}{\treewidth}%
\sbox{\treeboxone}{\usebox{\treebox}}%
\else\ifnum\value{branchcount}=2\sbox{\parentbox}{\ontop{#2}}%
\setlength{\branchwidthone}{\treewidthtwo}%
\addtolength{\branchwidthone}{\treeoffsetone}%
\addtolength{\branchwidthone}{-\treeshiftone}%
\addtolength{\branchwidthone}{-\treeoffsettwo}%
\setlength{\branchwidth}{\branchwidthone}%
\setlength{\daughteroffsetone}{\branchwidth}%
\addtolength{\daughteroffsetone}{-\branchwidthone}%
\addtolength{\daughteroffsetone}{-\treeshiftone}%
\setlength{\parentoffset}{-0.5\wd\parentbox}%
\addtolength{\parentoffset}{\treeoffsettwo}%
\addtolength{\parentoffset}{0.5\branchwidth}%
\setlength{\daughteroffset}{0in}%
\ifdim\parentoffset<0in%
\setlength{\daughteroffset}{-\parentoffset}%
\setlength{\parentoffset}{0in}\fi%
\setlength{\parentwidth}{\parentoffset}%
\addtolength{\parentwidth}{\wd\parentbox}%
\setlength{\treeoffset}{\daughteroffset}%
\addtolength{\treeoffset}{\treeoffsettwo}%
\setlength{\treewidth}{\wd\treeboxone}%
\addtolength{\treewidth}{\daughteroffsetone}%
\addtolength{\treewidth}{\treewidthtwo}%
\addtolength{\treewidth}{\daughteroffset}%
\ifdim\treewidth<\parentwidth\setlength{\treewidth}{\parentwidth}\fi%
\sbox{\treebox}{\begin{minipage}{\treewidth}%
\begin{flushleft}%
\hspace*{\parentoffset}\usebox{\parentbox}\\
{\setlength{\unitlength}{0.5\branchwidth}%
\hspace*{\treeoffset}\begin{picture}(2,0.5)%
\put(0,0){\line(2,1){1}}%
\put(2,0){\line(-2,1){1}}%
\end{picture}}\\
\vspace{-\baselineskip}
\hspace*{\daughteroffset}%
\makebox[\treewidthtwo][l]%
{\raisebox{-\ht\treeboxtwo}{\usebox{\treeboxtwo}}}%
\hspace*{\daughteroffsetone}%
\raisebox{-\ht\treeboxone}{\usebox{\treeboxone}}%
\end{flushleft}%
\end{minipage}}%
\setlength{\treeoffsetone}{\parentoffset}%
\addtolength{\treeoffsetone}{0.5\wd\parentbox}%
\setlength{\treeshiftone}{0pt}%
\setlength{\treewidthone}{\treewidth}%
\sbox{\treeboxone}{\usebox{\treebox}}\poptree%
\else\ifnum\value{branchcount}=3\sbox{\parentbox}{\ontop{#2}}%
\setlength{\branchwidthone}{\treewidthtwo}%
\addtolength{\branchwidthone}{\treeoffsetone}%
\addtolength{\branchwidthone}{-\treeshiftone}%
\addtolength{\branchwidthone}{-\treeoffsettwo}%
\setlength{\branchwidthtwo}{\treewidththree}%
\addtolength{\branchwidthtwo}{\treeoffsettwo}%
\addtolength{\branchwidthtwo}{-\treeshifttwo}%
\addtolength{\branchwidthtwo}{-\treeoffsetthree}%
\setlength{\branchwidth}{\branchwidthone}%
\ifdim\branchwidthtwo>\branchwidth%
\setlength{\branchwidth}{\branchwidthtwo}\fi%
\setlength{\daughteroffsetone}{\branchwidth}%
\addtolength{\daughteroffsetone}{-\branchwidthone}%
\addtolength{\daughteroffsetone}{-\treeshiftone}%
\setlength{\daughteroffsettwo}{\branchwidth}%
\addtolength{\daughteroffsettwo}{-\branchwidthtwo}%
\addtolength{\daughteroffsettwo}{-\treeshifttwo}%
\setlength{\parentoffset}{-0.5\wd\parentbox}%
\addtolength{\parentoffset}{\treeoffsetthree}%
\addtolength{\parentoffset}{\branchwidth}%
\setlength{\daughteroffset}{0in}%
\ifdim\parentoffset<0in%
\setlength{\daughteroffset}{-\parentoffset}%
\setlength{\parentoffset}{0in}\fi%
\setlength{\parentwidth}{\parentoffset}%
\addtolength{\parentwidth}{\wd\parentbox}%
\setlength{\treeoffset}{\daughteroffset}%
\addtolength{\treeoffset}{\treeoffsetthree}%
\setlength{\treewidth}{\wd\treeboxone}%
\addtolength{\treewidth}{\daughteroffsetone}%
\addtolength{\treewidth}{\treewidthtwo}%
\addtolength{\treewidth}{\daughteroffsettwo}%
\addtolength{\treewidth}{\treewidththree}%
\addtolength{\treewidth}{\daughteroffset}%
\ifdim\treewidth<\parentwidth\setlength{\treewidth}{\parentwidth}\fi%
\sbox{\treebox}{\begin{minipage}{\treewidth}%
\begin{flushleft}%
\hspace*{\parentoffset}\usebox{\parentbox}\\
{\setlength{\unitlength}{0.5\branchwidth}%
\hspace*{\treeoffset}\begin{picture}(4,1)%
\put(0,0){\line(2,1){2}}%
\put(2,0){\line(0,1){1}}%
\put(4,0){\line(-2,1){2}}%
\end{picture}}\\
\vspace{-\baselineskip}
\hspace*{\daughteroffset}%
\makebox[\treewidththree][l]%
{\raisebox{-\ht\treeboxthree}{\usebox{\treeboxthree}}}%
\hspace*{\daughteroffsettwo}%
\makebox[\treewidthtwo][l]%
{\raisebox{-\ht\treeboxtwo}{\usebox{\treeboxtwo}}}%
\hspace*{\daughteroffsetone}%
\raisebox{-\ht\treeboxone}{\usebox{\treeboxone}}%
\end{flushleft}%
\end{minipage}}%
\setlength{\treeoffsetone}{\parentoffset}%
\addtolength{\treeoffsetone}{0.5\wd\parentbox}%
\setlength{\treeshiftone}{0pt}%
\setlength{\treewidthone}{\treewidth}%
\sbox{\treeboxone}{\usebox{\treebox}}\poptree\poptree%
\else\ifnum\value{branchcount}=4\sbox{\parentbox}{\ontop{#2}}%
\setlength{\branchwidthone}{\treewidthtwo}%
\addtolength{\branchwidthone}{\treeoffsetone}%
\addtolength{\branchwidthone}{-\treeshiftone}%
\addtolength{\branchwidthone}{-\treeoffsettwo}%
\setlength{\branchwidthtwo}{\treewidththree}%
\addtolength{\branchwidthtwo}{\treeoffsettwo}%
\addtolength{\branchwidthtwo}{-\treeshifttwo}%
\addtolength{\branchwidthtwo}{-\treeoffsetthree}%
\setlength{\branchwidththree}{\treewidthfour}%
\addtolength{\branchwidththree}{\treeoffsetthree}%
\addtolength{\branchwidththree}{-\treeshiftthree}%
\addtolength{\branchwidththree}{-\treeoffsetfour}%
\setlength{\branchwidth}{\branchwidthone}%
\ifdim\branchwidthtwo>\branchwidth%
\setlength{\branchwidth}{\branchwidthtwo}\fi%
\ifdim\branchwidththree>\branchwidth%
\setlength{\branchwidth}{\branchwidththree}\fi%
\setlength{\daughteroffsetone}{\branchwidth}%
\addtolength{\daughteroffsetone}{-\branchwidthone}%
\addtolength{\daughteroffsetone}{-\treeshiftone}%
\setlength{\daughteroffsettwo}{\branchwidth}%
\addtolength{\daughteroffsettwo}{-\branchwidthtwo}%
\addtolength{\daughteroffsettwo}{-\treeshifttwo}%
\setlength{\daughteroffsetthree}{\branchwidth}%
\addtolength{\daughteroffsetthree}{-\branchwidththree}%
\addtolength{\daughteroffsetthree}{-\treeshiftthree}%
\setlength{\parentoffset}{-0.5\wd\parentbox}%
\addtolength{\parentoffset}{\treeoffsetfour}%
\addtolength{\parentoffset}{1.5\branchwidth}%
\setlength{\daughteroffset}{0in}%
\ifdim\parentoffset<0in%
\setlength{\daughteroffset}{-\parentoffset}%
\setlength{\parentoffset}{0in}\fi%
\setlength{\parentwidth}{\parentoffset}%
\addtolength{\parentwidth}{\wd\parentbox}%
\setlength{\treeoffset}{\daughteroffset}%
\addtolength{\treeoffset}{\treeoffsetfour}%
\setlength{\treewidth}{\wd\treeboxone}%
\addtolength{\treewidth}{\daughteroffsetone}%
\addtolength{\treewidth}{\treewidthtwo}%
\addtolength{\treewidth}{\daughteroffsettwo}%
\addtolength{\treewidth}{\treewidththree}%
\addtolength{\treewidth}{\daughteroffsetthree}%
\addtolength{\treewidth}{\treewidthfour}%
\addtolength{\treewidth}{\daughteroffset}%
\ifdim\treewidth<\parentwidth\setlength{\treewidth}{\parentwidth}\fi%
\sbox{\treebox}{\begin{minipage}{\treewidth}%
\begin{flushleft}%
\hspace*{\parentoffset}\usebox{\parentbox}\\
{\setlength{\unitlength}{0.5\branchwidth}%
\hspace*{\treeoffset}\begin{picture}(6,1)%
\put(0,0){\line(3,1){3}}%
\put(2,0){\line(1,1){1}}%
\put(4,0){\line(-1,1){1}}%
\put(6,0){\line(-3,1){3}}%
\end{picture}}\\
\vspace{-\baselineskip}
\hspace*{\daughteroffset}%
\makebox[\treewidthfour][l]%
{\raisebox{-\ht\treeboxfour}{\usebox{\treeboxfour}}}%
\hspace*{\daughteroffsetthree}%
\makebox[\treewidththree][l]%
{\raisebox{-\ht\treeboxthree}{\usebox{\treeboxthree}}}%
\hspace*{\daughteroffsettwo}%
\makebox[\treewidthtwo][l]%
{\raisebox{-\ht\treeboxtwo}{\usebox{\treeboxtwo}}}%
\hspace*{\daughteroffsetone}%
\raisebox{-\ht\treeboxone}{\usebox{\treeboxone}}%
\end{flushleft}%
\end{minipage}}%
\setlength{\treeoffsetone}{\parentoffset}%
\addtolength{\treeoffsetone}{0.5\wd\parentbox}%
\setlength{\treeshiftone}{0pt}%
\setlength{\treewidthone}{\treewidth}%
\sbox{\treeboxone}{\usebox{\treebox}}\poptree\poptree\poptree%
\else\ifnum\value{branchcount}=5\sbox{\parentbox}{\ontop{#2}}%
\setlength{\branchwidthone}{\treewidthtwo}%
\addtolength{\branchwidthone}{\treeoffsetone}%
\addtolength{\branchwidthone}{-\treeshiftone}%
\addtolength{\branchwidthone}{-\treeoffsettwo}%
\setlength{\branchwidthtwo}{\treewidththree}%
\addtolength{\branchwidthtwo}{\treeoffsettwo}%
\addtolength{\branchwidthtwo}{-\treeshifttwo}%
\addtolength{\branchwidthtwo}{-\treeoffsetthree}%
\setlength{\branchwidththree}{\treewidthfour}%
\addtolength{\branchwidththree}{\treeoffsetthree}%
\addtolength{\branchwidththree}{-\treeshiftthree}%
\addtolength{\branchwidththree}{-\treeoffsetfour}%
\setlength{\branchwidthfour}{\treewidthfive}%
\addtolength{\branchwidthfour}{\treeoffsetfour}%
\addtolength{\branchwidthfour}{-\treeshiftfour}%
\addtolength{\branchwidthfour}{-\treeoffsetfive}%
\setlength{\branchwidth}{\branchwidthone}%
\ifdim\branchwidthtwo>\branchwidth%
\setlength{\branchwidth}{\branchwidthtwo}\fi%
\ifdim\branchwidththree>\branchwidth%
\setlength{\branchwidth}{\branchwidththree}\fi%
\ifdim\branchwidthfour>\branchwidth%
\setlength{\branchwidth}{\branchwidthfour}\fi%
\setlength{\daughteroffsetone}{\branchwidth}%
\addtolength{\daughteroffsetone}{-\branchwidthone}%
\addtolength{\daughteroffsetone}{-\treeshiftone}%
\setlength{\daughteroffsettwo}{\branchwidth}%
\addtolength{\daughteroffsettwo}{-\branchwidthtwo}%
\addtolength{\daughteroffsettwo}{-\treeshifttwo}%
\setlength{\daughteroffsetthree}{\branchwidth}%
\addtolength{\daughteroffsetthree}{-\branchwidththree}%
\addtolength{\daughteroffsetthree}{-\treeshiftthree}%
\setlength{\daughteroffsetfour}{\branchwidth}%
\addtolength{\daughteroffsetfour}{-\branchwidthfour}%
\addtolength{\daughteroffsetfour}{-\treeshiftfour}%
\setlength{\parentoffset}{-0.5\wd\parentbox}%
\addtolength{\parentoffset}{\treeoffsetfive}%
\addtolength{\parentoffset}{2\branchwidth}%
\setlength{\daughteroffset}{0in}%
\ifdim\parentoffset<0in%
\setlength{\daughteroffset}{-\parentoffset}%
\setlength{\parentoffset}{0in}\fi%
\setlength{\parentwidth}{\parentoffset}%
\addtolength{\parentwidth}{\wd\parentbox}%
\setlength{\treeoffset}{\daughteroffset}%
\addtolength{\treeoffset}{\treeoffsetfive}%
\setlength{\treewidth}{\wd\treeboxone}%
\addtolength{\treewidth}{\daughteroffsetone}%
\addtolength{\treewidth}{\treewidthtwo}%
\addtolength{\treewidth}{\daughteroffsettwo}%
\addtolength{\treewidth}{\treewidththree}%
\addtolength{\treewidth}{\daughteroffsetthree}%
\addtolength{\treewidth}{\treewidthfour}%
\addtolength{\treewidth}{\daughteroffsetfour}%
\addtolength{\treewidth}{\treewidthfive}%
\addtolength{\treewidth}{\daughteroffset}%
\ifdim\treewidth<\parentwidth\setlength{\treewidth}{\parentwidth}\fi%
\sbox{\treebox}{\begin{minipage}{\treewidth}%
\begin{flushleft}%
\hspace*{\parentoffset}\usebox{\parentbox}\\
{\setlength{\unitlength}{0.5\branchwidth}%
\hspace*{\treeoffset}\begin{picture}(8,1)%
\put(0,0){\line(4,1){4}}%
\put(2,0){\line(2,1){2}}%
\put(4,0){\line(0,1){1}}%
\put(6,0){\line(-2,1){2}}%
\put(8,0){\line(-4,1){4}}%
\end{picture}}\\
\vspace{-\baselineskip}
\hspace*{\daughteroffset}%
\makebox[\treewidthfive][l]%
{\raisebox{-\ht\treeboxfour}{\usebox{\treeboxfive}}}%
\hspace*{\daughteroffsetfour}%
\makebox[\treewidthfour][l]%
{\raisebox{-\ht\treeboxfour}{\usebox{\treeboxfour}}}%
\hspace*{\daughteroffsetthree}%
\makebox[\treewidththree][l]%
{\raisebox{-\ht\treeboxthree}{\usebox{\treeboxthree}}}%
\hspace*{\daughteroffsettwo}%
\makebox[\treewidthtwo][l]%
{\raisebox{-\ht\treeboxtwo}{\usebox{\treeboxtwo}}}%
\hspace*{\daughteroffsetone}%
\raisebox{-\ht\treeboxone}{\usebox{\treeboxone}}%
\end{flushleft}%
\end{minipage}}%
\setlength{\treeoffsetone}{\parentoffset}%
\addtolength{\treeoffsetone}{0.5\wd\parentbox}%
\setlength{\treeshiftone}{0pt}%
\setlength{\treewidthone}{\treewidth}%
\sbox{\treeboxone}{\usebox{\treebox}}\poptree\poptree\poptree\poptree%
\else\typeout{QobiTeX warning--- Can't handle #1 branching}\fi\fi\fi\fi\fi}
\newcommand{\tree}{%
\usebox{\treeboxone}
\setlength{\treeoffsetone}{\treeoffsettwo}%
\sbox{\treeboxone}{\usebox{\treeboxtwo}}%
\poptree}

\makeindex

\begin{document}
\doublespacing
\pagestyle{empty}
\bibliographystyle{plain}
\include{titlepage}
\newpage
\include{dedicate}
\newpage
\include{ack}
\newpage
\chapter*{Abstract}

This research concentrates on the design and analysis of 
an algorithm referred to as
\index{Virtual Network Configuration}
\acl{VNC}\index{VNC|see{Virtual Network Configuration}}
which uses predicted future states of a system for faster network
configuration and management.
\acl{VNC}\index{VNC} is applied to the configuration of a wireless mobile 
\acl{ATM}\index{ATM} network.
\acl{VNC} is built on techniques from parallel discrete event
simulation\index{Distributed 
Simulation} merged with constraints from real-time 
systems\index{Real Time!systems}
and applied to mobile ATM\index{ATM} configuration and 
handoff\index{Handoff}. 

Configuration\index{Configuration} in a mobile network is
a dynamic and continuous process. 
Factors such as load\index{Load}, distance\index{Distance}, 
capacity\index{Capacity} and topology\index{Topology} are all
constantly changing in a mobile environment. The \acl{VNC}\index{VNC}
algorithm anticipates configuration changes and speeds the reconfiguration 
process by pre-computing and caching\index{Cache} results. \acl{VNC} 
propagates local prediction results throughout the \acl{VNC} enhanced
system. The \index{Global Positioning System}
\acl{GPS}\index{GPS|see{Global Positioning System}} is an enabling
technology for the use of \acl{VNC} in mobile networks
because it provides location 
information and accurate time for each node.

This research has resulted in well defined structures for the encapsulation 
of physical processes within \acl{LP}s and a generic library for enhancing 
a system with \acl{VNC}. Enhancing an existing system with
\acl{VNC} is straight forward assuming the existing physical processes
do not have side effects. 
The benefit of prediction\index{VNC!prediction} is gained at the 
cost of additional traffic and processing. This research includes an 
analysis of \acl{VNC} and suggestions for optimization of the \acl{VNC} 
algorithm and its parameters.

\newpage
\pagenumbering{roman}
\tableofcontents
\newpage
\listoftables
\newpage
\listoffigures
\newpage
\setcounter{page}{1}
\renewcommand{\thepage}{\arabic{page}}
\pagestyle{plain}
\newpage

\newif\ifisdraft
\isdraftfalse

%
%

\chapter{Thesis Overview}
\section{What is Virtual Network Configuration?}

This research concentrates on the design and analysis of 
an algorithm referred to as
\index{Virtual Network Configuration} 
\acl{VNC}\index{VNC}
which predicts local future states of a system and efficiently 
propagates the effects of state changes through the system. The 
focus of this research is on the use of VNC\index{VNC} for the 
configuration of a wireless mobile 
\acl{ATM}\index{ATM} \cite{Prycker} network.
Results from the area of \acl{PDES}\index{Distributed 
Simulation} and real-time systems\index{Real Time!systems}
are synthesized and applied to mobile \acl{ATM}\index{ATM} configuration 
and handoff\index{Handoff}. This is accomplished by propagating the
effects of predicted local changes through the network via a
parallel simulation technique. As a local predicted event is
propagated through the network, results are cached within each
\acl{LP} which will be needed once the predicted change occurs.

Configuration\index{Configuration} in a mobile network is
a dynamic and continuous process. 
Factors such as load\index{Load}, distance\index{Distance}, 
capacity\index{Capacity} and topology\index{Topology} are all
constantly changing in a mobile environment. The \acl{VNC}\index{VNC}
algorithm anticipates configuration changes
and speeds the reconfiguration process by pre-computing and 
caching\index{Cache} results. Configuration speedup occurs via 
three mechanisms: pre-computation and caching of events, concurrency, 
and the fact that it is possible for results to be generating
faster than the critical path through the system. This is
known as super-criticality\index{Super-Criticality} and will be
explained later.
The \index{Global Positioning System}
\acl{GPS}\index{GPS|see{Global Positioning System}} is an enabling
technology for the implementation of this algorithm because 
it provides location information and accurate time for each node.

The \acl{VNC} algorithm can be explained
by an example.  A mobile node's direction, velocity, bandwidth, number
of connections, past history and other factors can be used to approximate 
a new configuration sometime into the future. All actual configuration
processes can begin to work ahead in time to pre-compute configuration
values based on the position the remote node is
expected to be some point in the future. If the prediction is
incorrect, but within a given bound of the actual result, only some 
processing will have to be rolled back in time. For example, consider
beamformed network links. Beamforming is a technique used to confine
electro-magnetic radiation to a physically narrow path from the source
to the destination node. Beamforming allows other links of the same
frequency to be in operation at the same time and reduces the possibility of
the link being detected by undesired listeners. In a beamforming network,
the beamsteering\index{Beamforming} process results 
may have to be adjusted, but the topology\index{Topology} and many higher 
level requirements will remain correct. This working ahead and rolling back 
to adjust for error is a continuous process, depending on the tradeoff
between allowable risk\index{Risk} and distance into the future of
predictions.

This research has resulted in well defined fields
fields\index{VNC!fields} that are added to each existing
message and additional structures that are added to existing processes.
The benefit of prediction\index{VNC!prediction} is gained at the cost 
of additional traffic and processing. An analysis of the efficiency 
and optimization\index{Optimization} of \acl{VNC} is presented in this
thesis.
\subsection{Thesis Outline}

The research consists of three main tasks: algorithm design,
analysis, and implementation and experimental validation. Optimization
is an additional effort which permeates all of these tasks.
The \acl{VNC}\index{VNC!algorithm} algorithm 
is fully developed, formally described, and simulated. 
The goal of the simulation is to determine the general characteristics 
of this algorithm and its suitability for use in network configuration 
and other applications\index{Application}. A method of assigning 
probabilities to predicted events is considered and
optimizations\index{VNC!optimization} to the basic algorithm 
are examined. In the analysis task, analytical results which 
describe the performance-cost tradeoff are determined. 

Finally, in the third task of this research, 
the \acl{RDRN}\index{RDRN} prototype
is used to measure configuration and hand-off times
without \acl{VNC}\index{VNC} and compare these with the
predictive\index{Mobile Network!classification!predictive} 
\acl{VNC}\index{VNC} method. 
The \acl{NCP}\index{RDRN!network control protocol}
of \acl{RDRN} \cite{BushICC96} is enhanced so that the orderwire packet 
radios carry both real\index{VNC!real message} and virtual 
messages\index{VNC!virtual message}.
\subsection{Algorithm Design}

The effort to complete the algorithm design is pursued through two 
methods. A Maisie\index{Maisie} \cite{bagrodia} simulation of 
\acl{VNC}\index{VNC} 
has been developed to explore the effects of alternative methods of
implementing the \acl{VNC} algorithm.
A portable library written in C language which implements \acl{VNC} has 
been developed. The code was designed to allow any system to be easily
enhanced with the \acl{VNC} algorithm. The method of
prediction used in the driving process(es)\index{Driving Process} is not
of primary concern in this research. The amount of error\index{VNC!error},
defined as the difference between the predicted and real values, is of
concern, and this error probability is modeled. One of the first goals of 
the simulation is to examine the effects of overhead\index{Overhead} due 
to additional messages and rollback\index{Rollback}. The interaction of 
real\index{Real Message} and virtual message\index{Virtual Message} 
processing is of interest, particularly the method used for state adjustment
in the \acl{VNC}\index{VNC} protocol.
\subsection{Virtual Network Configuration Analysis}

Analytical results for the performance and costs of \acl{VNC}\index{VNC} 
are determined in the analysis section. Analytical 
results are determined for the speedup gained by using \acl{VNC}\index{VNC} 
as well as the overhead\index{Overhead} required. Speedup is the 
expected time predicted results, such as the \acl{RDRN} beam table, are 
available with \acl{VNC} divided by
the expected time required to obtain the results without \acl{VNC}.

Parameters such as lookahead ($\Lambda$) and tolerance ($\Theta$)
provide control over the acceptable risk\index{VNC!risk} and probability of 
error versus speed of configuration. 
Increasing $\Lambda$ will better utilize the processors, but
longer range predictions will be less accurate. Allowing more tolerance
will reduce the probability of rollback\index{Rollback} and increase speed,
but introduce a greater possibility of error. However, there exists
control and flexibility with this algorithm. Thus a relationship     
between $\Theta$, $\Lambda$, and 
costs including bandwidth are developed.
Also, an attempt to develop a formal description or calculus for
working with a \acl{VNC} system are undertaken using concepts from Petri
Net Theory \cite{Reisig85, Peterson81}.
\subsection{Implementation and Experimental Validation}

The goal of \acl{VNC} is to reduce the time required by each step
of the configuration process.
In order to verify the impact of \acl{VNC}, measurements are taken
of configuration times for the \acl{RDRN} prototype.
Configuration measurements include the time to establish the radio link
(beamforming\index{Beamforming}) up to and including network
level configuration.
The fact that each host in the \acl{RDRN} network has its own \acl{GPS} 
receiver greatly facilitates these measurements;
mechanisms such as the \acl{NTP} \cite{RFC958}
are not required. To perform these measurements, the \acl{GPS} time
is recorded on the source node before the message is sent, and again
on the destination node immediately after the message is received.

Another step in configuration is determining the \acl{ES} topology.
\acl{ES}s are intermediate nodes within the \acl{RDRN} which
have the capability of residing on the edge between fixed
and wireless networks \cite{BushRDRN}. 
\acl{ES}s serve as connections for \acl{RN}s which are end nodes.
\acl{ES}s also have the 
capability of networking together to form a wireless network. 
Once the \acl{ES} topology has been
determined, \acl{NCP} Messages are exchanged which allow the
beamformed radio link to be established. The protocol which uses the
wireless virtual fiber is the \acl{AHDLC} Layer \cite{sunthesis}. The 
\acl{AHDLC} Layer carries standard \acl{ATM} cells; much of the 
remaining configuration occurs as in a wired network. In other words, 
above the \acl{AHDLC} level, the wireless network acts like a wired network, 
except for the significant impact of mobility. 

Consideration has to be given to
the costs incurred by \acl{VNC} in additional orderwire bandwidth consumed
and additional CPU processing. Utilizing
CPU processing for prediction when that CPU power would otherwise be wasted 
is a benefit, as long as processing does not become a bottleneck.
However, the bandwidth used for configuration should be minimized. 

A successful implementation shows a speedup in configuration and 
handoff while not exceeding the available orderwire bandwidth or causing 
a bottleneck in CPU processing.
However, the impact of \acl{VNC} goes far beyond network configuration to
any real time system which benefits from future state information.
A subset of the \acl{RDRN}\index{RDRN} architecture
is enhanced with \acl{VNC}\index{VNC} and a comparison made with 
non-\acl{VNC}\index{VNC} configuration measurements. The \acl{NCP},
the configuration subsystem of \acl{RDRN} which
performs the configuration of the entire \acl{RDRN} system,
is enhanced with \acl{VNC} as a proto-type systems for this research.
\subsection{Complexity in Self-Predictive Systems}

The purpose of this section is to ``step out of the box''
and view from a few other perspectives the impossible fear of perfect
prediction that we would like \acl{VNC} to perform.
A perfectly tuned \acl{VNC} system would
have a tolerance for error ($\Theta$) of zero and $\forall t$
$GVT(t) > t$, where
$GVT(t)$ is the time into the future that the \acl{VNC} system has
predicted. This means that the system should always be able to
perfectly predict its future. However, as shown in G\"odel's Theorem
this cannot be accomplished. Thus, \acl{VNC} allows some prediction
error to be present by introducing the notion of tolerance.

Another interesting, though troublesome concept
which interferes with our knowledge of the future is Chaos. As an
example, imagine \acl{LP}s which perform even simple computations such as 
the recursion in Equation \ref{chaos}, which is chaotic for certain
values of $r$. Depending on 
the value of $r$ and the initial value of $X$, there are periods of 
stability, a bifurcation, or jumping among stable values at $r=3$, and 
completely unstable values in other iterations \cite{Jackson92}.

\begin{equation}
X_n = r X_{n-1} (1 - X_{n-1})
\label{chaos}
\end{equation}

Complexity theory deals with complicated systems with large numbers of
processes (agents) whose global behaviors cannot be predicted simply
from the rules of the agent interactions \cite{Darley96}, \cite{Lux95}.
Emergent systems lie at the edge of order and chaos and have resisted
yielding to any known form of analysis. A simple, introductory
chronicle of the history of the new field of emergent systems can
be found in \cite{Waldrop}. In \cite{Darley96}, it is
hypothesized that a true emergent system is one for which the optimal
minimum means of prediction is simulation. But this
author has not seen any mention of the effect of chaos and emergence
on the operation of the simulation mechanism itself.
An interesting tool for studying this may be the Swarm Simulation 
System \cite{Heibeler94} which is a simulation tool for 
studying systems with large numbers of agents interacting with each other 
within a dynamic environment. It has been used for studying emergence
in economic, social, molecular, and ecological systems.

In \cite{Kittock93}, emergence in game theory is modeled and studied.
Agents learn by receiving feedback from their environment, however,
their environment consists of other agents who behave in exactly the
same manner. The agents use relatively simple rules, but the behavior
of the entire system is complex and still under study. This again points
to the fact that a self predicting system composed of many \acl{LP}s
may also have complex behavior.

In \acl{VNC}, the configuration system is operating with imperfect
future position information. Such systems are analyzed in \cite{Huberman93}.
In \cite{Huberman93}, agents with imperfect information about their
state can result in oscillatory and chaotic behavior for the entire system.
Finally, \cite{Sawhill93} attempts to quantify when a complex system
becomes chaotic, exhibiting self-similarity. This section has highlighted 
the many subtle complexities involved in attempting to build a 
self-predicting system.
\chapter{Wireless Mobile Network Challenges}
\section{Wireless ATM\index{ATM!architecture} Architectures}

Interest in developing wireless mobile telecommunication networks is
increasing rapidly. There is expected to be a growing
market\index{Wireless Market} for wireless mobile computing based on the fact
that the mobile
cellular voice communications market has grown rapidly in recent years
and because of the ease and practicality offered by wireless mobile computing
to the consumer. There were 43 million new wireless communications
subscribers\index{Wireless Subscribers} as of 1996. It is predicted that
in 1997 there will be 60 million new subscribers which will surpass the
number of new fixed network subscribers for the first time in history
\cite{Ericsson}. As more corporations begin to notice the potential profit
in wireless mobile computing, they are attempting to obtain the results of
the research done in this area. Besides the purely market driven
requirements for mobile wireless computing, a wireless system can be rapidly
deployed in a disaster\index{Disaster} situation bringing valuable information
to rescue\index{Rescue} workers. A mobile wireless system can also be cheaply
established in rugged terrain\index{Rugged Terrain} or in developing
countries\index{Developing Countries} where
the cost of laying cable\index{Cable} would be prohibitive. Also, with a
wireless mobile network, businesses\index{Businesses} will be able to more
easily bring computing power to the location where they can gain the most
profit. Finally, mobile wireless computing will be useful for the military.

This section provides background in order to
understand the \acl{VNC} algorithm and the results of this research. 
This work is an application 
of a new algorithm to provide rapid wireless mobile \acl{ATM} 
configuration. We will begin with an overview of wireless \acl{ATM} 
architectures\index{ATM!wireless architecture}, 
because the manner in which the wireless links are implemented will impact 
the handoff\index{Handoff} performance of the predictive \acl{VNC} 
algorithm.

There appear to be two views on wireless \acl{ATM} 
architecture\index{ATM!wireless architecture}. In one
view, \acl{ATM} cells\index{ATM!cell} should be maintained end-to-end including 
transmission over the wireless portion of the network. The other view 
is that \acl{ATM} cells\index{ATM!cell}
need not be preserved and a more efficient form of packetization
should be used over the wireless network. The \acl{ATM} cells\index{ATM!cell} 
are then reconstructed from the wireless packetization method after
being received by the wireless destination.

Some thoughts on the latter view are provided in \cite{McTiffin}.
A \acl{MAC} protocol for wireless \acl{ATM}\index{ATM!wireless} is examined 
with a focus on \acl{CDMA}\index{CDMA} access.  
One of the earliest proposals for a wireless \acl{ATM}\index{ATM!architecture} 
Architecture is
described in \cite{Raychaudhuri}. This architecture appears to be
the model\index{Model} on which the \acl{ATM} Forum\index{ATM Forum} 
is basing its Wireless \acl{ATM}\index{ATM!wireless architecture} Architecture 
\cite{Suzuki}. It assumes a base station 
has an \acl{ATM} Network Interface Unit\index{ATM Network Interface Unit} (NIU) 
and a Personal Communications Network Network Interface Unit (PCN NIU)\index{Personal Communications Network Network Interface Unit}. Various 
alternatives for a wireless \acl{MAC}\index{Wireless MAC}
are discussed and a \acl{MAC} frame is proposed which contains sequence 
numbers, service type, and a \acl{CRC}\index{CRC}.
\index{Time of Expiry} \acl{TOE} scheduling policy is proposed 
as a means for improving real-time data\index{Data traffic!real-time} traffic 
handling. A related work which considers changes to Q.2931\index{Q.2931} 
\cite{Q2931} to support mobility is proposed in \cite{Yuan}. Q.2931 is the
ITU standard for ATM signaling.

In the \acl{RDRN} Project\index{RDRN} \cite{BushRDRN},
\acl{ATM} cells\index{ATM!cell} are transmitted over the wireless link
intact within the
\acl{AHDLC}\index{Adaptive HDLC} protocol.
A related paper analyzes fixed size \acl{ATM} cells\index{ATM!cell} in a mobile
environment \cite{mobatmbuf}. In \cite{mobatmbuf}, a stochastic\index{Stochastic}
analysis of \acl{ATM} buffer\index{ATM!buffer} fill distribution\index{Buffer Fill
Distribution} is extended to a mobile 
environment.  The performance of wireless networks with link level 
retransmissions is examined in \cite{DeSimone}. These results
indicate that application throughput improves with the presence
of the link level retransmission protocol only when the wireless 
packet error rate exceeds a certain threshold.

In any wireless architecture\index{ATM!wireless architecture}, 
the occurrence of a hand-off must not
interfere with \acl{ATM} cell order\index{ATM!cell order}. General 
handoff\index{Handoff} methods as well as some specific
to \acl{ATM}\index{ATM!handoff} will be discussed.
\section{Wireless Mobile Network Challenges}
\label{introduction}

There is currently intense interest in adding seamless
mobility to \acl{ATM}\index{ATM} networks; 
progress in that area is developing rapidly. However,
there are several problems integrating seamless mobility with \acl{ATM}
which are difficult to solve.
One of these problems is configuration in a rapidly changing network
such as that presented by a mobile environment. This problem becomes
especially acute in the inherently 
connection-oriented\index{Mobile Network!classification!connection-oriented} 
\acl{ATM}\index{ATM} environment.
An end-to-end connection must be established before any transfer of data
can take place and this setup time must be as short as possible yet
allow for a constantly changing topology\index{Topology}.

A large part of the difficulty in configuration is due to the difficulty 
caused by
the handoff\index{Handoff} of a mobile node from
one base station to another. In contrast to today's wireless mobile
cellular voice networks, an efficient signaling mechanism for
handoff in a seamless mobile \acl{ATM} network has not yet been standardized and
the effect of the handoff delay on \acl{ATM} and higher layer protocols can be
significant. 
This section provides an overview of 
some of the previous work related to the area of handoff\index{Handoff} for 
mobile \acl{ATM}\index{ATM} networks which tries to alleviate signaling traffic 
and maintain a seamless \acl{ATM}\index{ATM} connection. 
There are a variety of general methods for how and when handoff\index{Handoff}
can occur.
See Figure \ref{htypes} for general examples of how handoff\index{Handoff} 
can occur. The Break-Make method of handoff breaks the mobile node's connection 
to the current base station before making a connection to the new base station.
The Make-Break method of handoff is exactly the reverse of Break-Make. 
The Chaining method establishes a route through the previous base station 
then to the new base station after a handoff. Finally, either the Break-Make or 
Make-Break method can be used with the Chaining method resulting in a 
Combination method of handoff.
An example of determining when handoff\index{Handoff} should be initiated
is addressed in \cite{Santucci}. This algorithm
is based on a least squares estimate of path loss parameters assuming
location is known. It is shown that this technique reduces outages
and performs better than an estimation windows method.

\begin{table}[htbp]
\centering
\begin{tabular}{||l|l||} \hline
\textbf{Handoff Type} & \textbf{Description} \\ \hline \hline
Break-Make\index{Handoff!Break-Make} & Break current connection first \\ \hline
Make-Break\index{Handoff!Make-Break} & Establish new connection first \\ \hline
Chaining\index{Handoff!Chaining}   & Route through old connection \\ \hline
Combination\index{Handoff!Combination} & Combine chaining with first two methods \\ \hline
\end{tabular}
\caption{\label{htypes}General Handoff Methods.}
\end{table}

\section{Approaches to Network Mobility}

Mobile Networks are divided into two classes:
connectionless\index{Mobile Network!classification!connectionless}
and 
connection-oriented\index{Mobile Network!classification!connection-oriented}
as shown in Figure \ref{mobtaxtab}. 
Connectionless networks pass
datagrams without requiring any form of signaling to establish a
channel. Connection-oriented networks require an initial signaling protocol
to establish a communication channel. 

\begin{sidetable}
\begin{tabular}{lllll}
\textbf{Network Class} & \textbf{Sub Class} & \textbf{Sub Class} & \textbf{Description} & \textbf{Example} \\ \hline
\multicolumn{3}{l}{Connection-less}      & Handoffs require packet forwarding & \\ \hline
& \multicolumn{2}{l}{Registration Agent} & Packets forwarded by local agent & IP Mobility \cite{RFC2002}, CDPD \cite{CDPD} \\
\multicolumn{3}{l}{Connection-Oriented}  & Handoffs require call-setup & \\ 
& \multicolumn{2}{l}{Registration Agent} & Calls forwarded by local agent & AMPS \cite{AMPS}, GSM \cite{GSM} \\
& \multicolumn{2}{l}{Pivot Based}        & Distinct pivot BTS for handoff & \\ \hline
& & Tree Based         & Pre-established multicast connections & Acampora \cite{Acampora} \\
& & Peer Based         & Source and destination & ATM Forum \cite{Yuan} \\
& &                    & BTS negotiate handoff &  \\ \hline
& \multicolumn{2}{l}{Adaptive Path}      & Maintains optimal routes & \\
& \multicolumn{2}{l}{}                   & during handoff & \\ \hline
& & Predictive         & Protocol stack and routing  & VNC \cite{BushICC96} \\
& &                    & pre-configured by prediction 
                                                     & Full Mobility Architecture \cite{Liuphd} \\ \hline
& & Non-Predictive     & Routes optimized after handoff occurs & BAHAMA \cite{BAHAMA} \\
\end{tabular}
\caption{\label{mobtaxtab}Overview of Mobile Communications Network Mechanisms.}
\end{sidetable}

\begin{figure*}[htbp]
\centerline{\psfig{file=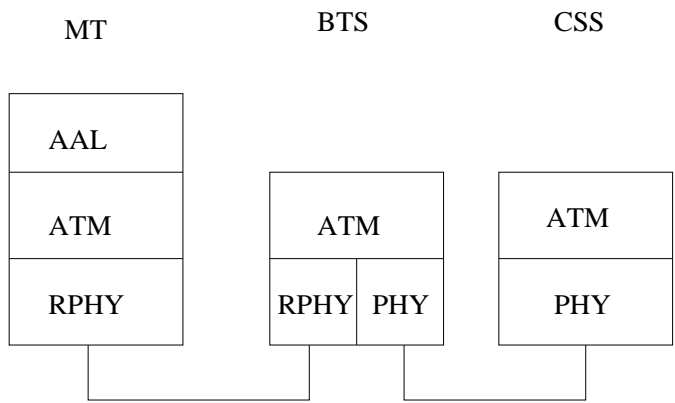,width=3.5in}} 
\caption{Basic Terminology.}                                                
\label{mobreq}                                     
\end{figure*}

In order to discuss the variety of disparate mechanisms proposed for
mobile networking, a common terminology is required. Throughout this
paper the acronym and terms defined in \cite{Hauwermeiren} are used;
\acl{MT}, \acl{BTS},
\acl{HO}, and \acl{CSS} are shown in Figure \ref{mobreq}.
The node where the old and
new connections meet is called the \acl{ANO} point. Bridging 
points are the nodes between which a connection change occurs. For
example, the bridging nodes in a simple hand-off would be the \acl{MT} and
the \acl{ANO}. Non-macro diversity is assumed throughout this paper, that is, only
one transmission path per data stream exists from source to destination at 
any time.

\subsection{Registration Agent-Based Mechanism}
\label{registration}

\begin{figure*}[htpb]
\centerline{\psfig{file=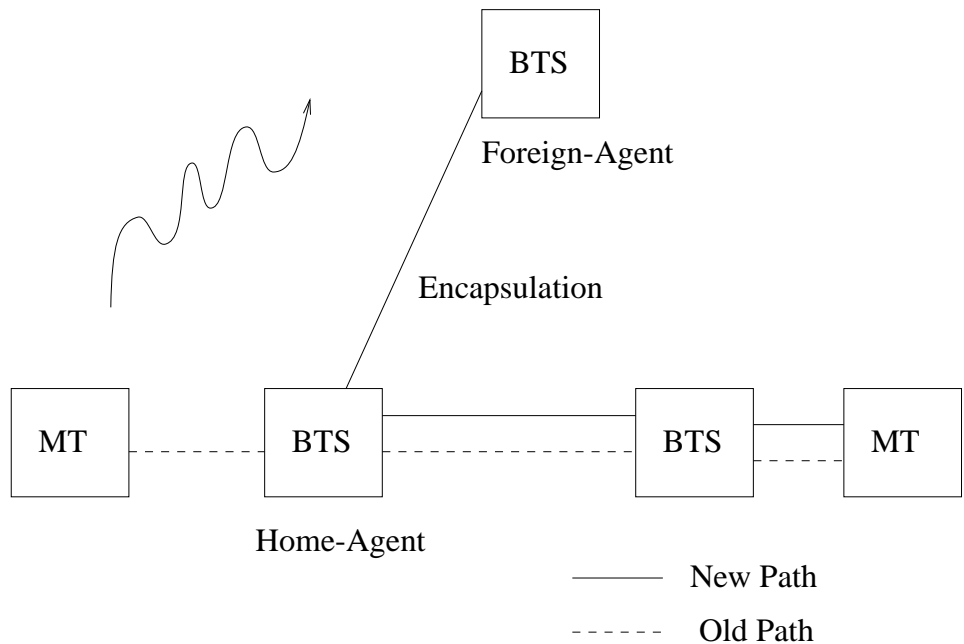,width=5.0in}}
\caption{Registration Agent-Based Mobility Method.}
\label{regag}
\end{figure*}

Figure \ref{regag} shows a hypothetical Registration Agent-Based mechanism
similar to Mobile-IP which is described in this section. The initial 
bridge points are the \acf{MT} and \acf{BTS} containing the Home Agent
with the Home Agent \acf{BTS} serving
as the \acf{ANO}. Routing optimizations can remove the home \acf{BTS} from 
the path, with the source and destination \acs{BTS}s as the bridge points. 
The Registration Agent-Based mechanisms contain a Home and Foreign
Agent such as in Mobile-IP\index{Mobile-IP} \cite{RFC2002}, explained 
next, or a \acf{HLR} and 
\acf{VLR} as in mobile cellular voice networks\index{Cellular Voice Network} 
\cite{AMPS} and \cite{GSM}. These agents maintain the information
concerning whether a mobile node is currently associated with its home 
network or is a visiting mobile node, that is, not currently in its home
network.

The IP Mobility Support\index{IP Mobility Support} Draft \cite{RFC2002} is a 
good example of 
the connectionless\index{Mobile Network!classification!connectionless} 
registration\index{Mobile Network!classification!registration} 
mechanism. It describes enhancements that
allow transparent routing of IP\index{IP} datagrams. This protocol uses an
agent registration\index{Mobile Network!classification!registration} process 
to maintain home and current 
location information. All packets are routed to the home address as in the 
fixed network IP\index{IP} protocol. The home agent has a 
registration\index{Mobile Network!classification!registration} table which 
indicates whether the packet needs to be tunneled to a new address
by encapsulating the packet within an IP packet bound for the new address, 
or to deliver the packet as in conventional fixed network IP\index{IP}. The 
mobile host registers with a foreign agent when it moves from its home
network to a new network. 
The foreign agent receives the packet tunneled to it and delivers the packet 
to the mobile host by de-encapsulating the packet and forwarding the packet
as in fixed network IP. IP Mobility Support\index{IP Mobility Support} allows 
for routers or entire networks to become mobile as a single unit. For example
a ship or airplane could contain a network of mobile hosts and routers.

A Registration Agent-Based method for data over connection-oriented           
voice networks is the \acf{CDPD} described in
\cite{CDPD} and \cite{Kunzinger} protocol. The \acl{CDPD} protocol operates 
much like the Mobile  
IP protocol for mobile cellular voice networks and is designed to co-exist    
with the \acf{AMPS}\cite{AMPS}. \acl{CDPD} uses registration and encapsulation 
to forward
packets to the current location of the mobile host just as in Mobile-IP.
\acl{CDPD} was developed independently of Mobile-IP. Although the \acl{CDPD} and    
Mobile-IP groups were aware of each others' existence, there has been     
surprisingly little interaction between the groups. This may be because      
the \acl{CDPD} development group is a closed group focused on developing a standard
to quickly meet commercial business requirements, whereas the \acf{IETF} group
has moved more slowly in attempts to find better solutions to the technical
and engineering problems involved. 

An example of a
connection-oriented\index{Mobile Network!classification!connection-oriented}
mechanism for voice networks is \acl{AMPS} \cite{AMPS}. \acl{AMPS} uses 
an \acf{HLR}\index{AMPS!home
location register} and \acf{VLR}\index{AMPS!visiting location register} 
to store the mobiles' \acf{MIN}\index{AMPS!mobile identification number}
and \index{AMPS!electronic serial number} \acf{ESN}.
The \acf{HLR} and \acf{VLR} reside on the 
\index{AMPS!mobile switching center} \acf{MSC} which is connected
to each mobile base station.
A mobile node can determine whether it is in its home coverage area or in a
new base station coverage area by comparing
the base station identification, \index{AMPS!station identification}
The \acf{SID}, advertised by the base station, with a previously stored value.
Registration and call routing\index{AMPS!call routing} are performed over \acf{SS7}.
Call setup is routed first to the destination mobile's home \acf{MSC}. The 
home \acf{MSC} will route the call to the \acf{MSC} which is currently serving 
the visiting mobile. If the visiting mobile node can be reached, the visiting 
mobile's \acf{MSC} will respond with a \acf{TDN}\index{AMPS!temporary directory number}.
If the \acf{MSC} cannot reach the visiting mobile node, the \acf{MSC} will respond with
a redirection indicating that the home \acf{MSC} must try another route or another
number.
\subsection{Tree-Based Mechanism}
\label{tree-based}

\begin{figure*}[htbp]
\centerline{\psfig{file=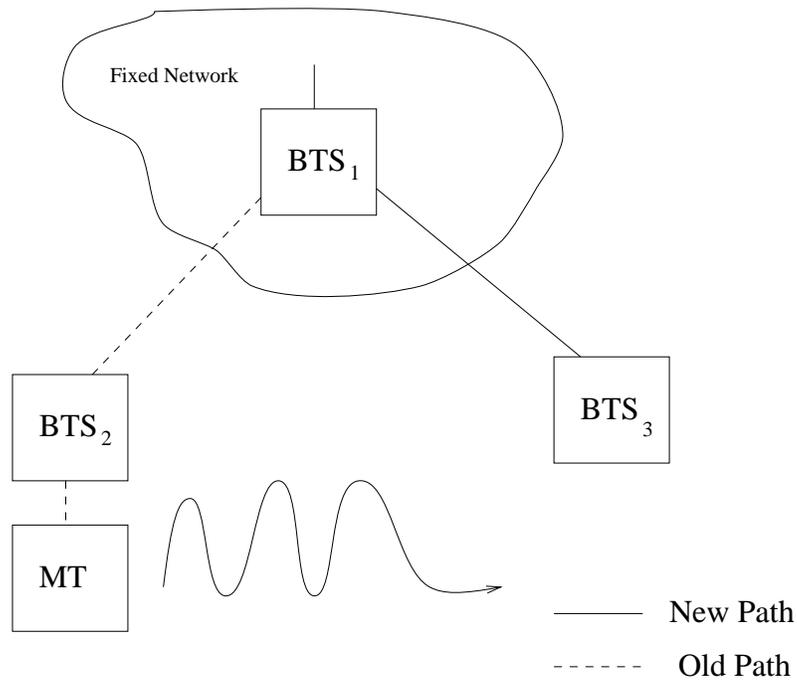,width=3.5in}}
\caption{Pivot-Based Mobility Method.}
\label{tree}
\end{figure*}

The Pivot-Based mobility mechanisms are characterized by a distinct
node which is chosen to act as a pivot. A new route is chosen from the
pivot node to the new base station after a handoff.
Figure \ref{tree} shows the general Pivot-Based mechanism. \acl{BTS} 1 is
the \acl{ANO} as the \acl{MT} moves from \acl{BTS} 2 to \acl{BTS} 3. In the Tree-Based mechanisms,
the \acl{ANO} initiates and manages the hand-off. In the Peer-Based mechanisms,
\acl{BTS} 2 and \acl{BTS} 3 have an equal role in negotiating the hand-off.

The simplest of the Pivot-Based methods is the tree-based method
which requires preallocation\index{Preallocation} of resources\index{Resource} but
does not require \acl{VC} setup and is thus efficient for performing hand-offs.
A good description of a
tree-based\index{Mobile Network!classification!tree-based} method is
described in \cite{Acampora}. In this method, the tree\index{Tree} is
formed by Virtual Paths\index{ATM!Virtual Path} from a root \acl{ATM} switch
in the fixed network. The leaves of the tree\index{Tree} are the base stations.
The \acl{ATM} switches\index{ATM!switch} within the tree\index{Tree} are enhanced
with translators
which monitor incoming cells\index{ATM!cell} and change routes based on
incoming cells\index{ATM!cell}.
A handoff\index{Handoff} is implied by cells\index{ATM!cell} arriving from a
given \acl{VC} and port; this causes the switch to route cells\index{ATM!cell}
along the new \acl{VC} and port without any additional signaling. This method
has been carried into more detail, such as considering \acl{QoS},
by \cite{Naghshineh}.

The mobile \acl{ATM} architecture described in \cite{Acampora} is based on a
{\em virtual connection tree\index{Virtual Connection Tree}}. The root
of the virtual connection tree is an \acl{ATM} switch in the wired network
whose leaves are the base stations. A mobile node will have a route
first through the virtual connection tree root and then to either
the wired network or back through the root to another mobile node.
There can be more than one virtual connection tree\index{Virtual
Connection Tree} throughout the network. When a mobile node joins
a particular virtual connection tree, virtual circuits are established
for that mobile node along every branch of the tree. These \acl{VC}s
remain established while the mobile node resides within the virtual 
connection tree. Thus when the mobile node performs a handoff\index{Virtual 
Connection Tree!handoff} from one base station to another within the same 
virtual connection tree\index{Virtual Connection Tree} signaling is not 
required to 
establish a new \acl{VC}. This works for \acl{ATM} cells traveling in the 
forward
direction relative to the mobile node, but cells in the reverse direction
will still arrive at the mobile node's previous destination. This
problem is solved by enhancing the virtual connection tree root \acl{ATM}
switch with a \acl{VC} translation device. When the first \acl{ATM} cell 
arrives at the
root with a new virtual circuit identifier, the root \acl{ATM} switch will
update the routing table to forward cells in the reverse direction properly.

The virtual connection tree method has the advantage of not requiring any 
changes to the \acl{ATM} cell structure or its interpretation. However, this 
method does have problems. Since a virtual connection 
tree\index{virtual connection tree} maintains \acl{VC}s for all mobile nodes to
every branch of the tree, there are many \acl{VC}s open which are not being
used. In addition this leaves all potential \acl{VC}s open on all base stations.
If many mobile nodes went to a single base station at the same time, they
would all be immediately accepted potentially causing the base station to 
be overloaded. 
\subsection{Peer-Based Mechanism}
\label{peer}

The Peer-Based method is included in the Pivot-Based mobility class
illustrated in Figure \ref{tree}.
The Peer-Based method assumes a fixed node which acts as a pivot, but it
differs from the Tree-Based methods in that the base station nodes to
which the mobile unit is associated before and after the hand-off may negotiate 
among themselves as peers in order to attempt to establish the same 
\acl{QoS} for the mobile unit.

Indirect-TCP (I-TCP) as described in \cite{Bakre} is an example of this in the 
connection-oriented TCP
environment. I-TCP requires \acl{MSR} which have the
ability to transfer images of sockets involved in an established TCP connection 
from one MSR to another. This transfer provides a seamless TCP connection
as the mobile host performs a hand-off. The transfer is transparent to both
the mobile node and a fixed network node. Clearly, one would expect a 
significant amount of overhead in copying the connected sockets from 
one \acl{MSR} to another. In addition, IP packets are lost during the 
hand-off due to the
routing update which is required after the socket transfer. The routing
update is required because the host at the opposite end
of the connection which did not hand-off \acl{FH} is connected to a 
new \acl{MSR}. Thus the FH and the new \acl{MSR} are the bridge nodes in this 
routing update. However, the results
reported in \cite{Bakre} indicate that I-TCP performed better than regular
TCP over Mobile-IP. This is explained as being primarily due to the TCP restart
delay incurred by regular TCP.

Another Peer-Based mobile \acl{ATM} mechanism which is analogous to the 
I-TCP method is the architecture
for mobility support in wireless \acl{ATM} Networks described in \cite{Yuan}.
In \cite{Yuan} additions to Q.2931 are proposed which allow the \acl{ATM} switch
before, and the new \acl{ATM} switch after a hand-off to negotiate a 
comparable \acl{QoS} for the mobile host and transfer state information in 
order to maintain cell order.
The mechanism described in \cite{Yuan} is currently a strong candidate for 
standardization by the \acl{ATM} Forum.
If the \acl{QoS} can be maintained through the destination \acl{ATM} switch, then the
hand-off is seamless\footnote{Contribution \cite{Yuan} only mentions that data link 
information is transferred between \acl{ATM} switches to maintain state information 
and that this guarantees in-sequence cell delivery. This would appear to 
require more explanation from the authors.}.
\subsection{Adaptive Path-Based Mechanism}
\label{adaptive}

The Adaptive 
Path-based\index{Mobile Network!classification!Adaptive Path-Based} 
mechanisms do not require resource\index{Resource} 
preallocation\index{Preallocation},
but require more complex modifications of the signaling protocol. The
routing is an integral part of the mechanism and maintains 
an optimal route between connections. 
There is no home or foreign agent; a distributed topology and routing 
protocol keep cell transfer tables accurate and consistent as topology 
changes.

An Adaptive Path-Based
method of hand-off for \acl{ATM} is described in \cite{BAHAMA, Veer, Veer97}. 
This system consists of \acl{PBS}\index{BAHAMA!Portable Base Station} 
and mobile users. \acl{PBS}s are base stations which perform \acl{ATM} 
cell\index{ATM!cell} \textbf{forwarding}, not \acl{VCI}/\acl{VPI} translation. 
This is a very important point, because it means that \acl{ATM} cell 
forwarding, based on \acl{VPI}, occurs much like IP packet forwarding in 
Mobile-IP. This method tries to blend the best of the connectionless 
mechanisms into the connection-oriented environment. \acl{PBS}s are 
connected via Virtual Path Trees\index{BAHAMA!Virtual Path Tree} which are 
preconfigured \acl{ATM} \acl{VP}. These trees\index{Tree} can change based on 
the topology\index{Topology} as described in the 
{\em Virtual Trees Routing Protocol}\index{BAHAMA!Virtual Trees Routing 
Protocol} \cite{VTRP}. Since Virtual Path Trees are pre-established,
no \acl{VP} setup is required;
individual connections are established by simply choosing available 
\index{Virtual Channel} \acl{VC} on the source and destination nodes. 
The fact that \acl{VP}s are maintained based on topology\index{Topology}, 
not call-setup, allows fast hand-off\index{Handoff} since connection routes 
do not need to be established with each call-setup. The \acl{VPI} denotes a 
\acl{PBS} address which is the root of
the \acl{VPI} tree\index{BAHAMA!tree} and a \acl{PBS} can belong to more than one 
tree\index{BAHAMA!tree}. A Homing Algorithm\index{BAHAMA!Homing Algorithm} 
maintains cell\index{ATM!cell} 
order without centralized control or re-sequencing at any point in the 
network. 
The Homing Algorithm\index{BAHAMA!Homing Algorithm} routes 
cells\index{ATM!cell} through the destination node's home \acl{PBS} regardless of
where the mobile node is currently located. The route will update to a more
optimal route when it is possible to do so without causing cell reordering.
This is done for \acl{ATM} cells\index{ATM!cell} in both directions on a 
channel. This algorithm is considered an adaptive 
path\index{Mobile Network!classification!Adaptive Path-Based} 
algorithm because both the Virtual Path Trees\index{BAHAMA!Virtual Path Tree} 
and the mobile node's home \acl{PBS} update gradually to result in more 
efficient routes.
\subsection{Predictive-Based Mechanisms}
\label{predictive}

Clearly, some type of prediction can be used in any mobile network
mechanism.
However, the Predictive-Based mechanism as used in this work refers to a 
mobile algorithm which makes use of predicted mobile location to 
pre-establish a significant portion of the post-handoff connection.
These mechanisms are located under the Adaptive Path-Based mechanisms
because Predictive Path-Based mechanisms can attempt to optimize
the route based on predicted topology.

Many recently proposed mobile networking architectures and protocols
involve predictive mobility management\index{Predictive Mobility Management}
schemes.
An optimization\index{Optimization} to a\index{Mobile-IP} Mobile IP-like
protocol using\index{Daedalus!IP-Multicast} IP-Multicast is
described in \cite{Seshan}. Hand-offs are anticipated and data is
multicast\index{Multicast} to nodes within the
neighborhood\index{Daedalus!neighborhood} of the predicted
handoff\index{Daedalus!handoff}.
These nodes intelligently buffer the data so that no matter where
the mobile host (MH) re-associates after a handoff\index{Daedalus!handoff},
no data will be lost.
Another example \cite{Liu} and
\cite{Liuphd} proposes deploying mobile floating agents\index{Agents!floating}
which decouple services and resources\index{Resource} from the underlying
network. These agents would be pre-assigned and pre-connected to predicted
user locations. 
Finally, \cite{BushICC96} discusses a Predictive-Based Mobile \acl{PNNI}. This
mechanism inserts a lookahead capability into the topology portion of 
fixed network \acl{PNNI} while maintaining as much of the fixed network \acl{PNNI}
standard as possible.
\section{Comparison of Mechanisms}
\label{comparison}

The applicability of these seamless mobile handoff mechanisms to \acl{ATM}
varies in the efficiency and the ability of each mechanism to maintain the 
\acl{ATM} protocol standard end-to-end. A Registration Agent-Based mechanism 
for \acl{ATM} at the cell level would require significant changes and overhead 
to the standard 
\acl{ATM} protocol in order to implement tunneling, however, the Adaptive Path-Based
mechanism, \cite{BAHAMA}, has found an interesting middle ground discussed later
in this section. The Tree-Based mechanism is 
efficient but it is static; routes do not adjust within the network to form 
the most efficient paths, base stations are assumed to be stationary, and 
all channels must be preconfigured for each mobile at each base station. 
Each base station will be required to allocate many more VCs than are
really necessary. An advantage of the tree-based method is that it maintains 
the \acl{ATM} protocol end-to-end.
The Adaptive Path-Based mechanism described previously, \cite{BAHAMA}, 
is more flexible than the Tree-Based method. It maintains Virtual Path
Trees which remain consistent with the changing topology.
However it requires
modifications to the protocol such as using the 
\acl{GFC} to contain a sequence number and an end-of-transmission
indicator within the packet. The Homing Algorithm
used by this adaptive path-based algorithm allows the routing to be adaptive
to form efficient paths; however, the paths adapt relatively slowly.
The reason for the slow adaptation is that all cells sent
before a handoff must be routed through the old \acl{PBS} in order
insure cell order. Finally, the mechanism in \cite{BAHAMA} requires
cell forwarding rather than switching and a set of signaling and control 
algorithms which are very different from the standards currently in use.
The Connection-Oriented Peer-Based method of mobile \acl{ATM} handoff described
in this paper, \cite{Yuan}, although
more flexible than the Tree-Based method, leaves some open questions regarding 
the overhead of the peer \acl{ATM} switch negotiation and the maintenance
of \acl{ATM} cell order. The Adaptive Path-Based method in \cite{BAHAMA} 
integrates topology and routing in a more fundamental manner and would
appear to handle the long term operation of the wireless network
more efficiently.
Also, the enhancement to the Q.2931 signaling protocol to
implement the Peer-Based handoff scheme does not handle route optimization.
Route optimization may be required as a final step to this method in order
to make it comparable with \cite{BAHAMA}.

The metrics of interest which are relatively easy to measure include the handoff 
delay and the additional network load due to signaling. Additional metrics which 
are less easy to quantify are the flexibility of the method in handling
all cases of mobility and how closely the method aligns with existing 
standards. The flexibility of the method involves such criteria as whether 
the method can handle every hand-off situation and whether the
method requires a fixed network connection or whether it can support a 
completely wireless environment. Much more experimental validation and 
measurement is required of these methods.

The contribution of this section has been to categorize 
mobile wireless \acl{ATM} configuration as well as to compare and contrast 
them. There is a trade-off between flexibility,
handoff delay, and maintenance of the \acl{ATM} protocol standard, and
amount of increase in signal load. 
The proposals range from the Tree-Based method which is fast but not
very flexible to the Adaptive Path and Peer-Based mechanisms which are 
more flexible, but require more overhead for each handoff. 

Putting aside the requirement for adhering to the \acl{ATM} protocol standard,
the best candidate at this time appears to be the Adaptive Path-Based
mechanism described in \cite{BAHAMA}. 
Although the Predictive-Based mechanisms may prove better, the
Predictive-Based mechanisms are behind \cite{BAHAMA} in actual
development and testing. This work on \acl{VNC}, a Predictive-Based mechanism,
will help to remedy that situation.
The Adaptive Path-Based mechanism by definition treats topology
and routing in a fundamental and efficient manner, not as 
additional steps separate from the mobile algorithm. 
Finally, \cite{BAHAMA}, applies techniques, namely cell forwarding, from 
the already proven connectionless approach of Mobile-IP to the 
connection-oriented environment of mobile \acl{ATM}. 

In \acl{RDRN}, the orderwire determines and controls handoff. The
\acl{NCP} controls the operation of the orderwire and is discussed
in more detail in the next section.
\section{Overview of the \acl{RDRN} System and Orderwire}
\label{ncppro}

It is important to understand the \acl{RDRN} system which the orderwire
serves and the details of the orderwire operation. A brief review
of the \acl{RDRN} system begins with
one of the earliest proposals for a wireless \acl{ATM}
architecture which is described in \cite{Raychaudhuri}. 
In \cite{Raychaudhuri}, various alternatives for a wireless 
\acl{MAC} are discussed and a \acl{MAC} frame is proposed. The \acl{MAC}
contains sequence numbers, service type, and a \acl{TOE}
scheduling policy as a means for improving real-time data traffic
handling. A related work which considers changes to Q.2931
\cite{Q2931} to support mobility is proposed in \cite{Yuan}.
A \acl{MAC} protocol for wireless \acl{ATM} is examined in \cite{McTiffin}
with a focus on \acl{CDMA} in which 
\acl{ATM} cells are not preserved allowing a more efficient form of 
packetization over the wireless network links. The \acl{ATM} cells
are reconstructed from the wireless packetization method after
being received by the destination. The \acl{RDRN} architecture maintains 
standard \acl{ATM} cells through the 
wireless links. Research work on wireless \acl{ATM} LANs have been described in 
\cite{Rednet} and \cite{SWAN}. The  mobile wireless \acl{ATM} \acl{RDRN} 
differs from these LANs because the \acl{RDRN} uses point-to-point radio 
communication over much longer distances.
The system described in \cite{BAHAMA} and \cite{Veer} consists of \acl{PBS} 
and mobile users. PBSs are base stations which perform 
\acl{ATM} cell switching and are connected via Virtual Path Trees which are 
preconfigured \acl{ATM} \acl{VP}. These trees can change based on 
the topology as described in the {\em Virtual Trees Routing Protocol} 
\cite{VTRP}. However, \acl{ATM} cells are forwarded along the Virtual Path
Tree rather than switched, which differs from the \acl{ATM} standard. An 
alternative mobile 
wireless \acl{ATM} system is presented
in this paper which consists of a mobile PNNI architecture based on a 
general purpose predictive mechanism known as Virtual Network Configuration 
that allows seamless rapid handoff.

The objective of the \acl{RDRN} effort 
is to create an \acl{ATM}-based wireless communication system that 
will be adaptive at both the link and network levels to allow for rapid 
deployment and response to a changing environment. The objective of the 
architecture is to use an adaptive point-to-point topology to 
gain the advantages of \acl{ATM} for wireless networks. A prototype of this 
system has been implemented and will be demonstrated over a wide area
network. The system adapts to its environment and can automatically arrange 
itself into a high capacity, fault tolerant, and reliable network. 
The \acl{RDRN} architecture is composed of two overlayed networks: 

\begin{itemize}
\item A low bandwidth, low power omni-directional network for location 
dissemination, switch coordination, and management which is the orderwire 
network described in this section,
\item A ``cellular-like'' system for multiple end-user access to the switch
using directional antennas for spatial reuse, and 
and a high capacity, highly directional, multiple beam network for 
switch-to-switch communication.
\end{itemize}

\ifisdraft
        \onecolumn
\fi
\begin{sidefigure}
        \centerline{\psfig{file=figures/scenario.eps,width=8in}}
        \caption{RDRN High-level Architecture.}
        \label{scenario}
\end{sidefigure}
\ifisdraft
        \twocolumn
\fi

\begin{figure*}[htbp]
        \centerline{\psfig{file=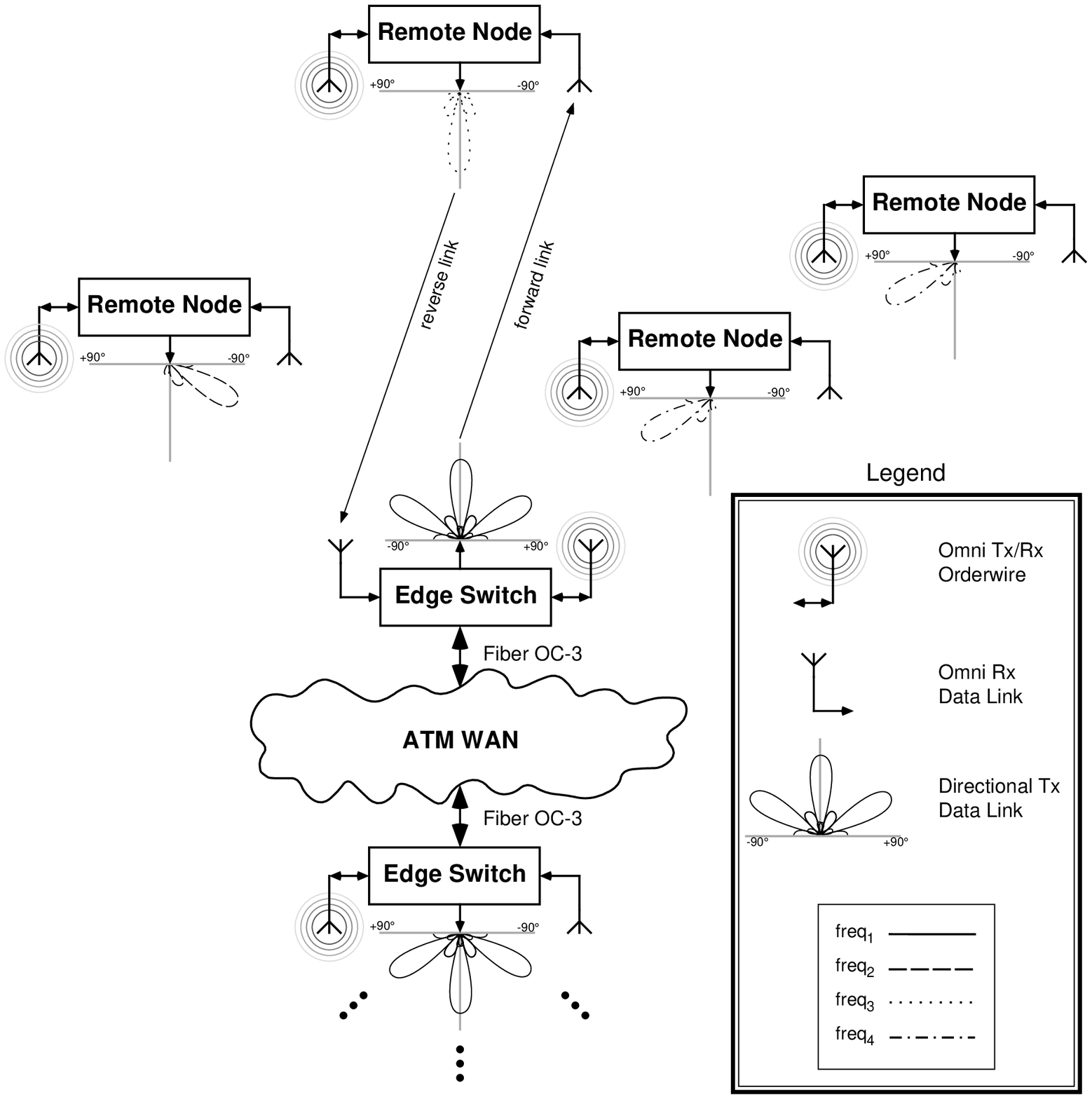,width=6.0in}}
        \caption{\acl{RDRN} Component Overview.}
        \label{hwowov}
\end{figure*}

The network currently consists of two types of nodes, \acl{EN}s and
\acl{RN}s as shown in Figure \ref{scenario}. \acl{EN}s were designed to
reside on the edge of a wired network and provide access to the wireless 
network; however, \acl{EN}s also have wireless links.
The \acl{EN} components include \acl{ES}es and optionally an \acl{ATM} switch, 
a radio handling the \acl{ATM}-based communications, a packet radio for 
the low speed orderwire running a protocol based on X.25 (AX.25), \acl{GPS} 
receiver, and a processor. Remote nodes (\acl{RN}) consist of 
the above, but do not contain an \acl{ATM} switch.
The \acl{EN}s and \acl{RN}s also include a phased array steerable antenna. 
The \acl{RDRN} uses
position information from the \acl{GPS} for steering antenna beams toward nearby 
nodes and nulls toward interferers, thus establishing the high capacity links
as illustrated in Figure \ref{hwowov}. Figure \ref{hwowov} highlights an
\acl{ES} (center of figure) with its omni-directional transmit and receive 
orderwire antenna and an omni-directional receive and directional transmit 
\acl{ATM}-based links. Note that two \acl{RN}s share the same $45^o$ beam from the
\acl{ES} and that four distinct frequencies are in use to avoid interference.
The decision involving which beams to establish and which frequencies
to use is made by the topology algorithm which is discussed in a later
section.

The \acl{ES} has the capability of switching \acl{ATM} cells among 
connected \acl{RN}s or passing the cells on to an \acl{ATM} switch to 
wire-based nodes. Note that the differences between an \acl{ES} and \acl{RN} 
are that the \acl{ES} performs switching and has the capability of higher 
speed radio links with other Edge Switches as well as connections to 
wired \acl{ATM} networks.

\begin{figure*}[htbp]
        \centerline{\psfig{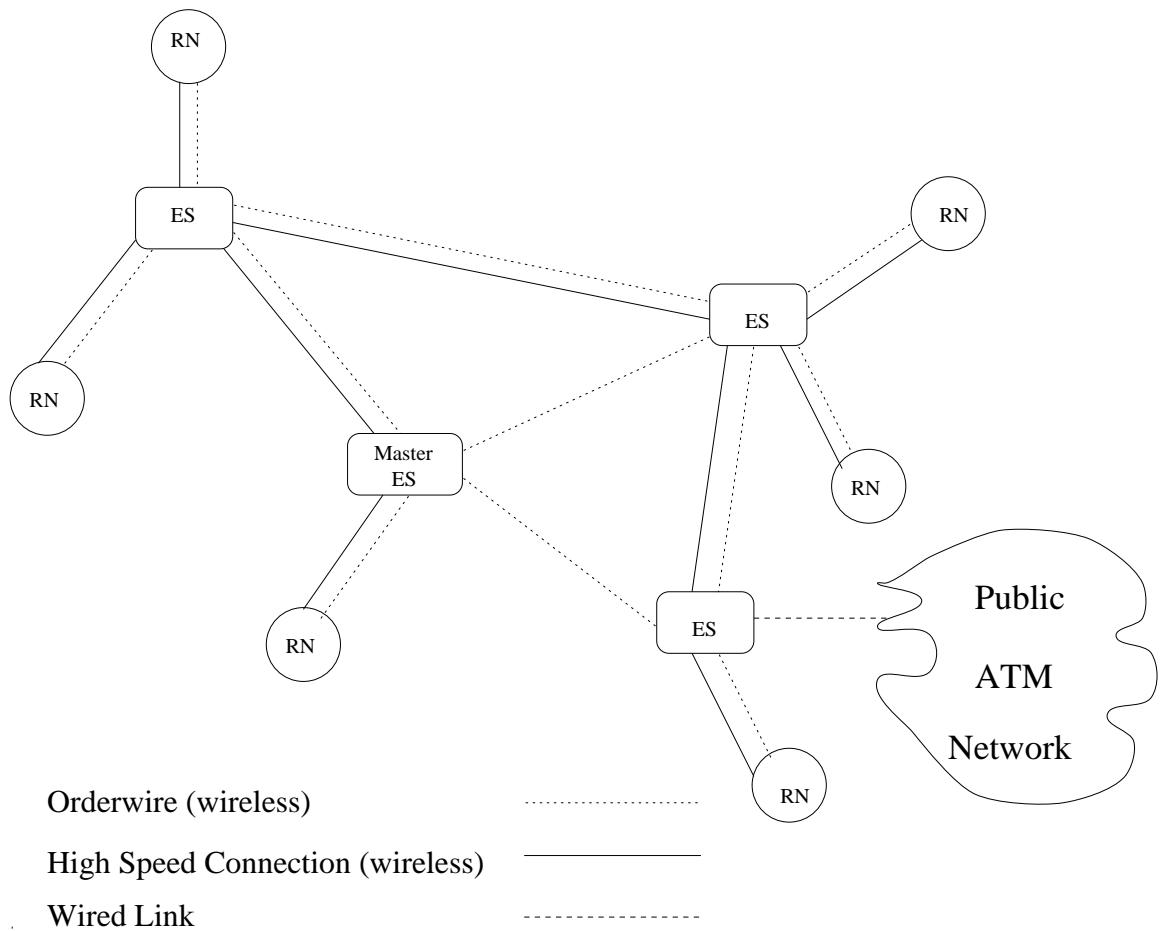}}
        \caption{Example Orderwire Topology.}
        \label{owov}
\end{figure*}

The orderwire network uses a low power, omni-directional channel, operating at 
19200 bps, for signaling, control, and communicating node locations to other 
network elements. The orderwire aids link establishment between the \acl{ES}es 
and between the \acl{RN}s and \acl{ES}es, tracking  remote nodes and determining 
link quality. The orderwire operates over packet radios and is part of the 
\acl{NCP}\footnote{The \acl{SNMP} \acl{MIB} for the \acl{NCP} operation as well as
live data from the running prototype \acl{RDRN} system can be
retrieved from \htmladdnormallink{http:/\-/\-www.ittc.ukans.edu/\-$\sim$sbush/\-rdrn/\-ncp.html}{http://www.ittc.ukans.edu/~sbush/rdrn/ncp.html}.}.
An example of the user data and orderwire network topology is shown in
Figure \ref{owov}. In this figure, an \acl{ES} serves as a link between
a wired and wireless network, while the remaining \acl{ES}es act as wireless
switches. The protocol stack for this network is shown in 
Figure \ref{hsstack}.

The focus of this section is on the \acl{NCP} and in particular on the orderwire 
network and protocols used to configure the network. This includes protocol layer 
configuration, link quality, hand-off, and host/switch assignment
along with information provided by the \acl{GPS} system such as position
and time. The details of the user data network will be covered in 
this paper only in terms of services required from, and interactions 
with, the \acl{NCP}. 

\begin{figure}[htbp]
        \centerline{\psfig{file=figures/protos.eps,width=4.0in}}
        \caption{Wireless \acl{ATM} Protocol Stack.}
        \label{hsstack}
\end{figure}

\subsection{Network Control Protocol}
  
At the physical level the orderwire is used to exchange
position, time and link quality information and to establish the wireless
\acl{ATM} connections. The process of setting up the wireless connections
involves setting up links between \acl{ES}es and between \acl{ES}es and 
\acl{RN}s. Error conditions are discussed at the end of this Section and
in Section \ref{perfemulation}.

The network has one Master \acl{ES}, which runs
the topology configuration algorithm \cite{cla} and distributes the resulting
topology information to all the connected \acl{ES}es over 
connection-oriented orderwire packet radio links. In the current prototype 
the connection-oriented link layer for the orderwire uses AX.25 
\cite{AX.25}. The term ``AX.25 connection-oriented''
in the following discussion means that the AX.25 Numbered Information Frames are
used. Broadcast mode in AX.25 uses Unnumbered Information Frames.
The Master \acl{ES} is the first active \acl{ES},
and any \acl{ES} has the capability of becoming a Master \acl{ES}.

The following description of the \acl{RDRN} Orderwire operation references
the state transitions labeled in Table \ref{ESFSM}.  The first \acl{ES} to 
become active broadcasts its callsign and start-up-time in a 
\textbf{MYCALL} packet and listens for responses from any other \acl{ES}es 
(Label I1 in Table \ref{ESFSM}). In this prototype system, the packet radio 
callsign is assigned by the FCC and identifies the radio operator. Since it 
is assumed that this is the first active \acl{ES}, there is no response within 
the given time period, $T$. At the end of $T$ seconds, the \acl{ES} rebroadcasts 
its \textbf{MYCALL} packet and waits another $T$ 
seconds. At the end of $2T$ seconds, if there are still no responses from other 
\acl{ES}es, the \acl{ES} activates and takes on the role of the Master \acl{ES} 
(Label I4 in Table \ref{ESFSM}). If the first two or more \acl{ES}es start up 
within $T$ seconds of each other, at the end of the interval $T$, the \acl{ES}es 
compare the start-up times in all the received
\textbf{MYCALL} packets and the \acl{ES} with the oldest start-up time
becomes the Master \acl{ES}. If the oldest two start-up times are equal,
the \acl{ES} with the lowest lexicographic callsign becomes the Master \acl{ES}.
In this system, accurate time stamps are provided 
by the \acl{GPS} receiver.

\begin{table*}[hbp]
\centering
\begin{tabular}{||l|l|l|l|l||}                                                                      \hline
\textbf{State}    & \textbf{Input Packet} & \textbf{Ouput Packet} & \textbf{Next State}  & \textbf{Label}          \\ \hline \hline
Initialization & None               & \textbf{MYCALL}   & Initialization    & I1 \\ \cline{2-5}
               & \textbf{MYCALL}   & \textbf{NEWSWITCH}& Initialization    & I2 \\ \cline{2-5}
               & \textbf{NEWSWITCH}& \textbf{SWITCHPOS}& Active       & I3 \\ \hline
(Master Only)  & timeout ($T$)      & \textbf{TOPOLOGY} & Active            & I4 \\ \cline{2-5}
               & \textbf{TOPOLOGY} & None               & Active            & I4 \\ \hline
Active         & \textbf{USER\_POS}& None               & RnUpdate          & A1 \\ \cline{2-5}
               & \textbf{TOPOLOGY} & None               & Active            & A2 \\ \hline
               & \textbf{MYCALL}   & None               & Initialization    & A3 \\ \cline{2-5}
               & \textbf{SWITCHPOS}& None               & Initialization    & A4 \\ \cline{2-5}
RnUpdate       & None               & None               & Active            & R1 \\ \cline{2-5}
               & None               & \textbf{HANDOFF}  & Active            & R2 \\ \hline
\end{tabular}
\caption{\label{ESFSM} Edge Switch Finite State Machine.}
\end{table*}

\begin{table*}[hbp]
\centering
\begin{tabular}{||l|l|l|l|l||}                                                                     \hline
\textbf{State}   & \textbf{Input Packet} & \textbf{Ouput Packet} & \textbf{Next State}  & \textbf{Label}          \\ \hline \hline
Initialization & None                   & None                   & Active         & C1 \\ \hline
Active         & \textbf{timeout}      & \textbf{USER\_POS}    & Active         & A1 \\ \cline{2-5}
               & \textbf{HANDOFF}      & None                   & Initialization & A2 \\ \hline
\end{tabular}
\caption{\label{RNFSM} Remote Node Finite State Machine.}
\end{table*}

Each successive \acl{ES} that becomes active broadcasts its callsign in a 
\textbf{MYCALL} packet (Label I1 in Table \ref{ESFSM}). Upon receipt of 
a \textbf{MYCALL} packet, the Master \acl{ES} extracts the callsign of the 
source, establishes an AX.25 connection-oriented link to the new \acl{ES} and 
sends it a \textbf{NEWSWITCH} packet (Label I2 in Table \ref{ESFSM}). 
On receipt of the \textbf{NEWSWITCH} packet over the AX.25 connection-oriented 
orderwire 
link, the \acl{ES} obtains its position from its \acl{GPS} receiver and sends 
its position to the Master \acl{ES} as a \textbf{SWITCHPOS} packet over the 
AX.25 connection-oriented orderwire link (Label I3 in Table \ref{ESFSM}). On receipt of 
a \textbf{SWITCHPOS} packet, the Master \acl{ES} records the position of the 
new \acl{ES} in its switch position table and runs the topology configuration 
algorithm \cite{cla} to determine the best possible interconnection of all 
the \acl{ES}es. The master then distributes the resulting information to all 
the \acl{ES}es in the form of a \textbf{TOPOLOGY} packet over the AX.25 connection-oriented 
orderwire links (Label I4 in Table \ref{ESFSM}). Each 
\acl{ES} then uses this information to setup the \acl{ES} links as specified by 
the topology algorithm. The Master \acl{ES} also distributes a copy of its 
switch position table to all the \acl{ES}es over the AX.25 connection-oriented 
orderwire links which can be used to configure \acl{RN}s as discussed below. 
The \acl{ES} can then use the callsign information in the switch position 
table to setup any additional AX.25 connection-oriented orderwire packet radio links 
corresponding to the \acl{ES} links required to exchange any link quality 
information. Thus, this scheme results in an AX.25 connection-oriented star network of 
orderwire links with the Master \acl{ES} at the center of the star and the 
AX.25 connection-oriented orderwire links between those \acl{ES}es which have
corresponding \acl{ES} links as shown 
in Figure \ref{owov}. The underlying AX.25 protocol generates error
messages when a frame has failed successful transmission after a
specified number of attempts. In the event of failure of the master node, 
which can be 
detected by listening for the AX-25 messages generated on node failure, the 
remaining \acl{ES}es exchange \textbf{MYCALL} packets (Label I1-I3 in Table 
\ref{ESFSM}), elect a new master node, and the network of \acl{ES}es is 
reconfigured using the topology configuration algorithm \cite{cla} (I4).

The \acl{ES}/\acl{RN} operation is described by Tables \ref{ESFSM} and 
\ref{RNFSM}. The packet contents are shown in Table \ref{prototab}.
Each \acl{RN} that becomes active obtains its position from its \acl{GPS} 
receiver and broadcasts its position as a \textbf{USER\_POS} packet 
over the orderwire network (Label A1 in Table \ref{RNFSM}). The timeout value 
in Table \ref{RNFSM} is the \textbf{USER\_POS} update period. The 
\textbf{USER\_POS} packet is received by all \acl{ES}es within range (Label A2 
in Table \ref{RNFSM}). Each \acl{ES} then computes the optimal grouping of \acl{RN}s 
within a beam via an algorithm described in \cite{ShaneBeam}.
The beamforming 
algorithm returns the steering angles for each of the beams originating from 
the \acl{ES} so that all the \acl{RN}s are covered. If the \acl{NCP}
determines that a beam and \acl{TDMA} time slot are available to support 
the new \acl{RN}, the \acl{ES} steers its beams so that all its connected 
\acl{RN}s and the new \acl{RN} are covered (Label R1 in Table \ref{RNFSM}). The 
\acl{ES} also records the new \acl{RN}'s position in its user position table, 
establishes a AX.25 connection-oriented orderwire link to the new \acl{RN}, and sends the 
new \acl{RN} a \textbf{HANDOFF} packet with link setup information indicating 
that the \acl{RN} is associated with it (Label R2 in Table \ref{RNFSM}). If the 
new \acl{RN} cannot be accommodated, the \acl{ES} sends it a \textbf{HANDOFF} 
packet with the callsign of the next closest \acl{ES}, to which the \acl{RN} 
sends another \textbf{USER\_POS} packet over a AX.25 connection-oriented orderwire link. 
This \acl{ES} then uses the beamform algorithm to determine if it can handle 
the \acl{RN}. 

This scheme uses feedback from the beamforming algorithm together with the 
distance information to configure the \acl{RN}. It should be noted that the 
underlying AX.25 protocol \cite{AX.25} provides error free transmissions over 
AX.25 connection-oriented orderwire links. Also the AX.25 connection-oriented 
orderwire link can be established from either end and the handshake 
mechanism for setting up such a link is handled by AX.25. If the \acl{RN} does 
not receive a \textbf{HANDOFF} packet within a given time it uses a retry 
mechanism to ensure successful broadcast of its \textbf{USER\_POS} packet.  

An AX.25 connection-oriented orderwire link is retained as long as a \acl{RN} 
is connected to a particular \acl{ES} and a corresponding high-speed link 
exists between 
them to enable exchange of link quality information. The link can be
removed 
when the mobile \acl{RN} migrates to another \acl{ES} in case of a hand-off. 
Thus at the end of this network configuration process, three overlayed networks 
are established; an orderwire network, an \acl{RN} to \acl{ES} network, and an 
\acl{ES} to \acl{ES} network. The orderwire network has links between the 
Master \acl{ES} and every other active \acl{ES} in a star configuration, links 
between \acl{ES}es connected by \acl{ES} links as well as links between 
\acl{RN}s 
and the \acl{ES}es to which they are connected, as shown in Figure \ref{owov}. 
Raw pipes for the user data links between \acl{RN}s and appropriate \acl{ES}es 
as well as for the user data links between \acl{ES}es are also established.
Any \acl{ES} which powers-up, powers-down, changes position, or becomes
unusable due to any malfunction causes the entire \acl{NCP} system to
reconfigure. Multiple \acl{NCP} groups will form when a set of nodes are
out of range of each other. \acl{ES} nodes which receive \textbf{NEWSWITCH}
packets from multiple Master \acl{ES}es will join the group whose Master
\acl{ES} has the oldest start time or which has the lowest lexicographic
callsign in case of equal start times. However, if digipeating is
enabled, any \acl{ES} node can act as a forwarder of messages for
nodes which are out of range, which would enable a single group to be
formed.

\begin{table*}[htbp]
\centering 
\begin{tabular}{||l|l||}                                         \hline
\textbf{Packet Types}  & \textbf{Packet Contents}             \\ \hline \hline 
MYCALL        & Callsign, Start-Up-Time                       \\ \hline 
NEWSWITCH     & {\em empty packet}                            \\ \hline 
SWITCHPOS     & \acl{GPS} Time, \acl{GPS} Position                        \\ \hline
TOPOLOGY      & Callsigns and Positions of each node          \\ \hline
USER\_POS     & Callsign, \acl{GPS} Time, \acl{GPS} Position              \\ \hline
HANDOFF       & Frequency, Time Slot, \acl{ES} \acl{GPS} Position         \\ \hline
\end{tabular}                                                           
\caption{\label{prototab} Network Control Protocol Packets.}         
\end{table*} 

\subsection{Orderwire Performance Emulation}
\label{perfemulation}


The purpose of this section is to illustrate the time required
by \acl{NCP} to perform configuration without \acl{VNC} using an 
emulation consisting of Maisie \cite{bagrodia} and the actual orderwire code.
The emulation of the orderwire systems satisfies several goals. It 
allows tests of configurations
that are beyond the scope of the prototype \acl{RDRN} hardware. Specifically,
it verifies the correct operation of the \acl{RDRN} \acl{NCP}
in a wide variety of situations. 
As an additional benefit, much of the actual orderwire code was used with
the Maisie emulation allowing further validation of that
code.
The \acl{ES} and \acl{RN} are modeled as a collection of 
Maisie entities. This 
is an emulation rather than a simulation because the Maisie code is 
linked with the working orderwire code and also with the topology 
algorithm. 
There is a Maisie entity for each major component of the \acl{RDRN} system
including the \acl{GPS} receiver, packet radio, inter-\acl{ES} links, \acl{RN} 
to \acl{ES} links and
the Master \acl{ES}, and \acl{RN} network configuration processors, as
well as other miscellaneous entities. 
The input parameters to the emulation are shown in Tables \ref{emin_move},
\ref{emin_time}, and \ref{emin_beam}.

\begin{table*}[htbp]
\centering
\begin{tabular}{||l|l||}                                           \hline
\textbf{Parameter }  & \textbf{Definition}         \\ \hline \hline
NumRN		& Number of Remote Nodes  \\ \hline
NumES 		& Number of Edge Switches  \\ \hline
ESDist 		& Inter Edge Switch spacing (forms rectangular area) \\	\hline
T 		& ES/ES MYCALL configuration time \\ \hline
maxV 		& Maximum \acl{RN} speed for uniform distribution \\ \hline
S 		& Time to wait between node initial startups \\ \hline
ESspd 		& Initial Edge Switch speed \\ \hline
ESdir 		& Initial Edge Switch direction \\ \hline
RNspd 		& Initial Remote Node speed \\ \hline
RNdir 		& Initial Remote Node direction \\ \hline
\end{tabular}
\caption{\label{emin_move} NCP Emulation Mobility Input Parameters.}
\end{table*}

\begin{table*}[htbp]
\centering
\begin{tabular}{||l|l||}                                           \hline
\textbf{Parameter }  & \textbf{Definition}         \\ \hline \hline
EndTime         & Emulation end time (tenths of seconds) \\ \hline
VCCallTime      & Inter High Speed Connection Setup Times \\ \hline
VCCallDuration  & High Speed Connection Life Times  \\ \hline
\end{tabular}
\caption{\label{emin_time} NCP Emulation Time Input Parameters.}
\end{table*}

\begin{table*}[htbp]
\centering
\begin{tabular}{||l|l||}                                           \hline
\textbf{Parameter}  & \textbf{Definition}         \\ \hline \hline
UseRealTopology & Connect to MatLab and run actual program \\ \hline
Rlink           & Maximum beam distance \\ \hline
Fmax            & Number of non-interfering frequency pairs \\ \hline
Imult           & Interference multiplier \\ \hline
Twidth          & Transmitting Beam width \\ \hline
Rwidth          & Receiving Beam width  \\ \hline
\end{tabular}
\caption{\label{emin_beam} NCP Emulation Beam Input Parameters.}
\end{table*}

The time between \acl{VC} requests from an \acl{RN} for connections over the inter-\acl{ES} antenna beams
is assumed to be Poisson. This represents \acl{ATM} \acl{VC} usage over the
physical link. The \acl{RN} will maintain a
constant speed and direction until a hand-off occurs, then a new
speed and direction are generated from a uniform distribution.
This simplifies the analytical computation. Note that \acl{NCP} packet transfer 
times as measured in Table \ref{timetab} are used here.

The architecture for the \acl{RDRN} link management and control is shown in
Figure \ref{ncsa}. The topology module executes only on the Master \acl{ES}
node. The remaining modules are used on all \acl{ES} nodes and \acl{RN}s.
The beamform module determines an optimal steering
angle for the given number of beams which connects all \acl{RN}s to be 
associated
with this \acl{ES}. It computes an estimated signal to noise interference 
ratio (SIR) and generates a table of complex weights which, once loaded, 
will control the beam formation. 
The connection table is used by the Adaptive \acl{HDLC} and
\acl{ATM} protocol stacks for configuration via the adaptation manager.

\ifisdraft
	\onecolumn
\fi
\begin{sidefigure}
        \centerline{\psfig{file=figures/linkd.eps,width=8in}}
        \caption{Network Control System Architecture.}
        \label{ncsa}
\end{sidefigure}
\ifisdraft
	\twocolumn
\fi

\begin{figure*}[htbp]
        \centerline{\psfig{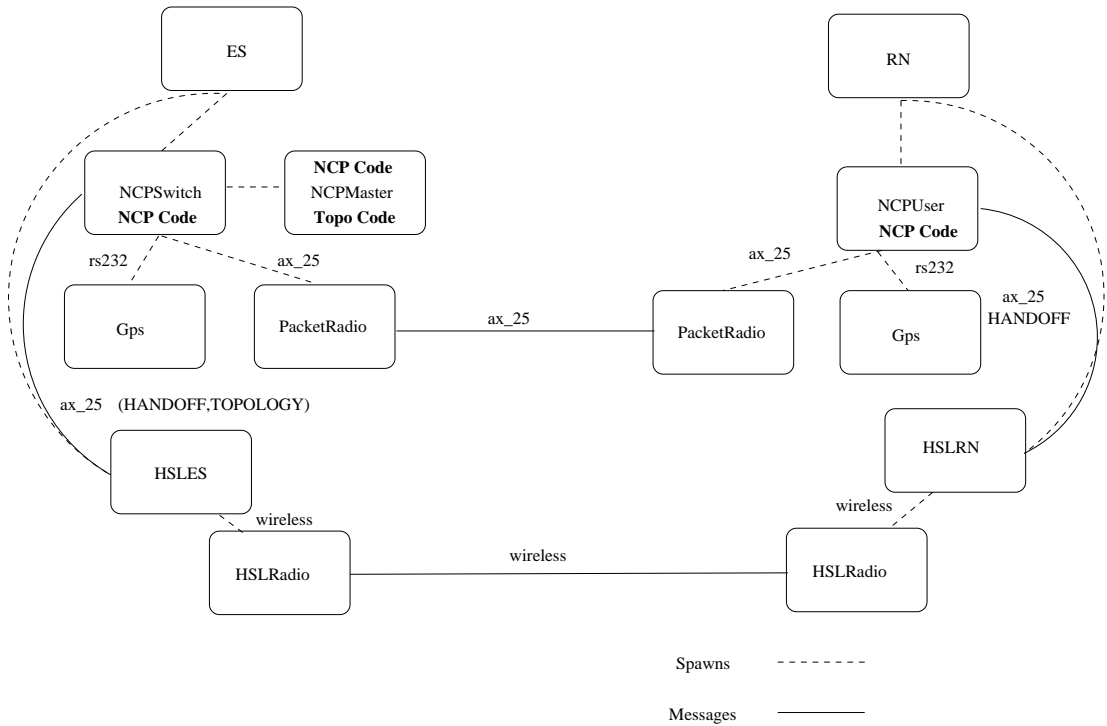}}
        \caption{Emulation Design.}
        \label{ncpem}
\end{figure*}

The emulation uses as much of the actual network control code as possible.
The packet radio driver and \acl{GPS} driver interact with Maisie rather than 
an actual packet radio and \acl{GPS} receiver. The remaining code is the actual
\acl{NCP} code. Figure \ref{ncpem} shows the structure of the Maisie entities. 
The entity names are shown in the boxes and the user defined Maisie message 
types are shown along the
lines. Direct communication between entities is represented as a solid line.
The dashed lines indicate from where Maisie entities are spawned. The parent
and child entities pass the message types indicated in Figure \ref{ncpem}.

The \acl{RN} entity which performs \acl{ATM} \acl{VC} setup
generates calls that the \acl{ES} node 
will attempt to accept.
The entity labeled HSLRN in Figure \ref{ncpem} represents the High Speed Link
\acl{RN} and the entity labeled HSLES in Figure \ref{ncpem} represents the
High Speed Link \acl{ES} entity. These are the radios upon which the
\acl{ATM} protocol is running.
If the \acl{EN} moves out of range or the 
\acl{ES} has no beam or slot available the setup is aborted. As the \acl{RN} 
moves,
the \acl{ES} will hand off the connection to the proper \acl{ES} based on 
closest distance between \acl{RN} and \acl{ES}. 

\subsection {Emulation Results}
\label{emres}

This section discusses some of the results from the emulation. 
Some of these results revealed problems which are not immediately
apparent from the \acl{FSM} in Tables \ref{ESFSM} and \ref{RNFSM}.
The emulation produces \acl{NCP} \acl{FSM}
output which shows the transitions based on the \acl{FSM}
in Tables \ref{ESFSM} and \ref{RNFSM}. The Finite State Machine output 
from the simulation provided an easy comparison with diagrams to insure 
correct operation of the protocol.

\subsubsection{Effect of Scale on \acl{NCP}}

The emulation was run to determine the effect on the \acl{NCP} as the number
of \acl{ES} and \acl{RN} nodes increased. The dominate component of the configuration
time is the topology calculation run by the \acl{ES} which is designated as the
master. Topology calculation involves searching through the problem space of
constraints on the directional beams for all feasible topologies and 
choosing an optimal topology from that set as described in \cite{cla}. 
The units on all values should be consistent with the GPS coordinate units, 
and all angles are assumed to be degrees. The beam constraint values used 
in this emulation are: maximum link distance 1000.0 meters, maximum 
frequencies 3, interference multiplier 1.0, transmit beam width 10.0 degrees, 
and receive beam width 10.0 degrees. These values were chosen for the
emulation before the antenna hardware had been built and the antenna
patterns measured. 

The topology calculation is performed
in MatLab and uses the MatLab provided external C interface. Passing
information through this interface is clearly slow, therefore
these results do represent the worst case bounds on the execution times 
of the prototype system.

A speedup will arise through the use of Virtual Network 
Configuration, which will provide a mechanism for predicting values 
and also allows processing to be distributed.
Another improvement which may be considered is to implement a 
hierarchical configuration. The network is partitioned into
a small number of clusters of nodes in such a way that nodes in each 
group are as close 
together as possible. The topology code is run as though these were
individual nodes located at the center of each group. This inter-group
connection will be added as constraints to the topology
computation for the intra-group connections. In this way the topology
program only needs to calculate small numbers of nodes which it does
relatively quickly.

\subsubsection{MYCALL Timer}

\begin{figure*}[htbp]
        \centerline{\psfig{file=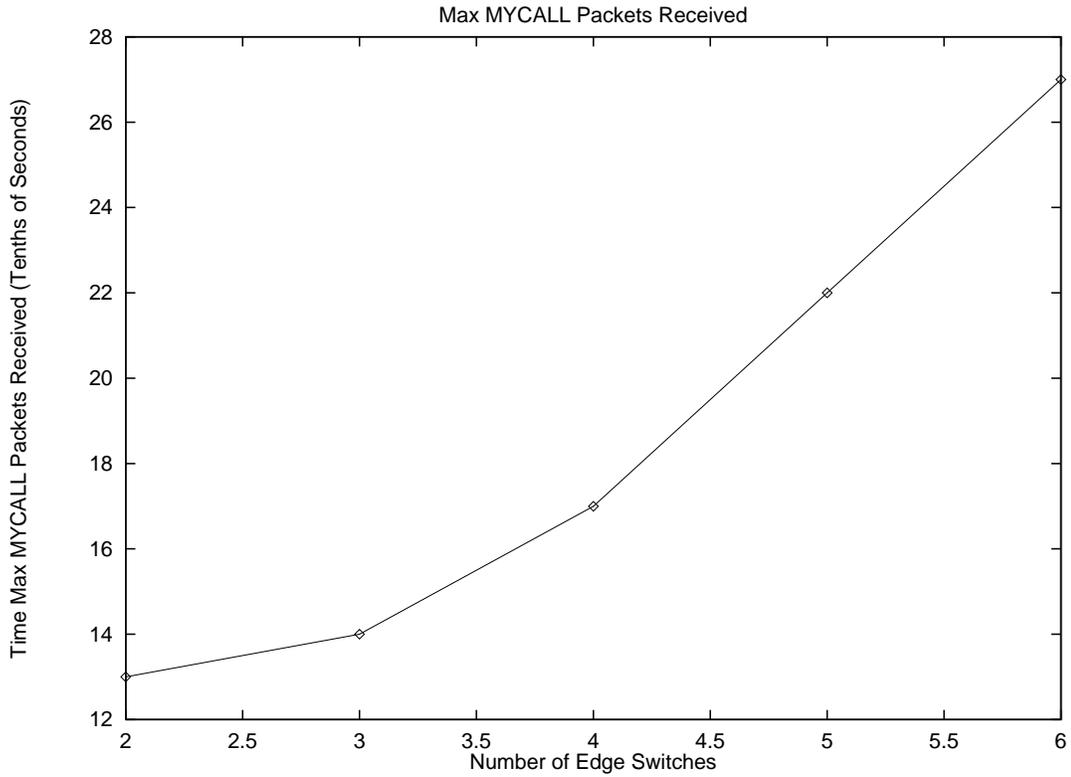,width=5.75in}}
        \caption{MYCALL Packets Received.}
        \label{mycalltime}
\end{figure*}

The MYCALL Timer, set to a value of $T$, controls
how long the system will wait to discover new \acl{ES} nodes before completing the
configuration. If this value is set too low, new \textbf{MYCALL} packets will
arrive after the topology calculation has begun, causing the system to 
needlessly reconfigure.
If the MYCALL Timer value is too long, time will be wasted, which will
have a large impact on a mobile \acl{ES} system. Table \ref{mycallsim} shows the
input parameters and Figure \ref{mycalltime} shows
the time required for all \textbf{MYCALL} packets to be received as a
function of the number of \acl{ES} nodes. These times are the 
optimal value of the MYCALL Timer as a function of the number of
ES nodes because the these times are exactly the amount of time 
required for all \acl{ES} nodes to respond. This time does not include the
topology calculation time which is currently significantly longer.
Clearly performance would benefit if these time consuming operations
could be precomputed.
In order to prevent the possibility of an infinite loop of reconfigurations
from occurring, an exponential increase in
the length of the \textbf{MYCALL} Timer value is introduced. 
As \textbf{MYCALL} packets arrive after
T has expired, the next configuration occurs with an increased value
of T. 

\begin{table*}[htbp]                           
\centering                                                      
\begin{tabular}{||l|l||}                              \hline
\textbf{Parameter }     & \textbf{Value}                 \\ \hline 
Number of RNs        & $0$                         \\ \hline 
Number of ESs        & $2$ thru $6$                \\ \hline
Inter-ES Distance    & $20$ m (65.62 ft)           \\ \hline 
T                    & $20$ s                      \\ \hline
Maximum Velocity     & $5$ m/s (11.16 mi/hr)       \\ \hline
Initial \acl{ES} Speed     & $0$ m/s                     \\ \hline
Initial \acl{ES} Direction & $0^o$                       \\ \hline
Initial \acl{RN} Speed     & N/A                         \\ \hline 
Initial \acl{RN} Direction & N/A                         \\ \hline
Call Inter-Arrival Time  & $1200$ s                    \\ \hline
VC Call Duration     & $600$ s                     \\ \hline
\end{tabular}
\caption{\label{mycallsim} MYCALL Timer Simulation Parameters.}
\end{table*}

\subsubsection{Link Usage Probability}

\begin{figure}[htbp]
        \centerline{\psfig{file=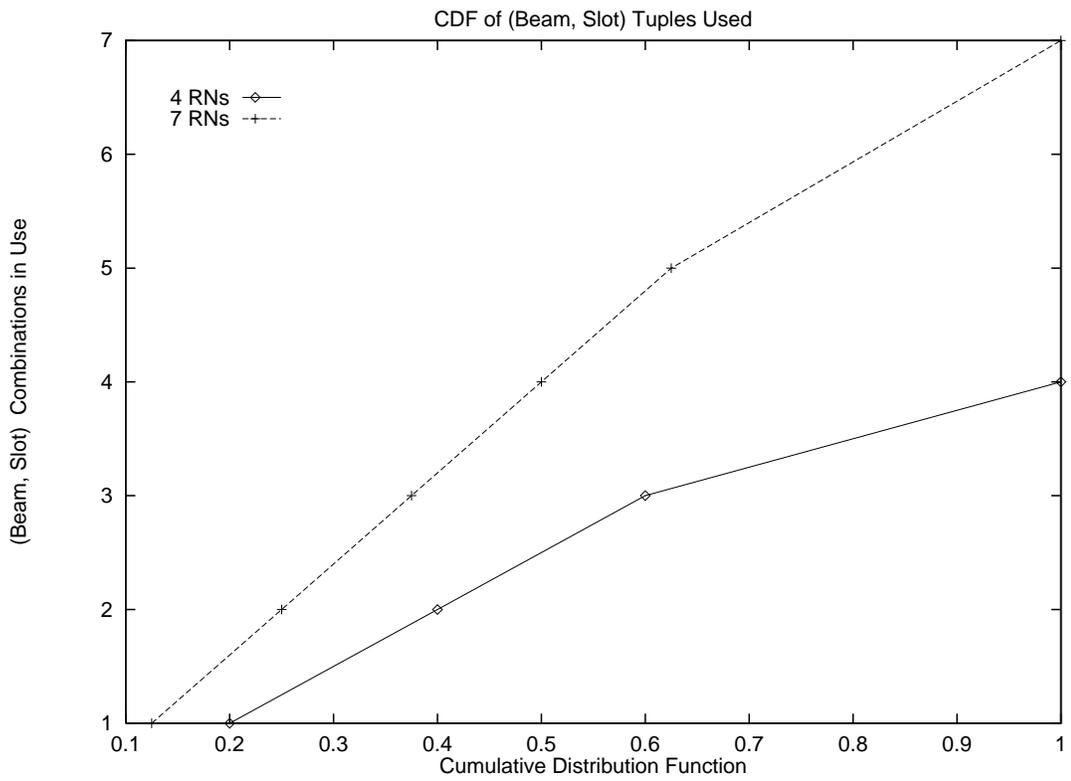,width=5.75in}}
        \caption{$(Beam, Slot)$ Usage.}
        \label{chanused}
\end{figure}

Multiple \acl{RN}s may share a single beam using Time Division Multiplexing
(TDMA) within a beam. The time slices are divided into slots, thus
a $(beam, slot)$ tuple defines a physical link. The emulation was run
to determine the probability distribution of links used as a function
of the number of \acl{RN}s. The parameters used in the emulation are shown in 
Table \ref{chanusedsim} the results of which indicate the number of links 
and thus the number of distinct $(beam, slot)$ tuples required. Figure 
\ref{chanused} shows the link usage cumulative distribution function
for 4 and 7 \acl{RN}s. The number of $(beam, slot)$ combinations
in use clearly increases with the number of \acl{RN}s as expected. 

\begin{table*}[htbp]                           
\centering                                                      
\begin{tabular}{||l|l||}                              \hline
\textbf{Parameter }     & \textbf{Value}          \\ \hline \hline 
Number of RNs        & $4$ and $7$          \\ \hline
Number of ESs        & $2$                  \\ \hline
Inter-ES Distance    & $20$ m (65.62 ft)    \\ \hline
T                    & $20$ s               \\ \hline
Maximum Velocity     & $5$ m/s (11.16 mi/hr)\\ \hline
Initial \acl{ES} Speed     & $0$ m/s              \\ \hline
Initial \acl{ES} Direction & $0^o$                \\ \hline
Initial \acl{RN} Speed     & $5$ m/s (11.16 mi/hr)\\ \hline 
Initial \acl{RN} Direction & $0^o$                \\ \hline
Call Inter-Arrival Time  & $1200$ s             \\ \hline
VC Call Duration     & $600$ s              \\ \hline
\end{tabular}
\caption{\label{chanusedsim} Link Usage Simulation Parameters.}
\end{table*}

\subsubsection{ES Mobility}

\begin{figure}[htbp]
        \centerline{\psfig{file=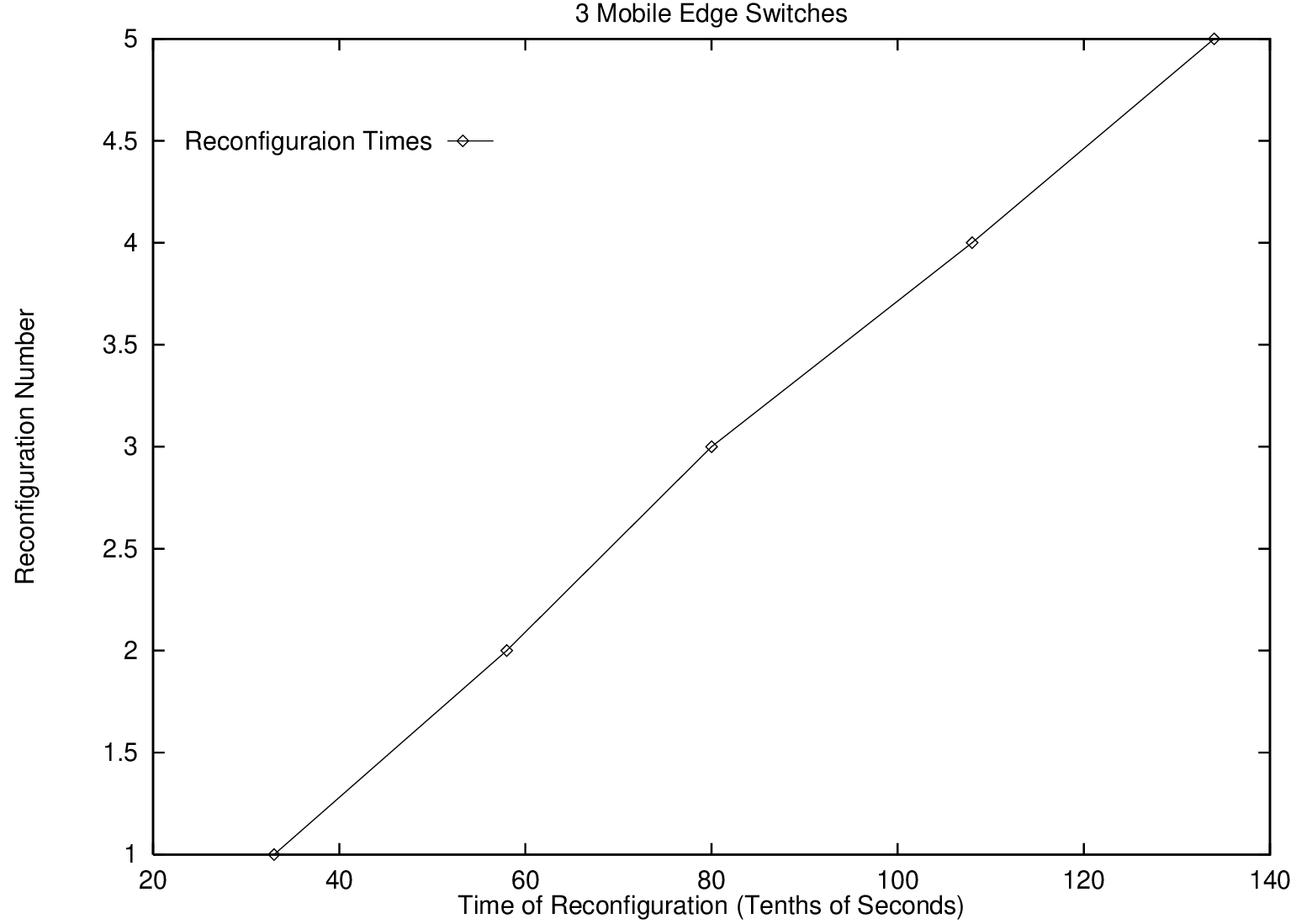,width=5.75in}}
        \caption{Mobile \acl{ES} Configuration Time.}
        \label{mobestime}
\end{figure}

ES mobility is a difficult problem because it requires many steps,
many of them very time consuming.
The parameters used in an emulation with mobile \acl{ES} nodes are shown in 
Table \ref{mobessim}.
As mentioned in the section on the MYCALL Timer, if a \textbf{MYCALL} packet
arrives after this timer has expired, a reconfiguration occurs. This could
happen due to a new \acl{ES} powering up or an \acl{ES} which has changed position.
Figure \ref{mobestime} shows the times at which reconfigurations occurred
in a situation in which \acl{ES} nodes were mobile.
Based on the state transitions generated from the emulation it is
apparent that the system is in a constant state of reconfiguration; no 
reconfiguration has time to complete before a new one begins. 
As \acl{ES} nodes move, the \acl{NCP} must notify \acl{RN}s associated with an \acl{ES} with the 
new position of the \acl{ES} as well as reconfigure the \acl{ES} nodes.
To solve this problem, a tolerance, which may be associated with the 
link quality, 
will be introduced which indicates how far nodes can move
within in a beam before the beam angle must be recalculated,
which will allow more time between reconfigurations.
It is expected that this tolerance in addition to Virtual Network 
Configuration will provide a solution to this problem.

\begin{table*}[htbp]                           
\centering                                                      
\begin{tabular}{||l|l||}                               \hline
\textbf{Parameter}       & \textbf{Value}                 \\ \hline \hline
Number of RNs          & $0$                         \\ \hline 
Number of ESs          & $3$                         \\ \hline 
Inter-ES Distance      & $20$ m (65.62 ft)           \\ \hline
T                      & $20$ s                      \\ \hline
Maximum Velocity       & $5$ m/s (11.16 mi/hr)       \\ \hline
Initial \acl{ES} Speed       & $1$ m/s (2.23 mi/hr)        \\ \hline
Initial \acl{ES} Direction   & $0^o$                       \\ \hline
Initial \acl{RN} Speed       & N/A                         \\ \hline
Initial \acl{RN} Direction   & N/A                         \\ \hline
Call Inter-Arrival Time    & $1200$ s                    \\ \hline
VC Call Duration       & $600$ s                     \\ \hline
\end{tabular}
\caption{\label{mobessim} Mobile ES Simulation Parameters.}
\end{table*}

\subsubsection{Effect of Communication Failures}

The emulation was run with a given probability of failure on each
Network Control Protocol packet type. The following results
are based on the output of the finite state machine (FSM) transition 
output of the emulation and an explanation is given for each case.

A dropped \textbf{MYCALL}
packet has no effect as long as at least one of the \textbf{MYCALL}
packets from each \acl{ES} is received at the master ES. This is the
only use of the AX.25 broadcast mode in the \acl{ES} configuration. The
broadcast AX.25 mode is a one time, best effort delivery;
therefore, \textbf{MYCALL} packets are repeatedly broadcast at the
NCP layer.

The Maisie emulation demonstrated that
a dropped \textbf{NEWSWITCH} packet caused the
protocol to fail. This is because the master \acl{ES} will wait until
it receives all \textbf{SWITCHPOS} packets from all \acl{ES} nodes for 
which it had received \textbf{MYCALL} packets. The \textbf{NEWSWITCH}
packet is sent over the AX.25 in connection-oriented mode, e.g. 
a mode in which corrupted frames are retransmitted.
A dropped \textbf{SWITCHPOS} packet has the same effect as a dropped
\textbf{NEWSWITCH} packet.
In order to avoid this situation, the \acl{NCP} will re-send the \textbf{NEWSWITCH}
if no response is received.

Finally, the Maisie emulation showed that a lost \textbf{TOPOLOGY} packet 
results in a partitioned network. The \acl{ES} which fails to receive the 
\textbf{TOPOLOGY} packet is not joined with the remaining \acl{ES} nodes; 
however, this \acl{ES} node continued to receive and 
process \textbf{USER\_POS} packets from all \acl{RN}s. It therefore attempts to 
form an initial connection with all \acl{RN}s.
The solution for this condition is not to allow \acl{RN} associations with
an \acl{ES} node until the \textbf{TOPOLOGY} packet is received. Because
\textbf{MYCALL} packets are transmitted via broadcast AX.25, each \acl{ES} node 
can simply count the number of \textbf{MYCALL} packets and estimate the 
time for the master \acl{ES} node to calculate the topology using the number of 
\textbf{MYCALL} packets as an estimate for the size of the network. 
If no \textbf{TOPOLOGY} 
packet is received within this time period, the \acl{ES} node retransmits its 
\textbf{SWITCHPOS} packet to the master \acl{ES} node in order to get a
\textbf{TOPOLOGY} packet as a reply. 

This section has provided a background on mobile wireless telecommunications
and \acf{RDRN}. In particular this section has focused on the \acf{NCP}.
\acf{VNC} will enhance the \acl{NCP} to provide speedup in configuration.
The next section introduces the parallel simulation algorithms from
which \acl{VNC} has been derived.
\section{Application of Parallel Simulation to \acl{VNC}}

A general solution to the challenge of mobile wireless \acl{ATM} network
configuration has been developed in the VNC algorithm\index{VNC!algorithm}. 
This algorithm is based on concepts for distributed\index{Distributed} and 
parallel simulation\index{Parallel Simulation}, which are classified in
this section. 

Figure \ref{simtax} shows a classification of parallel 
simulation\index{Parallel Simulation} taken from \cite{Lin90a}. In application 
distribution\index{Parallel Simulation!application level distribution},
the simulation is run independently with different parameter settings. 
In function level 
distribution\index{Parallel Simulation!function level distribution}, multiple 
processors work together on a single simulation execution.

\begin{sidefigure}

\leaf{\textbf{\small Support Function}\\ \textbf{Distribution}}

\leaf{\textbf{\small Synchronous}}

\leaf{\textbf{\small Conservative}}
\leaf{\textbf{\small Semi-Optimistic}}
\leaf{\textbf{\small Optimistic}}
\branch{3}{\textbf{\small Space Division}}

\leaf{\textbf{\small Time Division}}
\branch{3}{\textbf{\small Model Function}\\ \textbf{Distribution}}

\branch{2}{\textbf{\small Function-Level}\\ \textbf{Distribution}}

\leaf{\textbf{\small Applications-Level}\\ \textbf{Distribution}}
\branch{2}{\textbf{\small Parallel Discrete}\\ \textbf{Event Simulation}}

\tree

\caption{\label{simtax} Overview of Distributed Simulation Methods.}
\end{sidefigure}
\ifisdraft
	\twocolumn
\fi

Function level distribution can be further broken down into
model function\index{Parallel Simulation!model function distribution} 
and support function\index{Parallel Simulation!support function distribution} 
distribution. In support function\index{Parallel Simulation!support function
distribution} distribution, support 
functions\index{Parallel Simulation!function} such as random number
generation are run on separate processors. In model 
distribution\index{Parallel Simulation!model distribution}, individual 
simulation events are distributed\index{Parallel Simulation!distributed} on 
separate processors.

The model distribution\index{Parallel Simulation!model distribution} class 
can be further divided into synchronous\index{Parallel Simulation!synchronous}, 
asynchronous\index{Parallel Simulation!asynchronous} (space division)\index{Parallel Simulation!space division},
and time division\index{Parallel Simulation!time division} approaches.
In the synchronous\index{Parallel Simulation!synchronous} approach, 
events with the same timestamp are executed in parallel\index{Parallel Simulation!parallel}. 
The asynchronous\index{Parallel Simulation!asynchronous} 
approach allows greater parallelism\index{Parallel Simulation!parallelism} 
to occur. The model\index{Parallel Simulation!model} is partitioned into 
logical processes\index{Logical Process} 
which execute without regard to a common clock.
The time division approach partitions the model\index{Model} in simulation time.
Each time period is executed on a separate processor; an estimate
is made for the input at times not starting at zero. When the 
simulation completes, if results of consecutive time partitions
do not match, a fix-up phase is required. The asynchronous\index{Parallel 
Simulation!asynchronous} class is examined in more detail in Section 
\ref{vncorg}.
\section{Mobile Computing and Distributed Simulation}

The asynchronous\index{Parallel Simulation!asynchronous} approach is of 
most interest in this project because it has the capability of taking the
most advantage of parallelism. This method can be divided into 
optimistic\index{Parallel Simulation!optimistic}, 
semi-optimistic\index{Parallel Simulation!semi-optimistic}, and
conservative\index{Parallel Simulation!conservative} methods. These 
divisions reflect the amount of deviation allowed from a purely sequential 
simulation. Optimistic\index{Parallel Simulation!optimistic}
methods allow the most deviation from a sequential simulation and are
thus able to take the most advantage of parallelism\index{Parallelism}. 
The price paid for this is possible out of order message arrival which 
is corrected in by a rollback\index{Rollback} 
mechanism described in detail later. It is this 
rollback\index{Rollback} mechanism combined with connection-oriented\index{Mobile 
Network!classification!connection-oriented}, predictive mobile
network\index{Mobile Network!classification!predictive mobile network} 
mechanisms which enables the development of 
Virtual Network Configuration\index{Virtual Network Configuration}.
Changes to a local portion of the \acl{VNC} enhanced system are predicted
and the results are propagated throughout the system
allowing information needed by other elements of the system to be pre-computed 
and cached.

There are several benefits to be gained from the application of optimistic 
distributed simulation techniques to network configuration in a mobile wireless 
environment. Allowing the configuration processes to work ahead, generate
results, and cache them for future use provides speedup provided that the
predictions are accurate. As an additional benefit, optimistic distributed 
simulation has a potential to better utilize inherent parallelism thus 
speeding up operations. The parallelism referred to in this study is
among processors currently existing among multiple \acl{RDRN} nodes. However,
by using \acl{VNC}, predictive processes would already exist to take advantage 
of multiple processors if such processors were added in
the future. Optimistic distributed simulation has another characteristic 
known as super-criticality. 
The time required for a configuration result to be generated in a system 
partitioned into \acl{LP}s is measured; 
the longest path through which messages propagate is the bottleneck or 
``critical path''. Super-criticality 
results from the observation that with the lazy evaluation optimization, 
it is possible for the final result to be obtained in a time less than the 
critical path time. 
\section{Virtual Network Configuration Applications}
\label{VNC Examples}

Before going into further detail on \acl{VNC}\index{VNC} in the next section, 
consider how a predictive mechanism\index{Predictive Mobility 
Management!examples} can be used in a mobile wireless \acl{ATM}\index{ATM} network.
The following sections discuss application of \acl{VNC} at various layers of
a wireless network. This section clearly demonstrates the potential value for 
the \acl{VNC} algorithm developed in this thesis.

\subsection{Rapid Physical Layer Setup}

A virtual\index{VNC!virtual handoff message} handoff message causes 
physical layer states to be calculated and cached\index{Cache} prior to 
an actual handoff\index{Handoff}. Thus the only time required is the
time to access the cached information when a handoff occurs. An example 
of this would be
beamsteering calculations in the Rapidly Deployable Radio Network
Project\index{Rapidly Deployable Radio Network} \cite{BushRDRN}.

\subsection{Rapid Media Access Control (\acl{MAC}) Layer Setup}

Adaptive parameters\index{Adaptive Parameters} in the \acl{MAC} Layer will 
benefit from \acl{VNC}. For example the Adaptive HDLC\index{Adaptive HDLC} 
layer \cite{BushRDRN} in \acl{RDRN} will be able to pre-determine parameters such 
as the optimal frame length, modulation, and coding in advance of a 
handoff\index{Handoff}.

\subsection{Rapid \acl{ATM}\index{ATM!topology} Topology Calculation}

Knowledge of the future location of a mobile host allows the 
\acl{ATM}\index{ATM!topology} topology processes to configure prior to 
handoff\index{Handoff} taking place. Consider
the DEC Autonet \cite{Autonet} \cite{AutonetJ} which has switch specific 
features for topology\index{ATM!topology}
calculation such as Resilient Virtual Circuit\index{Resilient Virtual
Circuit} calculation and automatic topology\index{Topology} calculation.
These calculations can be time consuming. Consider
the Autonet topology\index{Topology} process which takes on 
the order of $58 + 3.34 d + 1.36 n + 0.315 d n$ 
milliseconds to determine the network topology, where $d$ is the network diameter 
and $n$ is the number of 
nodes. A 100 switch network could take over 0.5 seconds to reconfigure. 
This means the entire network would be down for half a second each time 
there was a handoff\index{Handoff}. 
In the best case using \acl{VNC}\index{VNC}, the entire topology\index{Topology} 
calculation is done 
with virtual messages\index{Virtual Message} and the only 
real-time\index{Real Time} component would be the time required to update the 
topology\index{Topology} from the \acl{VNC}\index{VNC} cache. 
Assuming this takes approximately 5 milliseconds, the speedup for a 100 node 
network would be 108 times faster with \acl{VNC}\index{VNC}.

\subsection{Rapid \acl{ATM}ARP\index{ATMARP} Server Updates and Configuration}

The \acl{ATMARP}\index{ATMARP} operation is 
described in \cite{RFC1577}. The purpose of \acl{ATMARP} is to provide the
\acl{ATM} address for a given network address. The configuration
of \acl{ATM}\index{ATMhost} hosts and \acl{ARP} servers\index{ARP server}
is considered implementation dependent.
\acl{VNC}\index{VNC} will allow the mobile host to be supplied with the 
location of the \acl{ATMARP}\index{ATMARP} server within its predicted new 
\acl{LIS}, before handoff\index{Handoff} takes place. This
saves the time taken by any implementation dependent protocol from
determining the \acl{ARP} server during handoff\index{Handoff}.

\subsection{Rapid NHRP\index{NHRP} Server 
Configuration\index{NHRP!server configuration} and Updates}

The Next Hop Resolution Protocol (NHRP\index{NHRP}) \cite{NBMA}
allows \acl{ATMARP}\index{ATMARP} to take place across Logical IP\index{IP} 
Subnets (LIS\index{LIS}). 
The configuration of NHRP\index{NHRP} servers may take place via 
NHRP\index{NHRP!Registration Packets} Registration Packets or manual configuration. 
NHRP\index{NHRP!server} servers can cache\index{Cache} the expected 
changes and have them take effect while the handoff\index{Handoff} is in 
progress, giving the appearance of immediate update.

\subsection{Rapid IP\index{IP} Route Calculation}

IP\index{IP} routing updates can take place prior to handoff\index{Handoff}, 
with results cached until handoff\index{Handoff} actually occurs. 
In \cite{Caceres}, handoff timing measurements were taken on a mobile
host connected to a 2 Mb/s WaveLAN product by NCR. The stationary hosts
connected to a 10 Mb/s Ethernet network. The hosts and base stations were
i486 processors with 330 MB hard disks, 16 MB of memory and they ran
BSD-Tahoe TCP with the Mobile IP software from Columbia University.
Even with instant notification of handoff\index{Handoff} provided 
when the mobile host crosses a cell boundary, 
it took approximately 0.05 seconds for the mobile host routing tables to 
update, and 0.15 seconds for the mobile support station routing tables 
to update. \acl{VNC}\index{VNC} could prepare the tables for update
before handoff\index{Handoff}, and assuming a 5 milliseconds to update 
the routing table from a local cache, this results in a speedup of 30 
times over non-\acl{VNC}\index{VNC} configuration. 

\subsection{Rapid Mobile-IP\index{Mobile-IP} Foreign 
Agent\index{Mobile-IP!Foreign Agent} Determination}

\begin{figure*}[htbp]
        \centerline{\psfig{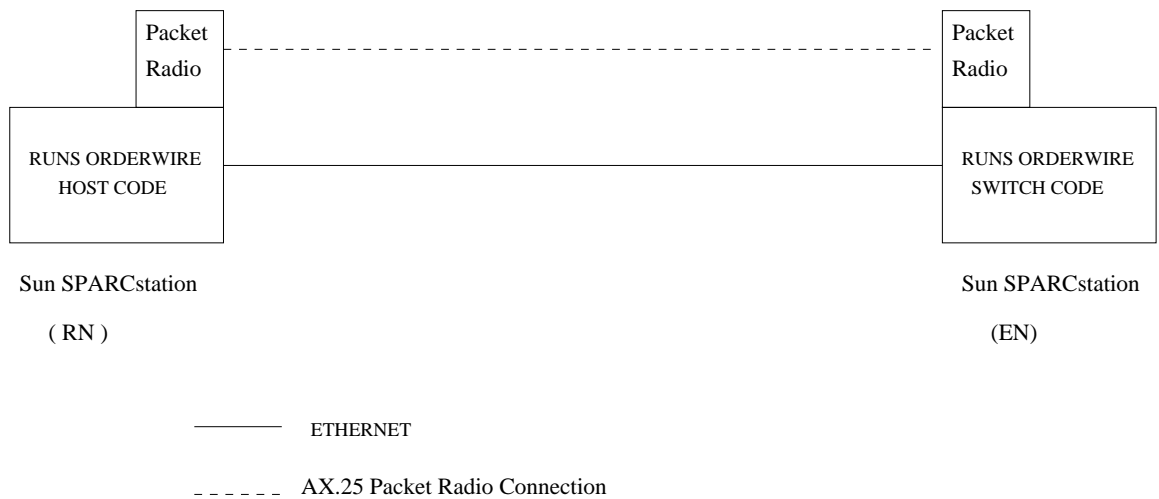}}
        \caption{Physical Setup for Packet Radio Timing.}
        \label{timepr}
\end{figure*}

Mobile-IP\index{Mobile-IP} configur\-ation results can be cached\index{Cache}
ahead of time. In the \acl{RDRN}
Project in particular, the foreign agent 
information is sent by the Network Configuration 
Protocol\index{RDRN!Network Control Protocol}
(NCP\index{NCP|see{Network Control Protocol}}). This results in
a savings of approximately 0.5 seconds as shown in Table \ref{timetab}
\cite{BushICC96}. These initial measurements were made on Sun Sparcs 
connected to Kantronics \acl{TNC}.
The experiments involved determining the time required to transmit and 
process each of the packet types listed in Table \ref{prototab} using the 
\acl{RDRN} packet radios. These times represent the time to packetize, transmit, 
receive, and depacketize each packet by the Network Control Protocol 
process. Figure \ref{timepr} illustrates the physical configuration used 
for the experiments involving the real packet radios. The results are 
presented in Table \ref{timetab}. 
Most of the overhead occurs during the initial system configuration which
occurs only once as long as \acl{ES}s remain stationary.
With regard to a handoff, the 473 millisecond time to transmit and process 
the handoff packet is on the same order of time as that required to compute 
the beam angles and steer the beams. The following sections provide an 
analysis and discuss the impact of scaling up the system on the configuration 
time.

\begin{table*}
\centering
\begin{tabular}{||l|c||}                    \hline
\textbf{Event}   & \textbf{Time (ms)}          \\ \hline \hline
USER\_POS     & 677                 \\ \hline
NEWSWITCH     & 439                 \\ \hline
HANDOFF       & 473                 \\ \hline
MYCALL        & 492                 \\ \hline
SWITCHPOS     & 679                 \\ \hline
TOPOLOGY      & 664                 \\ \hline
\end{tabular}
\caption{\label{timetab} Network Control Protocol Timing Results.}
\end{table*}

\subsection{Mobile \acl{PNNI}\index{PNNI}}

\begin{figure*}[htbp]
        \centerline{\psfig{file=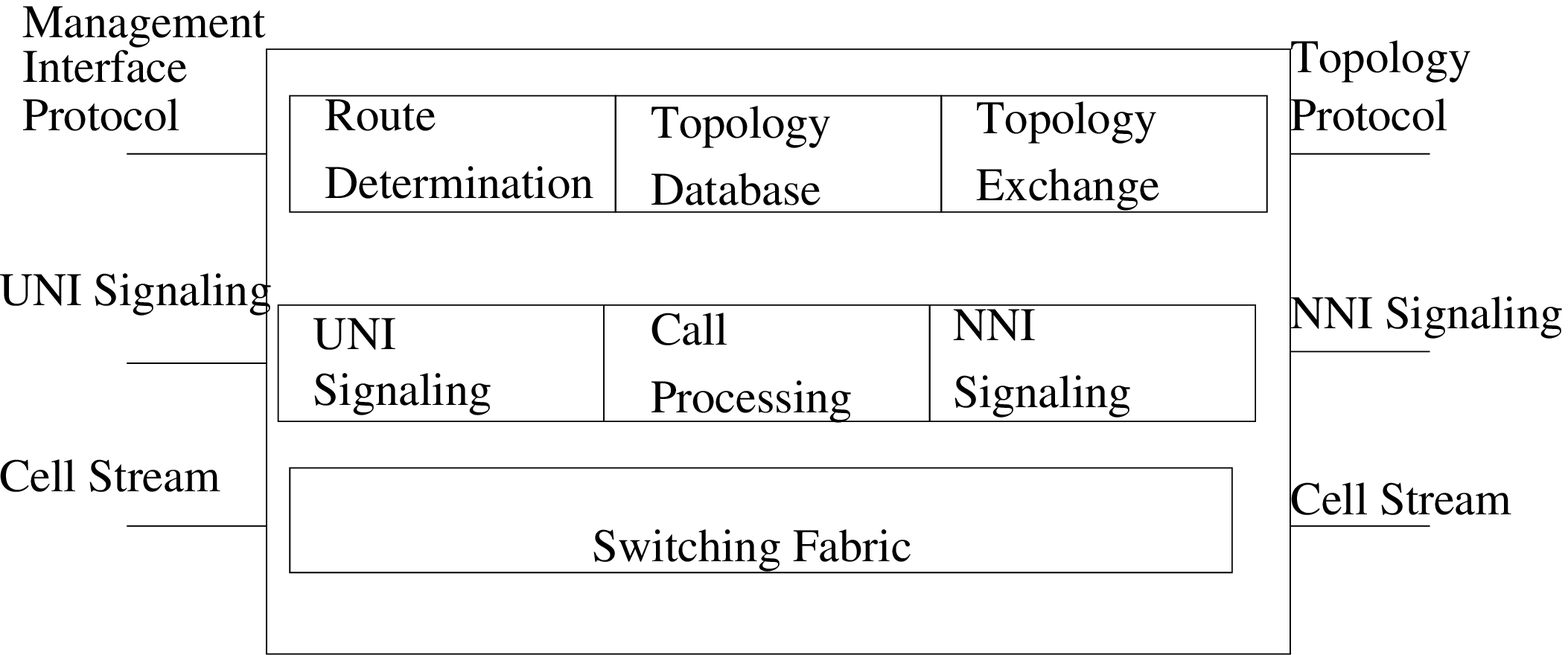,width=6.0in}}
        \caption{\acl{PNNI} Architecture Reference Model.}
        \label{pnniov}
\end{figure*}

\acl{VNC} can be used as a basis for a mobile Private Network-Network 
Interface (\acl{PNNI}) \cite{PNNI} implementation
with minimal changes to the evolving fixed network \acl{PNNI} standard.
Figure \ref{pnniov} shows a high level view of the \acl{PNNI} 
Architecture. Terminology used in the \acl{PNNI} Specification \cite{PNNI} is used 
here.

\begin{figure}[htbp]
        \centerline{\psfig{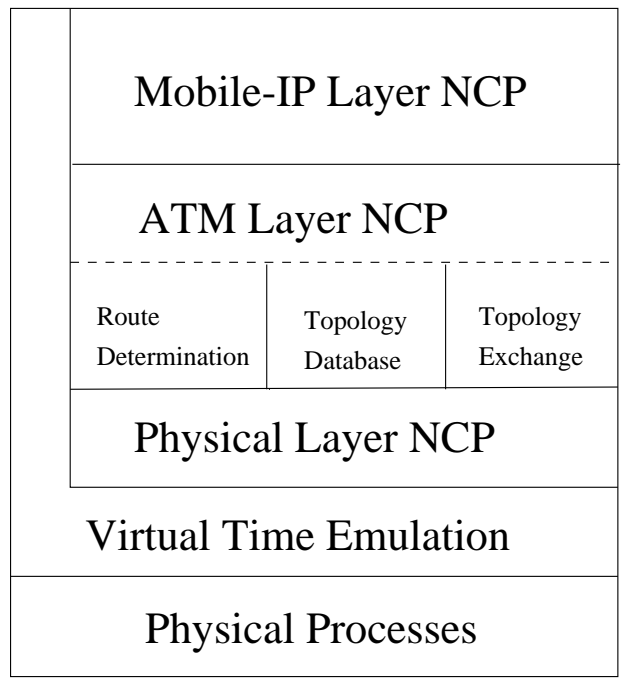}}
        \caption{Virtual Network Configuration Stack.}
        \label{vncstack}
\end{figure}

In this version of mobile \acl{PNNI}, the standard \acl{PNNI} route
determination, topology database, and topology exchange
would reside within the Network Control Protocol.
The NCP stack with \acl{VNC} is shown in Figure \ref{vncstack}.

The enabling mechanism for mobile \acl{PNNI} is the fact that \acl{VNC} will 
allow the NCP\index{Network Control Protocol}
to create a topology\index{PNNI!topology} which will exist after a hand-off 
occurs prior to the hand-off taking place.
This will cause \acl{PNNI} to perform its standard action of updating
its topology information immediately before the hand-off occurs.
Note that this is localized within a single \acl{PG}\index{Peer Group}.
A \acl{PG} in \acl{PNNI} is a set of \acl{LN}\index{Logical Node} 
grouped together for the purpose of creating a routing 
hierarchy\index{PNNI!routing hierarchy}. 
\acl{PNNI} topology messages are flooded among all Logical Nodes in a \acl{PG}. 
Note that an \acl{LN} may be
either a lowest level node\index{PNNI!lowest level node} or a Logical 
Group Node\index{Logical Group Node}. A Logical Group Node
may consist of many lowest level nodes treated together as a single node.
A Logical Group Node is represented by the \acl{PGL}\index{Peer Group Leader}.

The second enabling mechanism for mobile \acl{PNNI} is a change to the \acl{PNNI} 
signaling protocol. Standard \acl{PNNI} signaling\index{PNNI!signaling} is 
allowed to dynamically modify logical links
when triggered by a topology change. This will be similar to a
\textbf{CALL ABORT\index{CALL ABORT}} \acl{PNNI} signaling message
except that the ensuing \acl{PNNI} \textbf{RELEASE\index{RELEASE}} signaling 
messages will be contained within the scope of the Peer Group. This will
be called a \textbf{SCOPED CALL ABORT\index{SCOPED CALL ABORT}}.
When the topology changes due to an end system hand-off, a check will 
be made to determine which 
end system has changed Logical Nodes\index{Logical Node} \acl{LN}. 
An attempt is made to establish the same incoming \acl{VC}s 
at the new \acl{LN} as were at the original \acl{LN} and connections
are established from the new \acl{LN} to the original border 
\acl{LN}s\index{PNNI!border LN} of the \acl{PG}. 
A border node\index{PNNI!border node} in \acl{PNNI} has at least one link 
which crosses a Peer Group boundary.
This allows the end system to continue transmitting with the same \acl{VCI} as 
the hand-off occurs. 
The connections from the original \acl{LN}s to the border \acl{LN}s are released after 
the hand-off occurs. 
If the new \acl{LN} is already using a \acl{VCI} that was used at the original \acl{LN},
the HANDOFF packet will contain the replacement \acl{VCI}s to be used by
the end system.
There are now two branches of a logical link tree established  with
the border \acl{LN} as the root. After the hand-off takes place the old
branch is removed by the \textbf{SCOPED CALL ABORT}.

Note that link changes are localized to a single \acl{PG}.
The fact that
changes can be localized to a \acl{PG} greatly reduces the impact on
the network and implies that the mobile network should have many
levels in its \acl{PNNI} hierarchy\index{PNNI!hierarchy}.
In order to maintain cell order the new path within the \acl{PG} is chosen
so as to be equal to or longer than the original path based on
implementation dependent metrics.

\begin{figure*}[htbp]
        \centerline{\psfig{file=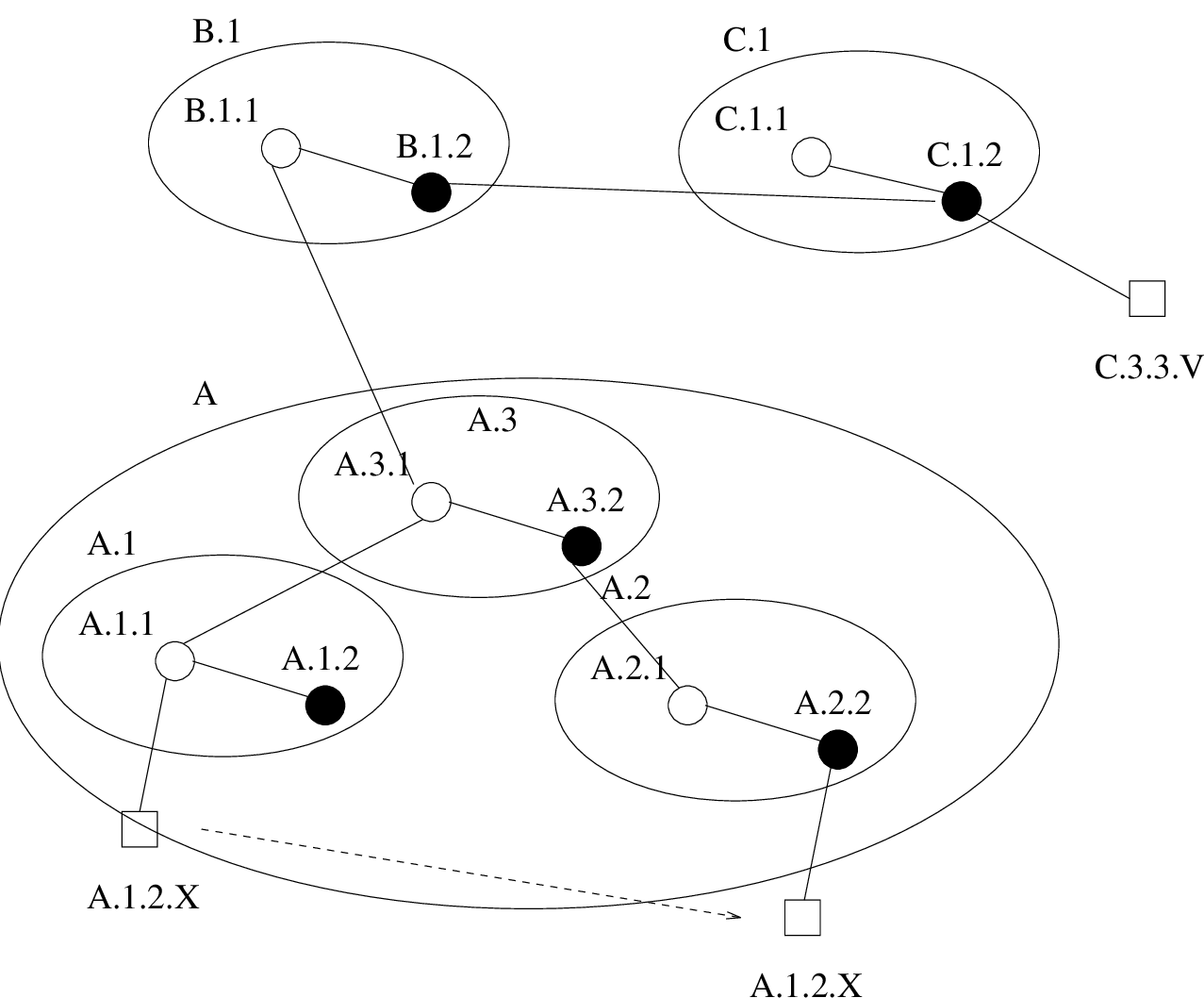,width=5.75in}}
        \caption{Mobile \acl{PNNI} Example.}
        \label{mobpnni}
\end{figure*}

An an example, consider the network shown in Figure \ref{mobpnni}. 
Peer Groups are enclosed in ellipses and the blackened nodes
represent the lowest level Peer Group Leader for each Peer Group.
End system A.1.2.X is about to hand-off from A.1.1 to A.2.2. The smallest 
scope which encompasses the old and new \acl{LN} is the \acl{LN} A.
A.3.1 is the outgoing border node for \acl{LN} A. A \textbf{CALL SETUP} uses
normal \acl{PNNI} operations to setup a logical link from A.3.1 to A.2.2.
After A.1.2.X hands off, a \textbf{SCOPED CALL ABORT} releases the 
logical link from A.3.1 to A.1.1.

\subsection{Rapid \acl{VC} Setup with Crankback\index{Crankback}}

An interesting feature of \acl{PNNI}\index{PNNI} involves 
crankback\index{Crankback}. 
Crankback is a mechanism for partially releasing a connection setup in
progress which has encountered a failure thus allowing the use of alternate
routing.
This is inherently built-in as a \acl{VNC} rollback\index{Rollback} within the
call setup process. The node in the route which 
detects the link failure will roll back to the time immediately prior to 
establishing the next hop in the route and as part of the rollback
mechanism, causes the predecessor node in the route to rollback and
attempt a different route.

\subsection{Rapid Dynamic Host Configuration}

The \acl{DHCP}\index{Dynamic Host Configuration Protocol} \cite{RFC1541} provides
for automatic configuration of a host from data stored on the network. \acl{DHCP} 
runs over the \acl{ATM}\index{ATM} link and thus must wait for the link to 
be established. A \acl{VNC} enabled \acl{DHCP} can prepare the contents of the host 
configuration messages ahead of time.

\subsection{Eliminate Unnecessary TCP\index{TCP-IP} Bandwidth Loss}

Knowing that a handoff\index{Handoff} is about to occur, source traffic
can be stopped using the \acl{ATM} \acl{ABR}\index{Available Bit Rate}
mechanism before the handoff\index{Handoff} and restarted afterwards.
If the state of the TCP\index{TCP-IP!timers} timers are preserved,
the Van Jacobson\index{Van Jacobson} TCP\index{TCP-IP} recovery mechanism
\cite{Jacobson}, which unnecessarily reduces bandwidth during a
handoff\index{Handoff}, can be avoided.

\subsection{Rapid Topology\index{ATM!topology} Update for Mobile Base Stations}

Although base stations which contain \acl{ATM}\index{ATM} switches 
generally remain stationary, there is no reason these base stations 
cannot be mobile. Configuration in this kind of environment is more 
challenging than a strictly mobile host environment as demonstrated
by the emulation results presented in Section \ref{emres} Figure
\ref{mobestime}. 
\acl{VNC}\index{VNC} provides a promising solution to this problem by
allowing the \acl{ES} topology to be precomputed and available for
use as soon as the \acl{ES} changes position.
\subsection{Proactive Network Management}

The trend in network management is to develop proactive network
management systems. These are management systems which ultimately
will notify network administration of potential faults
before they occur \cite{Schiaffino}. The interaction between such
a predictive network management system and a predictive mobile
network is an interesting case to consider. 

Imagine a network management station which uses \acl{MIB}\index{MIB} information
from either \acl{SNMP}\index{SNMP} or \acl{CMIP}\index{CMIP} to construct a 
model\index{Network Management!model} of the network which
can use provide alarms for both current and future events.
This is an intriguing application of \acl{VNC}\index{VNC} which is
simulated and discussed in Section \ref{vncsim}.
A network management station 
polls \cite{Takagi86} \cite{Steinber91} the managed objects for 
their current status which is a
current problem in network management. Polling too frequently
wastes bandwidth and degrades the performance of the system being
managed. \acl{VNC}\index{VNC} offers a solution because the predicted
events can be used as a guide to control the rate of polling.

Management entities rely upon standard mechanisms to obtain the state of
their managed entities in real time\index{Real Time}. These mechanisms 
use both solicited and unsolicited methods. The unsolicited method uses
messages sent from a managed entity to the manager. These messages are
called traps or notifications; the former are not acknowledged while
the latter are acknowledged. These messages are very similar to messages
used in distributed\index{Distributed Simulation} simulation algorithms; 
they contain a timestamp and
a value, they are sent to a particular destination, i.e. a management
entity, and they are the result of an event which has occurred.
 
Information requested by the management system from a particular managed
entity is solicited information. It also corresponds to messages
in distributed simulation\index{Distributed Simulation}. 
It provides a time and a value; however,
not all such messages are equivalent to messages in distributed 
simulation\index{Distributed Simulation}. These messages provide the management station 
with the current state of the managed entity, even though no event or change of
state may have occurred, or more than one event may have occurred. Designing
a manager which knows when best to request information is part of our goal.
 
We will assume for simplicity that each managed entity is represented by
a Logical Process (\acl{LP}\index{LP}). It would greatly 
facilitate system management if vendors provide not only the standards 
based \acl{SNMP}\index{SNMP} Management Information 
Base\index{Management Information Base} (\acl{MIB}), as they do now, 
but also a standard simulation code which models\index{Model} the entity 
or application behavior and can be plugged into the management system 
just as in the case with a \acl{MIB}\index{MIB|see{Management Information Base}}.

There are several parameters in this application which must
be determined. The first is how often this application should check 
the \acl{LP}\index{LP} to verify that past results match reality. This phase is 
called \textbf{Verification Query\index{Verification Query}} ($\Upsilon$).  
The optimum 
choice of verification query interval \index{Verification Query!verification 
query time} ($T_{query}$) is important because querying of entities 
should be minimized while still guaranteeing 
accuracy is maintained within some predefined tolerance, $\Theta$.
This tolerance could be set for each state variable or message value
sent from an \acl{LP}\index{LP}. 
 
The amount of time into the future which the emulation will attempt
to venture is another parameter which must be determined. This
lookahead sliding window of width $(t..\Lambda)$\index{$\Lambda$} where
$t$ is the current time, should be preconfigured
based on the accuracy required; the farther ahead the application 
attempts to go past real time\index{Real Time}, the more risk that 
is assumed.
 
\subsubsection{Optimum Choice of Verification Query Times}
 
One method of choosing the optimum verification query time
would be to query the entity based on the frequency of
the data monitored. Assuming the simulated data is
correct, query or sample in such a way as to perfectly reconstruct
the data, e.g. based on the maximum frequency component. A possible
drawback is that the actual data may be changing at a multiple
of the predicted value. The samples may appear to to be accurate when
they are invalid.
Some interesting points about this are that the error throughout the
simulated system may be randomized in such a way that errors among 
\acl{LP}s\index{LP} cancel. However, if the simulation is composed of many of 
the same class of \acl{LP}\index{LP}, the errors may compound rather than cancel
each other. See Section \ref{vncsim} on page \pageref{vncsim} for network
management simulation results.

\subsubsection{Predictive System Interaction}

There is an interesting interaction between the predictive management
system and the predictive mobile network. A predictive mobile network
such as the \acl{VNC} proposal for \acl{RDRN} \cite{BushICC96} will have results
cached in advance of use for many configuration parameters. 
These results should be part of the \acl{MIB}
for the mobile network and should include the predicted time of the
event which requires the result, the value of the result, and the
probability that the result will be correct at that time. Thus there
will be a triple associated with each predicted event: 
{\em (time, value, probability)}. Network management protocols, e.g. \acl{SNMP}
\cite{SNMP} and \acl{CMIP} \cite{CMIP}, include the time as part of 
the \acl{PDU}, however this time indicates only the real time the poll occurred.

A predictive management system could simply use \acl{LP}s to represent the
predictive mobile processes as previously described, however, this is
redundant since the mobile network itself has predicted events in
advance as part of its own management and control system. Therefore,
managing a predictive mobile network with a predictive network
management system provides an interesting problem in trying
to get the maximum benefit from both of these predictive systems.

Combining the two predictive systems in a low level manner, e.g. allowing 
the \acl{LP}s to exchange messages with each other, raises questions about 
synchronization between the mobile network and the management station. 
However, the predicted
mobile network results can be used as additional information to
refine the management system results. The management system will have
computed {\em (time, value, probability)} triples for each predicted result
as well. The final result by the management system would then be
an average of both triples weighted by the probability. An
additional weight may be added given the quality of either system. For
example the network management system might be weighted higher because
it has more knowledge about the entire network. Alternatively, the mobile 
network system may be weighted higher because the mobile
system may have better predictive capability for the detailed
events concerning handoff. Thus the two systems do not directly interact
with each other, but the final result is a combination of the results from
both predictive systems. A more complex method of combining results from these
two systems would involve a causal network such as the one described in 
\cite{Lehmann}.
Network management data freshness and other issues critical in a mobile 
environment are examined in \cite{Wu90hard} and \cite{Wu90consis}.
\chapter{VNC Algorithm Description}
\section{The Origin of Virtual Network Configuration}
\label{vncorg}

Recently proposed mobile networking architectures and protocols
involve predictive mobility management\index{Predictive Mobility Management}
schemes.  For example,	
an optimization\index{Optimization} to a\index{Mobile-IP} Mobile IP-like 
protocol using\index{Daedalus!IP-Multicast} IP-Multicast is
described in \cite{Seshan}. Hand-offs are anticipated and data is
multicast\index{Multicast} to nodes within the 
neighborhood\index{Daedalus!neighborhood} of the predicted 
handoff\index{Daedalus!handoff}.
These nodes intelligently buffer the data so that no matter where
the mobile host (MH) re-associates after a handoff\index{Daedalus!handoff}, 
no data will be lost.
Another example \cite{Liu}
\cite{Liuphd} proposes deploying mobile floating agents\index{Agents!floating}
which decouple services and resources\index{Resource} from the underlying 
network. These agents would be pre-assigned and pre-connected to predicted 
user locations.

One of the major contributions of this research is to recognize and
define an entirely new branch of the Time Warp Family Tree of algorithms.
\acl{VNC} integrates real and virtual time at a fundamental level allowing
processes to execute ahead in time. The \acl{VNC} 
algorithm must run in real-time, that is, with hard real-time constraints.
Consider the work leading towards the predictive \acl{VNC}\index{VNC} 
algorithm starting from 
a classic paper on synchronizing clocks in a distributed 
environment \cite{Lamport78}. A theorem from this paper 
limits the amount of parallelism in any 
distributed\index{Distributed Simulation} simulation algorithm:
\newtheorem{lamportr}{Rule}
\begin{lamportr}
If two events are scheduled for the same process, then the event with
the smaller timestamp must be executed before the one with the larger
timestamp.
\end{lamportr}
\begin{lamportr}
If an event executed at a process results in the scheduling of another
event at a different process, then the former must be executed before the
latter.
\end{lamportr}

A parallel simulation\index{Parallel Simulation} method, known as CMB (Chandy-Misra-Bryant),
which predates 
Time Warp\index{Time Warp} \cite{Jefferson82} is described in
\cite{Chandy79}. CMB is a conservative\index{Conservative}
algorithm which uses Null Messages to preserve message order and avoid 
deadlock.  Another method developed by the same author does not require 
Null Message\index{Null Message} overhead\index{Overhead},
but includes a central controller to maintain consistency and detect and
break deadlock\index{Deadlock}. There has been much research towards 
finding a faster algorithm and many algorithms claiming to be faster have 
compared themselves against the CMB method.

The basic Time Warp\index{Time Warp} Algorithm \cite{Jefferson82} was a major advance in 
distributed simulation\index{Distributed Simulation}. 
Time Warp is an algorithm used to speedup \acl{PDES} by taking advantage of 
parallelism among multiple processors. It is an optimistic method because
all messages are assumed to arrive in order and are processed as soon as
possible. If a message arrives out-of-order at an \acl{LP}, the \acl{LP} 
rollbacks back to a state which was saved prior to the arrival of the
out-of-order message. Rollback occurs by sending copies of all previously
generated messages as anti-messages. Anti-messages are exact copies of
the original message, except and anti-bit is set within the field of the message.
When the anti-message and real message meet, both messages are removed.
Thus, the rollback cancels the effects of out-of-order messages.
The rollback mechanism is a key part of \acl{VNC}, and algorithms which
improve Time Warp and rollback will also improve \acl{VNC}. 
There continues to be an explosion of new ideas and protocols for improving 
Time Warp\index{Time Warp}. An advantage to using a Time Warp\index{Time Warp} 
based algorithm is the ability to leverage future 
optimizations\index{Optimization}.
There have been many variations and improvements to this basic algorithm for
parallel simulation\index{Parallel Simulation}.
A collection of optimizations\index{Optimization} to 
Time Warp\index{Time Warp} is provided in \cite{Fujimoto}. 
The technical report describing Time Warp\index{Time Warp}
\cite{Jefferson82} does not solve the problem of determining Global Virtual 
Time\index{Global Virtual Time}, however an efficient algorithm for the 
determination of 
Global Virtual Time\index{Global Virtual Time} \index{Time Warp} is 
presented in \cite{Lazowaska90}. This algorithm does
not require message acknowledgments thus increasing the performance,
yet the algorithm works with unreliable communication links.

An analytical comparison of CMB\index{Chandy-Misra} and 
Time Warp\index{Time Warp} is the focus of
\cite{Lin90}. In this paper the comparison is done for the simplified
case of feed-forward and feedback networks. Conditions are developed
for Time Warp\index{Time Warp} to be conservative optimal. 
Conservative optimal\index{Conservative Optimal} means
that the time to complete a simulation is less than or equal to the
critical path \cite{Berry85}\index{Critical Path} through the 
event-precedence graph\index{Event-Precedence Graph} of a simulation.

A search for the upper bound of the performance of Time Warp\index{Time Warp} 
versus synchronous distributed\index{Distributed processing!synchronous} 
processing methods is presented 
in \cite{Felderman90}. Both methods are analyzed in a feed-forward network
with exponential processing times for each task. The analysis in
\cite{Felderman90} assumes that no Time Warp 
optimizations\index{Time Warp!optimization} 
are used. The result is that Time Warp\index{Time Warp} has an 
expected potential speedup\index{Time Warp!potential speedup} of no more 
than the natural logarithm of $P$ over the synchronous\index{Synchronous} 
method where $P$ is the number of processors.

A Markov Chain analysis model of Time Warp is given in \cite{Gupta91}.
This analysis uses standard exponential simplifying assumptions to 
obtain closed form results for performance measures such as fraction of 
processed events that commit, speedup, rollback recovery, expected length 
of rollback, probability mass function for the number of uncommitted
processed events, probability distribution function of the local
virtual time of a process, and the fraction of time the processors
remain idle. Although the analysis appears to be the most comprehensive
analysis to date, it has many simplifying assumptions such as no
communications delay, unbounded buffers, constant message population,
message destinations are uniformly distributed, and rollback takes no 
time. Thus, the analysis in \cite{Gupta91} is not directly applicable 
to the time sensitive nature of \acl{VNC}.

Further proof that Time Warp\index{Time Warp} out-performs 
CMB\index{Chandy-Misra} is provided in
\cite{Lipton90}. This is done by showing that there exists a simulation
model\index{Simulation!model} which out-performs 
CMB\index{Chandy-Misra} by exactly the number of processors used, 
but that no such 
model\index{Simulation!model} in which CMB\index{Chandy-Misra} 
out-performs Time Warp\index{Time Warp} by a factor of the number of 
processors used exists.

A detailed comparison of the CMB\index{Chandy-Misra} and 
Time Warp\index{Time Warp} methods are presented in \cite{Lin90a}. It is shown 
that Time Warp\index{Time Warp} 
out-performs conservative\index{Parallel Simulation!conservative} methods 
under most conditions. Improvements to Time Warp\index{Time Warp}
are suggested by reducing the overhead\index{Overhead} of state saving 
information and the introduction of a global virtual 
time\index{Global Virtual Time} calculation.
Simulation study results of Time Warp\index{Time Warp} are presented in 
\cite{Turnbull92}.
Various parameters such as communication delay\index{Delay!communications}, 
process delay\index{Delay!processing}, and
process topology\index{Topology} are varied and conditions under which 
Time Warp\index{Time Warp} and CMB\index{Chandy-Misra} perform 
best are determined.

\begin{figure*}[htbp]
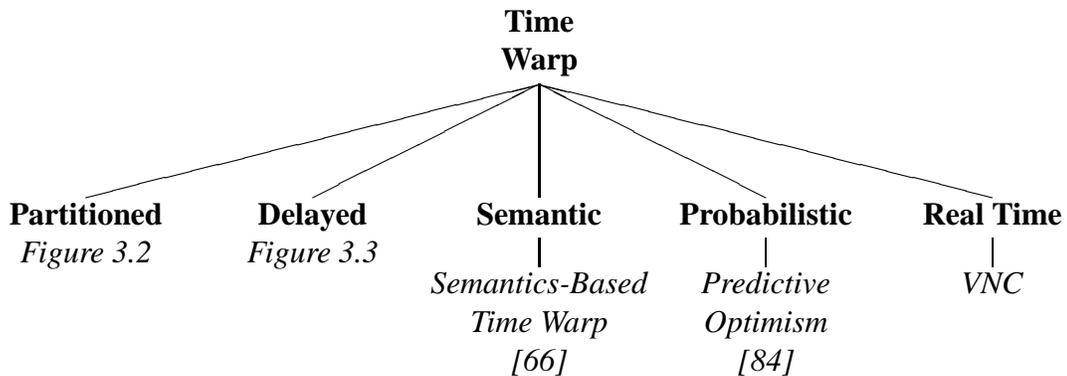

\leaf{\textbf{Partitioned} \\ \emph{Figure \ref{twpart}}}
\leaf{\textbf{Delayed} \\ \emph{Figure \ref{twdelay}}}
\leaf{\emph{Semantics-Based}\\ \emph{Time Warp} \\ \emph{\cite{Leong}}}
\branch{1}{\textbf{Semantic}}
\leaf{\emph{Predictive}\\ \emph{Optimism} \\ \emph{\cite{Nobl95}}}
\branch{1}{\textbf{Probabilistic}}
\leaf{\emph{VNC}}
\branch{1}{\textbf{Real Time}}
\branch{5}{\textbf{Time}\\ \textbf{Warp}}
\begin{center}
\tree
\end{center}
\caption{\label{twfam} Time Warp Family of Algorithms.}
\end{figure*}

The major contribution of this section is to recognize and             
define an entirely new branch of the Time Warp Family Tree of algorithms,
shown in Figure \ref{twfam}, which integrates real and virtual time at a 
fundamental level. The \acl{VNC} algorithm must run in real-time, that is, 
with hard real-time constraints.
Real-time constraints for a time warp simulation system are discussed
in \cite{Ghosh93}. The focus in \cite{Ghosh93} is the $R$-Schedulability
of events in Time Warp. Each event is assigned a real-time deadline 
($d_{E_{i,T}}$) for its execution in the simulation. $R$-Schedulability 
means that there exists a finite value ($R$) such that if each event's 
execution time is increased by $R$, the event can still be completed 
before its deadline.
The first theorem from \cite{Ghosh93} is that if there is no constraint on 
the number of such false events that may be created between any two successive
true events on an \acl{LP}, Time Warp cannot guarantee that a set of         
$R$-schedulable events can be processed without violating deadlines for
any finite $R$. 
There has been a rapidly expanding family of Time Warp 
algorithms focused on constraining the number of false events which are
discussed next.

\begin{figure*}[htbp]
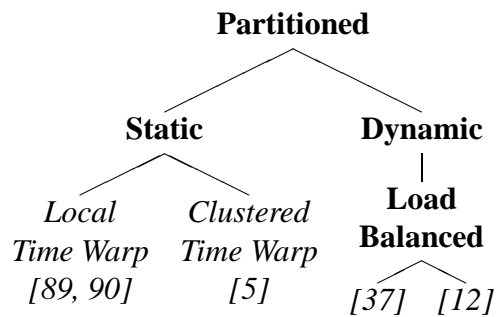

\leaf{\emph{Local}\\ \emph{Time Warp} \\ \emph{\cite{Raja93, Rajaei93}}}
\leaf{\emph{Clustered}\\ \emph{Time Warp} \\ \emph{\cite{Avril95}}}
\branch{2}{\textbf{Static}}
\leaf{\emph{\cite{Glaz93b}}}
\leaf{\emph{\cite{Boukerche94}}}
\branch{2}{\textbf{Load}\\ \textbf{Balanced}}
\branch{1}{\textbf{Dynamic}}
\branch{2}{\textbf{Partitioned}}
\begin{center}
\tree
\end{center}
\caption{\label{twpart}Partitioned Algorithms.}
\end{figure*}

Another contribution of this chapter is to
classify these algorithms as shown in Figures \ref{twfam}, \ref{twpart},
\ref{twdelay} and Table \ref{twfamtab}. 
Each new modification to the Time Warp mechanism attempts to improve 
performance by reducing the expected number of rollbacks. Partitioning 
methods attempt to divide tasks
into logical processes such that the inter-\acl{LP} communication is minimized.
Also included under partitioning are methods which dynamically move
\acl{LP}es from one processor to another in order to minimize load and/or
inter-\acl{LP} traffic. Delay methods attempt to introduce a minimal amount of
wait into \acl{LP}es such that the increased synchronization and reduced number 
of rollbacks more than compensates for the added delay. Many of the delay 
algorithms use some type of windowing method to bound the difference
between the fastest and slowest processes. The bounded sphere class
of delay mechanisms attempt to calculate the maximum number of nodes
which may need to be rolled back because they have processed messages
out of order. For example,
$S\downarrow(i, B)$ in \cite{Luba89} is the set of nodes effected
by incoming messages from node $i$ in time $B$ while $S\uparrow(i, B)$
is the set of nodes effected by outgoing messages from node $i$ in
time $B$. The downward pointing arrow in $S\downarrow(i, B)$ indicates
incoming messages while the upward pointing arrow in $S\uparrow(i, B)$
indicates outgoing messages. Another approach to reducing
rollback is to use all available semantic information within messages.  
For example, commutative sets of messages are messages which may be
processed out-of-order yet they produce the same result. Finally,
probabilistic methods attempt to predict certain characteristics of the
optimistic simulation, usually based on its immediate past history, and
take action to reduce rollback based on the predicted characteristic.
It is insightful to review a few of these algorithms because they not
only trace the development of Time Warp based algorithms but
because they illustrate the ``state of the art'' in preventing rollback, 
attempts at improving performance by constraining lookahead, partitioning
of \acl{LP}es into sequential and parallel environments, and the use of
semantic information. All of these techniques and more may be applied 
in the \acl{VNC} algorithm.

\begin{sidetable}
\begin{tabular}{lllll}
\textbf{Class} & \textbf{Sub Class} & \textbf{Sub Class} & \textbf{Description} & \textbf{Example} \\ \hline
\multicolumn{3}{l}{Probabilistic} & Predict next msg. arrival time & Predictive Optimism \cite{Leong} \\
\multicolumn{3}{l}{Semantic}      & Utilize msg contents to reduce rollback & Semantics Based Time Warp \cite{Leong} \\
\multicolumn{3}{l}{Partitioned}   & Inter-LP communication minimized & \\ \hline
& Dynamic         & & LPs change mode while executing & \\
& &  Load Balanced & LPs can migrate across hosts & \cite{Glaz93b}, \cite{Boukerche94} \\
& Static          & & LPs cannot change mode & Clustered Time Warp \cite{Avril95} \\ 
&                 & &  while executing             &
Local Time Warp \cite{Raja93, Rajaei93} \\ \hline
\multicolumn{3}{l}{Delayed}  & Delays added to reduce rollback & \\
& Windowed     & & Window based mechanisms reduce rollback & \\
& & Adaptive Window  & Window adapts to reduce rollback & Breathing Time Warp \cite{Stei93} \\
& & Fixed Window  & Window size does not adapt & Breathing Time Buckets \cite{Stei93} \\
& &               &                            & Moving Time Windows \cite{Madisetti87} \\ \hline
& Bounded Sphere & & Based on earliest time inter-LP & Bounded Lag \cite{Luba89b} \\
& &                  &  effects occur                      & WOLF \cite{Madisetti87, Soko90} \\ \hline
& Non-Windowed  & & A mechanism other than window to reduce & Adaptive Time Warp \cite{Ball} \\
&               & & rollback.                               & Near Perfect State Information \cite{Srini44} \\
\end{tabular}
\caption{\label{twfamtab} Time Warp Family of Algorithms.}
\end{sidetable}

\begin{sidefigure}
\leaf{\emph{WOLF} \\ \cite{Madisetti87, Soko90}}
\leaf{\emph{Bounded}\\ \emph{Lag} \\ \cite{Luba89b}}
\branch{2}{\textbf{Bounded} \\ \textbf{Sphere}}
\leaf{\emph{Breathing Time}\\ \emph{Buckets} \\ \cite{Stei93}}
\leaf{\emph{Moving Time}\\ \emph{Windows} \\ \cite{Madisetti87}}
\branch{2}{\textbf{Fixed}\\ \textbf{Window}}
\leaf{\emph{Breathing}\\ \emph{Time Warp} \\ \cite{Stei93}}
\branch{1}{\textbf{Adaptive}\\ \textbf{Window}}
\branch{2}{\textbf{Windowed}}
\leaf{\emph{Adaptive}\\ \emph{Time Warp} \\ \cite{Ball}}
\leaf{\emph{Near Perfect}\\ \emph{State Information} \\ \cite{Srini44}}
\branch{2}{\textbf{Non}\\ \textbf{Windowed}}
\branch{3}{\textbf{Delayed}}
\tree
\caption{\label{twdelay}Delaying Algorithms.}
\end{sidefigure}

The Bounded Lag algorithm \cite{Luba89b} for constraining rollback 
explicitly calculates, for each \acl{LP}, the earliest time that an event
from another \acl{LP} may affect the current \acl{LP}'s future. This calculation
is done by first determining the reachability sphere ($S\downarrow(i, B)$) 
which is the set of nodes which a message may reach in time $B$.
This depends on the minimum propagation delay of a message in simulation time
from node $i$ to node $j$ which is $d(i,j)$.
Once $S\downarrow(i, B)$ is known, the earliest time that node $i$
can be affected, $\alpha(i)$, is shown in Equation \ref{minaffect}
where $T(i)$ is the minimum message receive time in node $i$'s message 
receive queue. After processing all messages up to time $\alpha(i)$,
all \acl{LP}es must synchronize. 

\begin{figure*}
\begin{equation}
\alpha(i) = \min_{j \in S\downarrow(i, B) \land j \not= i}
            \{ d(j,i) + \min \{ T(j), d(i,j) + T(i) \} \}
\label{minaffect}
\end{equation}
\end{figure*}

The Bounded Lag algorithm is conservative because it 
synchronizes \acl{LP}es so that no message arrives out of order. The problem
is that a minimum $d(i,j)$ must be known and specified before the
simulation begins. A large $d(i,j)$ can negate any potential parallelism,
because a large $d(i,j)$ implies a large $\alpha(i)$ which implies
a longer time period between synchronizations. 
A filtered rollback extension to Bounded Lag is described in \cite{Luba89}.
Filtered Rollback allows $d(i,j)$ to be made arbitrarily small which
may possibly generate out of order messages. Thus the basic rollback
mechanism described in \cite{Jefferson82} is required. 

A thorough understanding of rollbacks and their containment are
essential for \acl{VNC}. In \cite{Luba89}, rollback cascades are
analyzed under the assumption that the Filtered Rollback mechanism is 
used. Rollback activity is viewed as a tree; a single rollback may
cause one or more rollbacks which branch-out indefinitely.
The analysis is based on a ``survival number'' of rollback
tree branches. The survival number is the difference between the
minimum propagation delay $d(j,i)$ and the delay in simulated
time for an event at node $i$ to effect the history at node $j$,
$t(i,j)$. Each generation of a rollback caused by an immediately
preceding node's rollback adds a positive or negative survival
number. These rollbacks can be thought of as a tree whose leaves
are rollbacks which have ``died out''. It is shown that it is possible
to calculate upper bounds, namely, infinite or finite number of 
nodes in the rollback tree.

A probabilistic method is described in \cite{Nobl95}. The concept
in \cite{Nobl95} is that optimistic simulation mechanisms are
making implicit predictions as to when the next message will arrive. 
A purely optimistic system assumes that if no message has arrived,
then no message \textbf{will} arrive and computation continues.
However, the immediate history of the simulation can be used to
attempt to predict when the next message will arrive. This information
can be used either for partitioning the location of the \acl{LP}es on
processors or for delaying computation when a message is expected
to arrive.

In \cite{McAffer90}, a foundation is laid for unifying conservative
and optimistic distributed simulation. Risk and aggressiveness are
parameters which are explicitly set by the simulation user. Aggressiveness
is the parameter controlling the amount of non-causality allowed
in order to gain parallelism and risk is the passing of such results
through the simulation system. Both aggressiveness and risk are
controlled via a windowing mechanism similar to the sliding
lookahead window of the \acl{VNC} algorithm.

A unified framework for conservative and optimistic simulation called
ADAPT is described in \cite{Jha94}. ADAPT allows the execution of a
``sub-model'' to dynamically change from a conservative to an optimistic 
simulation approach. This is accomplished by uniting conservative
and optimistic methods with the same Global Control Mechanism. The
mechanism in \cite{Jha94} has introduced a useful degree of flexibility
and described the mechanics for dynamically changing simulation approaches,
\cite{Jha94} does not quantify or discuss the optimal parameter settings
for each approach.

A hierarchical method of partitioning \acl{LP}es is described in 
\cite{Raja93, Rajaei93}. The salient feature of this algorithm 
is to partition \acl{LP}es into clusters. The \acl{LP}es operate as in
Time Warp. The individual clusters interact with each other
in a manner similar to \acl{LP}es. 

The \acl{CTW} is described in \cite{Avril95}. The \acl{CTW}
mechanism was developed concurrently but independently of \acl{VNC}.
This approach uses Time Warp between clusters of \acl{LP}es residing on
different processors and a sequential algorithm within clusters. This
is in some ways similar to the \acl{SLP} described later in \acl{VNC}. 
Since the
partitioning of the simulation system into clusters is a salient
feature of this algorithm, \acl{CTW} has been categorized
as a partitioned algorithm in Figure \ref{twpart}. One of the contributions
of \cite{Avril95} in \acl{CTW} is an attempt to efficiently
control a cluster of \acl{LP}es on a processor by means of the
\acl{CE}. The \acl{CE} allows the \acl{LP}es to behave as individual \acl{LP}es
as in the basic time warp algorithm or as a single collective \acl{LP}.
The \acl{CTW} algorithm is an optimization method for the \acl{VNC} \acl{SLP}es.

Semantics Based Time Warp is described in \cite{Leong}. In this algorithm,
the \acl{LP}es are viewed as abstract data type specifications. Messages
sent to an \acl{LP} are viewed as function call arguments and messages
received from \acl{LP}es are viewed as function return values. This allows
data type properties such as commutativity to be used to reduce rollback.
For example, if commutative messages arrive out-of-order, there is no
need for a rollback since the results will be the same. This is used in the
\textbf{MYCALL} Timeout phase of the \acl{NCP} which is transition labeled
I4 in Table \ref{ESFSM} on page \pageref{ESFSM}. The order in which
the initial \acl{ES} \textbf{MYCALL} messages are collected does not impact
the result of the configuration.

Another means of reducing rollback\index{Rollback}, in this case by     
decreasing the aggressiveness of Time Warp\index{Time Warp}, is given    
in \cite{Ball}. This scheme involves voluntarily suspending a        
processor whose rollback\index{Rollback} rate is too frequent because it 
is out-pacing its neighbors. \acl{VNC}\index{VNC} uses a fixed sliding 
window to control the rate of forward emulation progress, however, a 
mechanism based on those just mentioned could be investigated.

The \acl{NPSI} Adaptive Synchronization Algorithms for \acl{PDES} are 
discussed in \cite{Srini44} and \cite{Srini20}. The adaptive
algorithms use feedback from the simulation itself in order to adapt.
Some of the deeper implications of these types of
systems are discussed in Appendix \ref{philo}. The NPSI system requires
an overlay system to return feedback information to the \acl{LP}es.
The \acl{NPSI} Adaptive Synchronization Algorithm examines the system
state (or an approximation of the state) calculates an error potential
for future error, then translates the error potential into a value
which controls the amount of optimism.

Breathing Time Buckets described in \cite{stei92} is one of the simplest 
fixed window techniques. If there exists a minimum time interval between 
an each event and the earliest event generated by that event ($T$), then 
the system runs in time cycles of duration $T$. All \acl{LP}es synchronize 
after each cycle. The problem with this approach is that $T$ must exist 
and be known ahead of time. Also, $T$ should be large enough to allow a 
reasonable amount of parallelism, but not so large as to loose fidelity 
of the system results.

Breathing Time Warp (\cite{Stei93}) attempts to overcome the problems
with Breathing Time Buckets and Time Warp by combining the two 
mechanisms. The simulation mechanism operates in cycles which alternate
between a Time Warp phase and a Breathing Time Buckets phase. The 
reasoning for this mechanism is that messages close to \acl{GVT} are 
less likely to cause a rollback while messages with time-stamps far from 
\acl{GVT} are more likely to cause rollback. Breathing Time Warp also 
introduces the \emph{event horizon} which is the earliest time of the 
next new event generated in the current cycle. A user controlled
parameter controls the number of messages which are allowed to be
processed beyond \acl{GVT}. Once this number of messages is generated
in the Time Warp phase, the system switches to the Breathing Time
Buckets phase. This phase continues to process messages, but does
not send any new messages. Once the event horizon is crossed, processing
switches back to the Time Warp phase. One can picture the system
taking in a breath during the Time Warp phase and exhaling during the
Breathing Time Buckets phase.

An attempt to reduce roll-backs is presented in an algorithm 
called WOLF\index{WOLF} \cite{Madisetti87, Soko90}.
This method attempts to maintain a sphere of influence around each
rollback\index{WOLF!rollback} in order to limit its effects.

The Moving Time Window\index{Moving Time Window} \cite{Soko88, Soko90} 
simulation algorithm is an interesting alternative to 
Time Warp\index{Time Warp}. 
It controls the amount of aggressiveness in the system by means of a 
moving time window \acl{MTW}\index{MTW|see{Moving Time Windows}}. 
The trade-off in having no roll-backs in this algorithm is loss of 
fidelity\index{MTW!fidelity} in the simulation results. This could
be considered as another method for implementing the \acl{VNC} algorithm.

An adaptive simulation\index{Time Warp!adaptive simulation} application of 
Time Warp\index{Time Warp} is presented 
in \cite{Tinker}. The idea presented in this paper is to use 
Time Warp\index{Time Warp} to change the input parameters of a running 
simulation without having to restart the entire simulation. Also, it is 
suggested that events external to the simulation can be injected even 
after that event has been simulated.

Hybrid simulation and real system component models are discussed in
\cite{Bagrodia91}. The focus in \cite{Bagrodia91} is on \acl{PIPS}
Components of a 
performance specification for a distributed system are implemented
while the remainder of the system is simulated. More components are
implemented and tested with the simulated system in an iterative
manner until the entire distributed system is implemented. The
\acl{PIPS} system described in \cite{Bagrodia91} discusses using MAY or
Maisie as a tool to accomplish the task, but does not explicitly discuss
Time Warp.


The work in \cite{Ghosh93} provides some results relevant to 
\acl{VNC}. It is theorized that if a set of events
is $R$-schedulable in a conservative simulation, where $R \ge
\rho + c t + \sigma$ where $\rho$ is the time to restore an \acl{LP} state,
$c$ is the number of \acl{LP}es, $t$ is the time the simulation has been 
running, and $\sigma$ is the real time required to save an \acl{LP} state, 
then it can run to completion without 
missing any deadline by an \acl{NFT} Time Warp strategy 
with lazy cancellation.
\acl{NFT} Time Warp assumes that if an incorrect computation produces and 
incorrect event ($E_{i,T}$), then it must be the case that the correct
computation also produces an event ($E_{i,T}$) with the same timestamp
\footnote{This simplification makes the analysis in \cite{Ghosh93} 
tractable. This assumption also greatly simplifies the analysis
of \acl{VNC}. The \acl{VNC} algorithm is simplified because the state 
verification component of \acl{VNC} requires that saved states be 
compared with the real-time state of the process. This is done easily 
under the assumption that the $T$ (timestamp) values of the two events 
$E_{i,T_v}$ and $E_{i,T_r}$ are the same.}. This result shows that 
conditions exist in a Time Warp algorithm which guarantee that events 
are able meet a given deadline. This is encouraging for the \acl{VNC} 
algorithm since clearly events must be completed before real-time reaches 
the predicted time of the event for the cached results to be useful in 
\acl{VNC}.
Finally, this author has not been the only one to consider the use of
Time Warp\index{Time Warp} to speedup a real-time
process\index{Real Time!process}. In \cite{Tennenhouse}, the idea of
temporal decoupling is applied to a signal processing environment.
Differences in granularity of the rate of execution are utilized
to cache results before needed and to allocate resources\index{Resource}
more effectively. 

This section has shown the results of research into improving Time Warp,
especially in reducing rollback, as well as the limited results in applying 
Time Warp to real time systems. Improvements to Time Warp and the application
to real time systems are both directly applicable to \acl{VNC}.
Now consider the Virtual Network Algorithm in more detail.
\section{Virtual Network Configuration Description}

New terminology is introduced in the description of the 
\acl{VNC}\index{VNC} algorithm which follows and
terminology borrowed from previous distributed simulation algorithm 
descriptions has a slightly different meaning in \acl{VNC}\index{VNC}, 
thus it is important that terms be precisely defined.

The \acl{VNC}\index{VNC} algorithm encapsulates each \acl{PP}\index{PP} 
within an \acl{LP}\index{LP}. 
A \acl{PP}\index{PP} is nothing more than an executing task defined by 
program code. An example of a \acl{PP} is the \acl{RDRN} beam table computation
task. The beam table computation task generates a table of complex
weights which controls the angle of the radio beams based
on position input.
An \acl{LP}\index{LP} consists of the \acl{PP}\index{PP} and additional 
data structures and instructions which maintain message 
order and correct operation as the system executes ahead of real 
time\index{VNC!real time}. As an example, the beam table computation
\acl{PP} is encapsulated in an \acl{LP} which maintains generated beam
tables in its State Queue and handles rollback due to 
out-of-order input messages or out-of-tolerance real messages as
explained later. An \acl{LP}\index{LP} contains a 
\acl{QR}\index{QR|see{Receive Queue}}, 
Send Queue\index{Send Queue} (QS\index{QS|see{Send Queue}}), and 
State Queue\index{State Queue} (SQ\index{SQ|see{State Queue}}). 
The \acl{QR} maintains newly arriving messages in order by their Receive 
Time\index{Receive Time} (TR\index{TR|see{Receive Time}}). 
The QS maintains copies of previously sent messages in order of their send 
times. The state of an \acl{LP}\index{LP} is
periodically saved in the SQ. The \acl{LP}\index{LP} also contains its notion 
of time known as Local Virtual Time\index{Local Virtual Time}
(LVT\index{LVT|see{Local Virtual Time}}) and a 
Tolerance\index{Tolerance} ($\Theta$\index{$\Theta$|see{Tolerance}}).
Tolerance is the allowable deviation between actual and predicted
values of incoming messages. For example, when a real message enters the 
beam table computation \acl{LP} the position in the message value is compared
with the position which had been cached in the State Queue of the
\acl{LP}. If these values are out of tolerance, then corrective action
is taken in the form of a rollback as explained later.
Also, the Current State\index{Current State} 
(CS\index{CS|see{Current State}}) of an \acl{LP}\index{LP} is
the current state of the structures and \acl{PP} encapsulated within an
\acl{LP}\index{LP}.

The \acl{VNC}\index{VNC} system contains a notion of the complete system time 
known as Global Virtual Time\index{Global Virtual Time} 
(GVT\index{GVT|see{Global Virtual Time}}) and a sliding window of length 
Look Ahead\index{Look Ahead} time 
($\Lambda$\index{$\Lambda$|see{Look Ahead}}). There have been several 
proposals for efficient determination of \acl{GVT}, for example 
\cite{Lazowaska90}.
The \acl{GVT} algorithm in \cite{Lazowaska90} allows \acl{GVT} to be determined 
in a message-passing environment as opposed to the easier case of a shared 
memory environment. \acl{VNC} allows only message passing communication among
\acl{LP}es. The algorithm in \cite{Lazowaska90}  also allows normal processing 
to continue during the \acl{GVT} determination phase. \acl{GVT} is required
only for the purpose of throttling forward prediction in 
\acl{VNC}, that is, it governs how far into the future the system predicts. An 
alternative method for throttling the \acl{VNC} system makes use of the 
accurate time provided by the \acl{GPS} receiver and is discussed later in 
the analysis and implementation chapters.

\begin{figure*}[htbp]
        \centerline{\psfig{file=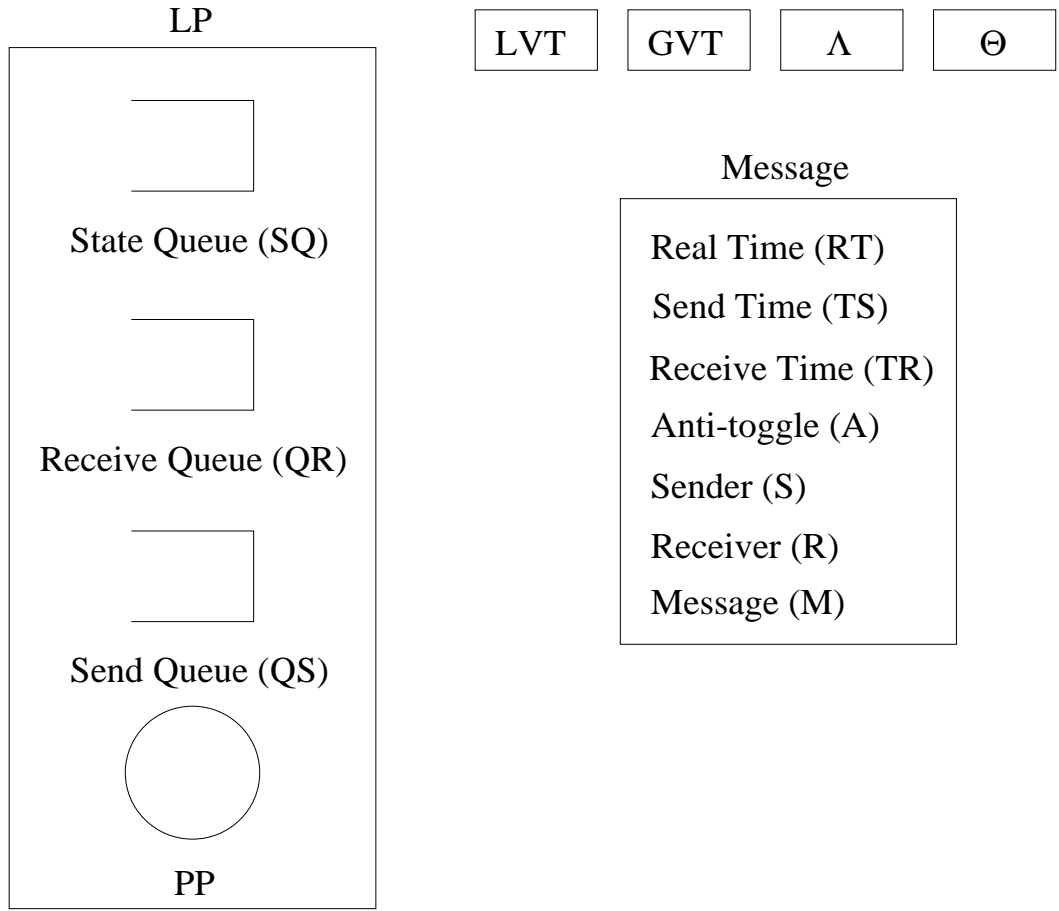,width=5.5in}}
        \caption{The VNC Logical Process and Message.}
        \label{vncspec}                                             
\end{figure*}

\acl{VNC} messages contain the Send Time\index{Send Time} 
(TS\index{TS|see{Send Time}}), 
Receive Time\index{Receive Time} (TR\index{TR|see{Receive Time}}), 
Anti-toggle\index{Anti-toggle} 
(A\index{A|see{Anti-toggle}}) and the actual message value itself (M). The 
TR is the time this message is predicted be valid at
the destination \acl{LP}\index{LP}. It is not the link transfer 
time from source to destination \acl{LP}.
The TS is the time this message was sent by the
originating \acl{LP}\index{LP}. The ``A'' field is the anti-toggle and is 
used for creating an anti-message\index{Anti-Message} to remove the effects
of false messages\index{False Message} as described
later. A message also contains a field for current 
Real Time\index{Real Time}
(RT\index{RT|see{Real Time}}). This is used to differentiate a real 
message\index{Real Message} from a virtual message\index{Virtual Message}.
A message which is generated and time-stamped with the current time is
called a real message\index{Real Message}. Messages which 
contain future event information and are time-stamped with a time greater 
than current time are called virtual messages\index{Virtual Message}. 
If a message arrives at an \acl{LP}\index{LP} out of order or with invalid 
information, it is called a false message\index{False Message}. 
A false message\index{False Message} will cause an 
\acl{LP}\index{LP} to rollback\index{Rollback}. The \acl{LP} structures and 
message fields are shown in Table \ref{lpspectab}, Table \ref{messpectab}
and in Figure \ref{vncspec}.

The \acl{VNC} algorithm requires a driving process\index{VNC!driving process}
to predict future events and inject them into the system. The driving 
process acts a source of virtual messages\index{VNC!virtual messages} 
for the \acl{VNC} system. All 
other processes react to virtual messages. For example, the 
\acl{GPS}\index{Global Positioning System}
receiver process runs in real-time\index{Real Time} providing
current time and location information
and has been  modified to inject future predicted time and location 
messages as well. 

A rollback is triggered either by messages arriving out of order at
the \acl{QR} of an \acl{LP} or by a predicted value previously
computed by this \acl{LP} which is beyond the allowable tolerance.
In either case, rollback\index{VNC!rollback} is a mechanism by 
which an \acl{LP}\index{LP} returns to a known correct state.
The rollback\index{Rollback} occurs in three phases. In the first phase, the 
\acl{LP}\index{LP} state
is restored to a time strictly earlier than the \acl{TR} of the
false message\index{False Message}. In the second phase, anti-messages are 
sent to cancel the effects of any invalid messages which had been generated before the
arrival of the false message\index{False Message}. An 
anti-message\index{Anti-Message} contains exactly the
same contents as the original message with the exception of an anti-toggle
bit\index{Anti-toggle!anti-toggle bit} which is set. When the 
anti-message\index{Anti-Message} and original message meet,
they are both annihilated\index{Annihilation}. The final phase consists 
of executing the \acl{LP}\index{LP} forward in time from its rollback\index{Rollback} 
state to the time the 
false message\index{False Message} arrived. No messages are canceled or sent 
between the time to which the \acl{LP}\index{LP} rolled back and the time of the 
false message\index{False Message}. 
Because these messages are correct, there is no need to cancel or re-send 
them. This increases performance, and it prevents
additional rollbacks. Note that another 
false message\index{False Message} or anti-message\index{Anti-Message} 
may arrive before this final phase has completed without causing problems.
The \acl{VNC} \acl{LP} has the contents 
shown in Table \ref{lpspectab}, the message fields are shown in 
Table \ref{messpectab}, and the message types are listed in 
Table \ref{mestypes} where $t$ is the real time at the receiving \acl{LP}.

\begin{table*}[htbp]
\centering
\begin{tabular}{||l|l||} \hline
\textbf{Structure} & \textbf{Description} \\ \hline \hline
Receive Queue (QR) & Ordered by message receive time (TR) \\ \hline
Send Queue (QS)    & Ordered by message send time (TS) \\ \hline
Local Virtual Time (LVT) & $LVT = \inf RQ$ \\ \hline
Current State (CS) & State of the logical and physical process \\ \hline
State Queue (SQ) & States (CS) are periodically saved \\ \hline
{\em Sliding Lookahead Window ($\Lambda$)} & $SLW = (t, t + \Lambda$] \\ \hline
{\em Tolerance ($\Theta$)} & Allowable deviation \\ \hline
\end{tabular}
\caption{\label{lpspectab}\acl{VNC} \acl{LP} Structures.}
\end{table*}

\begin{table*}[htbp]
\centering
\begin{tabular}{||l|l||} \hline
\textbf{Field} & \textbf{Description} \\ \hline \hline
Send Time (TS) & LVT of sending process when message is sent \\ \hline
Receive Time (TR) & Scheduled time to be received by receiving process \\ \hline
Anti-toggle (A) & Identifies message as normal or antimessage \\ \hline
Message (M) & The actual contents of the message  \\ \hline
{\em Real Time (RT)} & The GPS time at which the message originated \\ \hline
\end{tabular}
\caption{\label{messpectab}\acl{VNC} Message Fields.}
\end{table*}

\begin{table*}[htbp]
\centering
\begin{tabular}{||l|l||} \hline
Virtual Message & $RT > t$ \\ \hline
Real Message    & $RT \leq t$ \\ \hline
\end{tabular}
\caption{\label{mestypes}\acl{VNC} Message Types.}
\end{table*}

\subsection{Multiple Future Event Architecture}

It is possible to anticipate alternative future events using a direct
extension of the basic \acl{VNC} algorithm \cite{Tinker}. The driving process
generates multiple virtual messages, one for each possible future event with
corresponding probabilities of occurrence, or a ranking, for each
event. Instead of a single \acl{QR} for each \acl{LP}, multiple \acl{QR}s
for each version of an event are created dynamically for each LP.
The logical process can dynamically create \acl{QR}s for each event
and give priority to processing messages from the most likely versions'
\acl{QR}s. This enhancement to \acl{VNC} has not been implemented.
This architecture for implementing alternative futures, while a simple and 
natural extension of the \acl{VNC} algorithm, creates additional messages and
increases the message sizes. Messages require an additional field to
identify the probability of occurrence and an event identifier.

\subsection{Example of \acl{VNC} Operation}

\begin{figure*}[htbp]                                     
        \centerline{\psfig{file=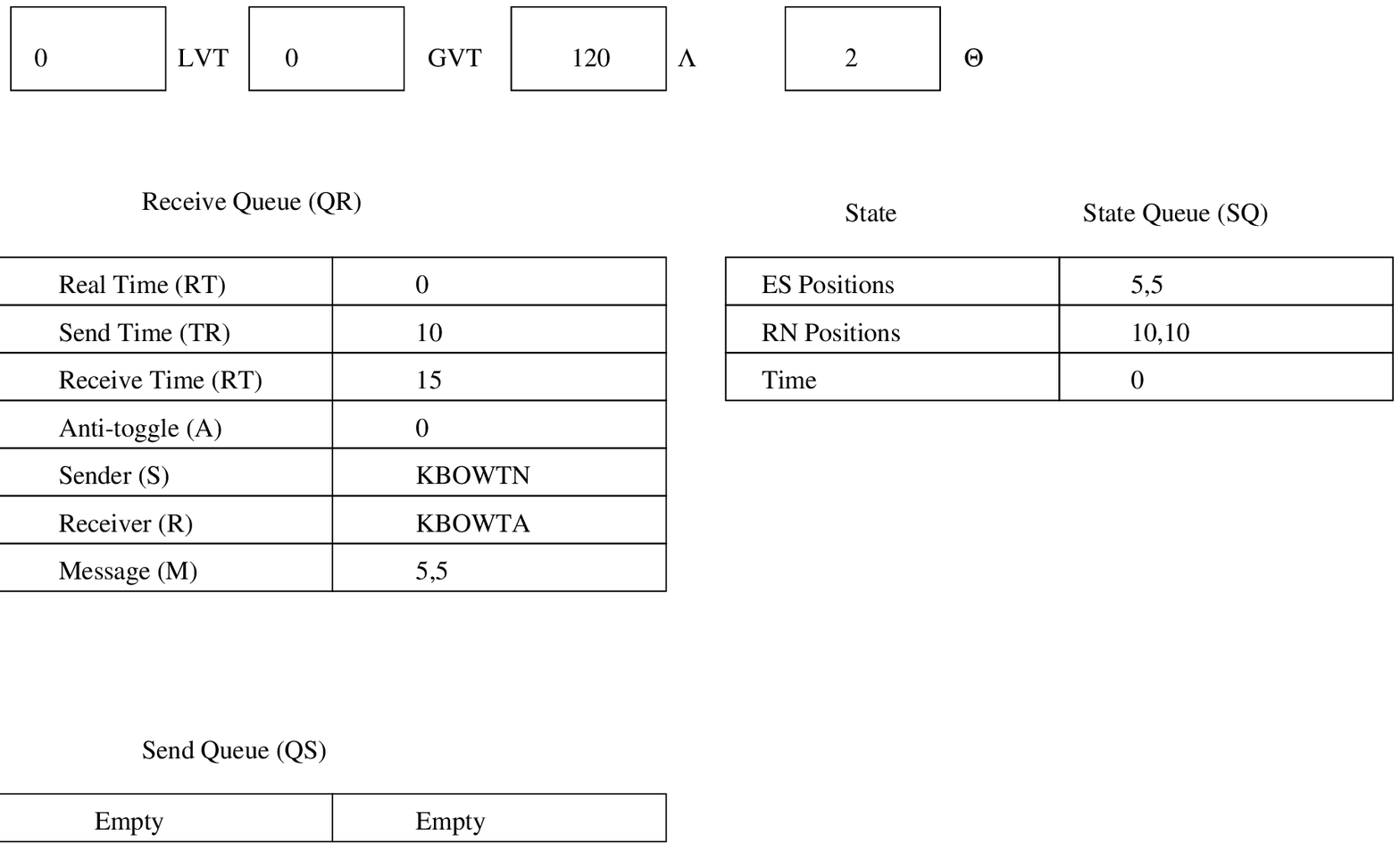,width=5.75in}}
        \caption{\acl{ES} Handoff \acl{LP}: First Virtual Message Arrives.}
        \label{vncex1}
\end{figure*}

\begin{figure*}[htbp]
        \centerline{\psfig{file=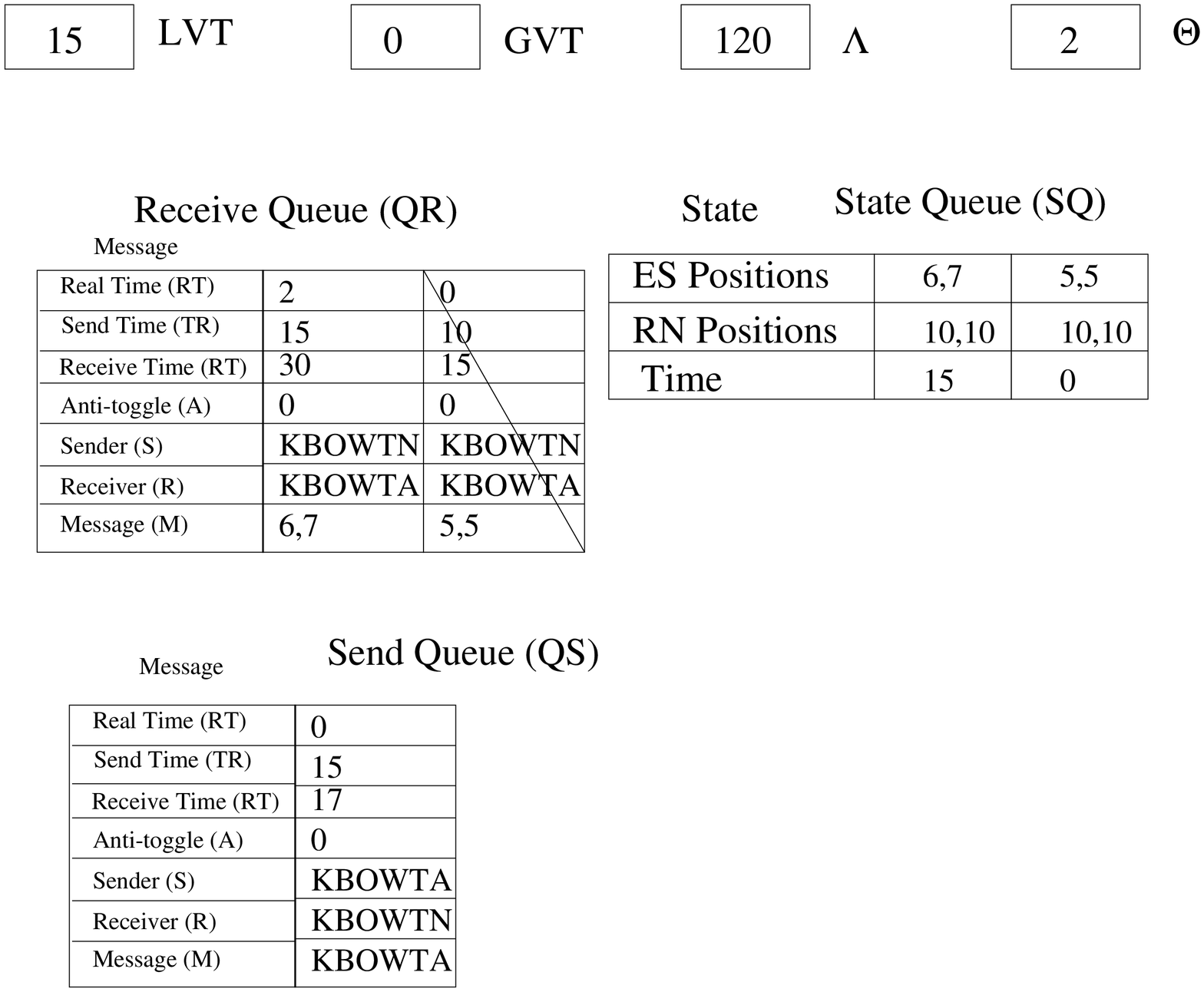,width=5.75in}}
        \caption{\acl{ES} Handoff \acl{LP}: Output Message Sent.}
        \label{vncex2}
\end{figure*}

\begin{figure*}[htbp]
        \centerline{\psfig{file=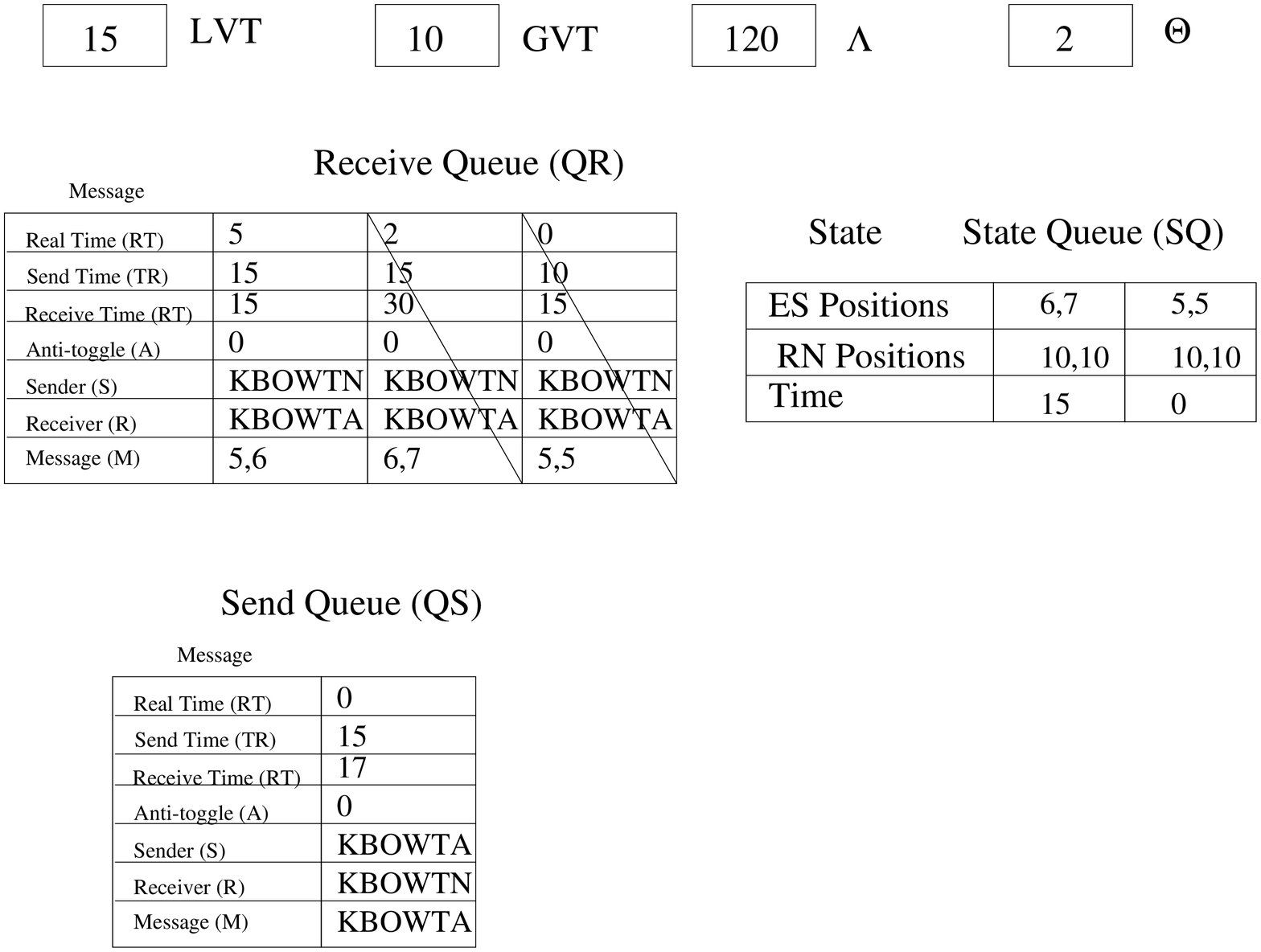,width=5.75in}}
        \caption{\acl{ES} Handoff \acl{LP}: First Real Message Arrives.}
        \label{vncex3}
\end{figure*}

\begin{figure*}[htbp]
        \centerline{\psfig{file=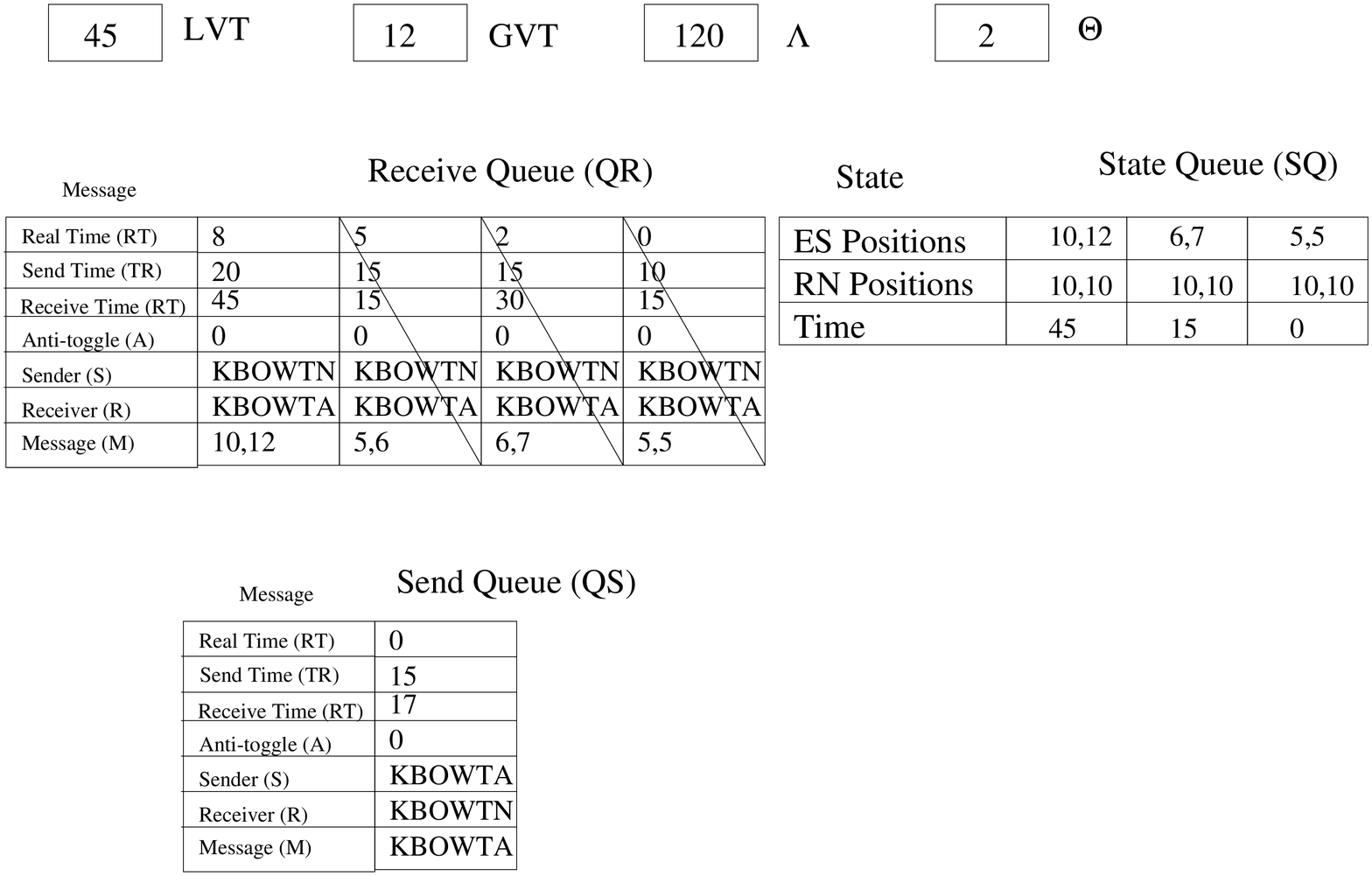,width=5.75in}}
        \caption{\acl{ES} Handoff \acl{LP}: Second Virtual Message Arrives.}
        \label{vncex4}
\end{figure*}

\begin{figure*}[htbp]
        \centerline{\psfig{file=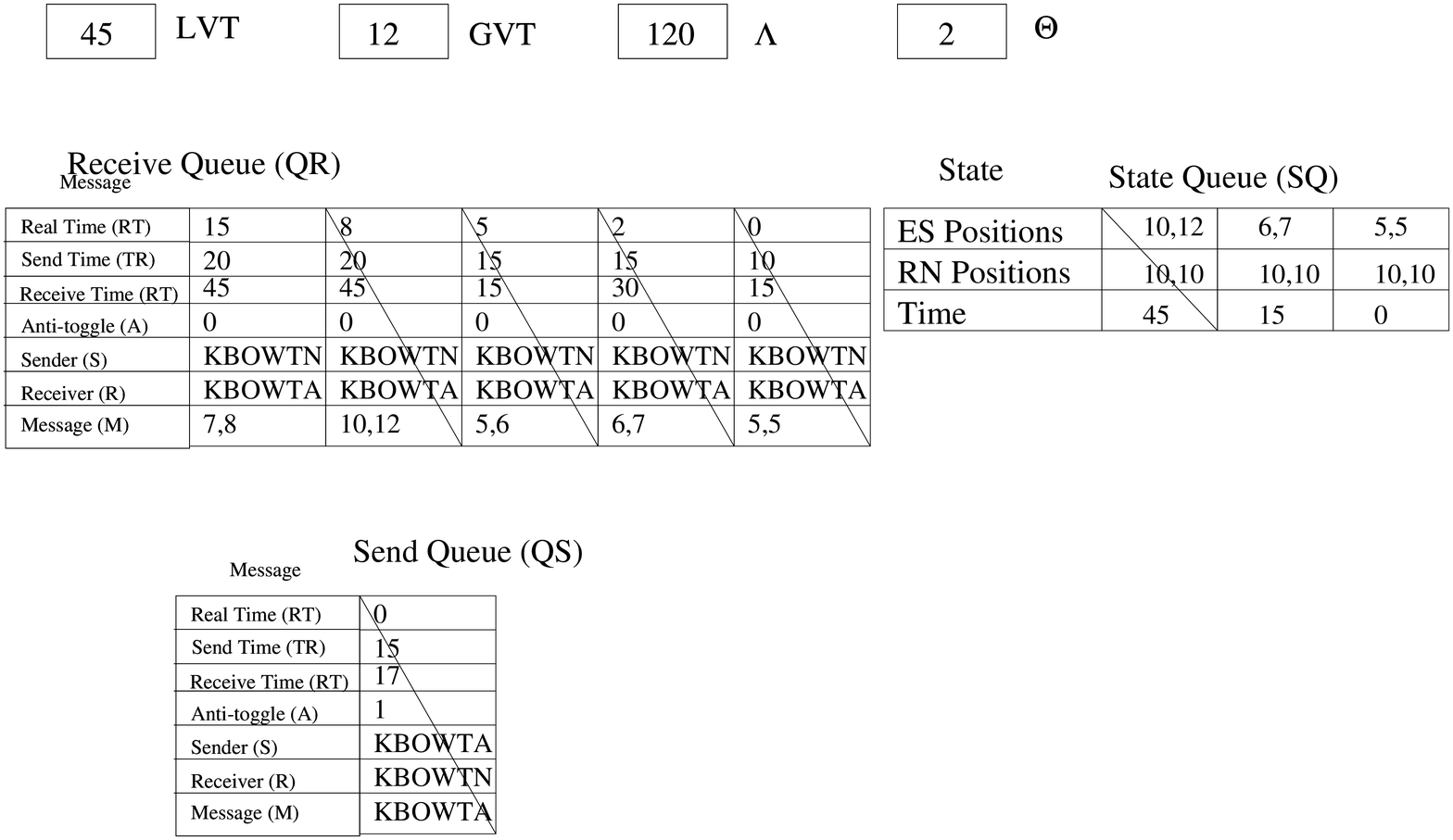,width=5.75in}}
        \caption{\acl{ES} Handoff \acl{LP}: Rollback Occurs.}
        \label{vncex5}
\end{figure*}

A specific simple example of the \acl{VNC} algorithm is shown in 
Figures \ref{vncex1} through \ref{vncex5}. The description of the
algorithm begins with a system which has just been powered up and
is generating real messages\index{Real Message}. The driving
process\index{VNC!driving process} will begin to
predict future events and inject virtual messages\index{Virtual Message}
based on those events. An \acl{LP}\index{LP} receives a message and must
first determine whether it is virtual\index{Virtual Message} or
real\index{Real Message}
by examining the RT field. If the message is a virtual 
message\index{VNC!virtual message},
the \acl{LP}\index{LP} compares the message with its \acl{LVT}\index{LVT} to
determine whether a rollback\index{Rollback} is necessary due to
an out of order message. If the message has not arrived in the past relative
to the \acl{LP}'s \acl{LVT}, the message then enters the \acl{QR} in order by 
\acl{TR}. 
The \acl{LP} takes the next message from the \acl{QR},
updates its \acl{LVT}\index{LVT}, and processes the message. If an outgoing
message is generated, a copy of the message is saved in the \acl{QS}\index{QS},
the \acl{TR}\index{TR} is set, and the
\acl{TS}\index{TS} is set to the current \acl{LVT}\index{LVT}.
The message is then sent to the destination \acl{LP}\index{LP}. If the
virtual message\index{Virtual Message}
arrived out of order, the \acl{LP}\index{LP} must rollback\index{Rollback}
as described previously. 
Figure \ref{vncex1} is an example of an \acl{ES} \acl{LP} which determines
handoff. Input messages arrive from \acl{LP}es on the \acl{RN} and enter \acl{LP}es
on the \acl{ES} which cause messages to arrive at the Handoff calculation \acl{LP}
shown in Figure \ref{vncex1}. At the start of the \acl{LP}'s execution, LVT 
and GVT are set to zero, Lookahead ($\Lambda$) is set to two minutes and a 
tolerance ($\Theta$) of 2 units of distance is allowed between predicted 
and actual location values for this process.

In Figure \ref{vncex2}, an output message is sent indicating to the \acl{RN}
that it should associate with this \acl{ES}. Also, the next virtual message 
has arrived, so the LVT will be set to the \acl{TR} of the virtual  message. 
The new \acl{LP} state is saved in the \acl{SQ}.

The first real message arrives in Figure \ref{vncex3} with an \acl{RT}
value of 5. The position in the message value (5,6) is compared with 
the value (5,5) which is the position in the \acl{SQ} with the closest
time at which the state was saved. It is found 
to be within the required tolerance, thus, no rollback occurs and no new 
messages need to be sent. Although the \acl{GVT} is not necessary for
\acl{VNC} operation as discussed in the next chapter, it is included
in this example. The \acl{GVT} is updated in Figure \ref{vncex3} to 
indicate the time that the entire system has predicted into the 
future.

Another virtual message arrives in Figure \ref{vncex4}. The message 
has arrived in proper order, therefore, the \acl{LVT} is updated, the 
message is processed and the state is saved. No output message was 
generated since no new \acl{ES}-\acl{RN} association was determined
by the \acl{PP} for this \acl{LP}.

An example of out of tolerance rollback is illustrated in Figure \ref{vncex5}.
A real message arrives and its message contents are compared with the 
closest saved state value. Because the message value is found to be out 
of tolerance, all state queue values with times
greater than the receive time of the real message are discarded. The
send queue message anti-toggle is set and the anti-message is sent.
The discarded states and messages are slashed in Figure
\ref{vncex5}. The rollback causes the \acl{LP} to go back to time 15 
because that is the time of the most recent saved state which is less
than the time of the out-of-tolerance message's \acl{TR}.

\begin{figure*}[htbp]                                                       
        \centerline{\psfig{file=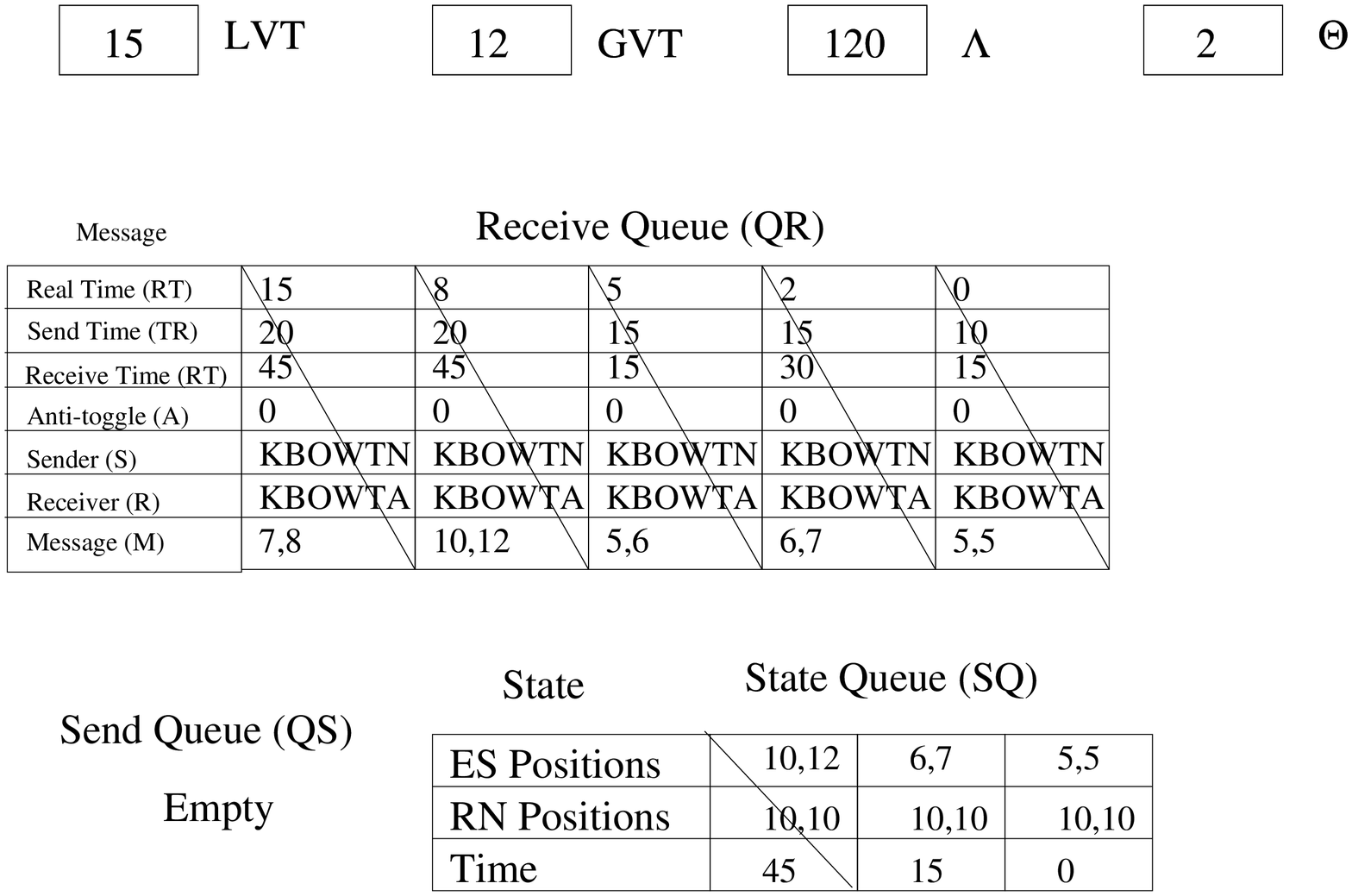,width=5.75in}}
        \caption{\acl{ES} Handoff \acl{LP}: After Rollback Occurs.}
        \label{vncex6}                                                        
\end{figure*}

After the rollback occurs, the \acl{LP} has the state shown in
Figure \ref{vncex6}. The \acl{LVT} has been adjusted to time 15
and the \acl{LP} continues normal processing. The algorithm is
presented in more detail in the next section.

\section{Pseudocode Specification for \acl{VNC}}

The \acl{VNC} algorithm requires both Driving Processes and Logical 
Processes. Driving Processes predict events
and inject virtual messages into the system. \acl{LP}s react
to both real and virtual messages. The \acl{VNC} Algorithm for 
a driving process is shown in Algorithm \ref{drivspec}. The
operation of the both the driving process and the logical process
repeat indefinitely.
If the Driving Process has not exceeded its lookahead time on
line \ref{checkdelay}, a new value $\Delta$ 
time units into the future
is computed by the function $C(t)$ and the result is assigned
to the message ($M$) and sent on line \ref{sendvm}. The receive time, 
which is the time
at which this message value is to be valid is assigned to ($M$) in 
line \ref{asstime}. 
Every $UpdatePeriod$ time units, real messages are generated
as shown in lines \ref{assval} and \ref{dsendm}.

\ifisdraft
	\onecolumn
\fi

\begin{algorithm}[htpb]
\algname{VncDriver}{$UpdatePeriod$, $\Lambda$}

\begin{algtab}
\algrepeat
	\alglabel{checkdelay}\algif{$GVT \le t + \Lambda$}
		\alglabel{predict} $M.val \leftarrow C(LVT+\Delta)$ \\
		\alglabel{asstime} $M.rt \leftarrow LVT+\Delta$ \\
		\alglabel{sendvm} Send(M) \\
	\algend
	\alglabel{checkupdate}\algif{$t \mod UpdatePeriod = 0$}
		\alglabel{assval} $M.val \leftarrow GPS.pos$ \\
		\alglabel{dsendm} Send(M)
\end{algtab}
\caption{\label{drivspec}VNC Driving Process Algorithm.}
\end{algorithm}

\ifisdraft
	\twocolumn
\fi

The \acl{VNC} Algorithm for an \acl{LP} is specified in 
Algorithm \ref{vncalg}. On line \ref{nextm}, $\inf$ is
infimum. On line \ref{checktol}, the
next message from the Receive Queue is checked to
determine whether the message is real. If the message
is real, line \ref{checktol} retrieves the state that was
saved closest to the receive time of the message and
checks whether the values of the saved state
are within tolerance. If the tolerance is exceeded, the
process will rollback. Also, if the message is received 
in the past relative to this process's LVT, the process
will rollback as shown in line \ref{checkrb}. 
In line \ref{commit}, the pre-computed and cached value in the
state queue is committed. Committing a value is an irreversible
action because it cannot be rolled back once committed. If the
process's LVT has not exceeded its lookahead time as determined in
line \ref{checkla}, then the virtual message is processed
in lines \ref{stateq} through \ref{lpend}.
The function $C_1(M, LVT)$ in line \ref{stateq} represents the
computation of the new state. The function $C_1(M, LVT)$ returns 
the state value for this \acl{LP} and updates the $LVT$ to the time 
at which that value is valid. The function $C_2(M, LVT)$ in line 
\ref{sendm} represents the computation of a new message value.

\ifisdraft
	\onecolumn
\fi

\begin{algorithm}[htbp]
\algname{Vnc}{$\Theta$, $\Lambda$}

\begin{algtab}
$LVT \leftarrow 0$ \\
$t \leftarrow GPS.t$ \\
\algrepeat
	\alglabel{nextm} $M \leftarrow \inf{M.tr \in QR}$ \\
	$CS(t).val \leftarrow C_1(M, t)$ \\
	\algifthen{\alglabel{checktol}($M.rt \le t) \land 
		(|SQ(t).val - \Theta| > CS(t).val)$}{Rollback()}
	\algifthen{\alglabel{checkrb} $M.rt < LVT$}{Rollback()} 
	\algifthen{\alglabel{commit} $M.rt \le t$}{Commit($SQ:SQ.t 
		= \approx M.rt)$}
	\algif{\alglabel{checkla} $LVT + \Lambda \le GVT$}
		\alglabel{stateq} $SQ.val \leftarrow C_1(M, LVT)$ \\
		$SQ.t \leftarrow LVT$ \\
		\alglabel{sendm} $M.val \leftarrow C_2(M, LVT)$ \\
		$M.rt \leftarrow LVT$ \\
		$QS \leftarrow M$ \\
		\alglabel{lpend} Send(M)
\end{algtab}
\caption{\label{vncalg}VNC Logical Process Algorithm.}
\end{algorithm}

\ifisdraft
	\twocolumn
\fi
\chapter{Algorithm Analysis}
\label{analysis}

This goal of this chapter is to analyze the performance of \acl{VNC}.
The characteristics of \acl{VNC} to be analyzed are speedup,
potentially wasted resources, and bandwidth overhead. 
Speedup is the ratio of the time required to perform an operation
without \acl{VNC} divided by the time required with \acl{VNC}.
Wasted resources are defined to be resources which are temporarily 
allocated but never used due to prediction inaccuracy. Bandwidth
overhead is the ratio of the amount of additional bandwidth required by
a \acl{VNC} system divided by the amount of bandwidth required by a none 
\acl{VNC} system.

Because the \acl{LP}s of a \acl{VNC} system are asynchronous, the \acl{LP}s
can take maximum advantage of parallelism. However, messages among processes 
may arrive at the destination process out-of-order. A Petri-Net model is used 
to quantify the amount of messages that arrive out-of-order for a particular 
\acl{VNC} system. 
Petri-Nets are commonly used for synchronization analysis where
``places'', usually shown as circles, represent entities such as
producers, consumers, or buffers, and ``transitions'', shown as squares
allow ``tokens'' shown as dots, to move from one place to another.
In this analysis, Petri-Net tokens represent \acl{VNC} messages and Petri-Net 
places represent \acl{VNC} \acl{LP}s. Characteristics of Petri-Nets are used 
to determine the likelihood of out-of-order messages.
The likelihood of the occurrence of out-of-order messages and out-of-tolerance 
messages is required by an equation that is developed in this chapter
to describe the speedup of \acl{VNC}. After analyzing the speedup, the
prediction accuracy and bandwidth are analyzed. The chapter concludes 
by considering enhancements and optimizations such as implementing 
multiple future events, eliminating the \acl{GVT} calculation, and 
elimination of real messages when they are not required.
\section{Utility Function Analysis}

Performance analysis of the \acf{VNC} algorithm must take into
account accuracy as well as speed. An inaccurate configuration can result
in committed resource\index{Resource}s which are never used and thus wasted, or in not
committing enough resource\index{Resource}s when needed thus causing a delay.
Unused resource\index{Resource} allocation\index{Allocation} must be minimized. Many of the mobile
wireless \acl{ATM} mechanisms previously mentioned depend on keeping 
resources permanently allocated, such as \acl{ATM} \acl{VC}s in \cite{Acampora}.
\acl{VNC} does not require permanent over-allocation of resource\index{Resource}s, however, 
the \acl{VNC} algorithm may make a false\index{False Message} prediction which \emph{temporarily}
establishes resource\index{Resource}s which may never be used. A \acl{VNC} system whose
tolerances are reduced in order to produce more accurate results
will have fewer unused allocated resource\index{Resource}s, however, the tradeoff 
is a reduction in speedup\index{Speedup}.

Equation \ref{vncutil} quantifies the 
advantage of using \acl{VNC} where $\eta$ is the expected speedup\index{Speedup} 
using \acl{VNC} over a non-VNC system, $\Phi_s$ is the marginal utility 
function of the configuration speed, and $\alpha$ is the expected quantity 
of wasted resource\index{Resource}s other than bandwidth, and 
$\Phi_w$ is the marginal utility function\index{Function} of the 
allocated but unused resource\index{Resource}. An example of a resource\index{Resource} which may be
temporarily wasted due to prediction error is a \acf{VC} which may
be established temporarily and not used.
The expected bandwidth overhead\index{Virtual Network Configuration!overhead} is 
represented by $\beta$ and $\Phi_b$ is the marginal utility
function of bandwidth. 

\begin{figure*}
\begin{equation}
U_{VNC} = \eta \Phi_s - \alpha \Phi_w - \beta \Phi_b
\label{vncutil}
\end{equation}
\end{figure*}

The marginal utility function\index{Function}s $\Phi_s$, $\Phi_w$ and $\Phi_b$ are subjective 
functions which describe the value of a particular service to the
user. The function\index{Function}s $\Phi_s$, $\Phi_w$ and $\Phi_b$ may be determined by 
monetary considerations and user perceptions.
The following sections develop propositions which describe the behavior
of the \acl{VNC} algorithm and from these propositions equations for $\eta$, 
$\alpha$ and $\beta$ are defined. 
\section{Petri-Net Analysis for Virtual\index{Virtual Network Configuration!verification} Network Configuration}
\label{pnet}

This goal of this section is to determine the 
probability of out-of-order messages arriving at an \acl{LP} and 
to determine the expected proportion of out-of-order messages
($E[X]$) and the probability of rollback\index{Rollback} due to out-of-order 
messages ($P_{oo}$). Another goal of this section is to develop 
a new and simpler approach to analyzing Time\index{Time} Warp based algorithms
in general and \acl{VNC} in particular.
The contribution of this section is unique because most current
optimistic synchronization analysis has been explicitly time-based
yielding limited results except for very specific cases.
The approach taken in this section is topological; timing is implicit
rather than explicit. A \acl{C/E} is used in this analysis because it 
is the simplest form of a Petri Net that is ideal for studying \acl{VNC} 
synchronization behavior. 

\begin{figure*}[htbp]                                                      
        \centerline{\psfig{file=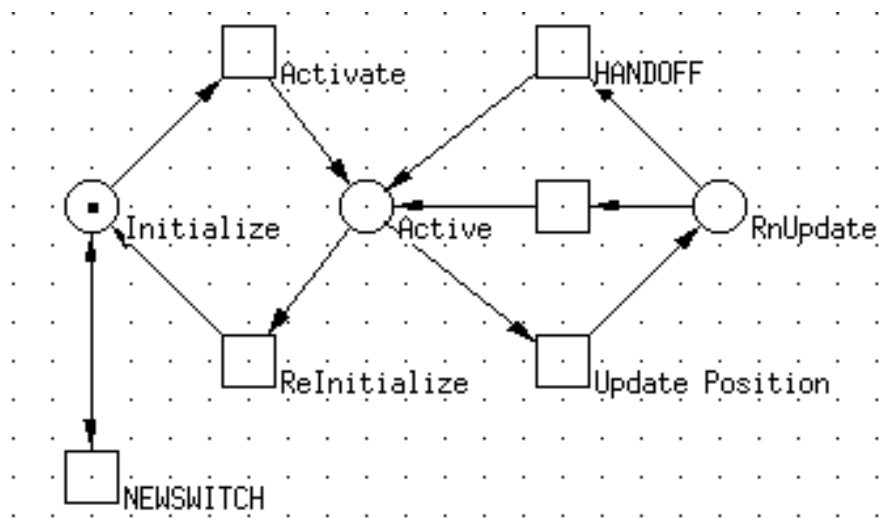,width=4.75in}}
        \caption{\label{escenet}Subset of NCP FSM as a C/E Net.}
\end{figure*}

A \acl{C/E} network consists of condition and transition elements that
contain tokens. Tokens reside in condition elements. When all condition 
elements 
leading to a transition element contain a token, several changes take 
place in the \acl{C/E} network. First, 
the tokens are removed from the conditions that triggered the event, the 
event occurs, and finally tokens are placed in all condition outputs from the 
transition that was triggered. Multiple tokens in a condition
and the uniqueness of the tokens is irrelevant in a \acl{C/E} Net. In this 
analysis, tokens represent virtual\index{Virtual Network Configuration!verification} messages, 
conditions represent processes, and transitions represent \acl{LP} 
interconnections. 
The notation from \cite{Reisig85} is used: $\Sigma=(B,E;F,C)$ is a \acl{C/E}
Net where 
$B$ is the set of conditions, $E$ is the set of transitions, and $F \subseteq 
(B \times E) \cup (E \times B)$ where $\cup$ is union and $\times$ is the 
cross product of all conditions and transitions. A marking is the set of
conditions containing tokens at any given time during \acl{C/E} operation
and $C$ is the set of all possible
sets of markings of $\Sigma$. The input conditions to a transition will
be written as ``$\petripre{e}$'' and the output conditions will be written as
``$\petripost{e}$''. Let $c \subseteq C$, then a
transition $e \in E$ is triggered when $\petripre{e} \subseteq (c \subseteq B)$ 
and $\petripost{e} \cap c = \emptyset$. If $c$ is the current set of enabled 
conditions and after the next transition ($e$) the new set of enabled 
conditions is $c'$, then this is represented more compactly as $c[e \rangle 
c'$. \acl{C/E} networks provide insight into liveness, isomorphism, 
reachability, a method for determining synchronous\index{Synchronous} behavior, and behavior 
based on the topology\index{Topology} of \acf{VNC} \acf{LP} communication. An example 
of a \acl{C/E} Net is shown in Figure \ref{escenet} for an \acl{ES}. The initial 
marking begins with a token in the Initialize condition. Receipt of a 
\textbf{MYCALL} packet causes the transmission of a \textbf{NEWSWITCH} 
message event to occur. After a Time\index{Time}out occurs, the \acl{ES} 
begins to handle \acl{RN} activity. Clearly, the \acl{C/E} network shown
in Figure \ref{escenet} is a subset of the \acf{FSM}
from Table \ref{ESFSM}. Every \acl{FSM} has an equivalent \acl{C/E} Net
\cite[p. 42]{Peterson81}.

Some common terminology and concepts are defined next that will be needed
for a topological analysis of \acl{VNC}. These terms and concepts are
introduced in a brief manner and build upon one another. Their relationship 
with \acl{VNC} will soon be made clear. The following notation is used:
``$\lnot$'' means ``logical not'', ``$\exists$'' means ``there exists'', 
``$\forall$'' means ``for each'', ``$\land$'' means ``logical and'', 
``$\lor$'' means ``logical or'', ``$\in$'' means that an element is a member 
of a set, ``$\equiv$'' means ``defined as'', and ``$\rightarrow$'' defines 
a mapping or function\index{Function}. Also, $a \prec b$ indicates an ordering 
between two elements, $a$ and $b$, such that $a$ precedes $b$ in some 
relation. ``$\Rightarrow$'' means ``logical implication'' and 
``$\leftrightarrow$'' means ``logical equivalence''.

A region of a particular similarity relation ($\cdot$) of $B \subseteq A$
means that $\forall a,b \in B : a \cdot b$ and $\forall a \in A :
a \not\in B \Rightarrow \exists b \in B : \lnot (a \cdot b)$. This means
that the relation is ``full'' on $B$ and $B$ is a maximal subset on that
the relation is full. In other words, a graph of the relation ($\cdot$) would
show $B$ as the largest fully connected subset of nodes in $A$.

\begin{figure*}[htbp]                                                     
        \centerline{\psfig{file=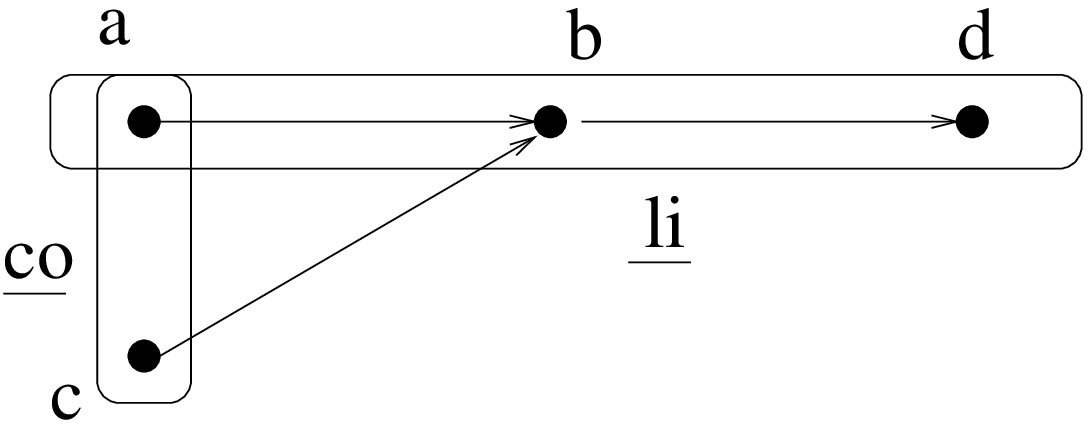,width=5.5in}}
        \caption{\label{kdense} Demonstration of \li and \co.}
\end{figure*}

Let ``\li'' represent a linear ordering such that $a \li b 
\leftrightarrow (a \prec b) \lor (b \prec a) \lor (a \equiv b)$. 
Let ``\co'' represent a concurrent ordering $a \co b \leftrightarrow 
\lnot (a \li b) \lor (a \equiv b)$. 
Figure \ref{kdense} illustrates a region of \co that contains
$\{a, c\}$ and region of \li that contains $\{a, b, d\}$ where $\{a, b, c, d\}$
represents \acl{LP}s and the relation is ``sends a message to''. 
Trivially, if every process in the
\acl{VNC} system is a region of $\li$ then regardless of how many 
driving\index{Driving Process} processes
there are, no synchronization is necessary since there exist no concurrent
processes.

Let $D$ is the set of driving\index{Driving Process} processes and $R$ be the
set of the remaining processes in the \acl{VNC} system. Then $D \prec R \leftrightarrow$ 
$\forall d \in D$ $\forall r \in R : (d \prec r) \lor (d \co r)$.
In order for the virtual\index{Virtual Network Configuration!verification} messages 
that originate from $D$ to be used, 
$D \prec R$ where $R$ are the remaining non-driving processes. 
The relation used here is again assumed to be ``sends a message to''.

\begin{figure*}[htbp]
        \centerline{\psfig{file=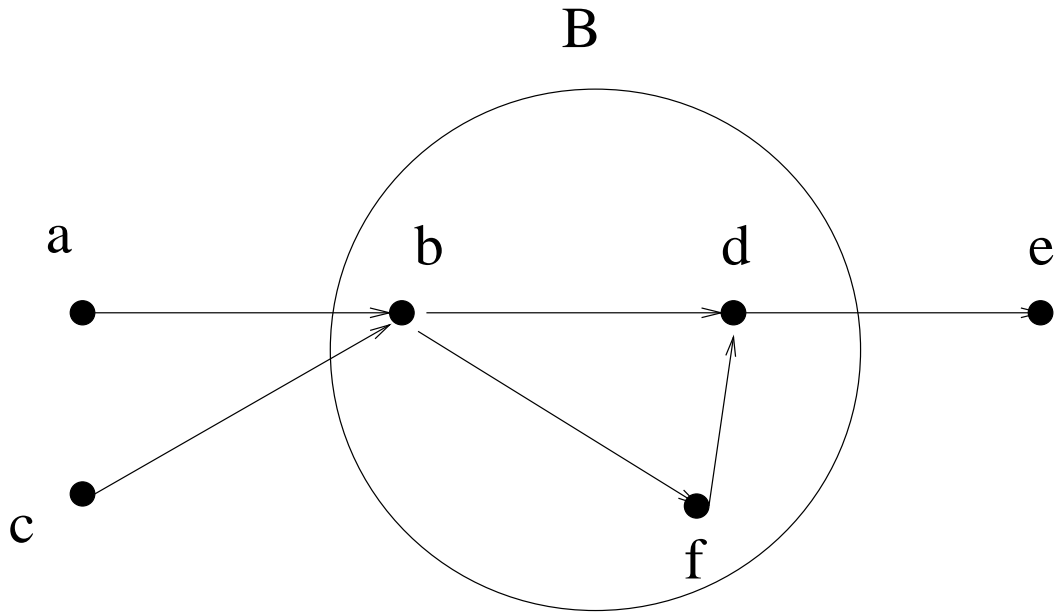,width=5.5in}}
        \caption{\label{extrema} Illustration of $B^-$ and $\overline{B}$.}
\end{figure*}

In the remaining definitions, let $A$, $B$, and $C$ be arbitrary sets where
$B \subseteq A$ used for defining additional operators.
Let $B \preceq C \equiv \forall b \in B \forall c \in C: b \prec c \lor b \co c$.
Let $B^- \equiv \{ a \in A | \{a\} \preceq B \}$ and 
$B^+ \equiv \{ a \in A | B \preceq \{a\} \}$ where $|$ means ``such that''. 
Also, let $\overline{B} \equiv \{ b \in B | \forall b' \in B: 
(b \co b') \lor (b \prec b') \}$ and $\underline{B} \equiv \{ b \in B | 
\forall b' \in B: (b \co b') \lor (b' \prec b) \}$. This is illustrated 
in Figure \ref{extrema} where all nodes are in the set $A$ and $B$ is the set 
of nodes that lie within the circle. 
$B^-$ is the set $\{a,b,c,d,f\}$ and $\overline{B}$ is the set $\{b\}$.

An occurrence network ($K$) is a \acl{C/E} network that is related to the
operation of a particular \acl{C/E} network ($\Sigma$). The
occurrence network ($K$) begins as an empty \acl{C/E} network;
conditions and events are added to $K$ as $\Sigma$ operates. Thus,
$K$ represents a particular sample of operation of $\Sigma$. 
There can be multiple events in $\Sigma$ which are capable of firing, but
only one event is chosen to fire, thus it is possible that a particular
$\Sigma$ will not always generate the same occurrence net ($K$) each time
it operates. Note
that $K$ has some special properties. The condition elements of $K$
have one and only one transition, because only one token
in $\Sigma$ may fire from a given condition. Also, $K$ is cycle
free because $K$ represents the operation of $\Sigma$.

A few more definitions are required before the relation described
above between $K$ and $\Sigma$ can be formally defined. This
relationship is called a Petri-Net process.
Once the Petri-Net process is defined, a
measure for the ``out-of-orderness'' of messages can be
developed based on synchronic distance.
A {\em line} is a subset that is a region of $\li$ and a {\em cut}
is a subset that is a region of $\co$. A slice (``$\slice$'') is a cut
of an occurrence network ($K$) containing condition elements and $\slice(K)$ is
the set of {\em all} slices of $K$. The region of \co shown in Figure
\ref{kdense} illustrates a cut where nodes represent conditions and the
relation defines an event from one condition to another in a \acl{C/E} Network.

A formal definition of the relation between an occurrence net and
a \acl{C/E} net is given by a Petri-Net process.
A Petri-Net process ($p$) is defined as a mapping from a network $K$ to a \acl{C/E} Network
$\Sigma$, $p : K \rightarrow \Sigma$, such that each slice of $K$ is mapped
injectively (one-to-one) into a marking and $(p(\petripre{t}) = \petripre{p(t)})
\land (p(\petripost{t}) = \petripost{p(t)})$. Also
note that $p^{-1}$ is used to indicate the inverse mapping of $p$. Think
of $K$ as a particular sample of the operation of a \acl{C/E} Network. A
\acl{C/E} Network can generate multiple processes
Another useful characteristic is whether a network is K-dense. A network
is K-dense if and only if every $\slice(K)$ has a non-empty
intersection with every region of $\li$ in $K$. This means that each
slice intersects every sequential path of operation. 

All of the preceding definitions have been leading towards the development
of a measure for the ``out-of-orderness'' of messages that is relatively
simple to calculate because it does not rely on explicit time values
or distributions. In the following
explanation, a measure is developed for the synchronization between events.
Consider $D_1$ and $D_2$ that
are two slices of $K$ and $M$ is a set of events in a \acl{C/E} Network. 

$\mu(M, D_1, D_2)$ is defined as $| M \cap D^+_2 \cap D^-_1 | - 
| M \cap D^-_1 \cap D^+_2 |$.
Note that $\mu(M, D_1, D_2) = - \mu(M, D_2, D_1)$. Thus $\mu(M, D_1, D_2)$
is a number that defines the number of events between two specific
slices of a net.

Let $(p:K \rightarrow \Sigma) \in \pi_{\Sigma}$ where $\pi_{\Sigma}$ is the 
set of all finite processes of $\Sigma$.
A term known as ``variance'' is defined that describes
the number of events across all slices of a net ($K$).
The variance of $T_{\Sigma}$ is $\nu(p, T_1, T_2) \equiv
\max\{\mu(p^{-1}(T_1), D_1, D_2) - \mu(p^{-1}(T_2), D_1, D_2) | D_1, D_2 
\in \slice(K)\}$. Also, note that $\nu(p, T_1, T_2) = \nu(p, T_2, T_1)$
where and $T_1, T_2 \subseteq T_{\Sigma}$. This defines a measure of the 
number of events across all slices of a net ($K$).

\begin{figure*}[htbp]
        \centerline{\psfig{file=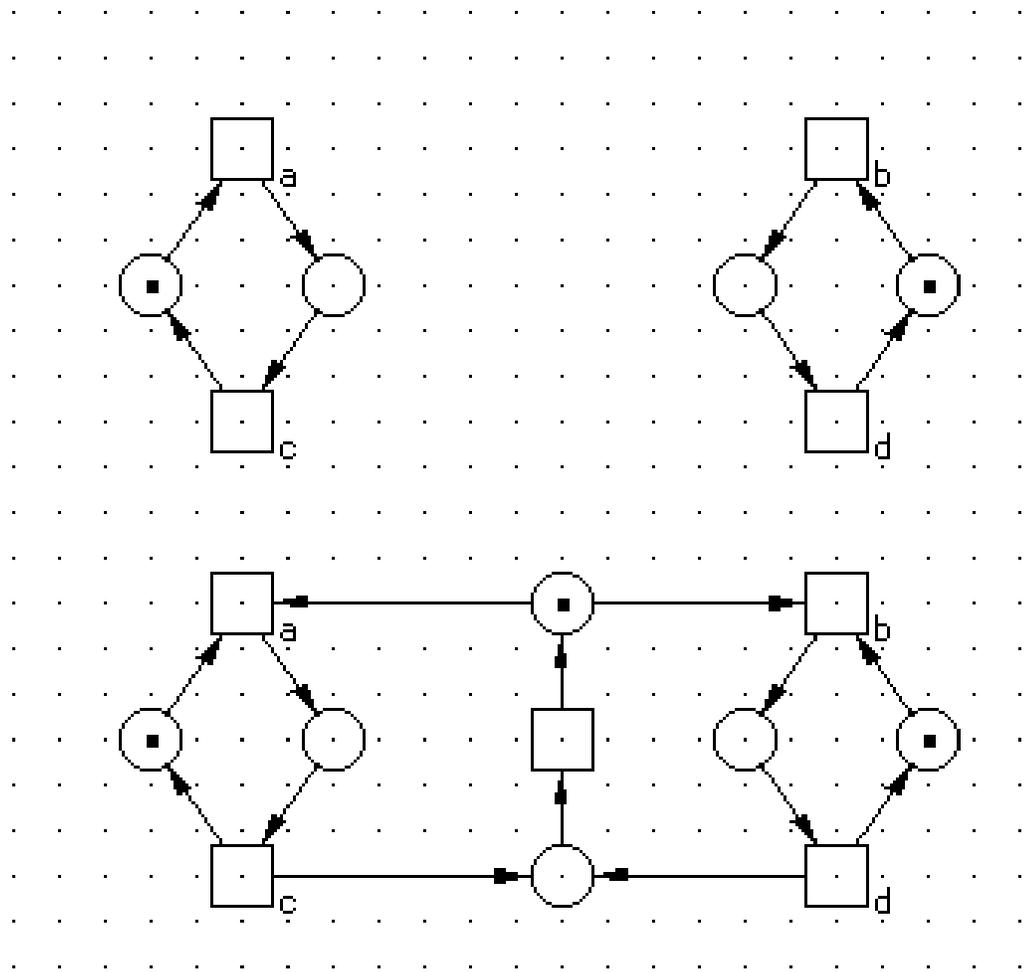,width=5.5in}}
        \caption{\label{synchdist} Example of Synchronic Distance.}
\end{figure*}

The synchronic distance ($\sigma(T_1, T_2) = \sup \{ \nu(p, T_1, T_2) 
| p \in \pi_{\Sigma} \})$ is the supremum of the variance in all finite
processes. This defines the measure of ``out-of-orderness'' across all 
possible $K$. By determining the synchronic distance, 
a measure for the likelihood of rollback\index{Rollback} in \acl{VNC} can be 
defined that is dependent on the topology\index{Topology} and is 
\textbf{independent of time}. Further details on synchronic distance
and the relation of synchronic distance to other measures of 
synchrony can be found in \cite{Voss}. A more intuitive method for
calculating the synchronic distance is to insert a virtual condition
into the \acl{C/E} net. This condition has no meaning or effect on
\acl{C/E} operation. It is allowed to hold multiple tokens
and begins with enough tokens so that it can emit a token whenever
a condition connected to its output transition is ready to fire. The
virtual condition has inputs from all members of $T_1$ and output 
transitions of all members of $T_2$. The synchronic distance is the 
maximum variation in the number of tokens in the virtual condition.
The greater the possibility of rollback\index{Rollback}, the larger
the value of $\sigma(T_1, T_2)$. A simple example in Figure \ref{synchdist}
intuitively illustrates what the synchronic distance means. Using the 
virtual condition method to calculate the synchronic
distance between $\{a, b\}$ and $\{c, d\}$ in the upper \acl{C/E} Network,
the synchronic distance is found to be two. 
By adding two more conditions and another transition to the \acl{C/E}
network, the synchronic
distance of the lower \acl{C/E} Network shown in Figure \ref{synchdist} is one. 
The larger the value of $\sigma(T_1, T_2)$ the less synchronized the events 
in sets $T_1$ and $T_2$.
If these events indicate message transmission, then the
less synchronized the events, the greater the likelihood that
the messages based on events $T_1$ and $T_2$ will be out-of-order. This
allows the likelihood of out-of-order message arrival at an \acl{LP} to 
be determined based on the inherent synchronization of a system.
However, a completely synchronized system will not gain the full
potential provided by optimistic parallel\index{Parallel} synchronization.

A P/T Network is similar to a \acl{C/E}
network except that a P/T Net allows multiple tokens in a 
place and multiple tokens may be required to cause a transition to fire. 
Places are defined by the set $S$ and transitions by the set $T$.
The operation of a P/T network can be described by a matrix. The
rows of the matrix represent places and the columns represent 
transitions. The last column of the matrix represents the 
current number of tokens in a place. Each element of the
matrix contains the number of tokens which either leave (negative
integer) or enter (positive integer) a place when the transition fires.
When a transition fires, the column corresponding to the
transition is added to the last column of the matrix. Thus, the
last column of the matrix changes as the number of nodes in each
place change.
The matrix representation of a P/T Network is shown in Matrix \ref{ptmatrix},
where $LP_n \in S$, $c_n \in T$ and $w_{i,j}$ is the weight or number of 
tokens required by link $j$ to fire or the number of tokens 
generated by place $i$. Note that $LP_n$ 
and $c_n$ bordering Matrix \ref{ptmatrix} indicate labels for rows and 
columns. Note also that there exists a duality between places and 
transitions such that places and transitions can be interchanged 
\cite[p. 13]{Peterson81}. 
P/T networks can be extended from the state representation of \acl{C/E}
networks to examine problems involving quantities of elements in
a system, such as producer/consumer problems.
The places in this analysis are analogous to \acl{LP}s because they
produce and consume both real and virtual messages.
Transitions in this analysis are analogous to connections between \acl{LP}s, 
and tokens to messages. The 
weight, or number of tokens, is $-w_{i,j}$ for outgoing tokens and $w_{i,j}$ 
for incoming tokens. The current marking, or expected value of the number 
of tokens held in each place is given in column vector $\vec{m_N}$. 
A transition to the next state is determined by $\vec{m_{N+1}} = \vec{m_N} + 
\vec{c_i}$ where $\vec{c_i}$ is the column vector of the transition that 
fired and $N$ is the current matrix index.

\ifisdraft
	\begin{table*}
\fi
\begin{equation}
\label{ptmatrix}
\centering
\textbf{$M_N$} =
    \bordermatrix{ & c_1    & c_2      & c_3     & ... & m_N \cr
              LP\index{LP}_1 & w_{1,1}& w_{1,2}  & w_{1,3} & ... & n_1 \cr
              LP\index{LP}_2 & w_{2,1}& w_{2,2}  & w_{2,3} & ... & n_2 \cr
              LP\index{LP}_3 & w_{3,1}& w_{3,2}  & w_{3,3} & ... & n_3 \cr
            \vdots & \vdots & \vdots   & \vdots  & \vdots & \vdots \cr}
\end{equation}
\ifisdraft
	\end{table*}
\fi

A global synchronic distance value $GSV = \max_{f_1, f_2 \subset T}
{\{\sigma(f_1, f2)\}}$ where $T$ consists of the set of all transitions.
The global synchronic distance is used to define a 
normalized synchronization measure. The global synchronization measure 
is the maximum synchronic distance in a P/T network and $\sigma_n(I_1,I_2) 
\in [0,1]$ is a normalized synchronization value shown in Equation 
\ref{normsig} where $\{I_n\}$ is a set of all incoming transitions
to a particular place.
A probability of being within tolerance is defined in vector 
$\vec{p}$ shown in Matrix \ref{pmatrix}. Each $LP_i$ along the
side of Matrix \ref{pmatrix} indicates a \acf{LP} and the 
$1 - P_{ot}$ along the top of Matrix \ref{pmatrix} indicates
$p_i$ values that are the individual probabilities that the
tolerance is not exceeded. The probability of out-of-tolerance 
rollback is discussed in more detail in Section \ref{stabsec}.
Let ($LP_i$, $c_j$) be the transition from $LP_i$ across
connection $c_j$.
After each transition of \textbf{$M_N$} from ($LP_i$, $c_j$), the
next value of $n_i$ which is the element in the $i^{th}$ row of the
last column of $M_N$ is ${\sigma_n(I_1,I_2) p_i}^{n_i}$. 

\begin{equation}
\sigma_n(I_1,I_2) = 1.0 - {\sigma(I_1,I_2) \over GSV}
\label{normsig}
\end{equation}

\ifisdraft
	\begin{table*}
\fi
\begin{equation}
\label{pmatrix}
\centering
\textbf{$\vec{p}$} =
    \bordermatrix{ & 1 - P_{ot} \cr                              
              LP\index{LP}_1 & p_1  \cr                                    
              LP\index{LP}_2 & p_2  \cr
              LP\index{LP}_3 & p_3  \cr
            \vdots & \vdots \cr}
\end{equation}
\ifisdraft
	\end{table*}
\fi

As $p_i^{n_i}$ approaches zero, the likelihood of an out-of-tolerance 
induced rollback\index{Rollback} increases. As $\sigma_n(I_1,I_2) p_i^{n_i}$ 
becomes very small, the likelihood of a rollback\index{Rollback} increases either 
due to violation of causality or an out-of-tolerance state value. The
$\sigma_n(I_1,I_2)$ value is treated as a probability because it has
the axiomatic properties of a probability. The axiomatic properties are 
that $\sigma_n(I_1,I_2)$ assigns a number greater than or equal to 
zero to each synchronic value, $\sigma_n(I_1,I_2)$ has the value 
of one when messages are always in order, and $\sigma_n(A) + \sigma_n(B)
= \sigma_n(A \cup B)$ where $A$ and $B$ are mutually exclusive sets 
of transitions.

\begin{figure*}[htbp]
        \centerline{\psfig{file=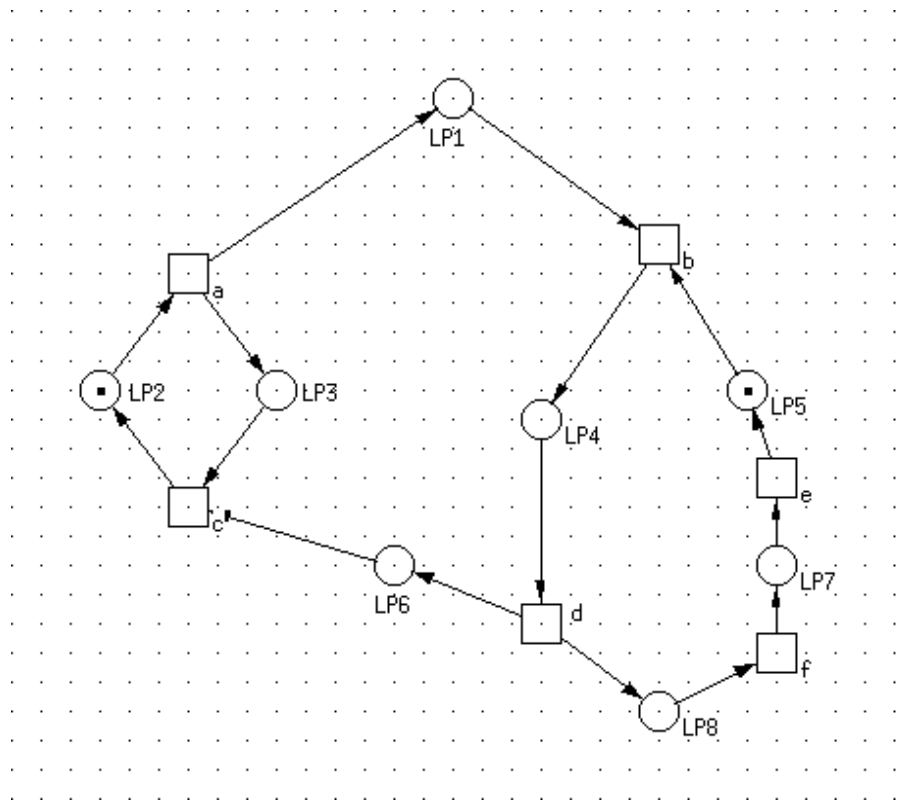,width=4.75in}}
        \caption{\label{synchex} Example of $P_{oo}$ Analysis.}      
\end{figure*}

A brief example is shown in Figure \ref{synchex}. The initial state
shown in Figure \ref{synchex} is represented in Matrix 
\ref{synchextab1}. The \acl{GSV} of this network is four.
The tolerance vector for this example
is shown in Vector \ref{pmatrixex}. Consider transition $a$ shown
in Figure \ref{synchex}; it
is enabled since tokens are available in all of its inputs. 
The element in the $\vec{p}$ column vector shown in Vector \ref{pmatrixex} 
is taken to the power of the corresponding elements of the column vector 
$\vec{a}$ in Matrix \ref{synchextab1} that are greater than zero 
($p_i^{n_i}$). This is the probability that all
messages passing through transition $a$ arrive within tolerance.
All columns of rows of $\vec{a}$ that are greater than zero
that have greater than zero values form the input set ($\{I_n\}$)
for $\sigma_n(I_1,I_2)$. Since transition $a$ has only one input,
$\sigma_n(\{a\})$ is one.
When transition $a$ fires, column vector $\vec{a}$ is added to column
vector $\vec{m_0}$ to generate a new vector $\vec{m_1}$. Matrix 
\ref{synchextab2} results after transition $a$ fires. Continuing
in this manner, Matrix \ref{synchextab3} shows the result after
transition $b$ fires. Since $\sigma_n(\{b\})$ is one, row
$LP_4$ of $\vec{m_2}$ is 0.3. 

\ifisdraft
	\begin{table*}
\fi
\begin{equation}
\label{synchextab1}
\centering
\textbf{$M_0$} =
    \bordermatrix{           & a  & b  & c  & d & e & f & m_0 \cr 
              LP\index{LP}_1 & 1  & -1 & 0  & 0 & 1 & 0 & 0   \cr
              LP\index{LP}_2 & -1 & 0  & 1  & 0 & 0 & 0 & 1   \cr
              LP\index{LP}_3 & 1  & 0  & -1 & 0 & 0 & 0 & 0   \cr
              LP\index{LP}_4 & 0  & 1  & 0  & -1& 0 & 0 & 0   \cr
              LP\index{LP}_5 & 0  & -1 & 0  & 0 & 1 & 0 & 1   \cr
              LP\index{LP}_6 & 0  & 0  & -1 & 1 & 0 & 0 & 0   \cr
              LP\index{LP}_7 & 0  & 0  & 0  & 0 & -1& 1 & 0   \cr
              LP\index{LP}_8 & 0  & 0  & 0  & 1 & 0 & -1& 0   \cr}
\end{equation}
\ifisdraft
	\end{table*}
\fi

\ifisdraft
	\begin{table*}
\fi
\begin{equation}
\label{pmatrixex}
\centering
\textbf{$\vec{p}$} =
    \bordermatrix{ & 1 - P_{ot} \cr
              LP\index{LP}_1 & 0.7  \cr
              LP\index{LP}_2 & 0.2  \cr
              LP\index{LP}_3 & 0.3  \cr
              LP\index{LP}_4 & 0.4  \cr
              LP\index{LP}_5 & 0.6  \cr
              LP\index{LP}_6 & 0.4  \cr
              LP\index{LP}_7 & 0.2  \cr
              LP\index{LP}_8 & 0.1  \cr}
\end{equation}
\ifisdraft
	\end{table*}
\fi

\ifisdraft
	\begin{table*}
\fi
\begin{equation}
\label{synchextab2}
\centering
\textbf{$M_1$} =
    \bordermatrix{           & a  & b  & c  & d & e & f & m_1 \cr
              LP\index{LP}_1 & 1  & -1 & 0  & 0 & 1 & 0 & 0.7 \cr
              LP\index{LP}_2 & -1 & 0  & 1  & 0 & 0 & 0 & 0   \cr
              LP\index{LP}_3 & 1  & 0  & -1 & 0 & 0 & 0 & 0.3 \cr
              LP\index{LP}_4 & 0  & 1  & 0  & -1& 0 & 0 & 0   \cr
              LP\index{LP}_5 & 0  & -1 & 0  & 0 & 1 & 0 & 1   \cr
              LP\index{LP}_6 & 0  & 0  & -1 & 1 & 0 & 0 & 0   \cr
              LP\index{LP}_7 & 0  & 0  & 0  & 0 & -1& 1 & 0   \cr
              LP\index{LP}_8 & 0  & 0  & 0  & 1 & 0 & -1& 0   \cr}
\end{equation}
\ifisdraft
	\end{table*}
\fi

\ifisdraft
	\begin{table*}
\fi
\begin{equation}
\label{synchextab3}
\centering
\textbf{$M_2$} =
    \bordermatrix{           & a  & b  & c  & d & e & f & m_2 \cr
              LP\index{LP}_1 & 1  & -1 & 0  & 0 & 1 & 0 & 0   \cr
              LP\index{LP}_2 & -1 & 0  & 1  & 0 & 0 & 0 & 0   \cr
              LP\index{LP}_3 & 1  & 0  & -1 & 0 & 0 & 0 & 0.3 \cr
              LP\index{LP}_4 & 0  & 1  & 0  & -1& 0 & 0 & 0.3 \cr
              LP\index{LP}_5 & 0  & -1 & 0  & 0 & 1 & 0 & 0   \cr
              LP\index{LP}_6 & 0  & 0  & -1 & 1 & 0 & 0 & 0   \cr
              LP\index{LP}_7 & 0  & 0  & 0  & 0 & -1& 1 & 0   \cr
              LP\index{LP}_8 & 0  & 0  & 0  & 1 & 0 & -1& 0   \cr}
\end{equation}
\ifisdraft
	\end{table*}
\fi

The analysis presented in this section reduces the time and 
topological complexities characteristic of more explicit time analysis 
methods to simpler and more insightful matrix manipulations. The
method presented in this section is used in the following section 
to determine the probability of rollback\index{Rollback} due to out-of-order
messages, $P_{oo} = 1 - \sigma_n(I_1,I_2)$. 

Also, the worst case proportion of out-of-order messages ($X$) 
can be calculated as follows. The synchronic distance
($\sigma(I_1,I_2)$) is a measure of the maximum difference
in the rate of firing among transitions. The maximum possible
rate that $\sigma(I_1,I_2)$ can occur is the rate of the slowest
firing transition in sets $I_1,I_2$. Thus, $E[X] \le 
\min_{\{Transition \in I_1,I_2\}}\{rate(Transition)\}$ where $rate(I)$
is the rate at that transition $I$ fires.

\subsection{P/T Analysis of VNC}

\begin{figure*}[htbp]
        \centerline{\psfig{file=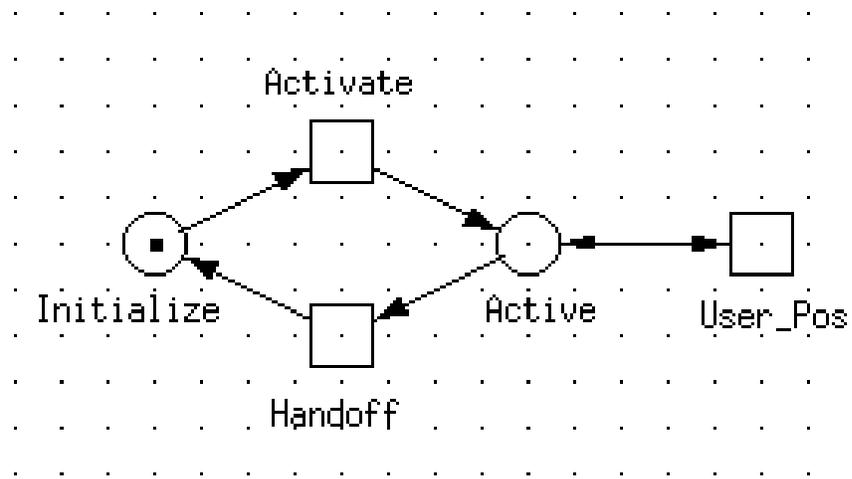,width=4.75in}}
\caption{\label{rncenet}RN NCP FSM as a C/E Net.}
\end{figure*}

\begin{figure*}[htbp]
        \centerline{\psfig{file=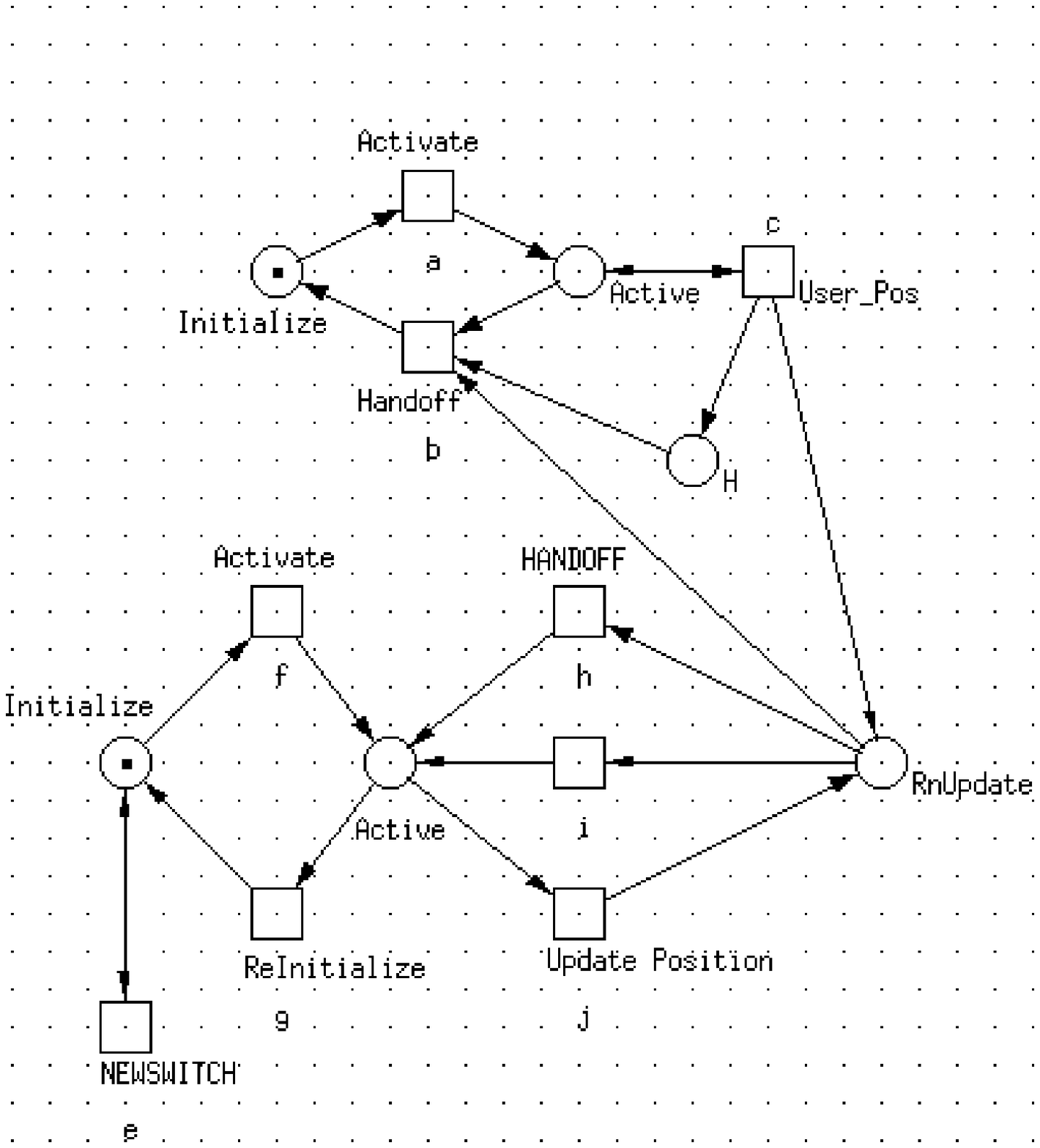,width=4.75in}}
        \caption{\label{rdrnpetr} Example of RDRN\index{RDRN} Causal Analysis.}
\end{figure*}

\begin{figure*}[htbp]
        \centerline{\psfig{file=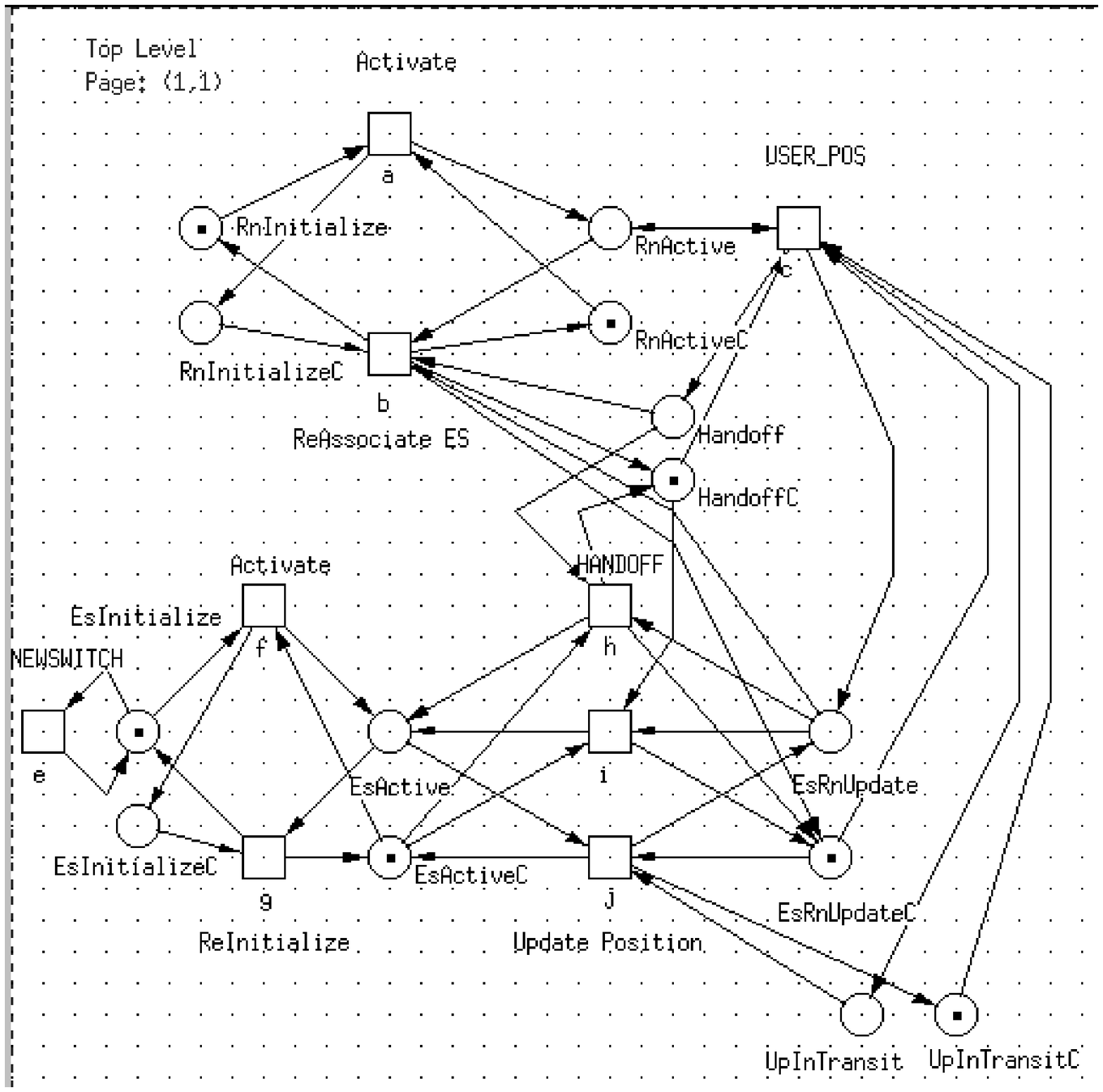,width=4.75in}}
        \caption{\label{rdrnprod} RDRN\index{RDRN} Causal Analysis with PROD.}
\end{figure*}

This section applies the synchronic measure analysis to the \acf{RDRN}
\acf{NCP}. Figure \ref{escenet} and Figure \ref{rncenet} show the 
\acl{C/E} Nets for an \acf{ES} and \acf{RN}. Figure \ref{rdrnpetr} shows 
a simple example of the Petri-Net analysis applied directly to the
\acl{NCP}. The net shown in Figure \ref{rdrnpetr} is a combination of
Figure \ref{escenet} and Figure \ref{rncenet} with a \emph{handoff}
place added to connect the two nets. 
The equivalent net shown in Figure \ref{rdrnprod} was analyzed using the 
PROD tool described in 
\cite{Varpaaniemi}. Note that complementary places exist in this net for 
two reasons. First, the tool that executes the reachability analysis 
assumes infinite capacity places. In order to implement single token 
capacity, the complement of each place was added. In addition the 
complement place provides the logical\index{Virtual Network Configuration!implementation} equivalent of a negative condition 
that is useful for controlling net behavior. 
Additional handoff\index{Handoff} places ($H$) can be added to hold $H$ tokens 
that represents an expected ratio of $H$ \textbf{USER\_POS}
packets for each \textbf{HANDOFF} packet. Using Figure \ref{rdrnpetr},
a relative measure for the out-of-orderedness of messages arriving
at each place can be determined. For example, messages arrive at
the \textbf{RnUpdate} place from transitions $j$ and $c$. The
synchronic distance between $j$ and $c$ is one, and the normalized
synchronic measure is $\sigma_n(\{j\},\{c\}) = 1 - \frac{1}{H}$.
Therefore these messages will arrive in order with a high likelihood if
$H$ is large. However,
messages arrive into the \textbf{Active} place of the \acl{ES} from
three different transitions, $f$, $h$, and $i$. Messages only arrive from
transition $f$ due to a re-initialization, re-initialization will only occur
with multiple \acl{ES}s and will not be considered in Figure \ref{rdrnpetr}.
That leaves transitions $h$ and $i$ that have a normalized 
synchronic value of $\sigma_n(\{h\},\{i\}) = 1 - \frac{(H-1)}{(H+1)}$. 
Thus, there is a high probability that a \textbf{HANDOFF} message may 
arrive out-of-order 
relative to a return to the \textbf{Active} place from an \textbf{RnUpdate},
because, except for the initial handoff\index{Handoff}, the ratio of 
\textbf{USER\_POS} to \textbf{HANDOFF} messages will usually be high.
Finally transitions $a$ and $c$ lead to the \textbf{Active} place for the 
\acl{RN} with a normalized synchronic value of $\sigma_n(\{a\},\{c\}) = 
1 - \frac{(H-1)}{(H+1)}$. This example has shown how synchronic distance 
is used to determine the likelihood of virtual\index{Virtual Network Configuration!verification} messages arriving out-of-order 
at an \acl{LP} and determines the rate of occurrence of out-of-order messages
($X$) that is simply the rate at that \textbf{HANDOFF} messages arrive
at \textbf{RnUpdate}. Therefore, $E[X]$ is less than or equal to the
rate at which \textbf{HANDOFF} messages arrive at \textbf{RnUpdate}.

\section{Expected Speed\index{Speed}up: $\eta$}

This section analyzes the primary benefit of \acl{VNC} which is
speedup. There are many factors which influence speedup\index{Speedup} including
out-of-order message probability, out-of-tolerance state value
probability, rate of virtual\index{Virtual Network Configuration!verification} messages entering the system, 
task execution time, task partitioning into \acl{LP}s, rollback\index{Rollback}
overhead, prediction accuracy as a function\index{Function} of distance into the
future which predictions are attempted, and the effect of 
parallelism and optimistic synchronization. All of these factors
are considered in this section beginning with a direct analysis using the 
definitions from optimistic simulation.

The definition of \acf{GVT} can be applied to determine the relationship
among expected task execution time ($\tau_{task}$), the real time at 
which the state was cached ($t_{SQ}$), and real\index{Real} time ($t$). 
Consider 
the value ($V_v$) which is cached at real time $t_{SQ}$ in the \acf{SQ}
resulting from a particular predicted event. 
For example, refer to Figures \ref{vncex4} through \ref{vncex6} and 
notice that state queue values may be repeatedly added and discarded
as \acl{VNC} operation proceeds in the presence of rollback\index{Rollback}.
As rollback\index{Rollback}s occur, values for a particular predicted event may 
change, converging to the real\index{Real} value ($V_r$).
For correct operation of \acl{VNC}, $V_v$ should approach $V_r$ as
$t$ approaches $GVT(t)$ where $GVT(t)$ is the
$GVT$ of the \acl{VNC} system at time $t$. Explicitly, this is
$\forall \epsilon > 0$ $\exists \delta > 0$ s.t. $|f(t) - f(GVT(t))| < \epsilon 
\Rightarrow 0 < |GVT(t) - t| < \delta$ where $f(t) = V_r$ and $f(GVT(t)) = V_v$.
$f(t)$ is the prediction function\index{Function} of a driving\index{Driving Process} 
process. The purpose and function\index{Function} of the driving\index{Driving Process} 
process has been explained in Section \ref{vncorg}. Because \acl{VNC} will
always use the correct value when the predicted time ($\tau$) equals the
current real time ($t$) and it is assumed that the predictions will become
more accurate as the predicted  time of the event approaches the current time, 
the reasonable 
assumption is made that $\lim_{\tau \rightarrow t} f(\tau) = V_v$.
In order for the \acl{VNC} system to always look ahead, 
$\forall t$ $GVT(t) \ge t$. This means that $\forall n \in \{LPs\}$ and
$\forall t$ $LVT_{lp_n}(t) \ge t$ and $\min_{m \in \{M\}} \{ m \} \ge t$ where
$m$ is the receive time of a message, $M$ is the set of messages in
the entire system and $LVT_{lp_n}$ is the \acl{LVT} of the $n^{th}$ \acl{LP}. 
In other words, the \acf{LVT} of each \acf{LP} must be greater than
or equal to real\index{Real} time and the smallest message \acl{TR} not yet
processed must also be greater than or equal to real\index{Real} time. The
smallest message \acl{TR} could cause a rollback to that time.
This implies that $\forall n,t$ $LVT_{dp_n}(t) \ge t$. In other words,
this implies that the \acf{LVT} of each driving\index{Driving Process} 
process must be greater than or equal to real\index{Real} time. An out-of-order
rollback\index{Rollback} occurs when $m < LVT\index{LVT}(t)$. The largest saved
state time such that $t_{SQ} < m$
is used to restore the state of the \acl{LP}, where $t_{SQ}$ is the real
time the state was saved. Then the expected task execution
time ($\tau_{task}$) can take no longer than $t_{SQ} - t$ to complete
in order for $GVT$ to remain ahead of real\index{Real} time. Thus, a constraint
between expected task execution time ($\tau_{task}$), state save time
($t_{SQ}$), and real\index{Real} time ($t$) has been defined. The next 
section considers the effect of out-of-tolerance state values on
the rollback\index{Rollback} probability and the concept of stability in \acl{VNC}.
\subsection{Stability}
\label{stabsec}

Stability in \acl{VNC} is related to the ability of the system to reduce the 
number of rollback\index{Rollback}s. An unstable system is one in which there 
exists enough rollback\index{Rollback}s to cause the system to take longer
than real\index{Real}-time to reach the end of the \acf{SLW}.
Rollback is caused by the arrival of a message which should have been        
executed in the past and by out-of-tolerance states. In either case, messages
which had been generated prior to the rollback\index{Rollback} are false\index{False Message} messages. Rollback\index{Rollback}   
is contained by sending anti-messages to cancel the effects of false\index{False Message} messages. 
The more quickly the anti-messages can overtake the effect of false\index{False Message} messages,
the more efficiently rollback\index{Rollback} is contained.

\begin{figure*}[htbp]
        \centerline{\psfig{file=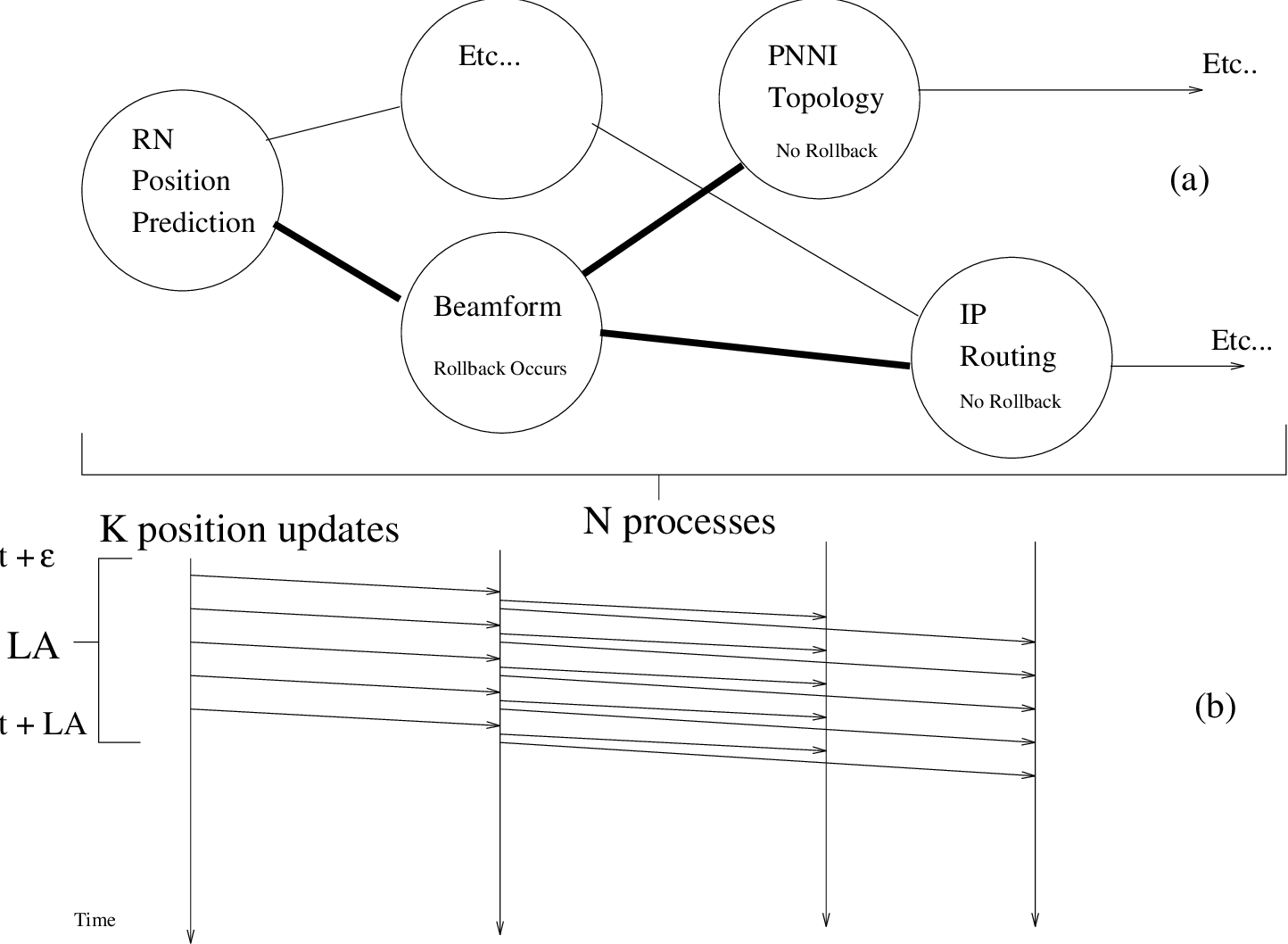,width=6in}}
        \caption{Relation Between Processes and Virtual\index{Virtual Network Configuration!verification} Messages.}
        \label{parms}
\end{figure*}

Consider a specific scenario in the \acl{RDRN} \cite{BushRDRN} of a single
configuration processor on the mobile host and a single configuration
processor on the base station.
All processing must occur in strict sequential order and the
only parallel\index{Parallel}ism occurs between the processing on the mobile host and
base station.
The simplified feasibility analysis\index{Feasibility analysis} presented
here assumes exponential processing times for each task.
The system is driven by the location predictions as shown in Figure
\ref{parms}. Time\index{Time} lines which indicate virtual\index{Virtual Network Configuration!verification} position updates beyond
current time ($t$) are drawn in the figure below the
corresponding processes. $\Lambda$\index{$\Lambda$} is the length of the
\acl{SLW}\index{VNC!sliding lookahead window} 
($SLW = (t,t+\Lambda]$)
which extends from the end of current time ($t+\epsilon$) to $t+\Lambda$
and contains $K$ position updates\index{RDRN!position updates}.
There are $N$ processes which process the $K$ inputs. Examples of these
processes are mobile host position prediction,
beamforming, \acl{PNNI}, IP\index{IP} Routing, \acl{ATMARP} caching
and Mobile IP\index{IP} configuration as shown in Figure \ref{parms}.

A cause of rollback\index{Rollback}s in \acl{VNC} is real\index{Real} messages which are out 
of tolerance\index{VNC!tolerance}. Those processes which require
a higher degree of tolerance are most likely to rollback\index{Rollback}. A worst case
probability of out-of-tolerance rollback\index{Rollback} for a single process, shown
in Equation \ref{cheb}, is based on Chebycheff's Inequality \cite{Papoulis}
from basic probability. The variance\index{Variance} of the
data is $\sigma^2$ and $\Theta$ is the acceptable 
tolerance\index{VNC!tolerance}
for a configuration process. Therefore, the performance gains of \acl{VNC} 
are reduced as a function\index{Function} of $P_{ot}$. At the cost of increasing the 
accuracy of the driving\index{Driving Process} process(es), that is, decreasing $\sigma^2$ in 
Proposition \ref{prb_hyp}, $P_{ot}$ becomes small thus increasing
the performance gain of \acl{VNC}.


\begin{hyp}
\label{prb_hyp}

The probability of rollback\index{Rollback} of an \acl{LP}\index{LP} is

\begin{equation}
P_{ot} \le \frac{\sigma^2}{\Theta^2}
\label{cheb}
\end{equation}

where $P_{ot}$ is the probability of out-of-tolerance rollback for
an \acl{LP}, $\sigma^2$ is the variance in the amount of error, and
$\Theta$ is the tolerance allowed for error.

\end{hyp}

If the $K$th position update predicts a handoff\index{Handoff}, then 
the total expected time for the prediction computation is $KN 
\frac{1}{\mu}$, where $N$ is the number of processes and $\frac{1}{\mu}$ 
is the expected exponential processing time per process and there
exists strict synchronization between processes. 

The expected time between rollback\index{Rollback}s for the \acl{VNC} system is critical
for determining its feasibility. The probability of 
rollback\index{VNC!rollback probability} for all processes is the
probability of out-of-order message occurrence and the probability of 
out-of-tolerance state values ($P_{rb} = P_{oo} + P_{ot}$). The 
received message rate per \acl{LP} is $R_m$ and there are $N$ \acl{LP}s.
The expected inter-rollback time for the system is shown in Equation 
\ref{erb}.


\begin{hyp}
\label{etrb_hyp}

The expected inter-rollback time is

\begin{equation}
\label{erb}
T_{rb} = \frac{1}{\lambda_{rb}} = \frac{1}{R_m N P_{rb}}
\end{equation}

where $T_{rb}$ is the expected inter-rollback time, $\lambda_{rb}$ is
the expected rollback rate, $R_{m}$ is the received message rate per
\acl{LP}, there are $N$ \acl{LP}s, and $P_{rb}$ is the probability of
rollback per process.

\end{hyp}

\subsection{Single Processor Logical Processes}

\begin{figure*}[htbp]
        \centerline{\psfig{file=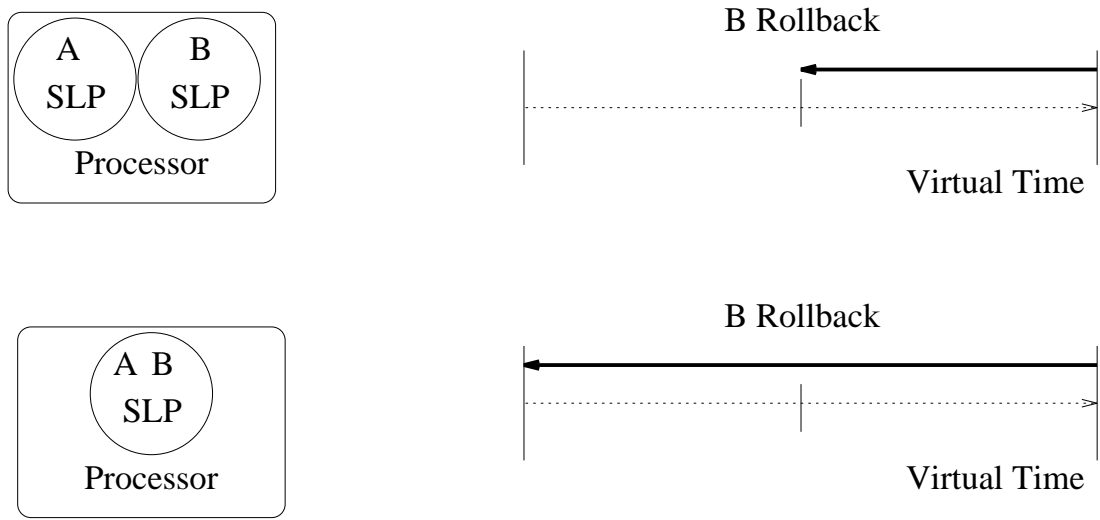,width=6in}}
        \caption{Single and Multiple Processor Logical Process System.}
        \label{slpcombo}
\end{figure*}

The contribution of this section is the analysis of single processor
logical process. Multiple \acl{LP}es on a single processor loose any gain in
concurrency since they are being served by a single processor, however, 
the \acl{LP}es maintain the \acl{VNC} lookahead if partitioned properly. The 
\acl{NCP}'s link management daemon code runs as a multiple process system
on single processors. In the current architecture\footnote{The 
revision control system identifier for \acl{NCP} code at the time this
document was written was \emph{esctrld.c,v 1.7 
1997/02/01 20:31:25} and \emph{rnctrld.c,v 1.8 1997/02/01 
20:31:25}.} each remote node link management daemon consists of three 
processes: a position prediction process, a control and \acf{SNMP}
agent process, and a beamform and stack activation process. Each edge switch
link management daemon consists of two processes: a control and \acl{SNMP}
agent process, and beamform and stack activation process.
Thus there are logical\index{Virtual Network Configuration!implementation} processes 
communicating on single processors and across multiple processors.
Multiple \acl{LP}es executing on a single
processor will not have speedup\index{Speedup} due to parallel\index{Parallel}ism because a single
processor must serve all the \acl{LP}es.
Because these \acl{LP}es reside on a single processor,
they are not operating in parallel\index{Parallel} as \acl{LP}es are assumed
to do in an optimistic system, thus
a new term needs to be applied to a task partitioned into \acl{LP}es on a
single processor. Each partition of tasks into \acl{LP}es on a single
processor is called a 
\acf{SLP}. In the upper portion of Figure \ref{slpcombo}, a task
has been partitioned into two logical\index{Virtual Network Configuration!implementation} processes. The same task exists in the
lower portion of Figure \ref{slpcombo} as a single \acl{LP}. If task B must
rollback because of an out-of-tolerance result, the entire single \acl{LP} must
rollback, while only the \acl{LP} for task B must rollback\index{Rollback} in the multiple
\acl{LP} case. Thus partitioning a task into multiple \acl{LP}es can save time
compared to a single \acl{LP} task.
Thus, \textbf{without considering parallel\index{Parallel}ism}, lookahead can be achieved by 
allowing the sequential system to work ahead while individual tasks 
within the system are allowed to rollback\index{Rollback}. Only tasks which deviate 
beyond a given preconfigured
tolerance are rolled back. Thus entire pre-computed and cached results 
are not lost due to inaccuracy, only parts of pre-computed results must 
be re-computed.
This section discusses single processor logical\index{Virtual Network Configuration!implementation} processes, the next section
reviews multiple processor logical\index{Virtual Network Configuration!implementation} processes, and a third section
discusses the combination of a single/multiple processor system.

\begin{figure*}[htbp]
        \centerline{\psfig{file=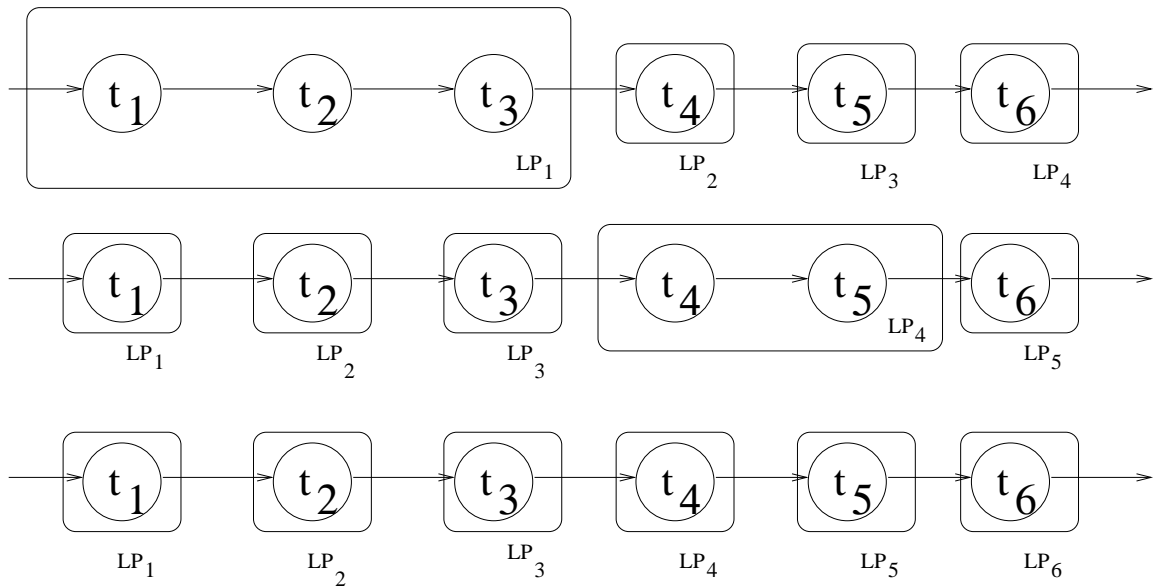,width=6in}}
        \caption{Possible Partitioning of Tasks into Logical Processes on a
                 Single Processor.}
        \label{vncpart}
\end{figure*}

Consider the optimal method of partitioning a single processor system into 
\acf{SLP}es in order to obtain speedup\index{Speedup} over a single \acl{LP} process. Assume 
$n$ tasks,
$task_1, ..., task_n$, with expected execution times of $\tau_1, ..., \tau_n$,
and that $task_n$ depends on messages from $task_{n-1}$ with a tolerance of 
$\Theta_n$. This is the largest error allowed in the 
input message such that the output is correct. Using the results from 
Proposition \ref{prb_hyp}, it is possible to 
determine a partitioning of tasks into logical\index{Virtual Network Configuration!implementation} processes such that speedup\index{Speedup} 
is achieved over operation of the same tasks encapsulated in a single 
\acl{LP}. Figure \ref{vncpart} shows possible groupings of the same set of 
six tasks into logical\index{Virtual Network Configuration!implementation} 
processes. It is hypothesized that the tasks most likely to 
rollback\index{Rollback} and which take the greatest amount of time to execute 
should be grouped together within \acl{SLP}es to minimize the
rollback time. 
There are $2^{n-1}$ possible groupings of tasks into \acl{SLP}es where $n$
is the number of tasks and message dependency among the tasks is maintained.
Those tasks least likely to rollback\index{Rollback} and those 
which execute quickly should be grouped within a single \acl{SLP}
to reduce the overhead\index{Virtual Network Configuration!overhead} of rollback\index{Rollback}. For example, if all the tasks 
in Figure \ref{vncpart} have an equal probability of rollback\index{Rollback} 
and $\tau_2 \gg \max \{ \tau_1, \tau_3, ... \}$ then the tasks should be 
partitioned such that $task_2$ is in a separate SLP:
$( task_1 | task_2 | task_3 ... task_n )$ where ``$|$'' indicates the 
grouping of tasks into sequential logical\index{Virtual Network Configuration!implementation} processes. 

\begin{figure*}[htbp]
        \centerline{\psfig{file=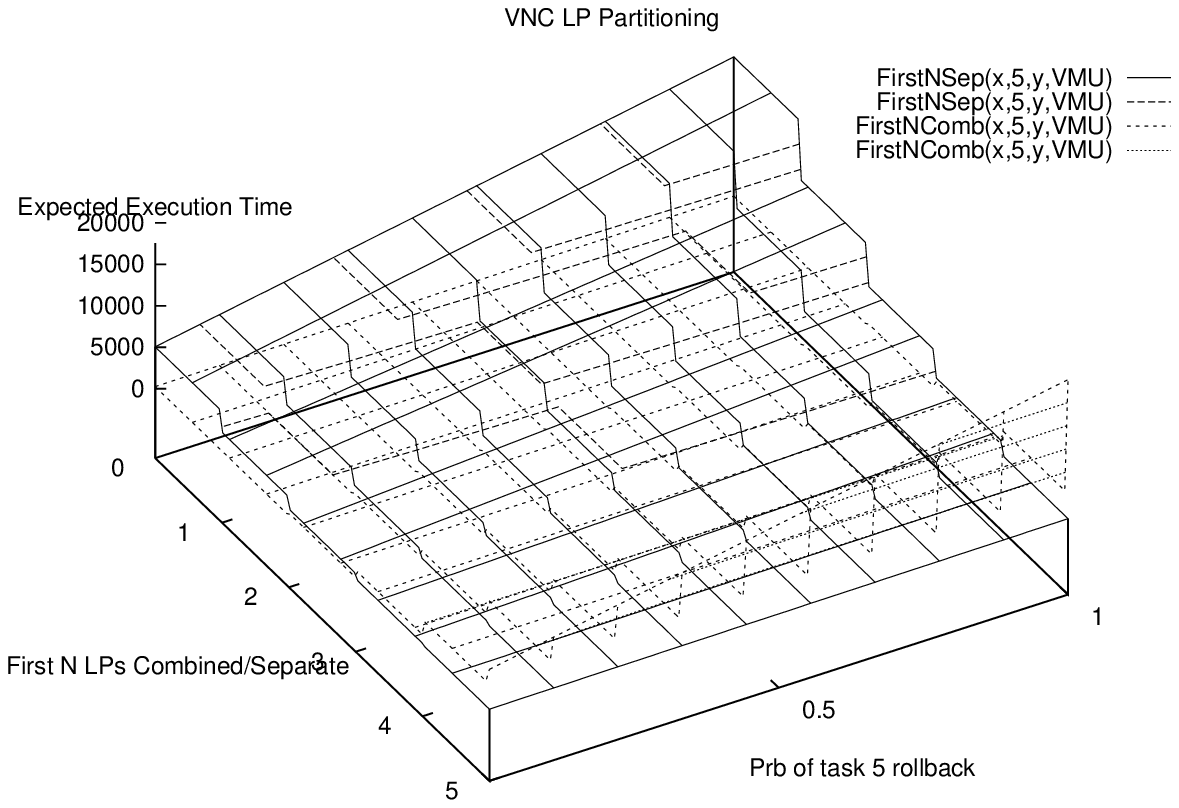,width=5.0in}}
        \caption{Optimal \acl{SLP} Partitioning.}
        \label{maxpart}                                         
\end{figure*}

For example, the expected \acl{LP} execution time for five tasks with 
equal probabilities of rollback\index{Rollback} of 0.1
are shown in Figure \ref{maxpart}. It is assumed that these tasks communicate
in order starting from Task 1 to Task 5 in order to generate a result. 
In Figure \ref{maxpart}, the x-axis indicates the boundary between
task partitions as the probability of rollback of task 5 is varied. 
With an x-value of 3, the solid surface
shows the expected execution time for the first three tasks combined
within a single \acl{LP} and the the remainder of the tasks encapsulated 
in separate \acl{LP}es. The dashed surface shows the first three tasks
encapsulated in separate \acl{LP}es and the remainder of the tasks 
encapsulated within an \acl{LP}. The graph in Figure \ref{maxpart} 
indicates a minimum for both curves when the high probability 
rollback\index{Rollback} tasks are encapsulated
in separate \acl{LP}es from the low probability of rollback\index{Rollback} 
tasks. As the probability of rollback\index{Rollback} increases, the 
expected execution time for all five processes is minimized when Task 5 
is encapsulated in a separate \acl{LP}.

\subsubsection{\acl{RDRN} \acl{NCP} Task Partition Analysis}

Consider an example from the \acf{RDRN} \acf{NCP} operation. Assume
the beamform table calculation time is exponentially distributed\index{Distributed} with
mean $1 \over \mu_{bT}$. Assume the stack activation operation
is also exponentially distributed\index{Distributed} with mean $1 \over \mu_{h}$.
Beamforming has a rollback\index{Rollback} probability of $P_{bT}$ and takes time
$\tau_{bT}$ to rollback\index{Rollback}. Also, stack activation time has a rollback\index{Rollback} 
probability of $P_{h}$ and takes time $\tau_{h}$ to rollback\index{Rollback}. If
both operations are encapsulated by a single logical\index{Virtual Network Configuration!implementation} process, then the
expected time of operation is shown in Equation \ref{splpcombined}. 
If each operation is encapsulated in a separate LP\index{LP}, then the expected time
is shown in Equation \ref{splpseparate}. Equations \ref{splpcombined}
and \ref{splpseparate} are formed by the sum of the expected time
to execute the task which is the first term and the rollback\index{Rollback} time
which is the second term. The probability of rollback\index{Rollback} in the combined
\acl{LP} is the probability that either task will rollback\index{Rollback}. Therefore,
the expected execution time of the tasks encapsulated in separate
\acl{LP}es is smaller since $\tau_{separate} < \tau_{combined}$.

\begin{figure*}
\begin{equation}
\tau_{combined} = ({1 \over \mu_{bT}} + {1 \over \mu_{h}}) + 
({1 \over \mu_{bT}} + {1 \over \mu_{h}})
(P_{bT} + P_{h}) (\tau_{bT} + \tau_{h})
\label{splpcombined}
\end{equation}
\end{figure*}

\begin{figure*}
\begin{equation}
\tau_{separate} = ({1 \over \mu_{bT}} + {1 \over \mu_{bT}} P_{bT} \tau_{bT}) +
({1 \over \mu{h}} + {1 \over \mu{h}} P_{h} \tau_{h})
\label{splpseparate}
\end{equation}
\end{figure*}

The grouping of tasks into
\acl{SLP}es can be done dynamically, that is, while the system is in 
operation. This dynamic adjustment is currently outside the scope
of this research but related to optimistic simulation load balancing 
\cite{Glaz93, Glaz93b}
and the recently developed topic of optimistic simulation dynamic 
partitioning \cite{Boukerche94, Konas95}.

\subsection{Single Processor Logical Process Prediction Rate}
\label{slprate}

The \acf{LVT} is a particular \acl{LP}'s notion of the current 
time. In optimistic simulation the \acl{LVT} of individual processes may be
different from one another and generally precede at a much faster
rate than real\index{Real} time. Thus, the rate at which a \acf{SLP} system can predict 
events (prediction rate) is the rate of change of the \acl{SLP}'s \acl{LVT} 
with respect to real\index{Real} time.
Assume a driving\index{Driving Process} process whose virtual\index{Virtual Network Configuration!verification} message generation rate is           
$\lambda_{vm}$. The \acl{LVT}
is increased by the expected amount $\Delta_{vm}$ every 
$\frac{1}{\lambda_{vm}}$ 
time units. The expected time spent executing the task is $\tau_{task}$.
The random variables $X$ and $Y$ are the proportion of messages which
are out-of-order and out-of-tolerance respectively.
The expected real\index{Real} time to handle 
a rollback\index{Rollback} is $\tau_{rb}$. Then the \acl{SLP}'s \acl{LVT} advances at the 
expected rate shown in Proposition \ref{splp_speed}. 


\begin{hyp}[Single Processor Logical Process Speed\index{Speed}]
\label{splp_speed}
The average prediction rate of a single logical\index{Virtual Network Configuration!implementation} processor system is

\begin{equation}
S_{cache} = {LVT \over t} = \lambda_{vm} (\Delta_{vm} - \slowSLPterm)
\label{splp_rate}
\end{equation}

where
the virtual message generation rate is $\lambda_{vm}$, the expected
lookahead per message is $\Delta_{vm}$, the proportion of out-of-order
messages is $X$, the proportion of out-of-tolerance messages is $Y$,
$\tau_{task}$ is the expected task execution time in real time,
$\tau_{rb}$ is the expected rollback overhead time in real time,
$LVT$ is the \acf{LVT}, and $t$ is real time.

\end{hyp}

\begin{figure*}[htbp]
        \centerline{\psfig{file=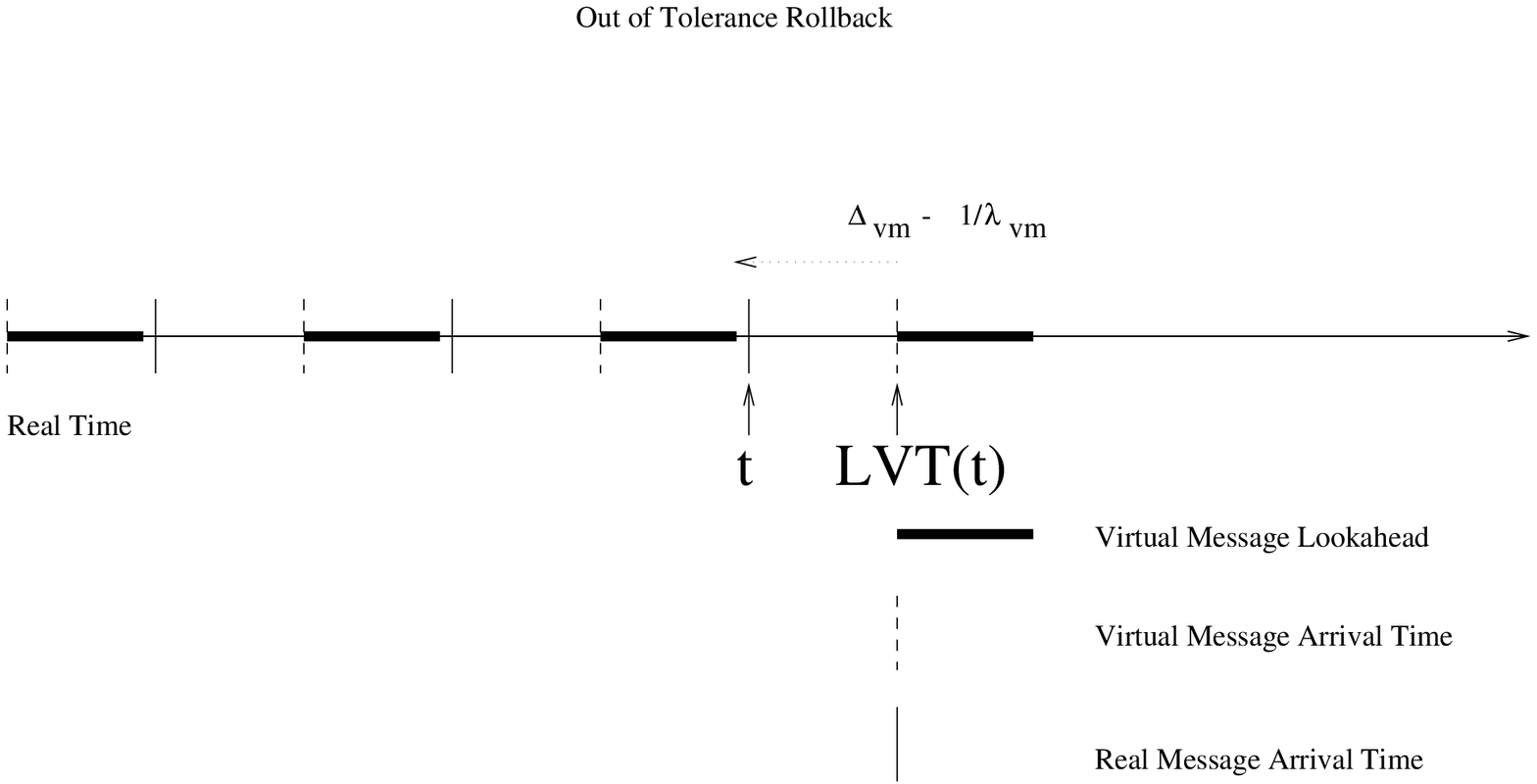,width=6.0in}}
        \caption{Out-of-Tolerance Rollback\index{Rollback}.}
        \label{troll}
\end{figure*}

In Proposition 
\ref{splp_speed} the expected lookahead per message ($\Delta_{vm}$) is
reduced by the real\index{Real} time taken to process the message ($\tau_{task}$).
The expected lookahead is also reduced by the time to re-execute the
task ($\tau_{task}$) and the rollback\index{Rollback} time ($\tau_{rb}$) times the
the proportion of occurrences of an out-of-order message ($E[X]$)
which results in the term $(\tau_{task} + \tau_{rb}) E[X]$.
Finally, the derivation of the ($\Delta_{vm} - {1 \over \lambda_{vm}}) 
E[Y]$ term is shown in Figure \ref{troll}. In Figure \ref{troll}, a real\index{Real}
message arrives at time $t$ which is labeled in Figure \ref{troll}. Note 
that real\index{Real} time $t$ and \acf{LVT} are both shown on the same time axis 
in Figure \ref{troll}. The current \acl{LVT} of the process is labeled at 
time LVT\index{LVT}($t$) in Figure \ref{troll}. The dotted line in Figure \ref{troll}
represents the time $\Delta_{vm} - \frac{1}{\lambda_{vm}}$ which
is subtracted from the \acl{LVT} when an out-of-tolerance
rollback occurs. The result of the subtraction of $\Delta_{vm} - 
\frac{1}{\lambda_{vm}}$ from the $LVT(t)$ results in the \acl{LVT} 
returning to real\index{Real} time as required by the algorithm. The virtual\index{Virtual Network Configuration!verification} message 
inter-arrival time is
$\frac{1}{\lambda_{vm}}$. Note that the ($\Delta_{vm} - {1 \over 
\lambda_{vm}}) E[Y]$ term causes the speedup\index{Speedup} to approach one based on 
the frequency of out-of-tolerance rollback\index{Rollback} ($E[Y]$).

In the specific case of the \acf{RDRN} \acf{RN} beamforming\index{Beamforming} \acl{SLP}
with no prediction error the expected prediction rate is 3.5 virtual\index{Virtual Network Configuration!verification} 
seconds per second given \oparms. The speedup\index{Speedup}
quantified in Equation \ref{splp_rate} is labeled $S_{cache}$
because there is also a synchronization speedup\index{Speedup}, $S_{parallel}$ which
is defined later.

\subsection{Sensitivity}
\label{sens}

If the proportion of out-of-tolerance messages, $Y$, cannot be reduced to zero,
the virtual\index{Virtual Network Configuration!verification} message
generation rates and expected virtual\index{Virtual Network Configuration!verification} message lookahead times can be
adjusted in order to improve speedup\index{Speedup}.
Given the closed form expression for \acl{VNC} speedup\index{Speedup} in Proposition
\ref{splp_speed}, it is important to determine the optimal values
for each parameter, particularly $\lambda_{vm}$ and $\Delta_{vm}$ and 
in addition, the sensitivity of each parameter.
Sensitivity information will indicate which parameters most affect the 
speedup. The parameters which most affect the speedup\index{Speedup} are the ones that 
will yield the best results if optimized.

A technique which optimizes a constrained objective function\index{Function} and
which also determines the sensitivity of each parameter within the
constraints is the Kuhn-Tucker\index{Kuhn-Tucker} method \cite[p. 314]{Luen}. The reason for
using this method rather than simply taking the derivative of
Equation \ref{splp_rate} is that the optimal value must reside
within a set of constraints. Depending on the particular application
of \acf{VNC}, the constraints may become more complex than those
shown in this example. The constraints for this example are
discussed in detail later. The sensitivity results appear as a by-product
of the Kuhn-Tucker\index{Kuhn-Tucker} method. The first order necessary conditions for 
an extremum using the Kuhn-Tucker\index{Kuhn-Tucker} method are listed in Equation \ref{kt}. 
The second 
order necessary conditions for an extremum are given in Equation \ref{sonc}
where L must be positive semi-definite over the active constraints and
$L$, $F$, $H$, and $G$ are Hessian\index{Hessian}s. The second
order sufficient conditions are the same as the first order necessary
conditions and the Hessian\index{Hessian} matrix in Equation \ref{sonc} is 
positive definite on the subspace $M = \{y:\nabla h(x) y = 0,
\nabla g_j(x) y = 0$ for all $j \in J\}$, where $J = \{j: g_j(x) = 0,
\mu_j \ge 0\}$. The sensitivity is determined by the Lagrange multipliers,
$\lambda^T$ and $\mu^T$. The Hessian\index{Hessian} of the objective function\index{Function} and
of each of the inequality constraints is a zero matrix, thus, the
eigenvalues $L$ in Equation \ref{sonc} are zero and the matrix
is clearly positive definite satisfying both the necessary and
sufficient conditions for an extremum.

\begin{figure*}
\begin{eqnarray}
\label{kt}
\nabla f(x^*) + \lambda^T \nabla h(x^*) + \mu^T \nabla g(x^*) = 0 & &\\ \nonumber
\mu^T g(x^*) = 0                                      & &\\ \nonumber
\mu \ge 0                                             & &\\ \nonumber
\end{eqnarray}
\end{figure*}

\begin{figure*}
\begin{equation}
\hess{L} = \hess{F} + \lambda^T \hess{H} + \mu^T \hess{G}
\label{sonc}
\end{equation}
\end{figure*}

The function\index{Function} $f$ in Equation \ref{kt} is the \acl{VNC} speedup\index{Speedup} given
in Equation \ref{splp_rate}. The matrix $h$ does not exist, because
there are no equality constraints, and the matrix $g$ consists of
the inequality constraints which are specified
in Equation \ref{gcon}. 
The bounds chosen in Equation \ref{gcon} are based on measurements from
a specific application, \acl{RDRN} configuration.
Clearly the upper bound constraints on $E[X]$ and 
$E[Y]$ are the virtual\index{Virtual Network Configuration!verification} message rate. The constraints for
$\tau_{task}$ and $\tau_{rb}$ are based on measurements of the
task execution time and the time to execute a rollback\index{Rollback}.
The maximum value for $\lambda_{vm}$ is determined by the rate
at which the virtual\index{Virtual Network Configuration!verification} message can be processed. Finally, the
maximum value for $\Delta_{vm}$ is determined by the required
caching period. If $\Delta_{vm}$ is too large, there may be no
state in the \acf{SQ} with which to compare an incoming
real message.

From inspection
of Equation \ref{splp_rate} and the constraint shown in Equation \ref{c1},
the constraints from \ref{gcon} are $\Delta_{vm} = 45.0$, $\tau_{task} = 5.0$, 
$\tau_{rb} = 1.0$, $E[X] = 0.0$, $E[Y] = 0.0$ which
results in the optimal solution shown in Equations \ref{sensol}.
The Lagrange multipliers $\mu_1$ through $\mu_6$ show that
$E[Y]$ $(-\mu_6 = -8.0)$, $\lambda_{vm}$ $(-\mu_1 = -40.0)$,
and $E[X]$ $(-\mu_5 = -1.2)$
have the greatest sensitivities. Therefore, reducing the out-of-tolerance
rollback has the greatest effect on speedup\index{Speedup}. In the next section
the effect of optimistic synchronization on speedup is derived.

\begin{eqnarray}
\label{c1} & \lambda_{vm} = {1 \over {\tau_{task} + (\tau_{task} + \tau_{rb}) 
  E[X] + \left(\Delta_{vm} - {1 \over \lambda_{vm}}\right) E[Y]}} & \\
\label{gcon} & 0.0 \le \lambda_{vm} \le {1 \over {\tau_{task} + (\tau_{task} 
+ \tau_{rb}) E[X] + \left(\Delta_{vm} - {1 \over \lambda_{vm}}\right) 
E[Y]}} & \\ 
           & 0.1 \le \Delta_{vm} \le 45.0 & \\
           & 5.0 \le \tau_{task} \le 10.0 & \nonumber \\
           & 1.0 \le \tau_{rb} \le 2.0    & \nonumber \\
           & 0.0 \le E[X] \le 1.0         & \nonumber \\
           & 0.0 \le E[Y] \le 1.0         & \nonumber \\
\label{sensol} & \lambda_{vm} = 1.0, \mu_1 = 40.0 & \\
               & \Delta_{vm} = 45.0, \mu_2 = 0.2  & \nonumber \\
               & \tau_{task} = 0.0, \mu_3 = 0.2   & \nonumber \\
               & \tau_{rb} = 0.0, \mu_4 = 0.0     & \nonumber \\
               & E[X] = 0.0, \mu_5 = 1.2          & \nonumber \\
               & E[Y] = 0.0, \mu_6 = 8.0          & \nonumber
\end{eqnarray}

\subsection{Sequential Execution Multiple Processors}

\begin{figure*}[htbp]
        \centerline{\psfig{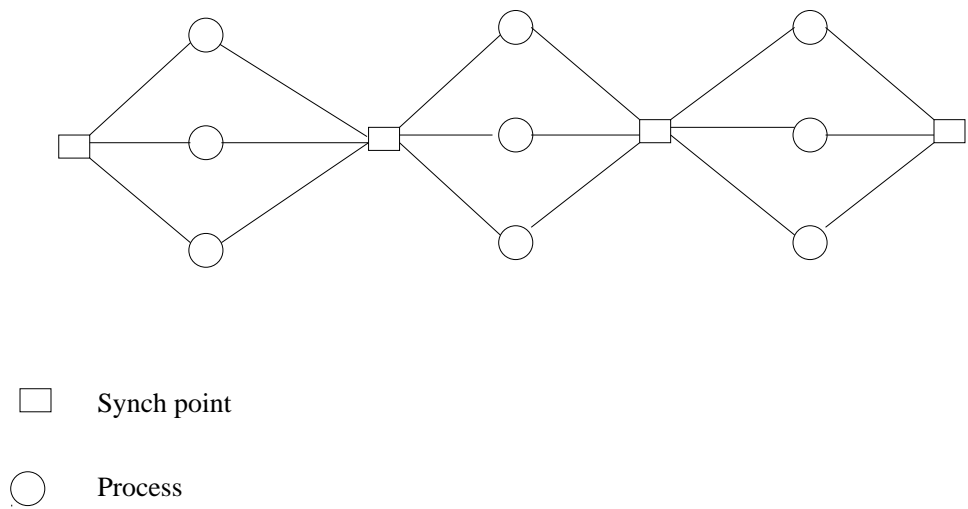}}
        \caption{Sequential Model of Operation.}
        \label{seqmodel}
\end{figure*}

A comparison of optimistic synchronization with sequential synchronization
has not been found in the literature because little work
has been done which combines optimistic synchronization with a
real time system with the exception of hybrid systems such as
the system described in \cite{Bagrodia91}. The hybrid system
described in \cite{Bagrodia91} is used as a design technique in which
distributed simulation \acf{LP}s are gradually replaced with
real system components allowing the emulated system to be
executed as the system is built. It does not focus on predicting
events as in \acl{VNC}.
This section examines sequential execution of tasks which corresponds 
with non-\acl{VNC} operation as shown in Figure \ref{seqmodel} in order 
to compare it with the \acl{VNC} algorithm in the next section.
As a specific example related to predicting handoff\index{Handoff}, consider the case 
of $K$ position updates with $P$ processes required for each handoff\index{Handoff} and 
each process has an exponential processing time with average $\frac{1}{\mu}$. 
In the sequential case, the expected completion time should be $K$ times the 
summation of $P$ exponential distributions. The summation of $P$ exponential 
distributions is a Gamma Distribution as shown in the sequential execution
pdf in Equation \ref{ppdf}. The average time to complete $K$ tasks is 
shown in Equation \ref{seqavg}.

\begin{equation}
f_T(x|P,\mu) = \left\{ 
	\begin{array}{ll} 
		{\mu^P \over \Gamma(P)} x^{P-1} \exp^{- \mu x} & x > 0 \\
		0 & x \le 0 
	\end{array} 
		\right.
\label{ppdf}
\end{equation}

\begin{equation}
T_{seq} = K \int_0^{\infty} x f_T(x|P, \mu) dx
\label{seqavg}
\end{equation}
\subsection{Asynchronous Execution Multiple Processors}

\begin{figure*}[htbp]                                                    
        \centerline{\psfig{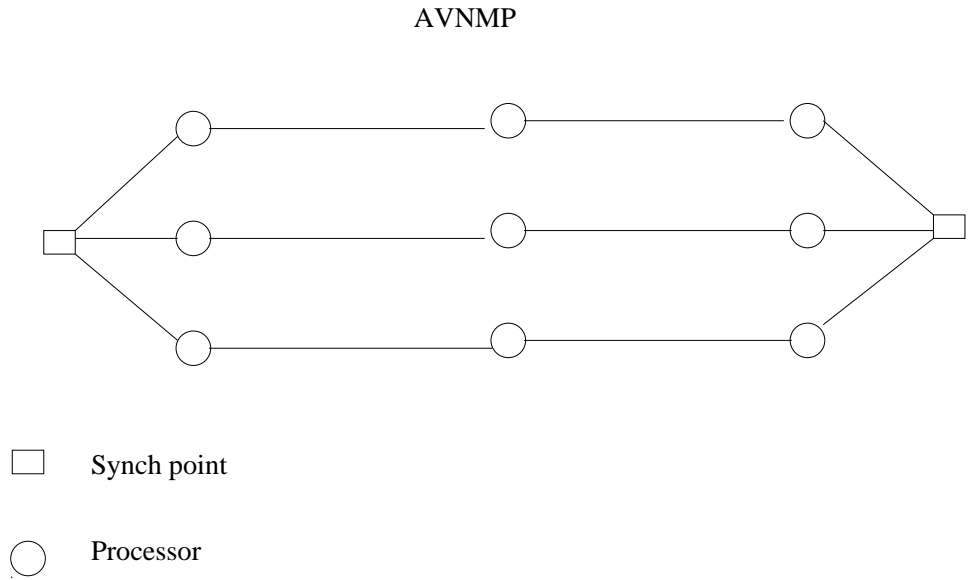}}  
        \caption{VNC Model of Parallelism.}   
        \label{vncmodel}
\end{figure*}

Assume that ordering of events is no longer a requirement. This represents
the asynchronous\index{Asynchronous} \acl{VNC} case and is shown in Figure \ref{vncmodel}. Note 
that this is the analysis of speedup\index{Speedup} due to parallel\index{Parallel}ism only, not the 
lookahead capability of \acl{VNC}.
This analysis of speedup\index{Speedup} assumes messages arrive in correct order and thus 
there is no rollback\index{Rollback}. However, this also assumes that there
are no optimization\index{Optimization} methods such as lazy cancellation.
Following \cite{Felderman90} the expected completion time is approximated 
by the maximum of $P$ $K$-stage Erlangs
where $P$ is the number of processes which can execute in parallel\index{Parallel} at each 
stage of execution. A $K$-stage Erlang model\index{Model} represents the total service
time as a series of exponential service times, where each service time
is performed by a process residing on an independent processor in this case.
There is no need to delay processing within the $K$-stage model\index{Model} because of
inter-process dependencies, as there is for synchronous\index{Synchronous} and sequential
cases.
For the specific case of \acf{RDRN} \acf{NCP} there would 
usually be two processes operating in parallel\index{Parallel}, the \acf{ES} and \acf{RN}. 
However, there can be three \acl{LP}s operating in parallel\index{Parallel} when operations 
between the Master \acl{ES}, and a particular associated \acl{ES} and \acl{RN} 
are required, or between an \acl{ES}, 
an \acl{RN}, and a destination \acl{ES} in a handoff\index{Handoff}. The most time consuming 
operations such as the \acl{ES} to \acl{ES} topology\index{Topology} calculation, if
partitioned in such a way as to minimize bandwidth overhead\index{Virtual Network Configuration!overhead}, can be executed 
in parallel\index{Parallel} on all \acl{ES}s.
Equation \ref{kstage} shows the pdf for a K-stage Erlang distribution.

\begin{equation}
f_T(x) = \frac{\mu e^{-\mu x}(\mu x)^{K-1}}{(K-1)!}
\label{kstage}
\end{equation}

As pointed out in \cite{Felderman90}, the probability that a $K$-stage Erlang
takes time less than or equal to $t$ is one minus the probability that
it takes time greater than $t$, which is simply one minus the probability
that there are $K$ arrivals in the interval $\left[0,t\right]$ from
a Poisson process at rate $\mu$. This result is shown in Equation \ref{simp1}.

\begin{figure*}
\begin{equation}
F_T(x) = 1 - e^{-\mu x}\sum_{i=0}^{K-1} \frac{(\mu x)^i}{i!}
\label{simp1}
\end{equation}
\end{figure*}

\begin{figure*}[htbp]
        \centerline{\psfig{file=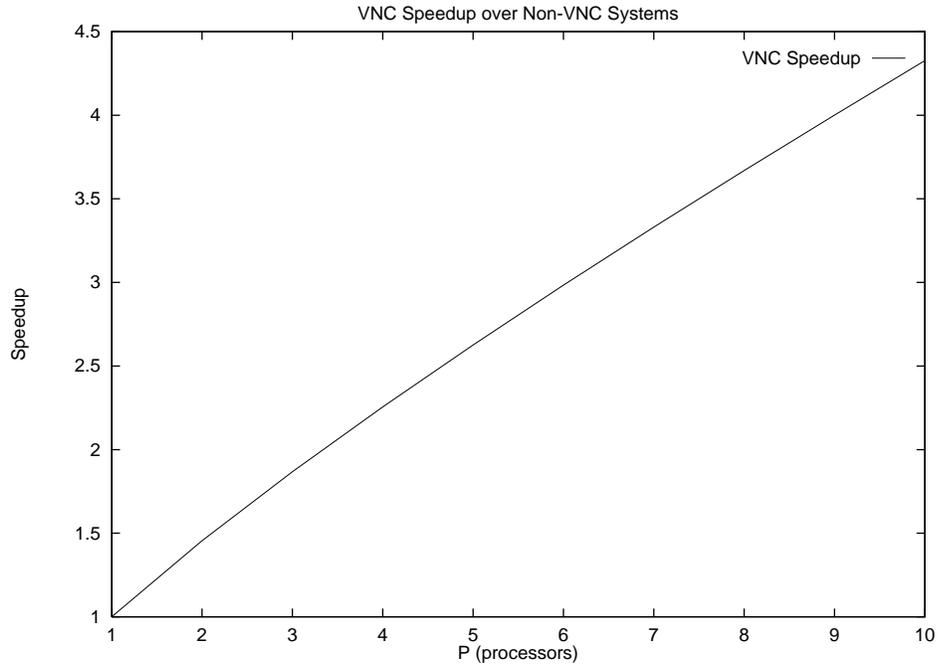,width=5.0in}}
        \caption{Speedup of VNC\index{VNC} over Non-VNC Systems Due to Parallelism.}
        \label{speedupgr}
\end{figure*}

The expected value is shown in Equation \ref{simp2}. This integral is
hard to solve with a closed form solution and \cite{Felderman90} instead tries
to find an approximate equation. This study attempts to be exact 
by using Equation \ref{simp2} and solving it numerically 
\cite[p. 378]{Kleinrock75}.
In Equation \ref{parspeedup} $S_{parallel}$ is the speedup\index{Speedup} of
optimistic synchronization over strictly sequential synchronization and
is graphed in Figure \ref{speedupgr} as a function\index{Function} of the
number of processors.
The speedup\index{Speedup} gained by parallel\index{Parallel}ism ($S_{parallel}$) augments the
speedup due to lookahead ($S_{cache}$) as shown in Equation \ref{l+p},
where $PR$ is the \acl{VNC} speedup\index{Speedup} and $X$ and $Y$ are random
variables representing the proportion of out-of-order and out-of-tolerance 
messages respectively.

\begin{figure*}
\begin{equation}
T_{async} = \int_{0}^{\infty} \left[ 1 - F_T(x) \right]  dx
\label{simp2}
\end{equation}
\end{figure*}

\begin{equation}
S_{parallel} = \frac{T_{seq}}{T_{async}}
\label{parspeedup}
\end{equation}

\begin{figure*}
\begin{eqnarray}
\label{l+p}
\lefteqn{PR_{X,Y} = } & & \\
& & \lambda_{vm} \left(\Delta_{vm} S_{parallel} - \tau_{task} -
   (\tau_{task} + \tau_{rb}) X - (\Delta_{vm}S_{parallel} - {1 \over 
   \lambda_{vm}}) Y\right) \nonumber
\end{eqnarray}
\end{figure*}

The \acl{RDRN} system consists of both \acl{SLP}s and LP\index{LP}s. Each \acf{ES} and
\acf{RN} executes network configuration processes (\acl{SLP}) and
these processors communicate between nodes (\acl{LP}). Thus, there is
the potential\index{Potential} for speedup\index{Speedup} via parallel\index{Parallel} processing.
The \acf{VNC} algorithm implementation\index{Virtual Network Configuration!implementation} is able to take advantage of both
\acl{SLP} lookahead without parallel\index{Parallel} processing and speedup\index{Speedup} due to parallel\index{Parallel}
processing because \acl{VNC} has been implemented on many nodes throughout the
network and each node has its own processor. Note that while
Clustered Time\index{Time} Warp \cite{Avril96}, which was developed concurrently
but independently of \acl{VNC}, uses a similar concept to \acf{SLP} and
\acf{LP}, it does not consider a real\index{Real}-time system as in \acl{VNC}. The
next section considers the prediction rate given the maximum lookahead
parameter $\Lambda$.
\subsection{Multiple Processor Logical Processes}

The goal of \acl{VNC} is to provide accurate predictions quickly
enough so that the results are available before they are 
needed. Without taking advantage of parallel\index{Parallel}ism,
a less sophisticated algorithm than \acl{VNC} could run ahead of 
real-time and cache results for future use such as the \acl{SLP}
system which assumes strict synchronization between processes
whose prediction rate is defined in Proposition \ref{splp_speed}. 
With such a simpler mechanism, $P_{oo}$ and $E[X]$ are always zero.
However, simply predicting and caching 
results ahead of time does not fully utilize inherent 
parallel\index{Parallel}ism in the 
system as long as messages between \acl{LP}s remain strictly synchronized.
Strict synchronization means that processes must wait until 
all messages are insured to be processed in order. 
Any speedup\index{Speedup} to be gained through parallel\index{Parallel}ism      
comes from the same mechanism as in optimistic parallel\index{Parallel} simulation; the
assumption that messages arrive in order by \acf{TR}, thus eliminating
unnecessary synchronization delay. However, messages arrive out-of-order
in \acl{VNC} for the following reasons. A general purpose system using 
the \acl{VNC} algorithm may have multiple driving\index{Driving Process} processes, each predicting
at different rates into the future. Another reason for out-of-order    
messages is that \acl{LP}s are not required to wait until processing completes
before sending the next message. For example, a process which loads beam table
information into memory need not wait for the load to complete before
sending a virtual\index{Virtual Network Configuration!verification} message. Also, processes may run faster for virtual\index{Virtual Network Configuration!verification} 
computations by allowing a larger tolerance. Finally, for 
testing purposes, hardware or processes may be 
replaced with simulated code, thus generating results faster than the 
actual process would.
Thus, although real\index{Real} and future time are working in parallel\index{Parallel} with strict
synchronization, no advantage is being taken of parallel\index{Parallel} processing.
This is demonstrated by the fact that, with strict
synchronization of messages, the same speedup\index{Speedup} ($S_{cache}$)
as defined in Proposition \ref{splp_speed} occurs regardless of
whether single processor or multiple processors are used.
What differentiates \acl{VNC} is the fact that it takes advantage 
of inherent parallel\index{Parallel}ism in the system than a sequential non-VNC 
pre-computation and 
caching method. Thus it is better able to meet the deadline imposed by 
predicting results before they are required. To see why this is true,
consider what happens as the overhead\index{Virtual Network Configuration!overhead} terms in Proposition 
\ref{splp_speed}, $\slowterm$, approach $\Delta_{vm}$.
The prediction rate becomes equal to real\index{Real}-time and can 
fall behind real\index{Real}-time as $\slowterm$ becomes larger. 
Optimistic synchronization helps to alleviate the
problem of the prediction rate falling behind real\index{Real}-time. Optimistic\index{Optimistic}
synchronization has another advantageous property, super-criticality.
A super critical system is one which can compute results faster
than the time taken by the critical path through the system. This 
can occur in \acl{VNC} using the lazy cancellation optimization\index{Optimization} as 
discussed in Section \ref{vncorg}. Super-criticality occurs when
task execution with false\index{False Message} message values generate a correct result.
Thus prematurely executed tasks do not rollback\index{Rollback} and a correct
result is generated faster than the route through the critical path.

The \acl{VNC} algorithm has two forms of speedup\index{Speedup} which need to
be clearly defined. There is the speedup\index{Speedup} in availability of
results because they have been pre-computed and cached. There is
also the speedup\index{Speedup} due to more efficient usage of parallel\index{Parallel}ism. The gain
in speedup\index{Speedup} due to parallel\index{Parallel}ism in \acl{VNC} can be significant
given the proper conditions. This can be achieved by neighboring \acf{RN}
and \acf{ES} nodes running on separate processors as they currently do,
rather than physical\index{Physical}ly adding multiple processors to each node.
In order that there be no confusion
as to which type of speedup\index{Speedup} is being analyzed, the speedup\index{Speedup} due
to pre-computing and caching results will be defined as
$S_{cache}$ and the speedup\index{Speedup} due to parallel\index{Parallel}ism 
will be defined as $S_{parallel}$.
Speedup due to parallel\index{Parallel}ism among multiple processors in \acl{VNC} will be 
gained from
the same mechanism which provides speedup\index{Speedup} in parallel\index{Parallel} simulation,
that is, it is assumed that all relevant messages are present and
will be processed in order by receive time. The method of maintaining
message order is optimistic in the form of rollback\index{Rollback}.
The following sections look at $S_{parallel}$ due to a multiprocessor
configuration system, such as the parallel\index{Parallel} operation of \acl{RN}
and \acl{ES} nodes. 
\subsection{\acl{VNC} Prediction Rate with a Fixed Lookahead}

\begin{figure*}[htbp]
        \centerline{\psfig{file=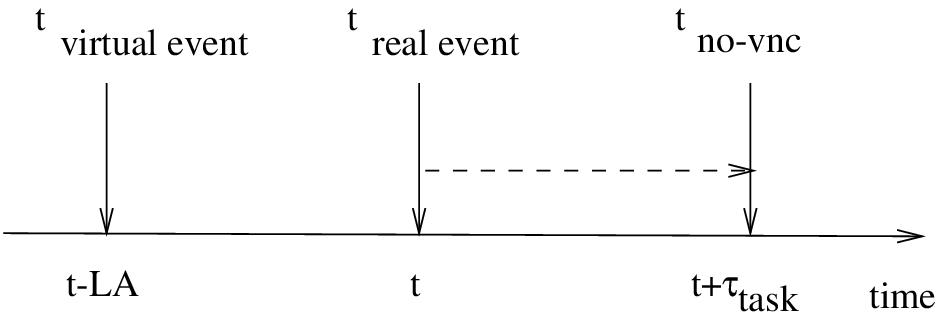,width=6.0in}}
        \caption{\acl{VNC} Cached before Real\index{Real Message} Event.}
        \label{cachesu1}
\end{figure*}

\begin{figure*}[htbp]                                                        
        \centerline{\psfig{file=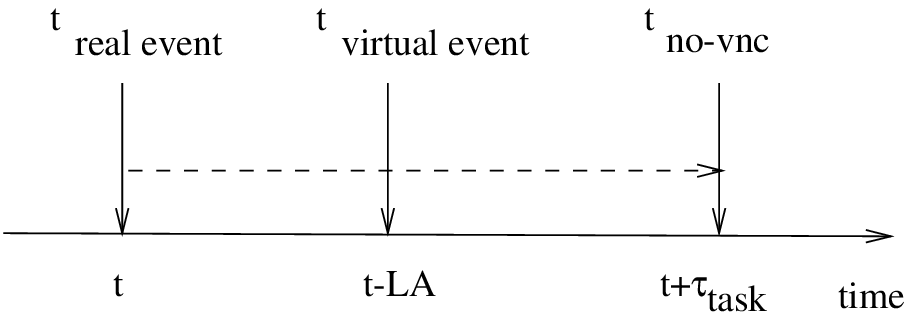,width=6.0in}}
        \caption{\acl{VNC} Cached later than Real\index{Real Message} Event.}
        \label{cachesu2}
\end{figure*}

\begin{figure*}[htbp]                                                        
        \centerline{\psfig{file=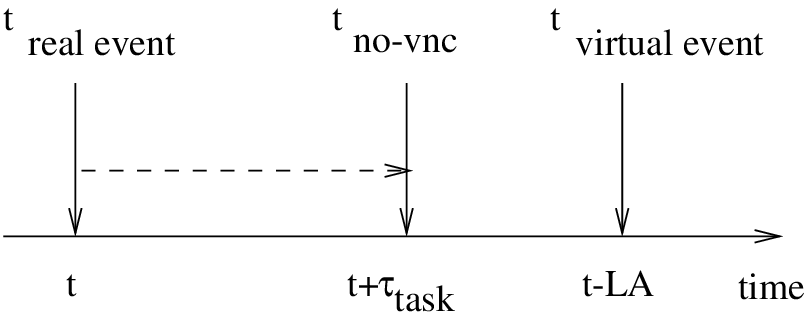,width=6.0in}}
        \caption{{VNC} Cached slower than Real\index{Real Message} Time\index{Time}.}
        \label{cachesu3}
\end{figure*}

There are three possible cases to consider when determining the
speedup of \acl{VNC} over non-lookahead sequential
execution. In this section we will determine the speedup\index{Speedup}
given each of these cases and their respective probabilities.
These cases are illustrated in Figures \ref{cachesu1}
through \ref{cachesu3}. The time that an event is predicted to
occur and the result cached is labeled $t_{virtual\ event}$, the 
time a real\index{Real} event occurs is labeled $t_{real\ event}$, and the 
time a result for the real\index{Real} event is calculated is labeled 
$t_{no-vnc}$. In \acl{VNC}, the 
virtual\index{Virtual Network Configuration!virtual event} event and its 
result can be cached before the 
real\index{Virtual Network Configuration!real event} event as shown in 
Figure \ref{cachesu1}, between the real\index{Real} event but before the 
real event result is calculated
shown in Figure \ref{cachesu2}, or after the real event result is 
calculated as shown in Figure \ref{cachesu3}. In each case, all events
are considered relative to the occurrence of the real\index{Real} event. It
is assumed that the real\index{Real} event occurs at time $t$. A random
variable called the lookahead ($LA$) is defined as $LVT - t$. The 
virtual event occurs at time $t - LA$. Assume that the task which must
be executed once the real\index{Real} event occurs takes $\tau_{task}$
time. Then without \acl{VNC} the task is completed at time
$t + \tau_{task}$.

The prediction rate has been defined in Equation \ref{l+p} and 
includes the time to predict an event and cache the result in the 
\acf{SQ}. Recall that in Section \ref{pnet} the expected value of $X$ 
has been determined based on the inherent synchronization of the \acl{LP}
topology. It was shown that $X$ has an expected value which varies with
the rate of hand-offs. Based on the topology of \acl{NCP} \acl{LP}s in Section 
\ref{pnet}, the virtual message path from the driving
process to the beamforming \acl{LP} is only one level. Thus, there are
no secondary rollbacks which could be generated. This means that the 
proportion of out-of-order rollback messages has a low variance. It
is clear that the
proportion of out-of-order messages is dependent on the \acf{LP}
architecture and the partitioning of tasks into \acl{LP}s. Thus, it is
difficult in an experimental implementation\index{Virtual Network 
Configuration!implementation} to vary $X$. It is easier to change 
the tolerance rather than change the \acl{LP} architecture to evaluate
the performance of \acl{VNC}. For these reasons, the analysis proceeds 
with $PR_{X,Y|X=E[X]}$.
Since the prediction rate is the rate of change of \acl{LVT} with
respect to time, the value of the \acl{LVT} is shown in Equation 
\ref{intlvt} where $C$ is an initial offset. This offset may occur 
because \acl{VNC} may begin running $C$ time units before or after 
the real\index{Real} system. Replacing $LVT$ in the definition of $LA$
with the right side of Equation \ref{intlvt} yields the Equation
for lookahead shown in Equation \ref{la}.
 
\begin{eqnarray}
\label{intlvt} \lefteqn{LVT_{X,Y|X=E[X]} = } & & \\
& \lambda_{vm} (\Delta_{vm} S_{parallel} - \tau_{task} - (\tau_{task} + \tau_{rb}) E[X] - & \nonumber \\
& (\Delta_{vm} S_{parallel} - {1 \over \lambda_{vm}}) Y) t + C & \nonumber \\
& \label{la} LA_{X,Y|X=E[X]} = (LVT_{X,Y|X=E[X]} - 1) t + C &
\end{eqnarray}

The probability of the event in which the \acl{VNC} result is cached
before the real\index{Real} event is defined in Equation \ref{pcache}. The probability 
of the event for which the \acl{VNC} result is cached after the real\index{Real} event 
but before the result would have been calculated in the non-\acl{VNC} system
is defined in Equation \ref{plate}. Finally, the probability of the event
for which the \acl{VNC} result is cached after the result would have been 
calculated in a non-\acl{VNC} system is defined in Equation \ref{pslow}.

\begin{eqnarray}
\label{pcache} P_{cache} = P[LA_{X,Y|X=E[X]} > \tau_{task}] \\
\label{plate} P_{late} = P[0 \le LA_{X,Y|X=E[X]} \le \tau_{task}] \\
\label{pslow} P_{slow} = P[LA_{X,Y|X=E[X]} < 0]
\end{eqnarray}

The goal of this analysis is to determine the effect of the proportion
of out-of-tolerance messages ($Y$) on the speedup\index{Speedup} of a \acl{VNC} system.
Hence we assume that the proportion $Y$ is a binomially distributed\index{Distributed} random 
variable with parameters $n$ and $p$ where $n$ is the total number 
of messages and $p$ is the probability of any single message being out of 
tolerance. It is helpful to simplify Equation \ref{la} by using
$\gamma_1$ and $\gamma_2$ as defined in Equations 
\ref{g1} and \ref{g2} in Equation \ref{sla}.

\begin{eqnarray}
\label{sla} LA_{X,Y|X=E[X]} = \gamma_1 - \gamma_2 Y & & \\
\label{g1} \gamma_1 = (\lambda_{vm} \Delta_{vm} S_{parallel} - \lambda_{vm} \tau_{task} - & & \\
& \lambda_{vm} (\tau_{rb} + \tau_{task}) E[X]) - 1) t + C & \nonumber \\
\label{g2} \gamma_2 = \lambda_{vm} (\Delta_{vm} S_{parallel} - {1 \over \lambda_{vm}} + \tau_{rb}) t & &
\end{eqnarray}

The early prediction probability as illustrated in Figure \ref{cachesu1} 
is shown in Equation \ref{fpcache}. The late prediction probability as
illustrated in Figure \ref{cachesu2} is shown in Equation \ref{fplate}. 
The probability for which \acl{VNC} falls behind real\index{Real} time as illustrated 
in Figure \ref{cachesu3} is shown in Equation \ref{fpslow}.
The three cases for determining \acl{VNC} speedup\index{Speedup} are thus
determined by the probability that $Y$ is greater or less than two
thresholds.

\begin{eqnarray}
\label{fpcache} & P_1(t) = P_{{cache}\ {X,Y|X=E[X]}} = P[Y < {{\gamma_1 - 
   \tau_{task}} \over \gamma_2}] & \\
\label{fplate} P_2(t) = \lefteqn{P_{{late}\ {X,Y|X=E[X]}} = } & & \\
& P[{{\gamma_1 - \tau_{task}} \over 
\gamma_2} \le Y \le {\gamma_1 \over \gamma_2}] & \nonumber \\
\label{fpslow} & P_3(t) = P_{{slow}\ {X,Y|X=E[X]}} = P[Y > {\gamma_1 \over 
\gamma_2}] & 
\end{eqnarray}

\begin{figure*}[htbp]
        \centerline{\psfig{file=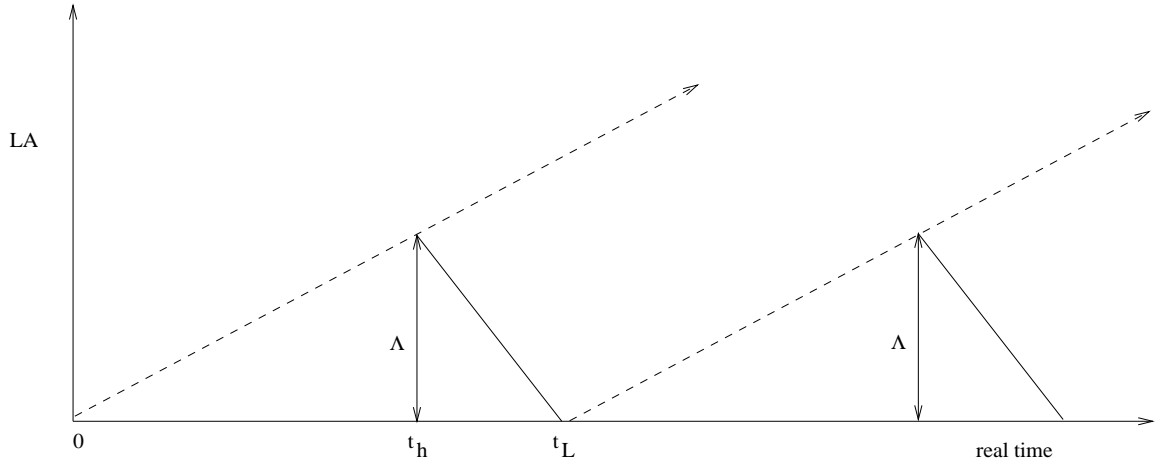,width=6.0in}}
        \caption{Lookahead with a Sliding Lookahead Window.}
        \label{lookahead}
\end{figure*}

The three probabilities in Equations \ref{fpcache}
through \ref{fpslow} depend on ($Y$) and real\index{Real} time because
the analysis assumes that the lookahead increases 
indefinitely which shifts the thresholds in such a manner as to 
increase \acl{VNC} performance as real\index{Real} time increases. However, the 
\acl{VNC} algorithm holds processing of virtual\index{Virtual Network Configuration!verification} messages once the 
end of the \acf{SLW} is reached. The hold time occurs when $LA = \Lambda$ 
where $\Lambda$ is the length of the \acl{SLW}. Once $\Lambda$ is reached, 
processing of virtual\index{Virtual Network Configuration!verification} messages is discontinued until real\index{Real}-time reaches 
\acl{LVT}. The lookahead versus real\index{Real} time including the effect of the \acl{SLW} 
is shown in Figure \ref{lookahead}. The dashed arrow represents the 
lookahead which increases at rate $PR$. The solid line returning to
zero is lookahead as the \acl{LP} delays. Because the curve in
Figure \ref{lookahead} from $0$ to $t_L$ repeats indefinitely, only the area
from $0$ to $t_L$ need be considered. For
each $P_i(t)$ $i=1,2,3$, the time average over the lookahead time ($t_L$) 
is shown 
by the integral in Equation \ref{ala}.

\begin{equation}
\label{ala}
P_{X,Y|X=E[X]} = {1 \over t_L} \int_0^{t_L} P_i(t) dt
\end{equation}

\begin{figure*}[htbp]
        \centerline{\psfig{file=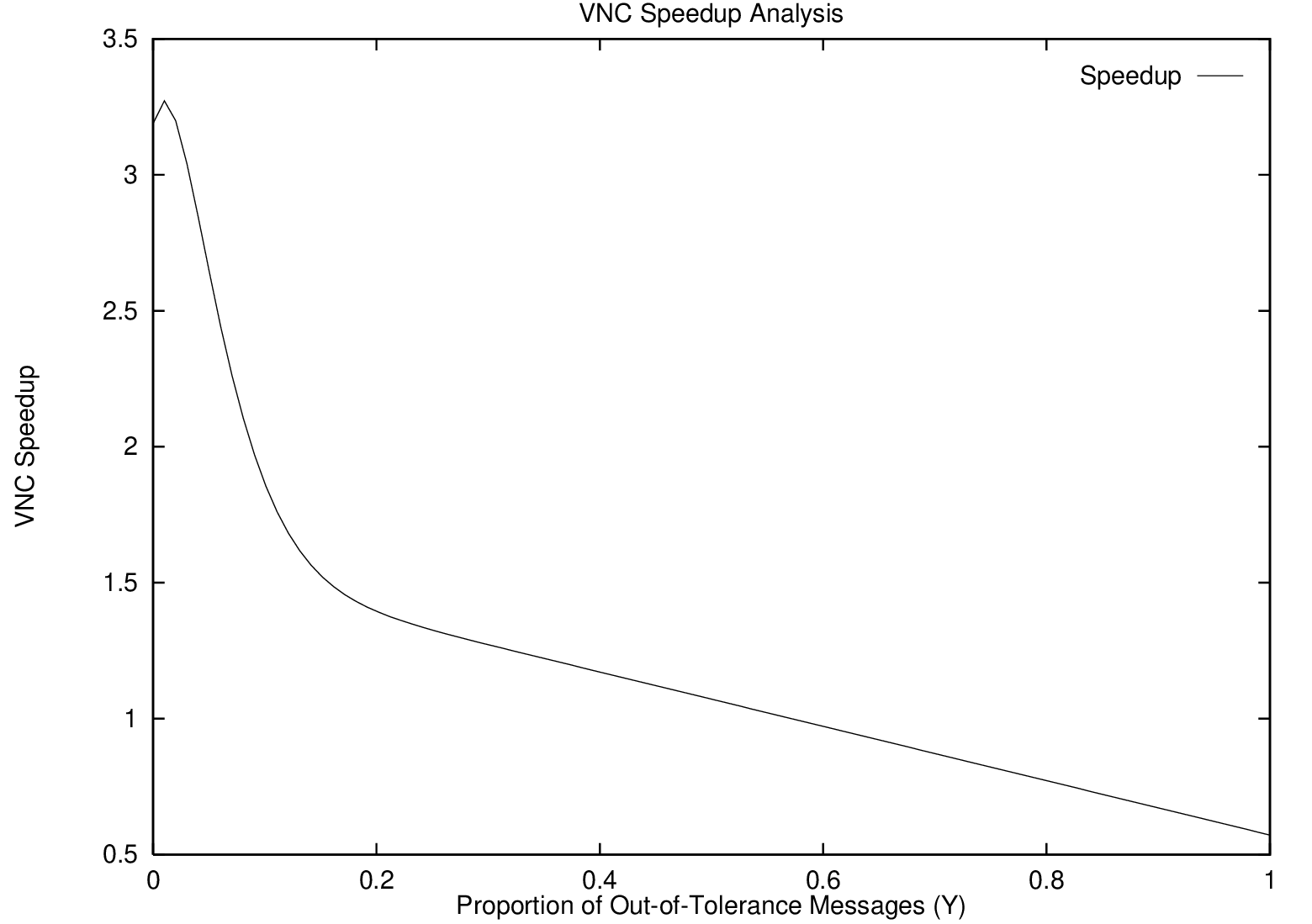,width=6.0in}}
        \caption{VNC Speed\index{Speed}up.}
        \label{speedup}
\end{figure*}

The probability of each of the events shown in Figures
\ref{cachesu1} through \ref{cachesu3} is multiplied by the speedup\index{Speedup}
for each event in order to derive the average speedup\index{Speedup}. For the 
case
shown in Figure \ref{cachesu1}, the speedup\index{Speedup} ($C_r$) is provided by the time
to read the cache over directly computing the result. For the remaining
cases the speedup\index{Speedup} is $PR_{X,Y|X=E[X]}$ which has been defined as 
${LVT_{X,Y|X=E[X]} \over t}$ as shown in Equation \ref{eta}.
The analytical results for speedup\index{Speedup} are graphed in Figure \ref{speedup}.
A high probability of out-of-tolerance rollback\index{Rollback} in Figure 
\ref{speedup} results in a speedup\index{Speedup} of less than one. Real\index{Real Message} messages
are always processed when they arrive at an \acl{LP}. Thus, no matter
how late \acl{VNC} results are, the system will continue to run near
real time.  However, when \acl{VNC} results are very late due to a high
proportion of out-of-tolerance messages, the \acl{VNC} system is slower than 
real time because out-of-tolerance rollback overhead\index{Virtual Network Configuration!overhead} processing occurs.
Anti-messages must be sent to correct other \acl{LP}s which
have processed messages which have now been found to be out of tolerance from 
the current \acl{LP}. This causes the speedup\index{Speedup} to be less than one when the 
out-of-tolerance probability is high. Thus, $PR_{X,Y|X=E[X]}$ will be less than 
one for the ``slow'' shown in Figure \ref{cachesu3}.

\begin{table*}
\begin{eqnarray}
\label{eta}
\lefteqn{\eta \equiv } \\
& & P_{{cache}\ {X|X=E[X]}} C_r + (P_{{late}\ {X|X=E[X]}} + 
P_{{slow}\ {X|X=E[X]}}) PR_{X,Y|X=E[X]} \nonumber
\end{eqnarray}
\end{table*}

\section{Prediction Accuracy: $\alpha$}
\label{pacc}

Accuracy is the ability of the system to predict future events. A higher
degree of accuracy will result in more ``cache hits'' of the predicted
state cache information. Smaller tolerances should result in greater
system accuracy, but this comes at the cost of a reduction in speedup\index{Speedup}.

Assume for simplicity that the effects of non-causality are negligible 
for the analysis in
this section. The effects of causality are discussed in more detail
in Section \ref{nogvt}. An \acl{LP}\index{LP} may deviate from the real\index{Real} object 
it represents because either the \acl{LP}\index{LP} does not accurately represent the actual 
entity or because events outside the scope of the predictive\index{Mobile Network!predictive} system 
may effect the entities being managed. Ignore events outside the
scope of the predictive\index{Mobile Network!predictive} system for this analysis and consider only the
deterministic error 
from inaccurate prediction of the driving\index{Driving Process} process. The error is defined
as the difference between an actual message value at the current time 
($v_t$) and a message value which had been predicted earlier ($v_{t_p}$). 
Thus the \textbf{M}essage \textbf{E}rror is $ME = v_t - v_{t_p}$. Virtual\index{Virtual Network Configuration!verification} 
message
values generated from a driving\index{Driving Process} process may contain some error.
It is assumed that the error in any output message generated by a process 
is a function\index{Function} of any error in the input message and the amount of time it 
takes to process the message. A larger processing time increases the
chances that external events may have changed before the processing
has completed.

\begin{figure*}[htbp]
        \centerline{\psfig{file=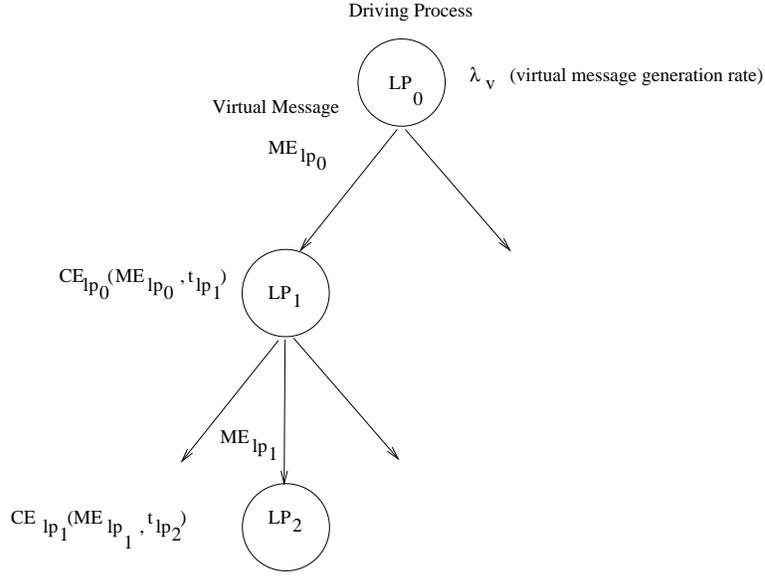,height=3in}}
        \caption{Accumulated Message Value Error.}
        \label{drivproc}
\end{figure*}

Two function\index{Function}s of total \textbf{AC}cumulated message value error 
($AC(\cdot)$) in a predicted result are described by Equations \ref{acn} 
and \ref{actau} and are illustrated in Figure \ref{drivproc}. 
$ME_{lp_0}$ is the amount of error in the value of the virtual\index{Virtual Network Configuration!verification} message 
injected into the predictive\index{Mobile Network!predictive} system by the driving\index{Driving Process} process ($lp_0$). The error 
introduced into the value of the output message produced by the computation 
of each \acl{LP}\index{LP} is represented by the \textbf{C}omputation \textbf{E}rror function\index{Function}
$CE_{lp_n}(ME_{lp_{n - 1}}, t_{lp_n})$. The real\index{Real} time 
taken for the $n^{th}$ \acl{LP}\index{LP} to generate a message is $t_{lp_n}$.
The error accumulates in the \acf{SQ} at each node by the amount
$CE_{lp_n}(ME_{lp_{n - 1}}, t_{lp_n})$ which is a function\index{Function} of
the error contained in the input message from the predecessor \acl{LP}
and the time to process that
message. Figure \ref{drivproc} shows a driving\index{Driving Process} process ($DP$) generating
a virtual\index{Virtual Network Configuration!verification} message which contains prediction error
($ME_{lp_0}$). The virtual\index{Virtual Network Configuration!verification} message with prediction error ($ME_{lp_0}$)
is processed by node $LP_1$ in $t_{lp_1}$ time units resulting
in an output message with error, $ME_{lp_1} = CE_{lp_0}(ME_{lp_0}, 
t_{lp_1})$.

\begin{hyp}
\label{ac_hyp}

The accumulated error in a message value is Equation \ref{acn} and
Equation \ref{actau}.

\begin{eqnarray}
\label{acn} AC_n(n) = \sum_{i = 1}^N CE_{lp_i}(ME_{lp_{i-1}}, t_{lp_i}) \\
\label{actau} AC_t(\tau) = \lim_{\sum t_{lp_i} \rightarrow \tau}
\sum_{i = 1}^{lpn} CE_{lp_i}(ME_{lp_{i-1}}, t_{lp_i})
\end{eqnarray}

where $CE_{lp_i}$ is the computational error added to a virtual
output message value, $ME_{lp_i}$ is the virtual message input
error, and $t_{lp_i}$ is the real time taken to process a virtual
message.

\end{hyp}

As shown in Proposition \ref{ac_hyp},
$AC_n(n)$ is the total accumulated error in the virtual\index{Virtual Network Configuration!verification} message output by
the $n^{th}$ \acl{LP}\index{LP} from the driving\index{Driving Process} process. $AC_t(\tau)$ is the accumulated
error in $\tau$ real\index{Real} time units from the generation of the initial virtual\index{Virtual Network Configuration!verification} 
message from the driving\index{Driving Process} process. Equation \ref{actau} is 
$\lim_{\sum t_{lp_i} \rightarrow \tau} \sum_{i = 1}^{lpn} AC_n(n)$, 
where $lpn$ is the number of \acl{LP}\index{LP} computations in time $\tau$. In other
words, $AC_t(\tau)$ is the error accumulated as messages pass
through $lpn$ \acl{LP}s in real\index{Real} time $\tau$.
For example, if a prediction result is generated
in the third \acl{LP}\index{LP} from the driving\index{Driving Process} process, then the total accumulated
error in the result is $AC_n(3)$. If 10 represents the number of time units
after the initial message was generated from the driving\index{Driving Process} process then
$AC_t(10)$ would be the amount of total accumulated error in the result.
A cache hit occurs when $|AC_t(\tau)| \le \Theta$,
where $\Theta$ is the tolerance associated with the last \acl{LP}\index{LP} required
to generate the final result.
Equations \ref{acn} and \ref{actau} provide a means of representing
the amount of error in a \acl{VNC} generated result. Once an event
has been predicted and results pre-computed and cached, it would
be useful to know what the probability is that the result has been
accurately calculated, especially if any results are committed before
a real\index{Real} message arrives. The out-of-tolerance check and rollback\index{Rollback} does not
occur until a real\index{Real} message arrives. If a resource\index{Resource}, such as a \acl{VC},
is established ahead of time based on the predicted result, then
this section has defined $\alpha = P[|AC_t(\Lambda)| > \Theta]$
where $\Theta$\index{$\Theta$} is the tolerance associated with the last 
\acl{LP}\index{LP} required to generate the final result.

\subsection{\acl{RDRN} Error Accumulation}

\begin{figure}[htbp]
        \centerline{\psfig{file=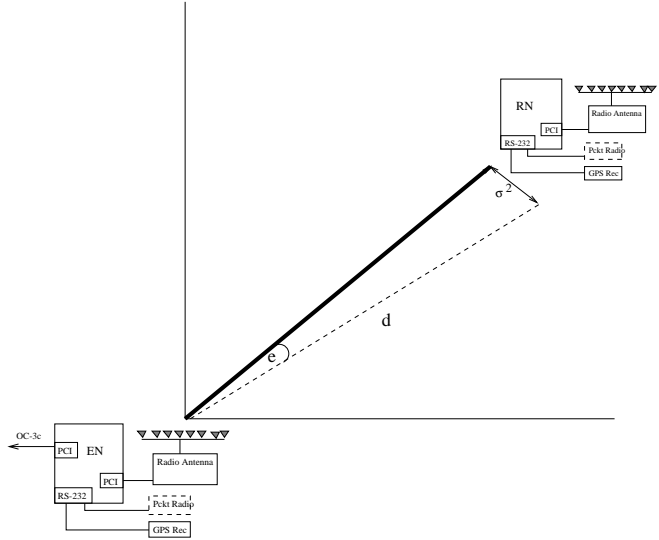,width=3.4in}}
        \caption{Beam Steering Variance.}
        \label{beamerr}
\end{figure}

In the \acl{RDRN} \acl{NCP} implementation\index{Virtual Network Configuration!implementation}, no network resource\index{Resource}s are 
committed, except for pre-loading the beam table, until the real\index{Real} message for 
an event is received. Assume the beamform table
has an estimated download time of five milliseconds and a
one degree tolerance for error in position. In the
implementation used for the experimental validation, there is only 
one level of error propagation, and thus no error accumulation. Thus,
$\alpha$ is shown in Equation \ref{impalpha} as illustrated in
Figure \ref{beamerr} where $d$ is the 
distance between the nodes, $\sigma^2$ is the variance in location
prediction accuracy, and $\Theta$ is the tolerance. Angle $e$ in
Figure \ref{beamerr} is the variance of error in the beam steering
angle due to the variance in the Cartesian co-ordinates provided
by the prediction process.

\begin{table*}
\begin{equation}
\alpha \equiv Pr\left[ \tan^{-1}\left( \sigma \over d \right) > \Theta \right]
\label{impalpha}
\end{equation}
\end{table*}
\section{Bandwidth: $\beta$}

\begin{figure*}[htbp]
        \centerline{\psfig{file=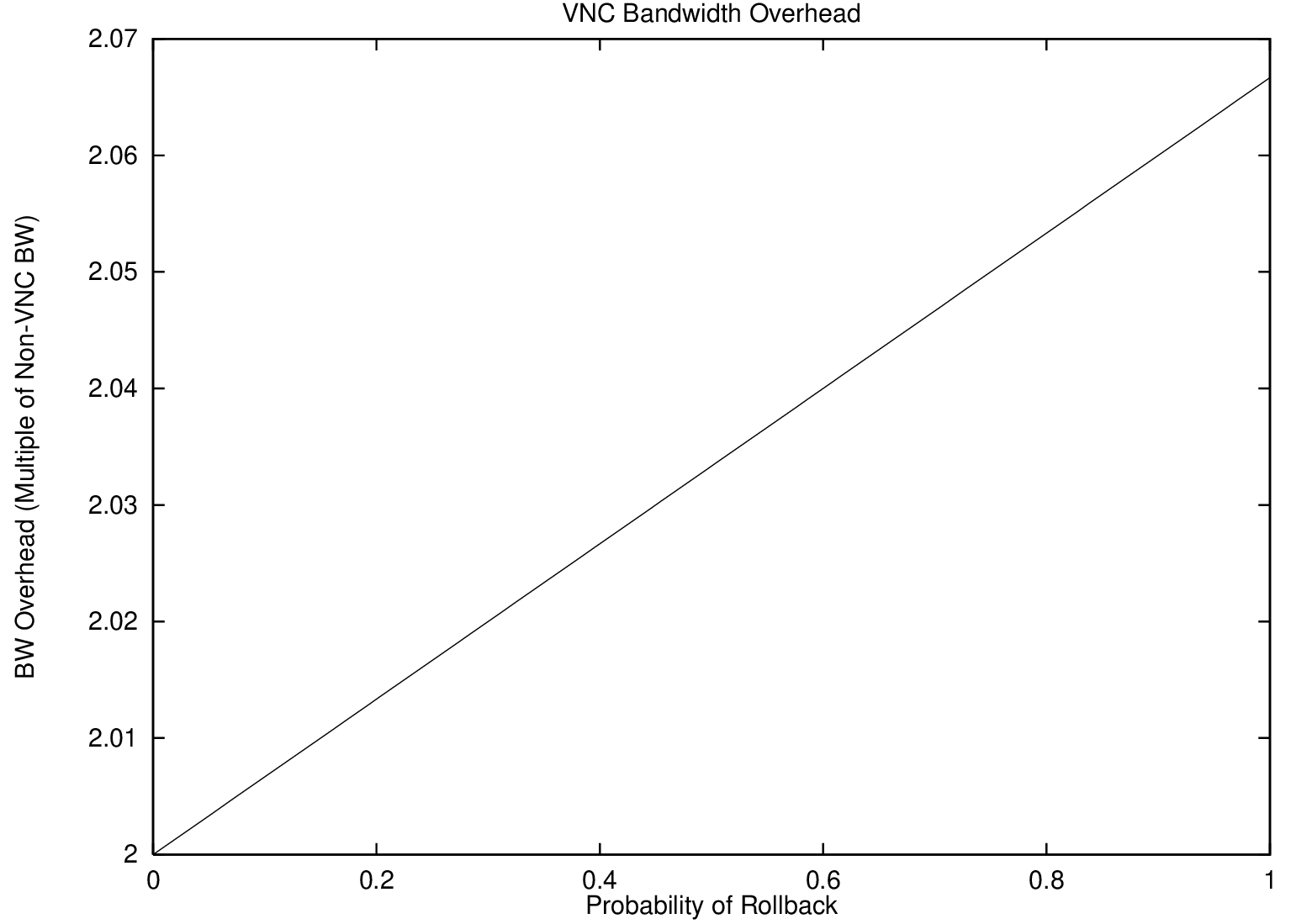,width=5.0in}}
        \caption{VNC Bandwidth Overhead.}
        \label{bwover}
\end{figure*}

\begin{figure*}[htbp]
        \centerline{\psfig{file=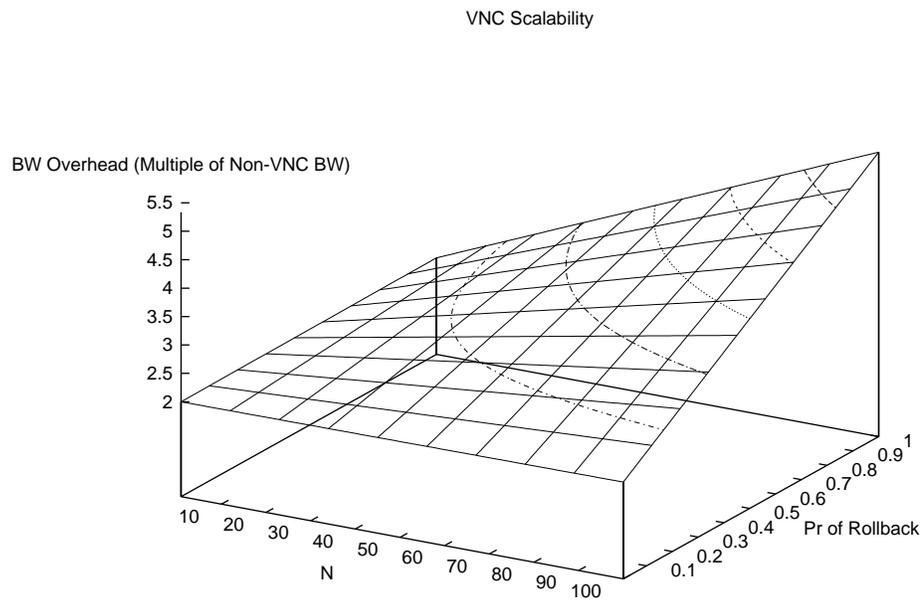,width=5.0in}}
        \caption{VNC Scalability.}
        \label{bwscale}                                           
\end{figure*}

The amount of overhead\index{Virtual Network Configuration!overhead} in bandwidth required by \acl{VNC} is
due to virtual\index{Virtual Network Configuration!verification} and anti-message load. With perfect prediction 
capability, there should be exactly one virtual\index{Virtual Network Configuration!verification} message from the 
driving process for each real\index{Real} message. The inter-rollback time,
$\frac{1}{\lambda_{rb}}$, has been determined in Proposition 
\ref{etrb_hyp} Equation \ref{erb}. 
Virtual messages are arriving and generating new messages at
a rate of $\lambda_v$. Thus, the worst case expected number of 
messages in the \acl{QS}
which will be sent as anti-messages is $\frac{\lambda_v}{\lambda_{rb}}$ 
when a rollback\index{Rollback} occurs.
The bandwidth overhead\index{Virtual Network Configuration!overhead} is shown in Equation \ref{boverhead} where
$\lambda_v$ is the virtual\index{Virtual Network Configuration!verification} message load, $\lambda_r$ is the real\index{Real}
message load, and $\lambda_{rb}$ is the expected rollback\index{Rollback} rate.
The bandwidth overhead\index{Virtual Network Configuration!overhead} as a function\index{Function} of rollback\index{Rollback} rate
is shown in Figure \ref{bwover}.
Scalability in \acl{VNC} is the rate at which the 
proportion of rollback\index{Rollback}s increases as the number of nodes increases.
The graph in Figure \ref{bwscale} illustrates the tradeoff between the 
number of \acl{LP}s and the rollback\index{Rollback} rate given \aparms\ 
and $Rm = {2 \over 30\mbox{\ ms}}$. The rollback\index{Rollback} rate in
this graph is the sum of both the out-of-order and the out-of-tolerance 
rollback rates. 

\begin{hyp}
\label{bw_hyp}

The expected bandwidth overhead is

\begin{equation}
\label{boverhead}
\beta = {{\frac{\lambda_v}{\lambda_{rb}} + \lambda_{v}} + \lambda_r \over \lambda_r}
\end{equation}

where $\lambda_{rb}$ is the expected rollback rate, 
$\lambda_{v}$ is the expected virtual message rate, 
and $\lambda_r$ is the expected real message rate.

\end{hyp}


\section{Performance Analysis}
\label{perfanal}

\begin{figure*}[htbp]
        \centerline{\psfig{file=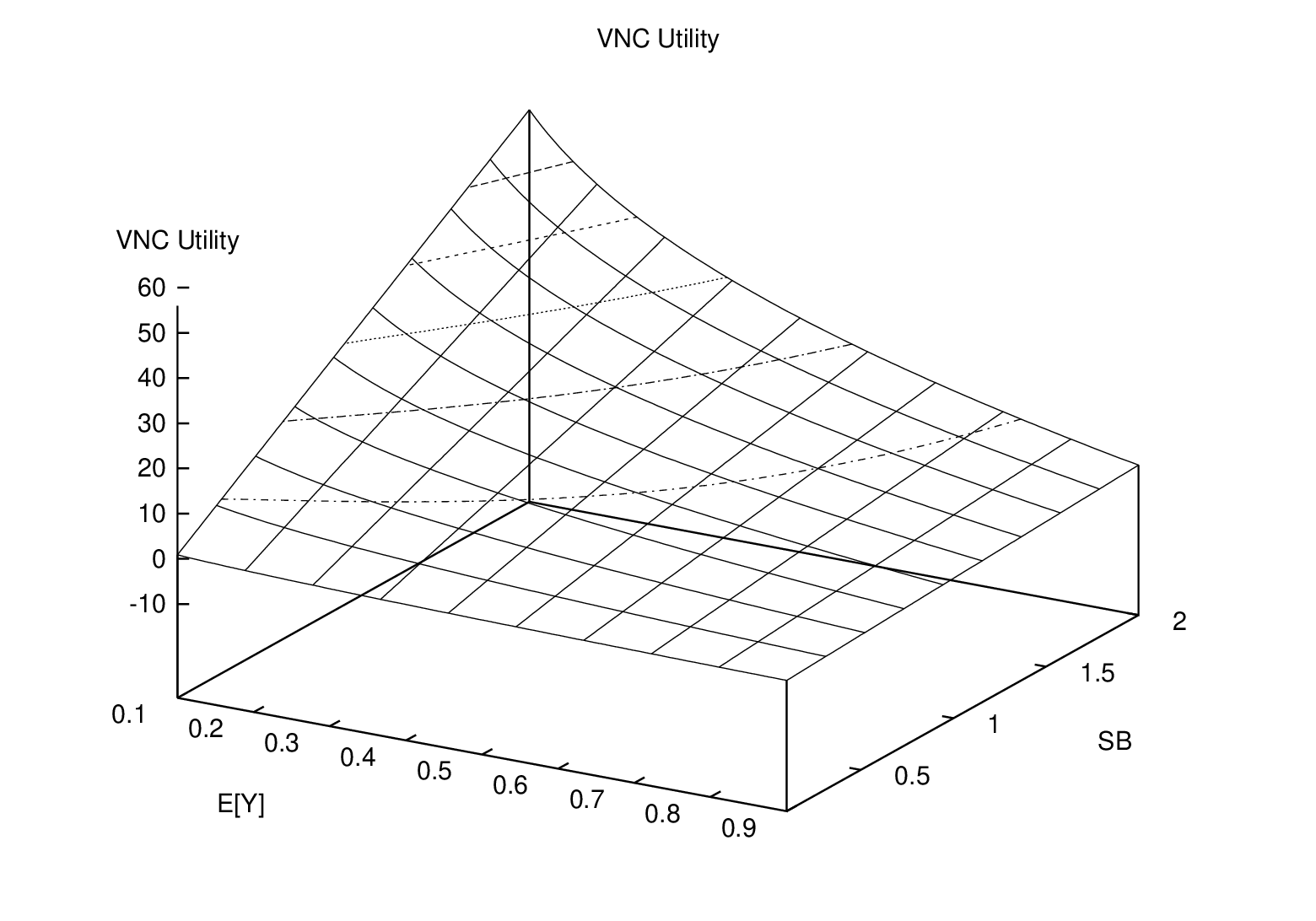,width=5.0in}}
        \caption{Overhead versus Speed\index{Speed}up as a Function of Probability of Rollback\index{Rollback}.}
        \label{overspeed}
\end{figure*}

Equation \ref{vncutil1} shows the complete \acl{VNC} performance utility.
The surface plot showing the utility of \acl{VNC} 
as a function\index{Function} of the proportion of out-of-tolerance messages is shown
in Figure \ref{overspeed} where $\Phi_s$, $\Phi_w$, $\Phi_b$ are one
and \aparms. The 
wasted resource\index{Resource}s utility is not included in Figure \ref{overspeed} 
because as discussed in Section \ref{pacc} there is only one level
of message generation and thus no error accumulation.
The y-axis is the relative marginal utility of speedup\index{Speedup} over reduction
in bandwidth overhead\index{Virtual Network Configuration!overhead} $SB = \frac{\Phi_s}{\Phi_b}$. Thus if bandwidth 
reduction is much more important than speedup\index{Speedup}, the utility is low and 
the proportion of rollback\index{Rollback} messages would have to be kept below 0.3 per 
millisecond in this case. However, if speedup\index{Speedup} is the primary desire 
relative to bandwidth, the proportion of out-of-tolerance rollback\index{Rollback} 
message values can be as high as 0.5 per millisecond. If the 
proportion of out-of-tolerance messages becomes too high, the utility 
becomes negative because prediction time begins to fall behind real\index{Real} time.

\begin{figure*}[htbp]
        \centerline{\psfig{file=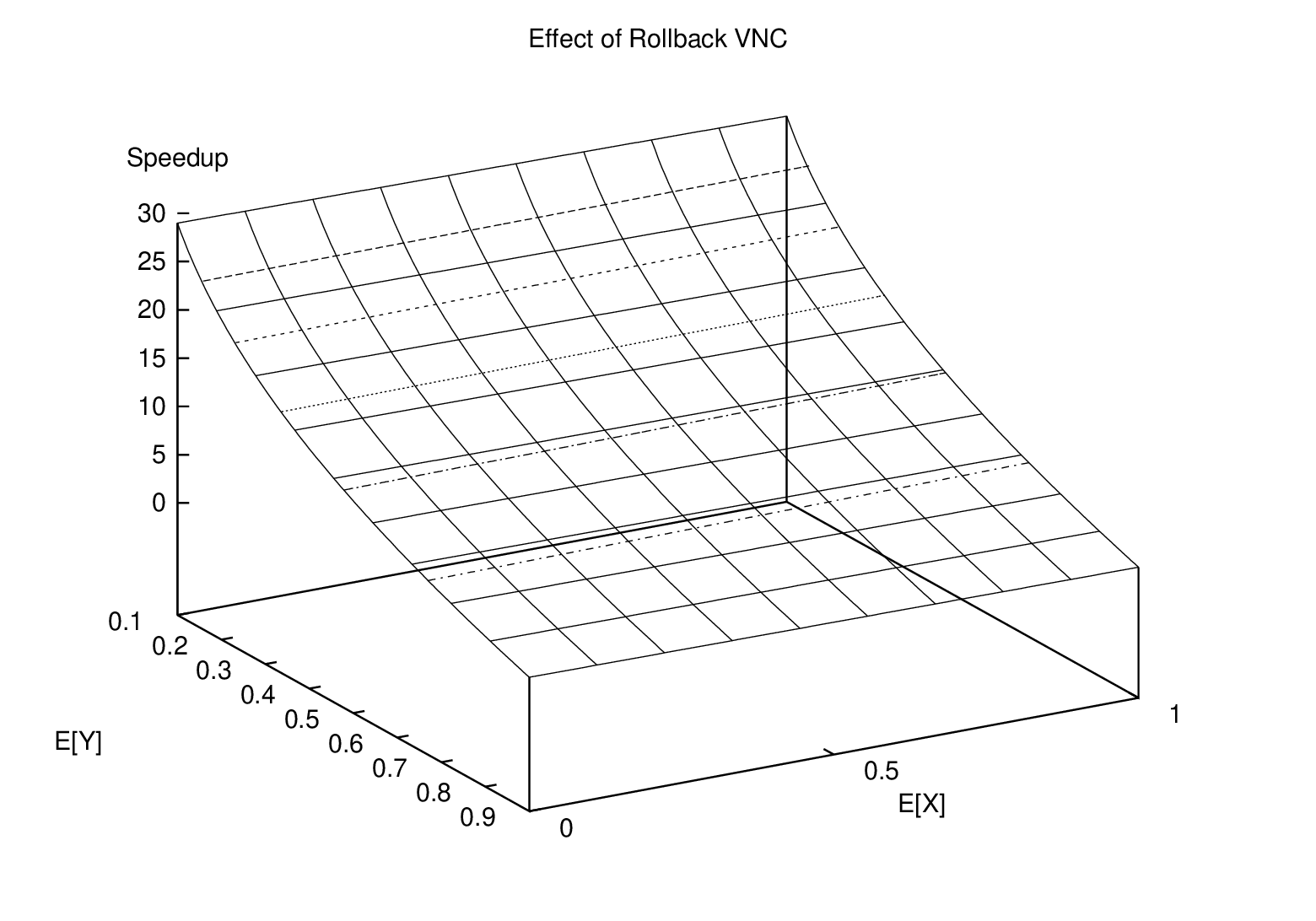,width=5.0in}}
        \caption{Effect of Non-Causality and Tolerance\index{Tolerance} on Speed\index{Speed}up.}
        \label{rollbacks}
\end{figure*}

The effect of the proportion of out-of-order and out-of-tolerance 
messages on \acl{VNC} speedup\index{Speedup} is shown in Figure \ref{rollbacks}. 
This graph shows that out-of-tolerance rollback\index{Rollback}s have a greater impact 
on speedup\index{Speedup} than out-of-order rollback\index{Rollback}s. The reason for the greater impact 
of the proportion of out-of-tolerance messages is that such rollback\index{Rollback}s 
caused by such messages always cause a process to rollback\index{Rollback} to real\index{Real} time. 
An out-of-order rollback\index{Rollback} only requires the process to rollback\index{Rollback} to the 
previous saved state.

\begin{figure*}[htbp]
        \centerline{\psfig{file=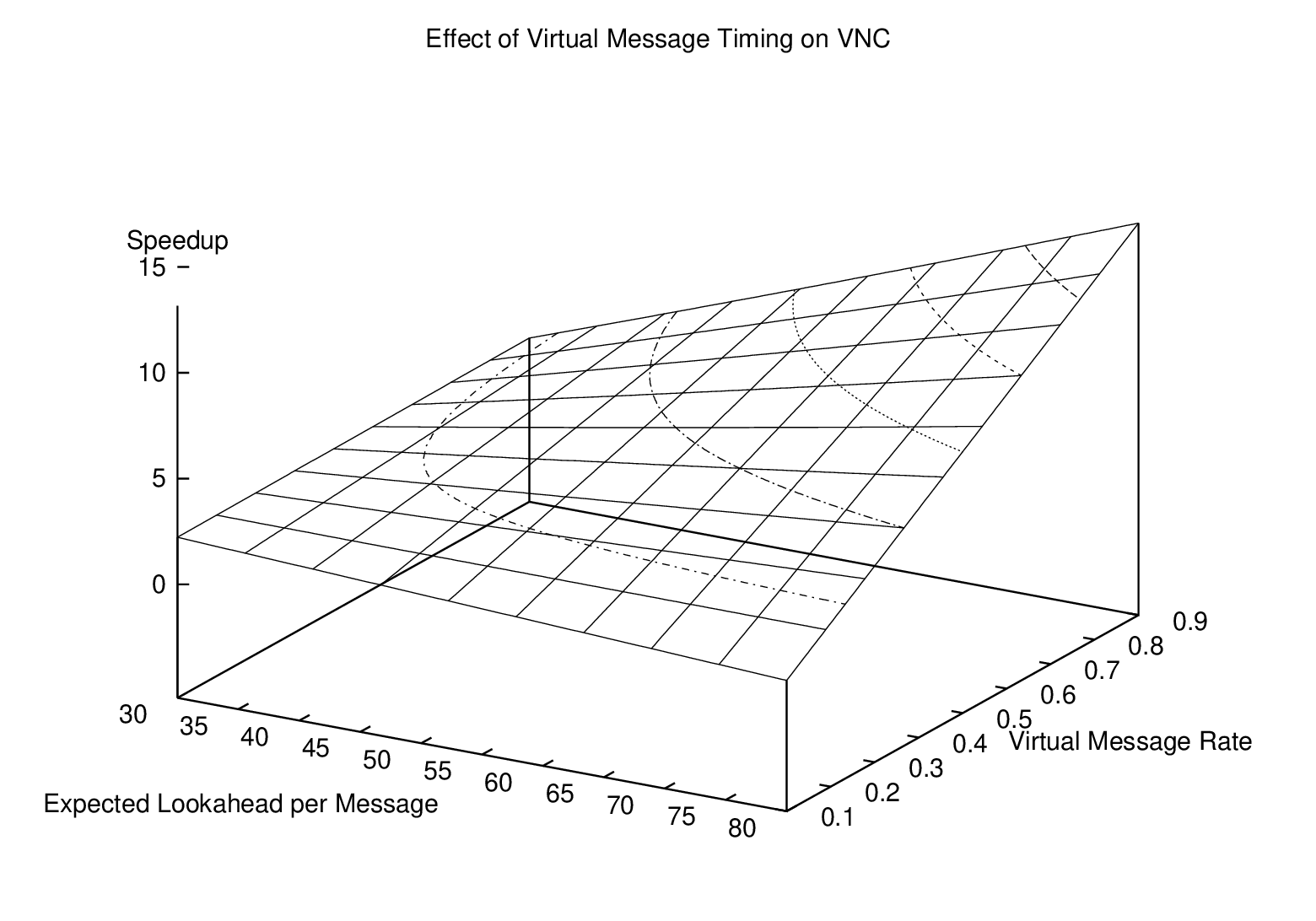,width=5.0in}}
        \caption{Effect of Virtual\index{Virtual Network Configuration!verification} Message Rate and Lookahead on Speed\index{Speed}up.}
        \label{vmrate}
\end{figure*}

Figure \ref{vmrate} shows the effect of the proportion of virtual\index{Virtual Network Configuration!verification}
messages and expected lookahead per virtual\index{Virtual Network Configuration!verification} message on speedup\index{Speedup}.
This graph is interesting because it shows how the proportion of virtual\index{Virtual Network Configuration!verification}
messages injected into the \acl{VNC} system and the expected lookahead
time of each message can affect the speedup\index{Speedup}. 
The real\index{Real} and virtual\index{Virtual Network Configuration!verification} message rates are $\frac{0.1}{\mbox{ms}}$, 
$Rm = {2 \over 30\mbox{\ ms}}$, \aparms.

\begin{figure*}
\begin{eqnarray}
\label{vncutil1}
U_{VNC} & = & \left( P_{{cache}\ {X|X=E[X]}} C_r + \right. \\ \nonumber
        &   & \left. (P_{{late}\ {X|X=E[X]}} + P_{{slow}\ {X|X=E[X]}}) PR_{X,Y|X
=E[X]} \right) \Phi_s - \\ \nonumber
        &   & P[|AC_t(\Lambda)| > \Theta] \Phi_w  - \\ \nonumber
        &   & \left({\frac{\lambda_v}{\lambda_{rb}} + \lambda_{v}}
                   \over \lambda_r\right) \Phi_b \nonumber
\end{eqnarray}
\end{figure*}

\section{Aspects of \acl{VNC} Performance}

The following sections discuss other aspects and optimization\index{Optimization}s
of the \acl{VNC} algorithm including handling multiple future events
and the relevance of \acf{GVT} to \acl{VNC}. Since all possible
alternative events cannot be predicted, 
only the most likely events are predicted in \acl{VNC}. However, knowledge 
of alternative events with a lower probability of occurrence will allow 
the system to prepare more intelligently.

Another consideration is the calculation of \acl{GVT}. This requires 
bandwidth and processing overhead\index{Virtual Network Configuration!overhead}. A discussion of the need for \acl{GVT} 
and an alternative to using \acl{GVT} which makes use of the accurate time
provided by the \acl{GPS} receiver is discussed in this section.
Finally, a bandwidth optimization\index{Optimization} is suggested in which real\index{Real} packets may
be sent less frequently.
\subsection{Multiple Future Events}

The architecture for implementing alternative futures 
discussed in Section\ref{vncorg}, while a simple and
natural extension of the \acl{VNC} algorithm, creates additional messages and 
increases message sizes. Messages require an additional field to 
identify the probability of occurrence and an event identifier. 
However, the \acl{VNC} tolerance\index{Tolerance} is shown
to provide consideration of events which fall within the tolerances
$\Theta_n$ where $n \in N$ and $N$ is the number of \acl{LP}s\index{LP}. 

The set of possible futures at time $t$ is represented by the set $E$.
A message value generating an event occurring in one of the possible 
futures is represented by $E_{val}$. As messages propagate through
the \acl{VNC} system, there is a neighborhood around each message value
defined by the tolerance ($\Theta_n$). However, each message value
also accumulates error ($AC_n(n)$). Let the neighborhood ($E_\Delta$)
be defined such that $E_\Delta \le |\Theta_n - AC_n(n)|$ for each $n 
\in \{\acl{LP}s\}$.  Thus, 
$|E_\Delta + AC_n(n)| \le {\min_{n \in N}} \Theta_n$ defines a valid prediction.
The infinite set of events in the neighborhood $E_\Delta \le |\min_{n \in N} 
\Theta_n - AC_n(n)|$ are valid. Therefore, multiple future events 
which fall within the bounds of the tolerances reduced by any accumulated
error can be implicitly considered.

\subsection{Global Virtual\index{Virtual Network Configuration!verification} Time\index{Time}}
\label{nogvt}

In order to maintain the lookahead ($\Lambda$), for the entire configuration
system, it is necessary to know how far into the future the system is 
currently predicting. The purpose of \acf{GVT} is to determine $\Lambda$ where
$\Lambda$ is used to stop the \acl{VNC} system from looking ahead once the 
system has predicted up to the lookahead time. This helps maintain 
synchronization and saves processing and bandwidth since it is not 
necessary to continue the prediction process indefinitely into the future, 
especially since the prediction process is assumed to be less accurate the 
further it predicts into the future.

Distributed simulation mechanisms require \acl{GVT} in order to determine
when to commit events. This is because the simulation cannot rollback\index{Rollback}
beyond \acl{GVT}. In \acl{VNC}, event results are assumed to be cached 
before real\index{Real} time reaches the \acl{LVT} of an \acl{LP}. The only purpose for 
\acl{GVT} in \acl{VNC} is to act as a throttle on computation into the future. 
Thus, the complexity and overhead\index{Virtual Network Configuration!overhead}
required to accurately determine the \acl{GVT} is unnecessary in \acl{VNC}. 
In the \acl{VNC} system implemented for \acl{RDRN}, while the \acl{LVT} of 
an \acl{LP} is greater than $t + \Lambda$, the \acl{LP} does not process 
virtual messages. An accurate time $t$ is provided by the \acl{GPS} receiver.
This has worked well in providing the necessary throttle for \acl{LP} 
processing. It is knowledge of the accurate \acl{GPS} time that eliminates
the need for \acl{GVT}.

\subsection{Real Message Optimization\index{Optimization}}
\label{noreal}

\begin{figure*}[htbp]
        \centerline{\psfig{file=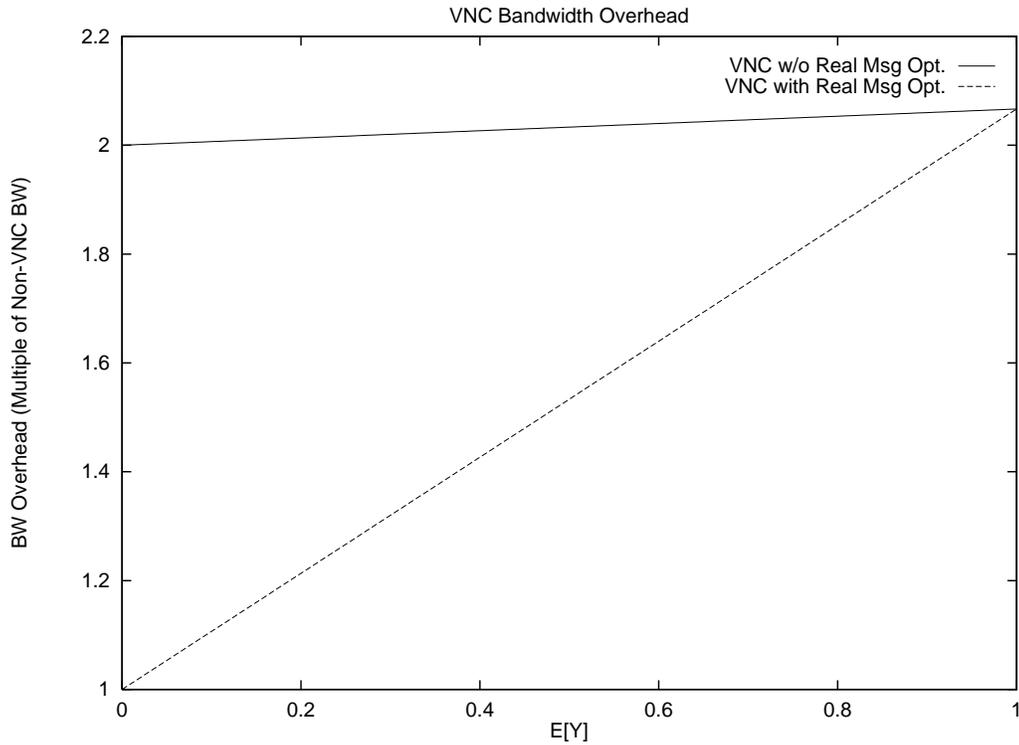,width=5.5in}}
        \caption{Bandwidth Overhead Reduction.}
        \label{bwreduce}
\end{figure*}

Real messages are only used in the \acl{VNC} algorithm as 
a verification\index{Virtual Network Configuration!verification} that a prediction has been accurate
within a given tolerance. The driving\index{Driving Process} process
need not send a real\index{Real} message if the virtual\index{Virtual Network Configuration!verification} messages
are within tolerance of the lowest tolerance within
the path of a virtual\index{Virtual Network Configuration!verification} message. This
requires that the driving\index{Driving Process} process have knowledge of the
destination processes' tolerance. The driving\index{Driving Process} process
will have copies of previously sent messages in its
send queue.
If real\index{Real} messages are only sent when an out-of-tolerance
condition occurs, then the orderwire load can be reduced
by up to 50\%. Figure \ref{bwreduce} compares the
orderwire load with and without the real\index{Real} message optimization\index{Optimization}.

\section{Analysis Summary}

The performance analysis of \acl{VNC} has quantified the costs
versus the speedup\index{Speedup} provided by \acl{VNC}. The costs have been
identified as the additional bandwidth and possible wasted
resources due to inaccurate prediction. Since the \acl{VNC}
algorithm combines optimistic synchronization with a real\index{Real}
time system, the probability of non-causal message order was
determined. A new approach using Petri-Nets and synchronic distance
determined the likelihood of out-of-order virtual\index{Virtual Network Configuration!verification} messages.
This new Petri-Net approach was applied to the \acl{RDRN} \acl{NCP}
to determine the ability of \acl{NCP} to enforce causality in
the virtual\index{Virtual Network Configuration!verification} messages. Other factors such as the distribution of 
partitioned tasks into \acl{LP}s were discussed.

The speedup\index{Speedup} was defined as the expected rate of change of the 
\acl{LVT} with respect to real\index{Real} time. The speedup\index{Speedup} was quantified
and a sensitivity analysis revealed the parameters most
affecting speedup\index{Speedup}. 
The bandwidth was quantified based on the probability of rollback\index{Rollback}
and the expected rollback\index{Rollback} rate of the \acl{VNC} system.
A general analysis of the accumulated error of the \acl{VNC} system
followed with the probability of error in the \acl{RDRN} \acl{NCP}.
Finally, the consideration of alternative future events, the
relevance of \acl{GVT}, and a bandwidth optimization\index{Optimization} technique 
were discussed.
The next section describes the experimental validation method
and compares the results with the analysis in this section.
\chapter{Experimental Validation}
\section{\acl{VNC} Experimental Validation Overview}

The experimental validation of \acl{VNC} examines simulations
and actual implementation\index{Virtual Network Configuration!implementation}s of the \acl{VNC} algorithm
in order to determine the speedup\index{Speedup} ($\eta$) and
bandwidth overhead\index{Virtual Network Configuration!overhead} ($\beta$) of the \acl{VNC} algorithm. 
The experimental validation is also an existence proof which
demonstrates the feasibility of the \acl{VNC} concepts.
For the \acl{RDRN}, all cached results
from the \acl{VNC} implementation\index{Virtual Network Configuration!implementation} are based directly on position and
the amount of position error is a controlled variable, thus there is
no accumulated error.

The implementation\index{Virtual Network Configuration!implementation} of the driving\index{Driving Process} process is critical 
for \acl{VNC}, therefore this chapter begins with a discussion of an
ideal implementation\index{Virtual Network Configuration!implementation} of the driving\index{Driving Process}
process for \acl{RDRN} position prediction. This
indicates the accuracy that can be expected from a position prediction
process. However, the implementation\index{Virtual Network Configuration!implementation} of the driving\index{Driving Process} process for
the experimental validation in this chapter simulates \acl{GPS} input. 
By simulating the \acl{GPS}\index{GPS}, the amount of error in virtual\index{Virtual Network Configuration!verification} messages can 
be accurately controlled.

\acl{VNC} has been experimental validated in a mobile network
configuration environment and in a simulated predictive\index{Mobile Network!predictive} network 
management environment. In the mobile network environment, a portion
of the \acl{RDRN} \acl{NCP} has been enhanced with \acl{VNC}. Specifically,
the task which processes \textbf{USER\_POS} messages pre-computes
and caches beam tables. In the  simulated predictive\index{Mobile Network!predictive} network
management environment, a standards based management system enhanced
with \acl{VNC} is simulated with Maisie\index{Maisie} \cite{bagrodia}.
\section{The Driving Process}

The \acl{VNC} driving\index{Driving Process} process for mobile systems will require accurate position
prediction. Previous mobile host location prediction algorithms have 
focused on an aggregate view of mobile host location prediction, primarily
for such purposes as base-station channel assignment and base-station
capacity planning. Examples are a fluid flow model\index{Model} \cite{Thomas} and 
the method of Hong and Rappaport \cite{Hong}. A location prediction
algorithm accurate enough for individual mobile host prediction
has been developed in \cite{liumma}. A brief overview of the algorithm
follows because the algorithm in \cite{liumma} is an ideal example
of a driving\index{Driving Process} process for \acl{VNC} and demonstrates the speedup\index{Speedup} that
\acl{VNC} is capable of providing with this prediction method. The algorithm 
allows individual mobile hosts to predict their future movement based on 
past history and known constraints in the mobile host's path. 

All movement ($\{M(k,t)\}$) is broken into two parts, regular\index{Motion!regular} and random\index{Motion!random} 
motion. A Markov model\index{Motion!markov model} is formed based on past history of regular and 
random motion and used to build a prediction mechanism for future movement 
as shown in Equation \ref{markovm}. The regular movement is identified by 
$S_{k,t}$ where $S$ is the state (geographical cell\index{Cell} area) identified by 
state index $k$ at time $t$ and the random movement is identified similarly 
by $X(k,t)$. $M(k,t)$ is the sum of the regular and random movement.

\begin{figure*}
\begin{equation}
\{M(k,t)\} = \{S_{k,t} | k \le K, t \in T\} + \{X(k,t) | k \le K, t \in T\}
\label{markovm}
\end{equation}
\end{figure*}

The mobile host location prediction algorithm in \cite{liumma} determines
regular movement as it occurs, then classifies and saves each regular 
move as part of a movement track or movement circle. A movement circle
is a series of position states which lead back to the initial state
while a movement track\index{Movement Track} leads from one point to another distinct point. A 
movement circle\index{Movement Circle} can be composed of movement tracks. Let $M_c$ denote a 
movement circle and $M_t$ denote a movement track. Then Equation 
\ref{randm} shows the random portion of the movement.

\begin{figure*}
\begin{equation}
\{X(k,t)\} = \{M(k,t)\} - (\{M_c(k,t) | k \le K, t \in T\} + 
\{M_t(k,t) | k \le K, t \in T\})
\label{randm}
\end{equation}
\end{figure*}

The result of this algorithm is a constantly updating model\index{Motion!model} of past
movement classified into regular and random movement. The proportion
of random movement to regular movement is called the randomness factor.
Simulation of this mobility algorithm in \cite{liumma} indicates a 
prediction efficiency of 95\%. The prediction efficiency is defined 
as the prediction accuracy rate over the regularity factor. The prediction 
accuracy rate is defined in \cite{liumma} as the probability of a correct
prediction. The regularity factor is the proportion of regular states,
$\{S_{k,t}\}$, to random states $\{X(k,t)\}$. 
The theoretically optimum line in \cite[p. 143]{liumma}, may have been better 
labeled the deterministic line. The deterministic line is an upper 
bound on prediction performance for all {\em regular} movement. The 
addition of the random portion of the movement may increase or decrease 
actual prediction results above or below the deterministic line.
A theoretically optimum (deterministic)
prediction accuracy rate is one with a randomness factor of zero and
a regularity factor\index{Motion!regularity factor} of one. The algorithm in \cite{liumma} does 
slightly worse than expected for completely deterministic regular 
movement but it improves as movement becomes more random. 
As a prediction algorithm for \acl{VNC}, a state as defined in \cite{liumma}
is chosen such that the area of the state corresponds exactly to the
\acl{VNC} tolerance, then based on the prediction accuracy rate in the
graph shown in \cite[p. 143]{liumma}
the probability of being out of tolerance is less than 30\% if the 
random movement ratio is kept below 0.4. An out-of-tolerance 
proportion of less than 30\% where virtual\index{Virtual Network Configuration!verification} messages are transmitted at 
a rate of $\lambda_{vm} = 0.03$ per millisecond results in a significant 
speedup as shown in Section \ref{perfanal}.

In the \acl{VNC} architecture, each node could independently run this 
predictive location algorithm in order to predict its own future location and 
notify the edge switch via a virtual\index{Virtual Network Configuration!verification} orderwire \textbf{USER\_POS} network 
control message as shown in Table \ref{prototab}. 
In the current version of the orderwire, a much simpler linear
prediction is used based on the last known location and speed. For
testing the \acl{VNC} concept, it is only necessary to have a
controlled amount of error for the predictive\index{Mobile Network!predictive} system. The same
code used to simulate the next location is used to predict the
next location with a controlled amount of error.

\section{A \acl{VNC} Enhanced \acl{RDRN} Orderwire Implementation\index{VNC!Implementation}}
\label{vncowi}

\begin{figure*}[htbp]
	\centerline{\psfig{file=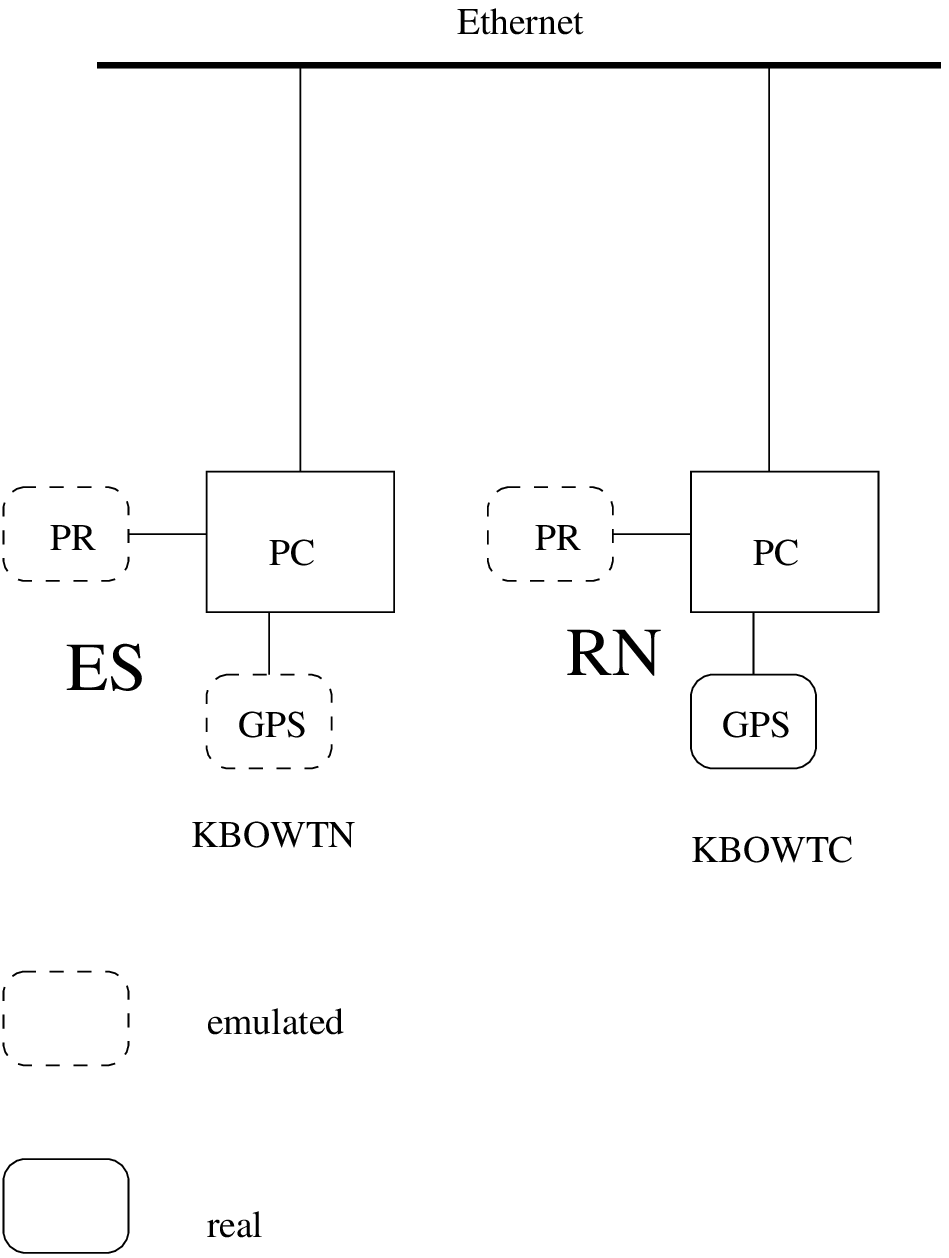,height=4.0in}}
        \caption{\acl{VNC} Test Environment.}
        \label{vnctestenv}
\end{figure*}

This section presents the results of performance measurements of 
the \acl{VNC} algorithm. The \acl{VNC} algorithm has been implemented 
for \acl{RDRN}. 
Measurements have been taken in the environment shown in Figure 
\ref{vnctestenv}. As mentioned previously, \acf{GPS} location is
simulated such that \acl{RN}s move in a linear path where the initial
direction is chosen at random. The complete \acf{NCP} code is used
in this emulation environment except for the 
simulated \acl{GPS} receiver which replaces the actual \acl{GPS} driver 
interface code. In addition the beam table creation and download
time is replaced with a delay which can be changed to examine
the effect of task execution time on \acl{VNC} performance.
The beam steering\index{Beamforming} code includes beam steering optimization\index{Optimization} 
\cite{ShaneBeam} between the \acl{ES} and \acl{RN} nodes. Thus, most
of the \acl{NCP} code is exercised in this emulation.

\begin{figure*}[htbp]
        \centerline{\psfig{file=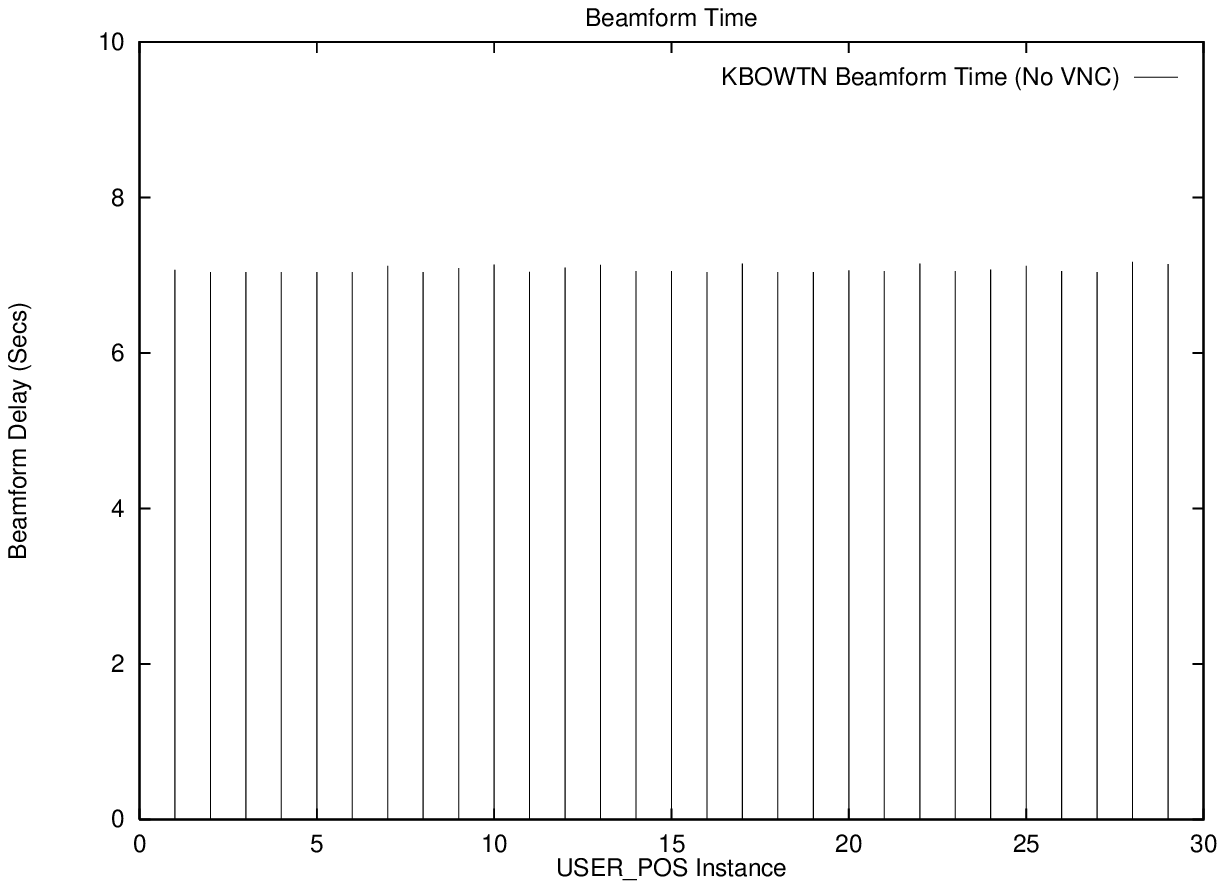,width=5.0in}}
        \caption{Edge Switch Beamforming\index{Beamforming} Time\index{Time} without VNC\index{VNC}.}
        \label{novnctestbeamtime}
\end{figure*}

The details of the \acl{NCP} operation have been discussed in Section
\ref{ncppro}. A position update message (\textbf{USER\_POS}) is
sent by \acl{RN} nodes to \acl{ES} nodes in order to update the \acl{RN} 
location. The first measurement is the time to process a position update 
message. 
This time begins from the receipt of an \acl{RN} position update message 
and ends when the \acl{RN} position update message has been processed. 
This may include the creation and download of a new beam table. When 
\acl{VNC} is disabled, the resulting beam creation\index{Beamforming} times are shown in Figure 
\ref{novnctestbeamtime}. The y-axis in Figure \ref{novnctestbeamtime}
plots the time required to generate a beam table. The x-axis identifies
a particular \textbf{USER\_POS} message. There is not a significant
variation in the beamform delay.

\begin{figure*}[htbp]
        \centerline{\psfig{file=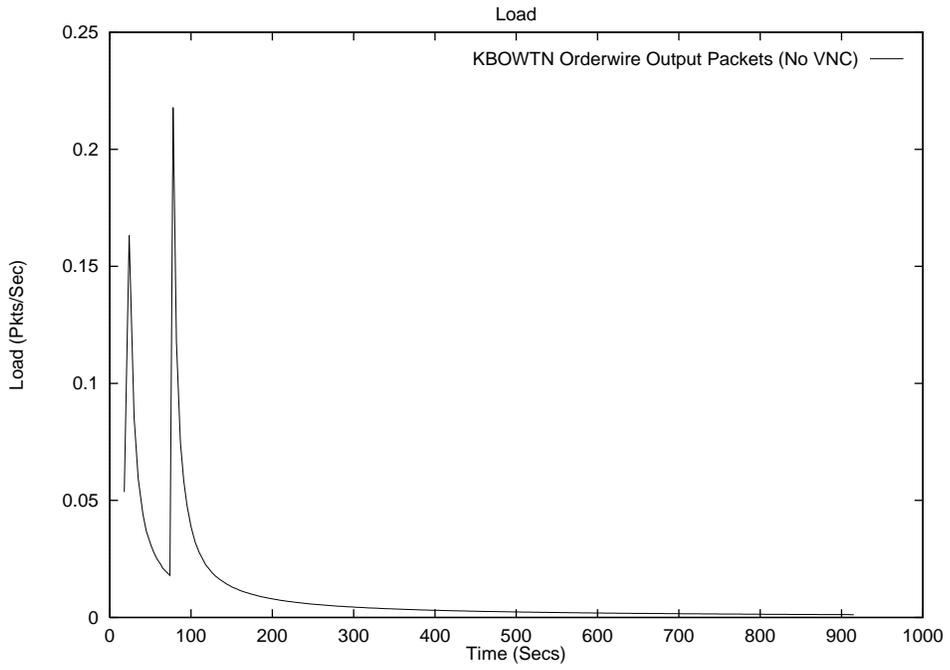,width=5.0in}}
        \caption{Edge Switch Orderwire Load without VNC\index{VNC}.}
        \label{novnctestload}
\end{figure*}

\begin{figure*}[htbp]
        \centerline{\psfig{file=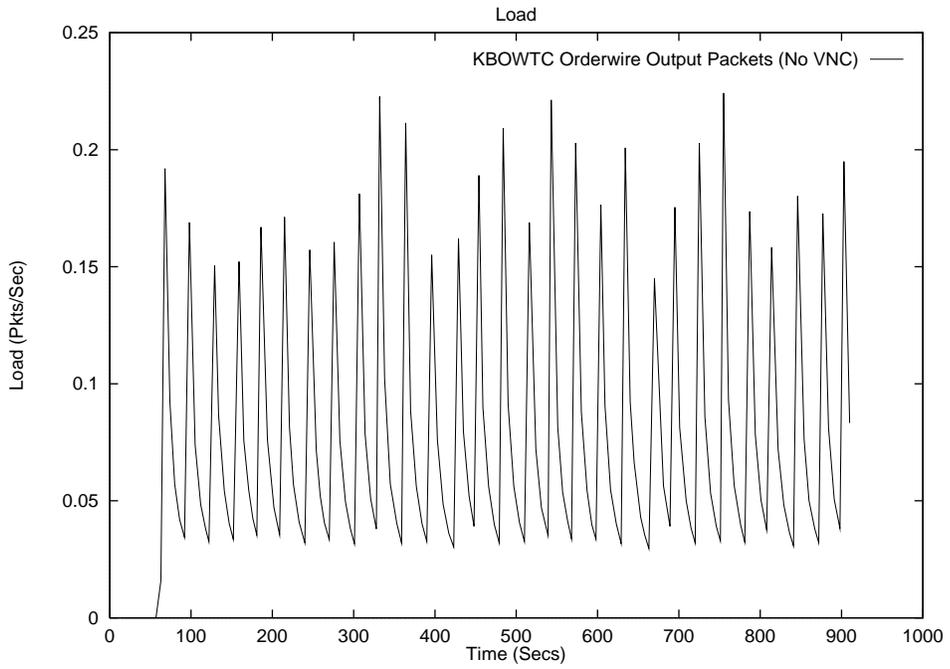,width=5.0in}}
        \caption{Remote Node Orderwire Load without VNC\index{VNC}.}
        \label{novncrntestload}
\end{figure*}

The orderwire network load as a function\index{Function} of time as measured on the \acl{ES} 
without \acl{VNC} is shown in 
Figure \ref{novnctestload} and the orderwire load due to the
\acl{RN} is shown in Figure \ref{novncrntestload}.
The load is normalized in terms of \acl{DES}\index{DES|see{Data Encryption Standard}} encoded packets per 
second and is averaged over one second intervals. The 
traffic on the orderwire consists of the initial \acl{ES} configuration 
followed by the \acl{RN} position update messages and handoff\index{Handoff} messages. The 
initial load shown in Figure \ref{novnctestload} is caused by the \acl{ES} 
configuration which quickly tapers off as \acl{ES} configuration ends and 
\textbf{HANDOFF} messages may be transmitted.

\begin{figure*}[htbp]
        \centerline{\psfig{file=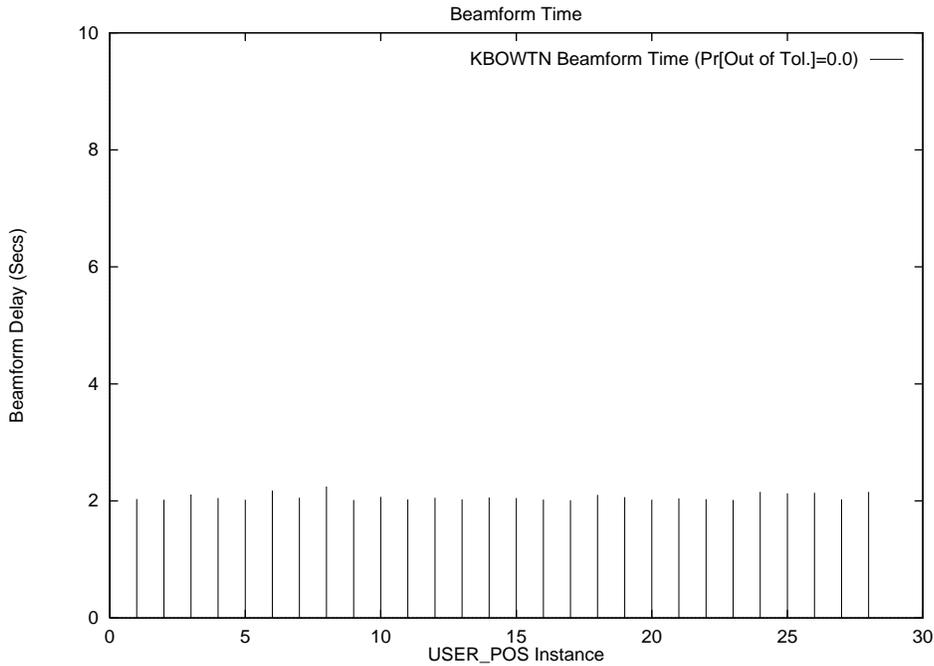,width=5.0in}}
        \caption{ES Beam formation Time\index{Time} with VNC\index{VNC} and No Prediction Error.}
        \label{vnc0testbeamtime}
\end{figure*}

\begin{figure*}[htbp]
        \centerline{\psfig{file=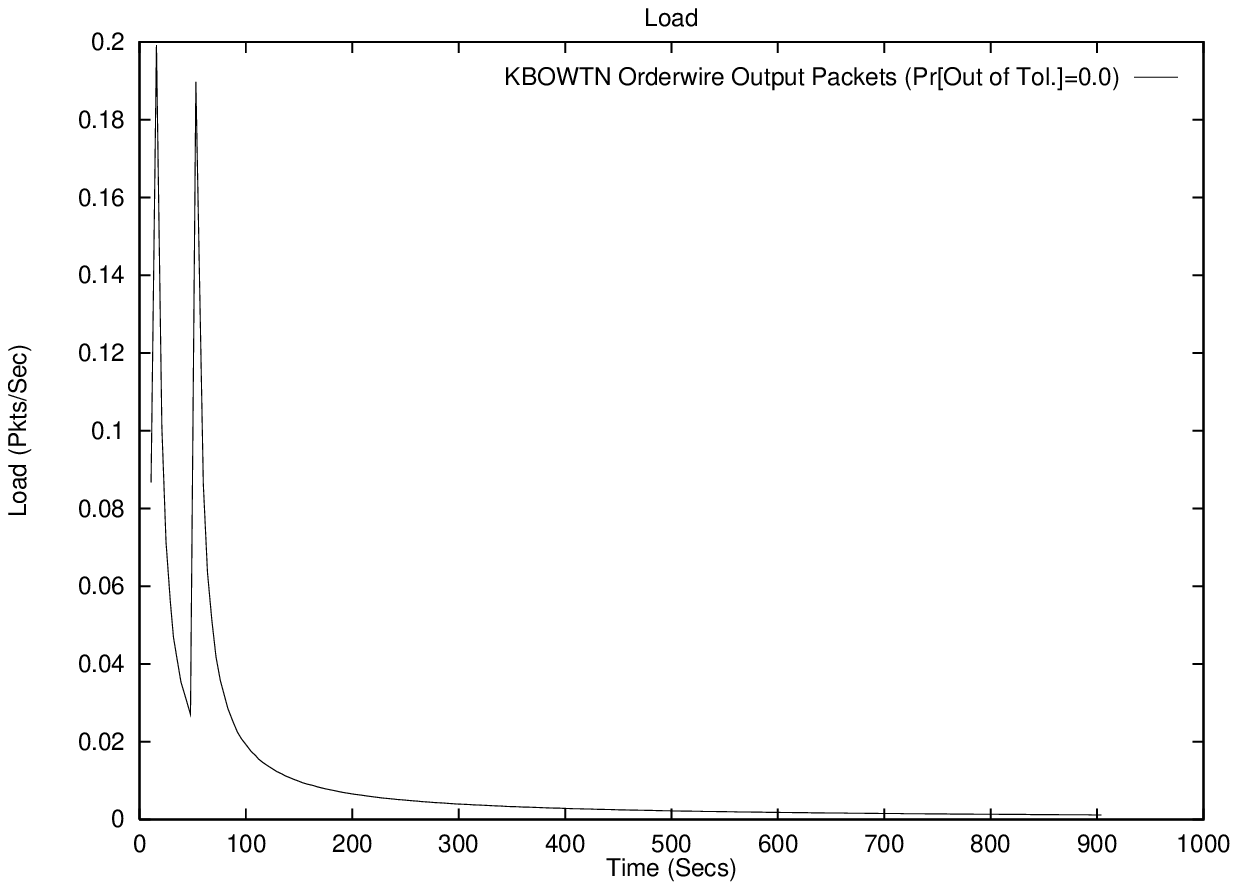,width=5.0in}}
        \caption{ES Orderwire Load with VNC\index{VNC} and No Prediction Error.}
        \label{vnc0testload}
\end{figure*}

\begin{figure*}[htbp]
        \centerline{\psfig{file=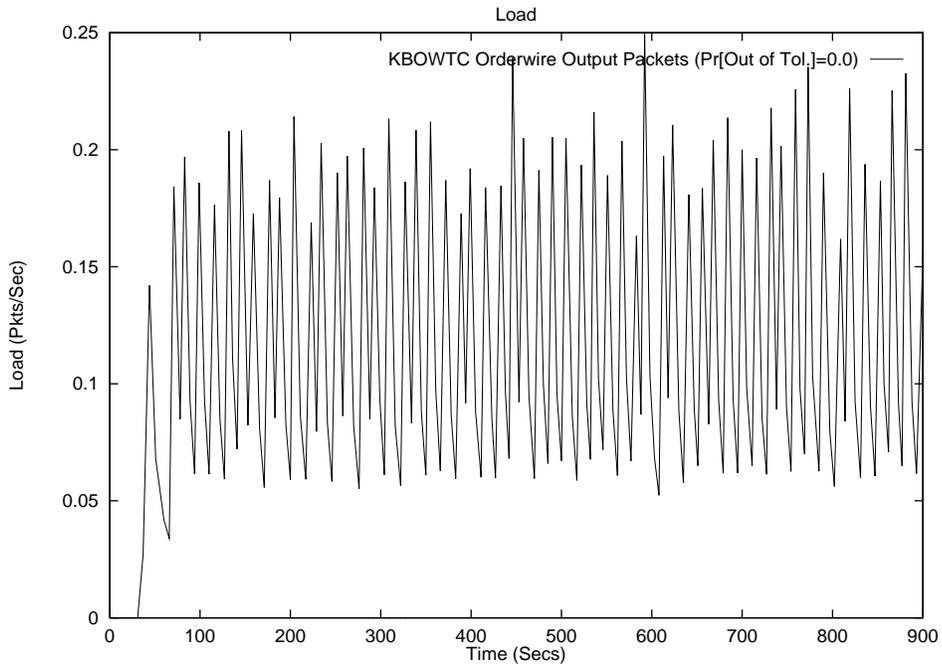,width=5.0in}}
        \caption{Remote Node Orderwire Load with VNC\index{VNC} and No Prediction Error.}
        \label{vnc0rntestload}
\end{figure*}

\begin{figure*}[htbp]
        \centerline{\psfig{file=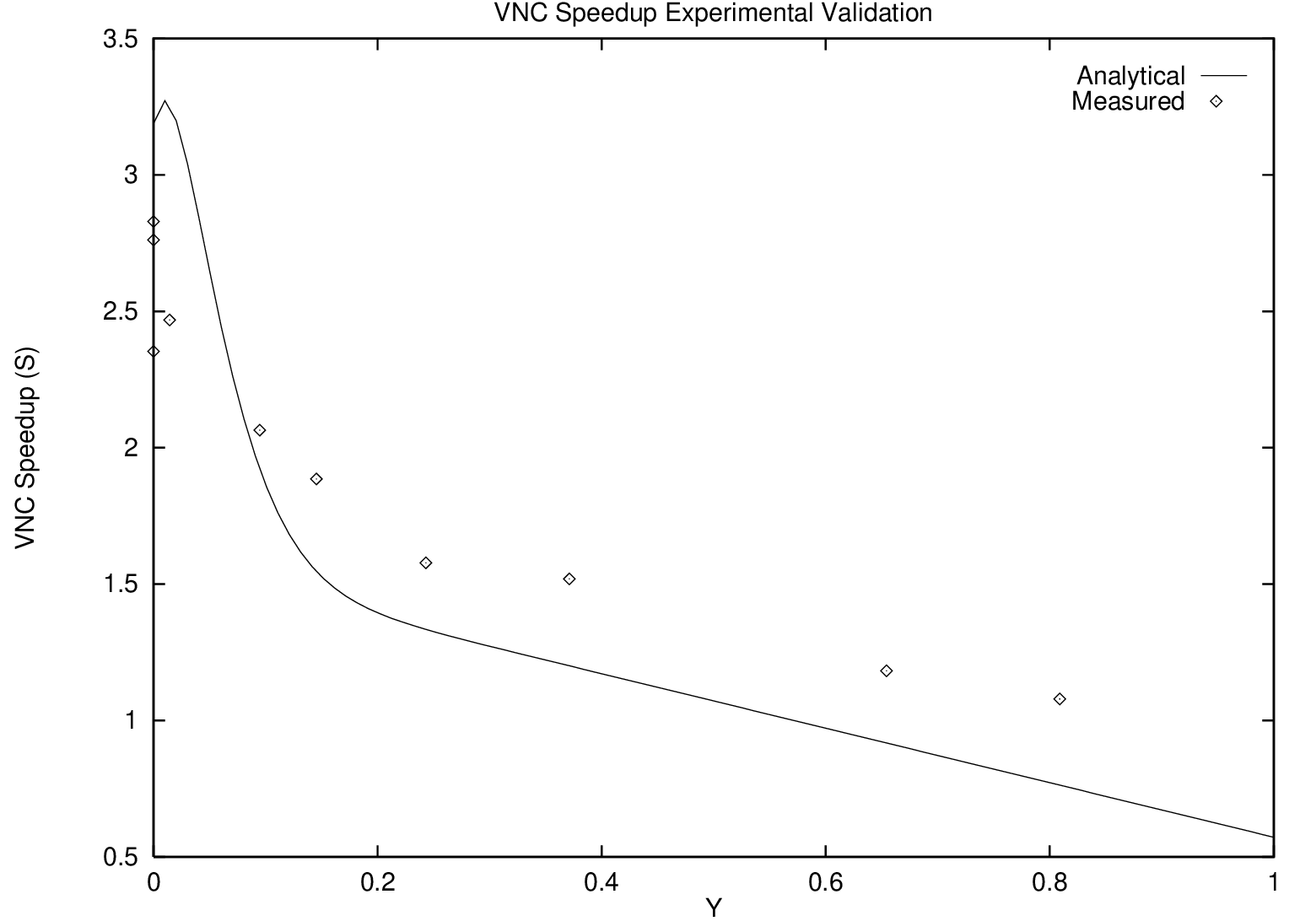,width=5.0in}}
        \caption{ES Beam Creation Speed\index{Speed}up.}
        \label{esspeedup}
\end{figure*}

The beam creation\index{Beamforming} time is 3.5 times faster if the beam results have 
been cached, thus the speedup\index{VNC!speedup} as determined from the analysis is shown 
in Figure \ref{esspeedup}.
The expected beam creation time without \acl{VNC} was 8.0 seconds. With
perfect prediction, the expected beam creation time with \acl{VNC} was
2.29 seconds for a speedup\index{Speedup} of 3.5 times. The measured results are plotted
with Figure \ref{esspeedup} in Figure \ref{esspeedup}.
The \acl{VNC} algorithm using perfect position prediction generated the
beamform delay shown in Figure \ref{vnc0testbeamtime}. The beam creation
time has been significantly reduced. Figure \ref{vnc0testload} shows the 
\acl{VNC} orderwire traffic load from the \acl{ES} and Figure 
\ref{vnc0rntestload} shows the load from the \acl{RN}. The load is 
approximately double the load shown in Figure \ref{novnctestload}.


\begin{figure*}[htbp]
        \centerline{\psfig{file=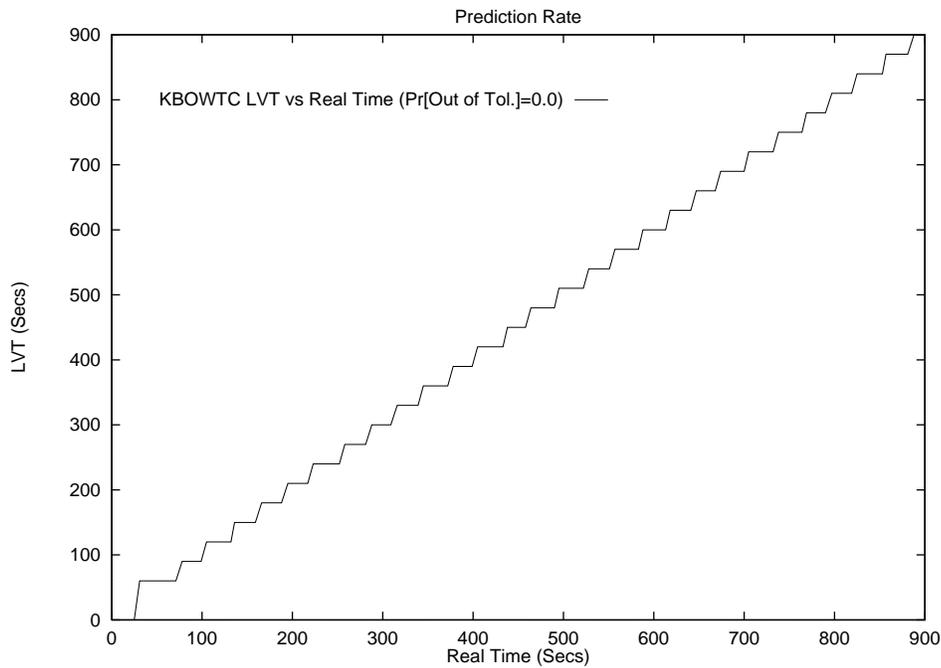,width=5.0in}}
        \caption{Remote Node LVT with VNC and No Prediction Error.}
        \label{rnvnc0testlvt}
\end{figure*}

\begin{figure*}[htbp]                                                  
        \centerline{\psfig{file=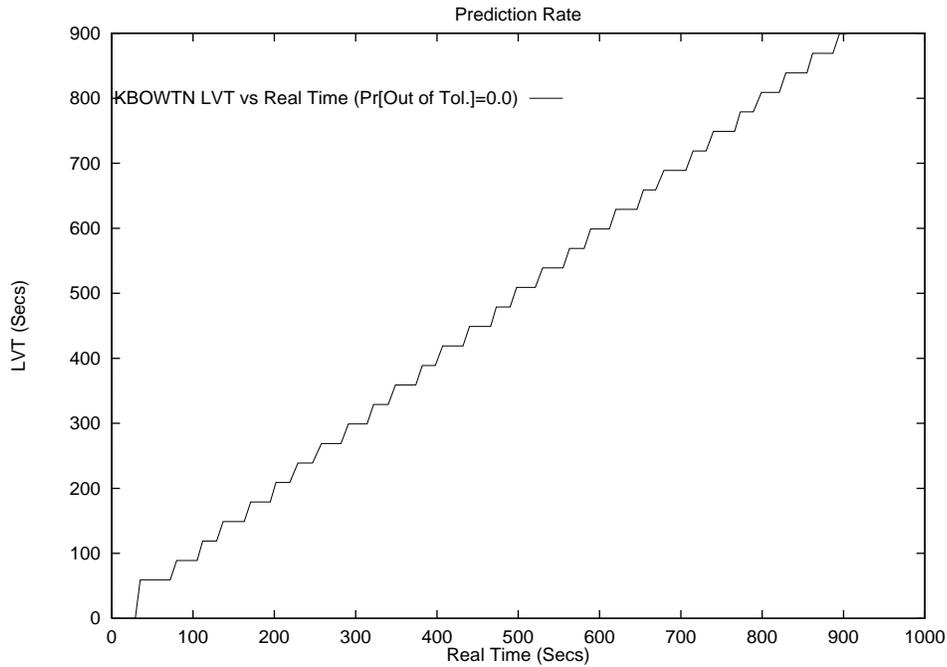,width=5.0in}}
        \caption{ES LVT with VNC and No Prediction Error.}      
        \label{esvnc0testlvt}
\end{figure*}

Figure \ref{rnvnc0testlvt} shows the \acl{RN} \acl{LVT} as a function\index{Function} of time.
This figure shows the \acl{LVT} quickly reaching the 30 second lookahead
and delaying until the end of the lookahead window as expected. The 
prediction rate shown in Figure \ref{rnvnc0testlvt} is 1.23 virtual\index{Virtual Network Configuration!verification} seconds 
per second as determined from the analysis in Section \ref{slprate} given
\aparms.
Figure \ref{esvnc0testlvt} shows the \acl{ES} \acl{LVT} as a function\index{Function} of time.
The \acl{ES} \acl{LVT} is driven by the \acf{TR} of messages entering its
input queue from the \acl{RN}. Thus the \acl{ES} \acl{LVT}\index{LVT} is similar to
the \acl{RN} \acl{LVT}.

\begin{figure*}[htbp]
        \centerline{\psfig{file=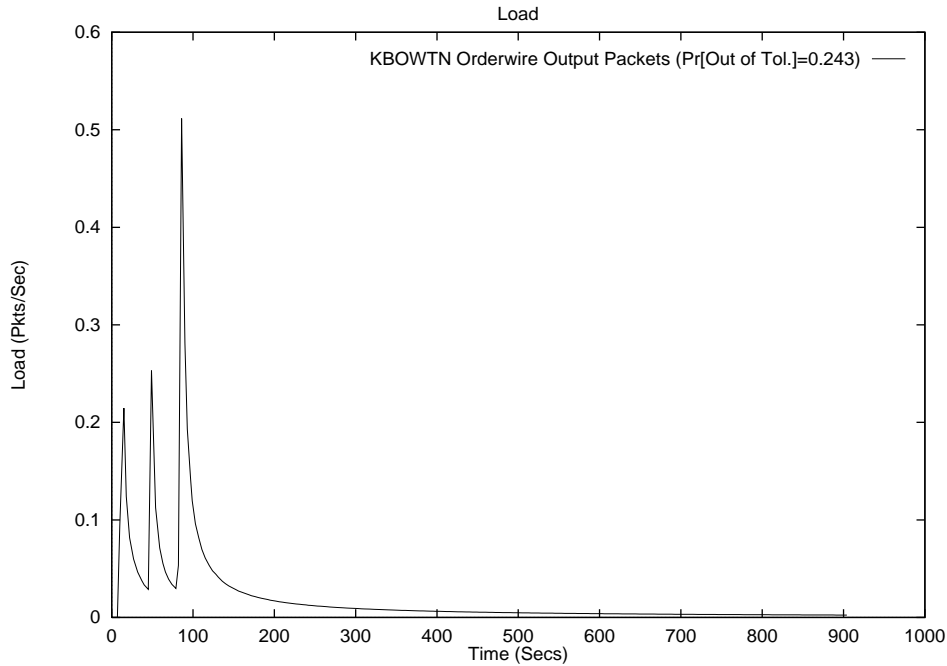,width=5.0in}}
        \caption{ES Orderwire Load with VNC and Large Prediction Error.}
        \label{vnc150testload}
\end{figure*}

\begin{figure*}[htbp]
        \centerline{\psfig{file=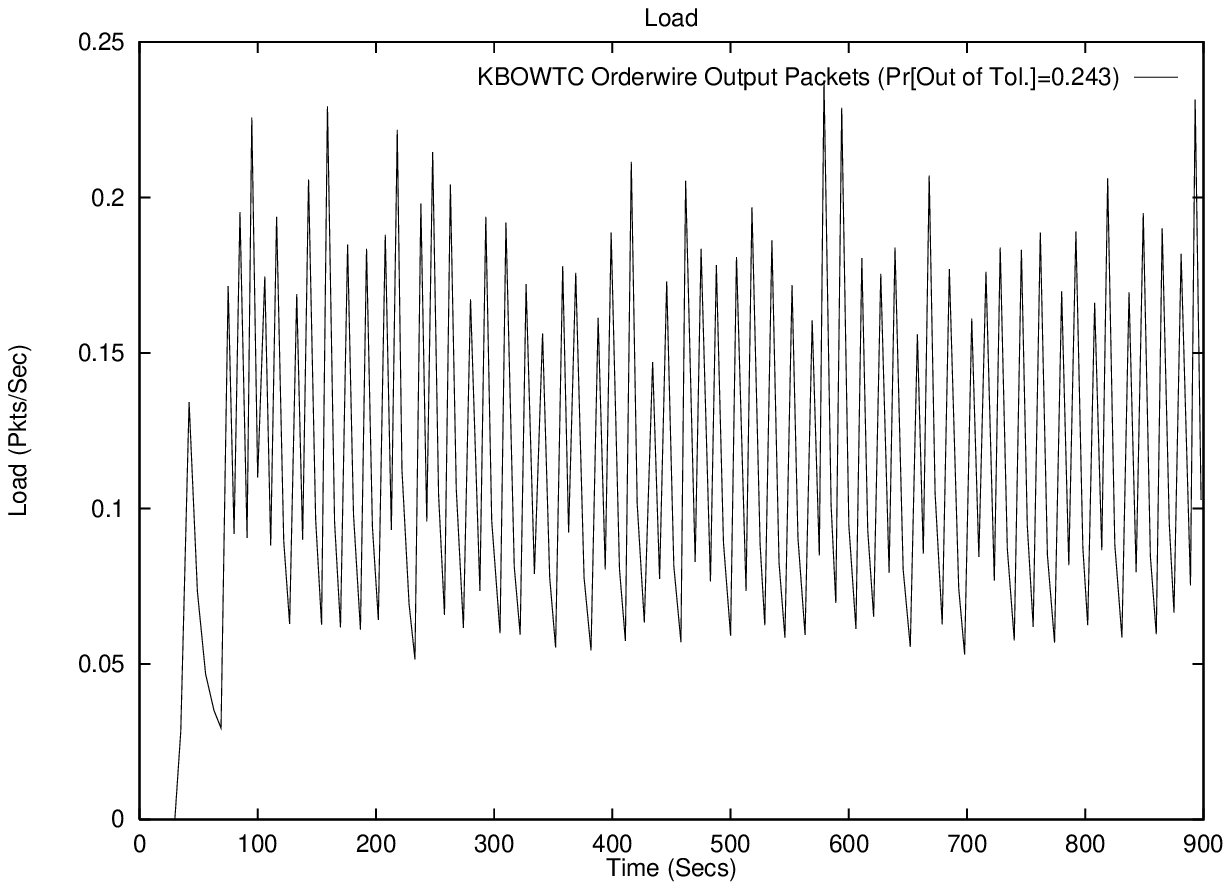,width=5.0in}}
        \caption{RN Orderwire Load with VNC and Large Prediction Error.}
        \label{vnc150rntestload}
\end{figure*}

The next set of results show the performance of the \acl{VNC} algorithm 
as the prediction algorithm becomes less perfect. The predicted values 
are degraded in
the driving\index{Driving Process} process by adding an exponentially distributed\index{Distributed} error to the
predicted values. As discussed in the analysis section,
rollback will increase the load on the orderwire and increase 
the user position message processing time.
Figure \ref{vnc150testload} shows the \acl{ES} orderwire load under the 
presence of rollback\index{VNC!rollback} and Figure \ref{vnc150rntestload} shows the \acl{RN} 
load. 

Compare the \acl{ES} load without \acl{VNC} Figure \ref{novnctestload},
the \acl{ES} load with \acl{VNC} and no prediction error in Figure 
\ref{vnc0testload}, and the \acl{ES} load with \acl{VNC} and large prediction 
error \ref{vnc150testload}. The first peak in all the \acl{ES} figures
is the initial \textbf{MYCALL} packet broadcast upon startup. The next
peak is the initial \textbf{HANDOFF} packet sent from the \acl{ES} to the
\acl{RN}. Notice that the handoff\index{Handoff} occurs earlier in Figure \ref{vnc0testload}
than it does in Figure \ref{novnctestload}. This due to the speedup\index{Speedup} gained
via \acl{VNC}. However, in Figure \ref{vnc150testload}, a third peak
occurs which is a \textbf{HANDOFF} anti-message. The \textbf{HANDOFF} 
anti-message is sent because the initial location prediction was out
of tolerance. Once the handoff\index{Handoff} location has been corrected by the
\textbf{HANDOFF} anti-message, no additional messages are sent from
the \acl{ES}.

\begin{figure*}[htbp]
        \centerline{\psfig{file=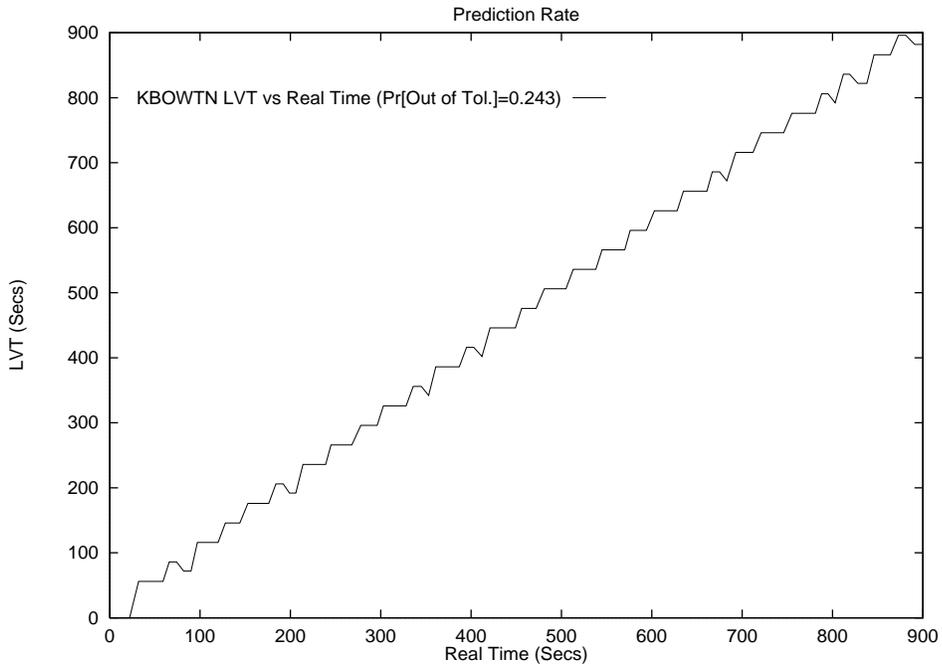,width=5.0in}}
        \caption{ES LVT with VNC and Large Prediction Error.}
        \label{esvnc150testlvt}
\end{figure*}

\begin{figure*}[htbp]
        \centerline{\psfig{file=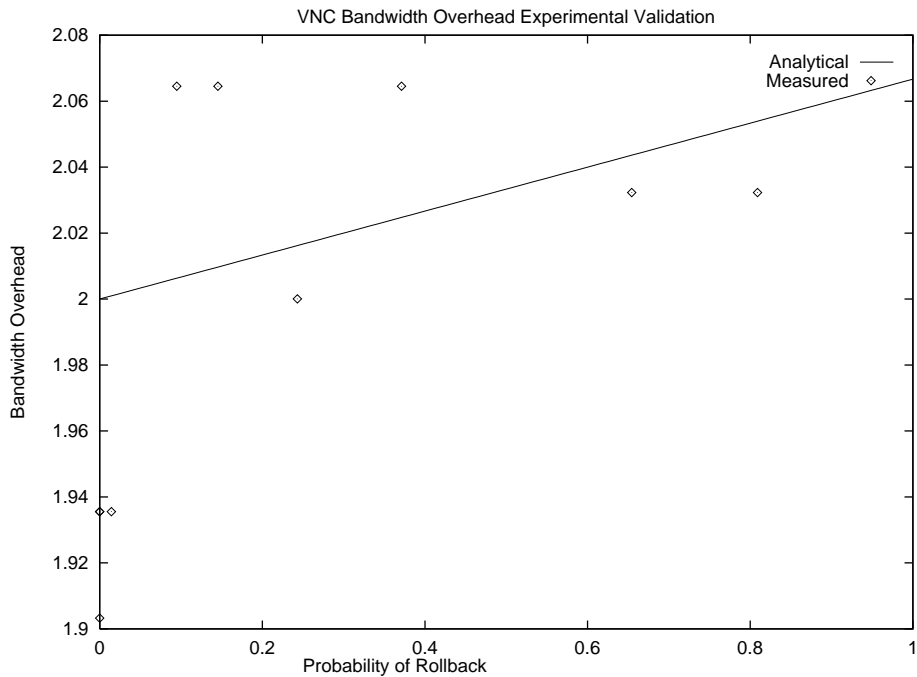,width=5.0in}}
        \caption{VNC Bandwidth Overhead.}
        \label{vncbo}
\end{figure*}

Figure \ref{vncbo} shows the analytical and measured bandwidth overhead\index{Virtual Network Configuration!overhead} as
the prediction error increases. The bandwidth overhead\index{Virtual Network Configuration!overhead} increases by a
small amount above twice the non-\acl{VNC} bandwidth as expected in
the analysis.
Figure \ref{esspeedup} shows the analytical and measured speedup\index{Speedup} in the 
processing of a \textbf{USER\_POS} packet as the prediction error increases. 
The measured results have a slightly lower speedup\index{Speedup} for low out-of-tolerance
message probability and higher speedup\index{Speedup} than the analytical results for
higher out-of-tolerance message probabilities. There are several possible
explanations for the inaccuracy. First, the emulation was done on a shared
processor with many other unrelated processes with different processor 
scheduling priorities. It is possible that these other processes have 
influenced the real\index{Real} time results. Second, the beam table creation task 
measured in this experimental implementation\index{Virtual Network Configuration!implementation} uses the \acl{VNC} time warp 
mechanism across two different processors. The simplified speedup\index{Speedup} analysis 
for the contribution of time warp to \acl{VNC} speedup\index{Speedup} may not be precise
enough. Finally, the analysis assumed an average 
time to perform the standard rollback\index{Rollback} ($\tau_{rb}$) operations such as 
restoring the \acl{LP} state to a previously saved state from the \acf{SQ} 
and modifying the \acl{LVT}. The actual time to perform a rollback\index{Rollback} may be 
smaller than the value used in the analysis. Over estimating $\tau_{rb}$
would have little or no effect when $Y$ is small, but it would have a
larger effect as $Y$ increases which appears to be the case in 
Figure \ref{esspeedup}.

\section{\acl{VNC} \acl{NCP} Experimental Implementation\index{VNC!Implementation} Summary}

The \acl{VNC} algorithm is most sensitive to the proportion of 
messages which cause a rollback\index{Rollback} rate. The dominate cause of rollback\index{Rollback} 
is the proportion of messages which cause an out-of-tolerance rollback\index{Rollback} 
($Y$) as determined in Section \ref{sens}. Clearly minimizing $Y$ has 
the greatest effect in optimizing the operation.
However, if ($Y$) cannot be minimized, the virtual\index{Virtual Network Configuration!verification} message
generation rates and expected virtual\index{Virtual Network Configuration!verification} message lookahead times can be
adjusted as discussed in Section \ref{perfanal}
and shown in Figure \ref{vmrate} in order to maximize the speedup\index{Speedup}.

The speedup\index{Speedup} provided by \acl{VNC} has been shown to be significant
in the analysis and the experimental validation results. There
are two contributing factors to the speedup\index{Speedup} in \acl{VNC}. The
primary source of speedup\index{Speedup} is the computing ahead and caching of
information required in the future. Another source of speedup\index{Speedup}
in \acl{VNC} comes from the fact that optimistic advantage
is taken of parallel\index{Parallel} processing available to the system.

The bandwidth overhead\index{Virtual Network Configuration!overhead} as a function\index{Function} of rollback\index{Rollback} rate
and the number of \acl{LP}s is shown in Figure \ref{bwscale}.
This graph illustrates the trade-off between the number of \acl{LP}s
and the rollback\index{Rollback} rate. The rollback\index{Rollback} rate in
this graph is the sum of both the out-of-order and the out-of-tolerance
rollback rates. The dominate contribution to the amount
of overhead\index{Virtual Network Configuration!overhead} is the rollback\index{Rollback} rate and not the number of
\acl{LP}s.
The dominate overhead\index{Virtual Network Configuration!overhead} required by \acl{VNC} is additional bandwidth.
A \acl{VNC} system which has perfectly adjusted parameters and
perfect prediction will require twice the bandwidth of the same
system without \acl{VNC}. As the rollback\index{Rollback} rate increases,
more bandwidth is required. The rate of increase in bandwidth is
shown in Figure \ref{bwscale}.

The experimental implementation\index{Virtual Network Configuration!implementation} has shown the speedup\index{Speedup} of \acl{VNC}
to decrease as a function\index{Function} of out-of-tolerance messages at the rate 
determined by the analysis. This is shown in Figure \ref{esspeedup}. 
Also, the bandwidth overhead\index{Virtual Network Configuration!overhead} has been
shown to increase at a very slight rate as a function\index{Function} of 
the proportion of out-of-tolerance messages as shown in Figure 
\ref{bwover}.

In the next section, the \acl{VNC} algorithm is used to enhance the 
operation of a predictive\index{Mobile Network!predictive} network management application. The predictive\index{Mobile Network!predictive}
capability of \acl{VNC} is utilized in the predictive\index{Mobile Network!predictive} management application,
however, the overhead\index{Virtual Network Configuration!overhead} measurement is different as discussed in Section
\ref{vncsim}.
\section{Predictive Network Management Simulation Results}
\label{vncsim}

While the goal of \acl{VNC} in the predictive\index{Predictive Network Management!predictive} management environment is
to enable prediction just as in the \acl{RDRN} environment, the focus
on the overhead\index{Virtual Network Configuration!overhead} is different. As discussed later in this section, the 
goal in the predictive\index{Predictive Network Management!predictive} management system is to reduce the verification\index{Virtual Network Configuration!verification} 
query overhead\index{Virtual Network Configuration!overhead}. The predictive\index{Predictive Network Management!predictive} management system is assumed to run on a
separate processing system, thus the message overhead\index{Virtual Network Configuration!overhead} among \acl{LP}s, 
is not as critical as in \acl{RDRN}.
In the \acl{RDRN} implementation\index{Virtual Network Configuration!implementation}, there is no verification\index{Virtual Network Configuration!verification} query and the 
inter-\acl{LP} message is crucial since the transmission medium between 
\acl{LP}s is a low bandwidth packet radio system.
Another measurement of interest for the \acl{VNC} enhanced predictive\index{Predictive Network Management!predictive} 
management environment is the prediction rate and the effects of out 
of tolerance and out of order rollback\index{Rollback}. Therefore, the metrics that 
this section focuses on are verification\index{Virtual Network Configuration!verification} query overhead\index{Virtual Network Configuration!overhead} and prediction 
rate.

Systems management means the management of 
heterogeneous subsystems of network devices, processing platforms,
distributed applications\index{Applications}, and other components found in
communications and computing environments. 
Current system management relies on presenting a model\index{Model} to the user of the
managed system which should accurately reflect the current
state of the system and should ideally be capable of predicting the 
future health of the system.
System management relies on a combination of asynchronous\index{Asynchronous}ly generated
alerts and polling to determine the health of a system
\cite{Steinber91}.

The management application presents state information such as link state, 
buffer fill and packet loss to the user in the form of a model\index{Model} \cite{Sylor}.
The model\index{Model} can be as simple as a passive display of nodes on a screen or
a more active model\index{Model} which allows displayed nodes to change color based 
on state changes, or react to user input by allowing the user to manipulate 
the nodes which causes values to be set on the managed entity. This model\index{Model} 
can be made even more active by enhancing it with predictive\index{Predictive Network Management!predictive} capability \cite{BushJNSM}.
This enables the management system to manage itself, for example, to optimize 
its polling rate. The two major management protocols, \acl{SNMP} \cite{SNMP} 
and \acl{CMIP} \cite{CMIP}, allow the management station to poll a managed 
entity to determine its state. In order to accomplish real\index{Real}-time and 
predictive network management in an efficient manner, the model\index{Model} should
be updated with real\index{Real}-time state information when it becomes available,
while other parts of the model\index{Model} work ahead in time. Those objects working 
ahead of real\index{Real}-time can predict future operation so that system management 
parameters such as polling times and thresholds can be dynamically adjusted 
and problems can be anticipated. The model\index{Model} will not deviate too far from 
reality because those processes which are found to deviate beyond a certain 
threshold will be rolled back, as explained in detail later.

One goal of this research is to minimize polling overhead\index{Virtual Network Configuration!overhead}
in the management of large systems \cite{Takagi86}.
Instead of basing the polling rate on the characteristics of the
data itself, the entity is emulated some time into the future in
order to determine the characteristics of the data to be polled.
Polling is still required with this predictive\index{Predictive Network Management!predictive} network management
system in order to verify the accuracy of the emulation. 

\section{Predictive Standards-Based Network Management Information}

Management information from standards-based managed
entities must be mapped into this predictive\index{Predictive Network Management!predictive} network management system.
Network management systems rely upon standard mechanisms to obtain the state of 
their managed entities in near real\index{Real}-time. These mechanisms, SNMP\index{SNMP}
\cite{SNMP} and CMIP\index{CMIP} \cite{CMIP} for example, use both solicited
and unsolicited methods. The unsolicited method uses
messages sent from a managed entity to the manager. These unsolicited
messages are
called traps or notifications; the former are not acknowledged while
the latter are acknowledged. These messages are very similar to messages
used in distributed\index{Distributed} simulation algorithms; they contain a timestamp and 
a value, they are sent to a particular destination, that is, a management
entity, and they are the result of an event which has occurred.

Information requested by the management system from a particular managed
entity is solicited information. It also corresponds to messages 
in distributed\index{Distributed} simulation. It provides a time and a value; however,
not all such messages are equivalent to messages in distributed\index{Distributed}
simulation and required in a predictive\index{Predictive Network Management!predictive} management system. These messages 
provide the management station with the 
current state of the managed entity, even though no change of 
state may have occurred or multiple state changes may have occurred. 
The design of a management system which requests information on the
state of its managed entities at the optimum time has always been a problem
in network management. If requested too frequently, bandwidth is wasted,
if not requested frequently enough, critical state change information
will be missed.

We will assume for simplicity that each managed entity is represented in
the predictive\index{Predictive Network Management!predictive} management system by a \acl{LP}. It would greatly 
facilitate system management if vendors provide not only the standards based 
SNMP Management Information Base (MIB) as they do now, but also a standard 
simulation code which model\index{Model}s the entity or application behavior and can be 
plugged into the management system just as in the case with a MIB. Vendors 
should have model\index{Model}s of their devices readily available from product 
development.

\section{Characteristics of the Predictive\index{Predictive Network Management!predictive} Network Management System}

There are two types of false\index{False Message} messages generated in this predictive\index{Predictive Network Management} 
network management system; those produced by messages arriving in the past 
Local Virtual\index{Virtual Network Configuration!verification} Time\index{Time} (LVT) of an \acl{LP} and those produced because the \acl{LP} is
generating results which do not match real\index{Real}ity. If rollback\index{Rollback}s occur for
both reasons the question arises as to whether the system will be
stable. 
An unstable system is one in which there
exists enough rollback\index{Rollback}s to cause the system to take longer
than real\index{Real}-time to reach the end of the \acf{SLW}.
A stable system is able to make reasonably accurate predictions far 
enough into the future to be useful.
An unstable system will have its performance degraded by rollback\index{Rollback}s
to the point where it is not able to predict ahead of real\index{Real}-time.
Initial results shown later indicate that predictive\index{Predictive Network Management!predictive} 
network management systems can be stable.

There are several parameters in this predictive\index{Predictive Network Management!predictive} network management system 
which must be determined. The first is how often the predictive\index{Predictive Network Management!predictive} network 
management system should check the \acl{LP} to verify that past results match 
reality. There are two conditions which cause \acl{LP}s in the system to 
have states which differ from the system being managed and to
produce inaccurate predictions. The first is that the predictive\index{Predictive Network Management!predictive}
model which comprises an \acl{LP} is most likely a simplification of the 
actual managed entity and thus cannot model\index{Model} the entity with perfect
fidelity. The second reason is that events outside the scope of the
model may occur which lead to inaccurate results. However, a benefit
of this system is that it will self-adjust for both of these conditions.

The optimum choice of verification\index{Virtual Network Configuration!verification} 
query time, $T_{query}$, is important because querying entities is
something the predictive\index{Predictive Network Management!predictive} management system should minimize while still 
guaranteeing that the
accuracy is maintained within some predefined tolerance ($\Theta$). 
For example, the network management station may predict user location
as explained later. If the physical\index{Physical} layer attempts spatial 
reuse via antenna beamforming\index{Beamforming}
techniques as in the RDRN\index{RDRN} project, then there is an acceptable amount
of error in the steering angle for the beam and thus the node location
is allowed a tolerance.
The tolerances could be set for each state variable or message value 
sent from a \acl{LP}. State\index{State}\index{State} verification\index{Virtual Network Configuration!verification} can be done in one of at least
two ways. The \acl{LP} state can be compared with previously saved states
as real\index{Real} time catches up to the saved state times or output message
values can be compared with previously saved output messages in the
send queue.
In the prototype implemented for this predictive\index{Predictive Network Management!predictive} network management 
system state verification\index{Virtual Network Configuration!verification} is done based on states saved in the state
queue.  This implies that all \acl{LP} states must be saved from the \acl{LP} LVT\index{LVT} 
back to the current time.

The amount of time into the future which the emulation will attempt
to venture is another parameter which must be determined. This 
\acl{SLW} width ($\Lambda$) should be preconfigured 
based on the accuracy required; the farther ahead this predictive\index{Predictive Network Management!predictive} 
network management system attempts to go 
past real\index{Real} time, the more risk that is assumed.

\subsection{Optimum Choice of Verification\index{Virtual Network Configuration!verification} Query Time\index{Time}s}

As previously stated, the prototype system performs the 
verification based on the states in the state queue.

One method of choosing the verification\index{Virtual Network Configuration!verification} query time would be to query 
the entity based on the frequency of 
the data we are trying to monitor. Assuming the simulated data is
correct, query or sample in such a way as to perfectly reconstruct
the data, for example, based on the maximum frequency component of the monitored
data. A possible
drawback is that the actual data may be changing at a multiple
of the predicted rate. The samples may appear to to be accurate when
they are invalid.

\subsection{Verification Tolerance\index{Tolerance}}

The verification\index{Virtual Network Configuration!verification} tolerance ($\Theta$) is the amount of difference 
allowed between the \acl{LP} state and the actual entity state.
A large tolerance decreases the number of false\index{False Message} messages and rollback\index{Rollback}s,
thus increasing performance and requiring fewer queries, but allows
a larger probability of error between predicted and the actual
values will cause rollback\index{Rollback}s in each \acl{LP} at real\index{Real} times of $t_{vfail}$
from the start of execution of each \acl{LP}.

The error throughout the simulated system may be randomized in such a way 
that errors among \acl{LP}s cancel. However, if the simulation is composed of 
many of the same class of \acl{LP}, the errors may compound rather than cancel
each other. The tolerance of a particular \acl{LP} ($\Theta_{lp_n}$) will be 
reached in time $t_{vfail_n} = \{ \mbox{lub\ } \tau \mbox{ s.t. } 
AC_t(\tau) > \Theta_{lp_n} \}$.
The verification\index{Virtual Network Configuration!verification} query period ($\Upsilon$) should be
less than or equal to $t_{vfail_n}$ in order to maintain accuracy within 
the tolerance.

The accuracy of any predicted event must be quantified. This could be
quantified as the probability of occurrence of a predicted event. The 
probability of occurrence will be a function\index{Accuracy!function} of the verification\index{Virtual Network Configuration!verification} tolerance, 
the time of last rollback\index{Rollback} due to verification\index{Virtual Network Configuration!verification} error, the error between the 
simulation and actual entity, and the \acl{SLW}. Every 
\acl{LP} will be in exact alignment with its \acl{PP} as a result of a 
state verification\index{Virtual Network Configuration!verification} query. This occurs every $T_{query} = t_{vfail}$ time 
units. 

\subsection{Length of Lookahead Window}

The length of the lookahead window ($\Lambda$) should be as large 
as possible while maintaining the required accuracy. The total error is 
also a function\index{Total Error!function} of the chain of messages which lead to the state in 
question.
Thus the farther ahead of real\index{Real}-time the predictive\index{Predictive Network Management!predictive} network management 
system advances,
$t_{ahead} = GVT\index{GVT} - t_{current-time}$, the greater the number of 
messages before a verification\index{Virtual Network Configuration!verification} query can be made and the 
greater the error. The maximum error is $AC_t(\Lambda)$.

\subsection{Calibration Mode of Operation}

It may be helpful to run the predictive\index{Predictive Network Management!predictive} network management system in a mode 
such that error between the actual entities and the predictive\index{Predictive Network Management!predictive} network 
management system are measured. This error information can be used during 
the normal predictive\index{Predictive Network Management!predictive} mode in order to help set the above parameters. 
This begins to remind one of back propagation in a neural network, that is,
the predictive\index{Predictive Network Management!predictive} network management system automatically adjusts parameters 
in response to real\index{Real} output in order to become more accurate. 
This calibration mode could be part of normal operation. The error can be
tracked simply by keeping track of the difference between the 
simulated messages and the result of verification\index{Virtual Network Configuration!verification} queries. 

\section{Model and Simulation}

A simulation of \acl{VNC} applied to a predictive\index{Predictive Network Management!predictive} management system
was implemented with Maisie\index{Maisie} \cite{bagrodia}. 
Maisie\index{Maisie} is the simulation environment used here.
Its suitability for this has been demonstrated in the RDRN\index{RDRN}
network management and control design and development and in
\cite{Short} to develop a mobile wireless network
parallel simulation\index{Parallel Simulation} environment. 

\subsection{Verification Query Rollback\index{Rollback} Versus Causality Rollback\index{Rollback}}

Verification query\index{State Adjustment} rollback\index{Rollback}s are the 
most critical part of the predictive\index{Predictive Network Management!predictive} management system. They are handled 
in a slightly different fashion 
from causality\index{VNC!causality} failure rollback\index{Rollback}s. A state verification\index{Virtual Network Configuration!verification} 
failure causes the \acl{LP} state to be corrected at the time of the state 
verification which failed. 
The state, $S_{v}$, has been obtained from the actual device from 
the verification\index{Virtual Network Configuration!verification} query at time $t_{v}$.
The \acl{LP} rolls back to exactly $t_{v}$ with state, $S_{v}$.
States greater than $t_{v}$ are removed from the state 
queue\index{State Queue}. 
Anti-messages\index{Anti-Message} are sent from the output message queue 
for all messages
greater than $t_{v}$. The \acl{LP} continues forward execution from this 
point.
Note that this implies that the message and state queues cannot
be purged of elements which are older than the GVT\index{GVT}. Only elements which
are older than real\index{Real} time can be purged.

\subsection{The Predictive\index{Predictive Network Management!predictive} Management System Simulation}

\begin{figure*}[htbp]
\centerline{\psfig{file=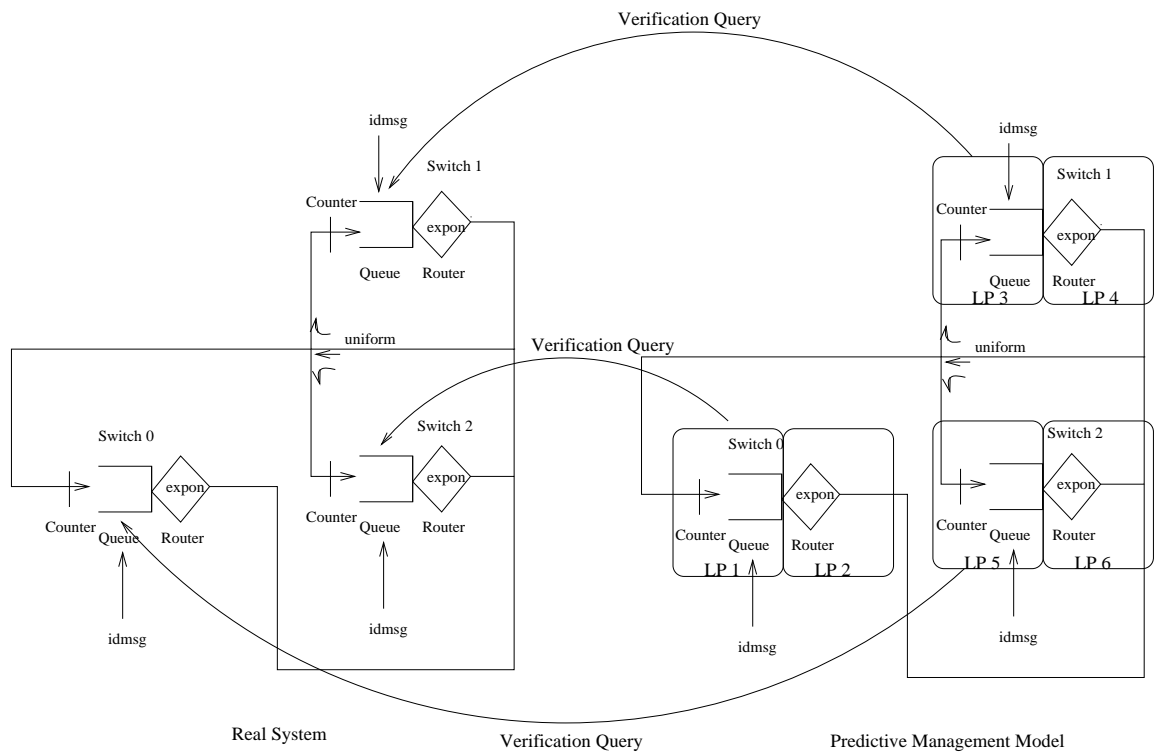,width=6.0in}}
\caption{Initial Feasibility Network Model}
\label{qrsim}
\end{figure*}

A small closed queuing network with FCFS servers
is used as the target system in this study. Figure \ref{qrsim} shows the
real system to be managed and the predictive\index{Predictive Network Management!predictive} management model\index{Model}. In this
initial feasibility study, the managed system and the predictive\index{Predictive Network Management!predictive} management
model are both model\index{Model}ed with Maisie\index{Maisie}. The verification\index{Virtual Network Configuration!verification} query between the
real system and the management model\index{Model} are explicitly illustrated in
Figure \ref{qrsim}.

The system consists of three switch-like entities, each switch contains a 
single queue and switches consisting of 10 exponentially distributed\index{Distributed} 
servers which must sequentially service each packet. A mean service time of 
10 time units is assumed. The servers represent the link rate. The packet is 
then forwarded with equal probability to another switch, including itself. 
Each switch is a driving\index{Driving Process} process; the switches forward real\index{Real} and virtual\index{Virtual Network Configuration!verification} 
messages. The cumulative number of packets which have entered each switch 
and queue is the state. This is similar to SNMP\index{SNMP} \cite{SNMP} statistics 
monitored by SNMP\index{SNMP} Counters, for example, the \textbf{ifInOctets} counter in
MIB-II interfaces \cite{RFC1156}.

Both real\index{Real} and virtual\index{Virtual Network Configuration!verification} messages contain the time at which service ends.
The switches are fully connected. An initial message 
enters each queue upon startup to associate a queue with its switch.
This is the purpose of the \textbf{idmsg} which enters the queues in
Figure \ref{qrsim}. The predictive\index{Predictive Network Management!predictive} system parameters 
are more compactly identified as a triple consisting of
Lookahead Window Size (seconds),  Tolerance\index{Tolerance} (counter value), 
and Verification\index{Virtual Network Configuration!verification} Query Period (seconds) in the form $(\Lambda, 
\Theta, \Upsilon)$. The effect of these parameters are examined on
the system of switches previously described.
The simulation was run with the following triples:
$(5,10,5)$, $(5,10,1)$, $(5,3,5)$, $(400,5,5)$. The graphs which follow
show the results for each triple. 

\begin{figure*}[htbp]
\centerline{\psfig{file=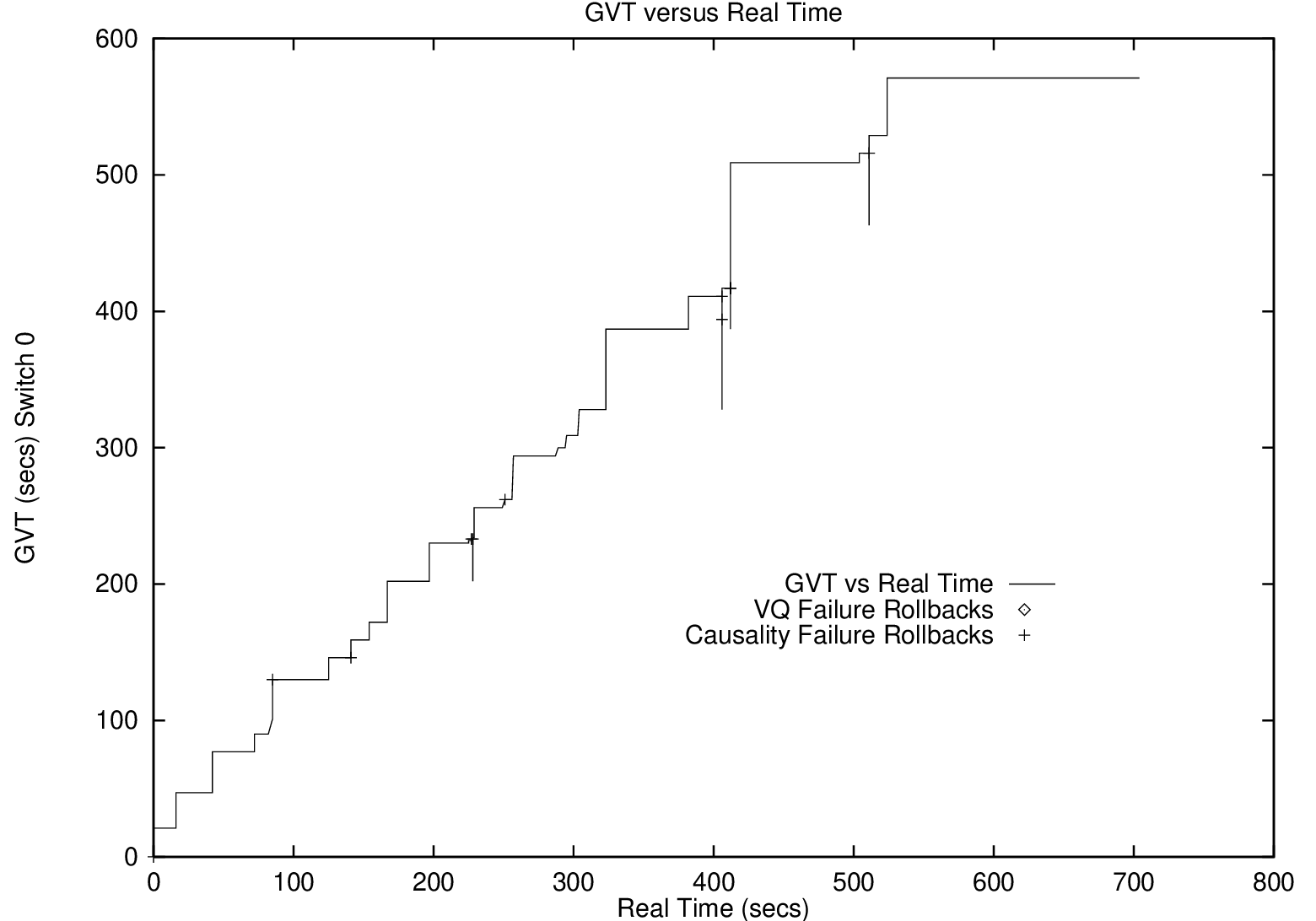,width=6.0in}}
\caption{Rollbacks Due to State\index{State}\index{State} Verification\index{Virtual Network Configuration!verification} Failure (5, 10, 5)}
\label{rb5105}
\end{figure*}

\begin{figure*}[htbp]
\centerline{\psfig{file=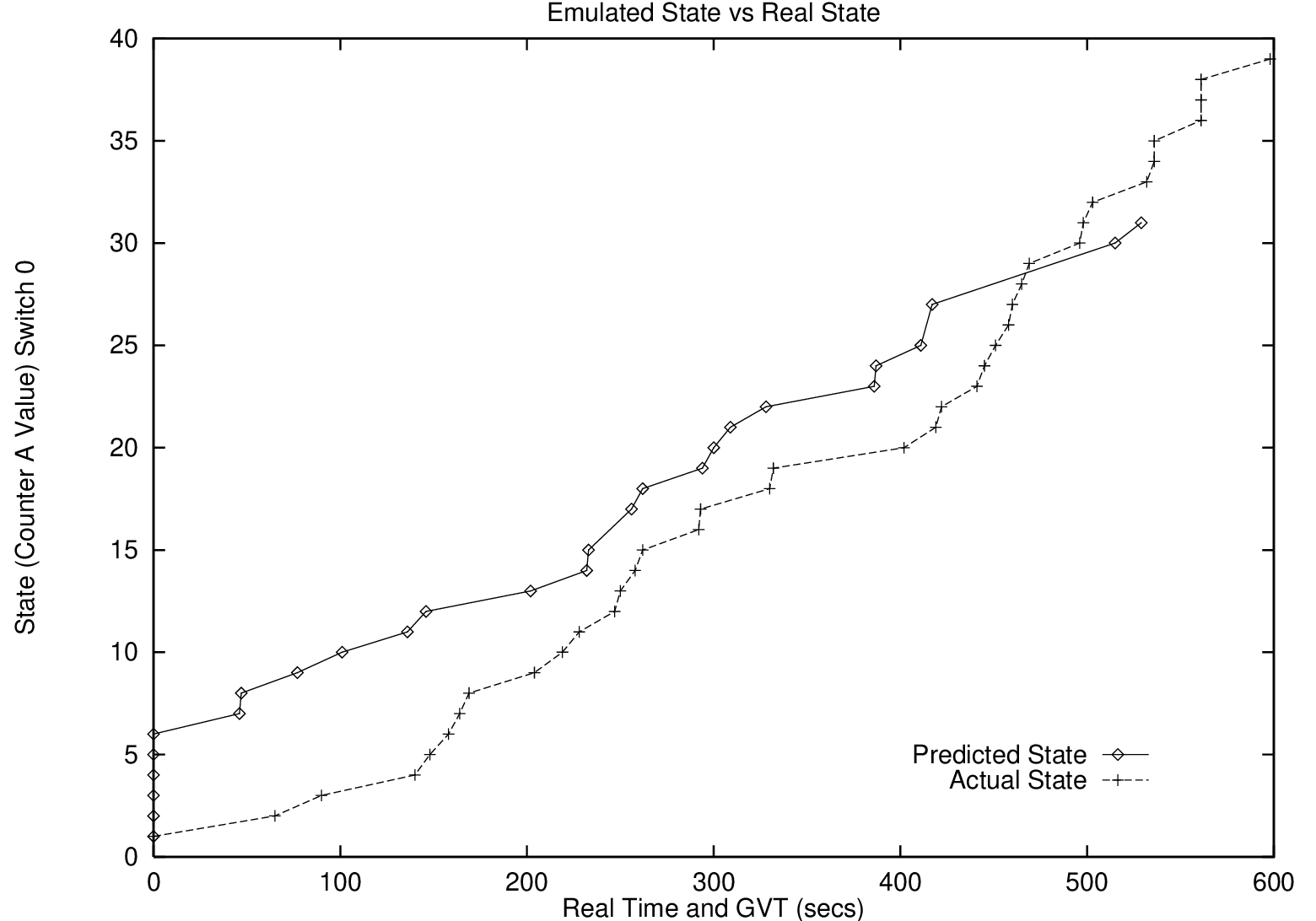,width=6.0in}}
\caption{State (5, 10, 5)}
\label{s5105}
\end{figure*}

The first run parameters were $(5, 10, 5)$. There were no state 
verification rollback\index{Rollback}s although there were some causality induced rollback\index{Rollback}s
as shown in Figure \ref{rb5105}. GVT\index{GVT} increased almost instantaneously versus 
real time; at times the next event far exceeded the look-ahead window. This
is the reason for the nearly vertical jumps in the GVT\index{GVT} as a function\index{GVT!function} of 
real-time graph as shown in Figure \ref{rb5105}. The state graph for this
run is shown in Figure \ref{s5105}.

In the initial implementation\index{Virtual Network Configuration!implementation}, state verification\index{Virtual Network Configuration!verification} was performed in the
\acl{LP} immediately after each new message was received. However, the probability
that an \acl{LP} had saved a future state, while processing at its LVT\index{LVT},
with the same state save time as the time at which a real\index{Real} message arrived was low. 
Thus, there was frequently nothing with which to 
compare the current state in order to perform the state verification\index{Virtual Network Configuration!verification}.
However, it was observed that the predictive\index{Predictive Network Management!predictive} system was simulating up to the lookahead 
window very quickly and spending most of its time holding, during which 
time it was doing nothing. The implementation\index{Virtual Network Configuration!implementation} was modified so that each entity 
would perform state verification\index{Virtual Network Configuration!verification} during its hold time\index{VNC!hold time}.
This design change better utilized the processors and resulted in
more accurate alignment between the actual and logical\index{Virtual Network Configuration!implementation} processes.

\begin{figure*}[htbp]
\centerline{\psfig{file=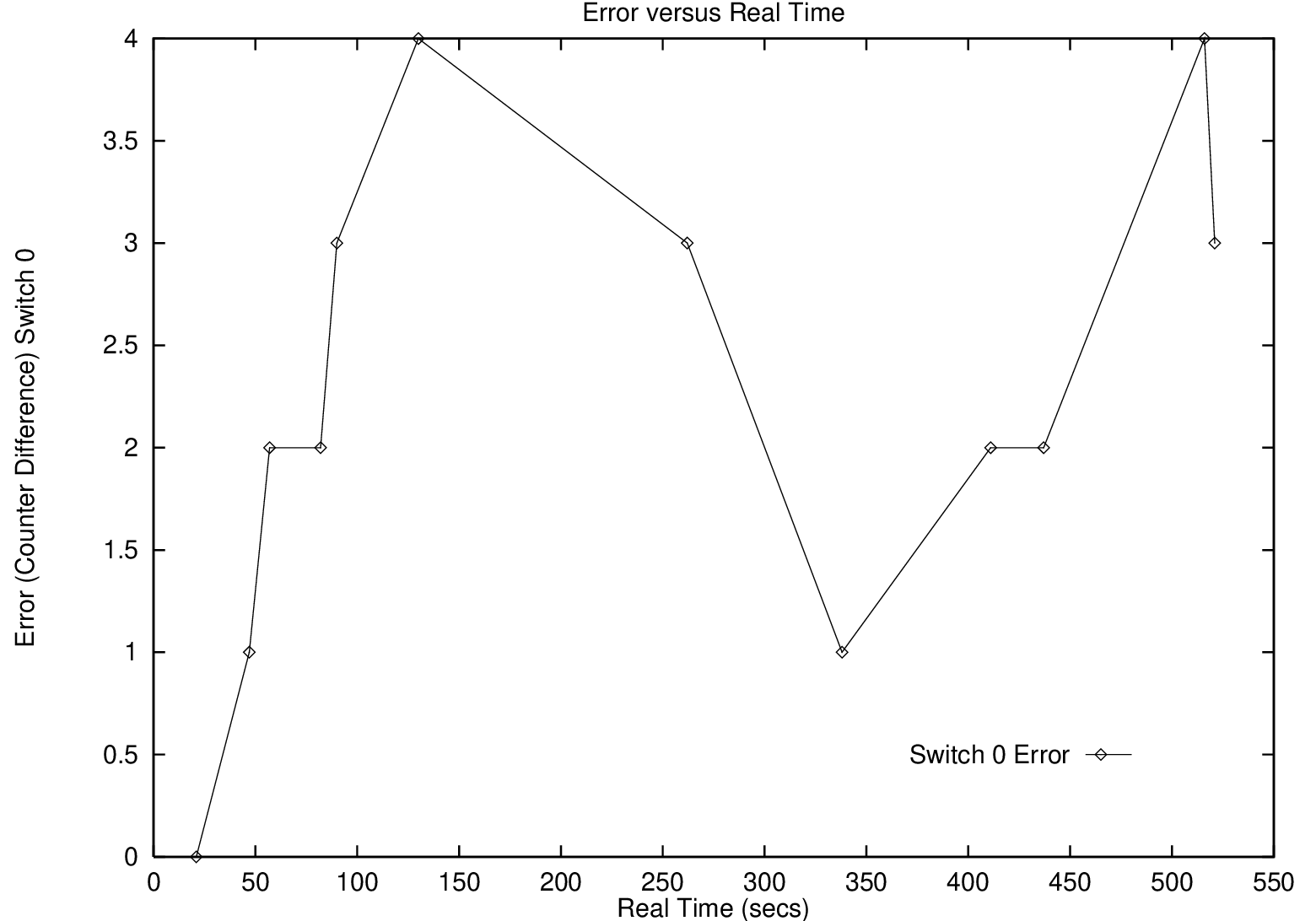,width=6.0in}}
\caption{Amount of Error (5, 10, 5)}
\label{e5105}
\end{figure*}

\begin{figure*}[htbp]
\centerline{\psfig{file=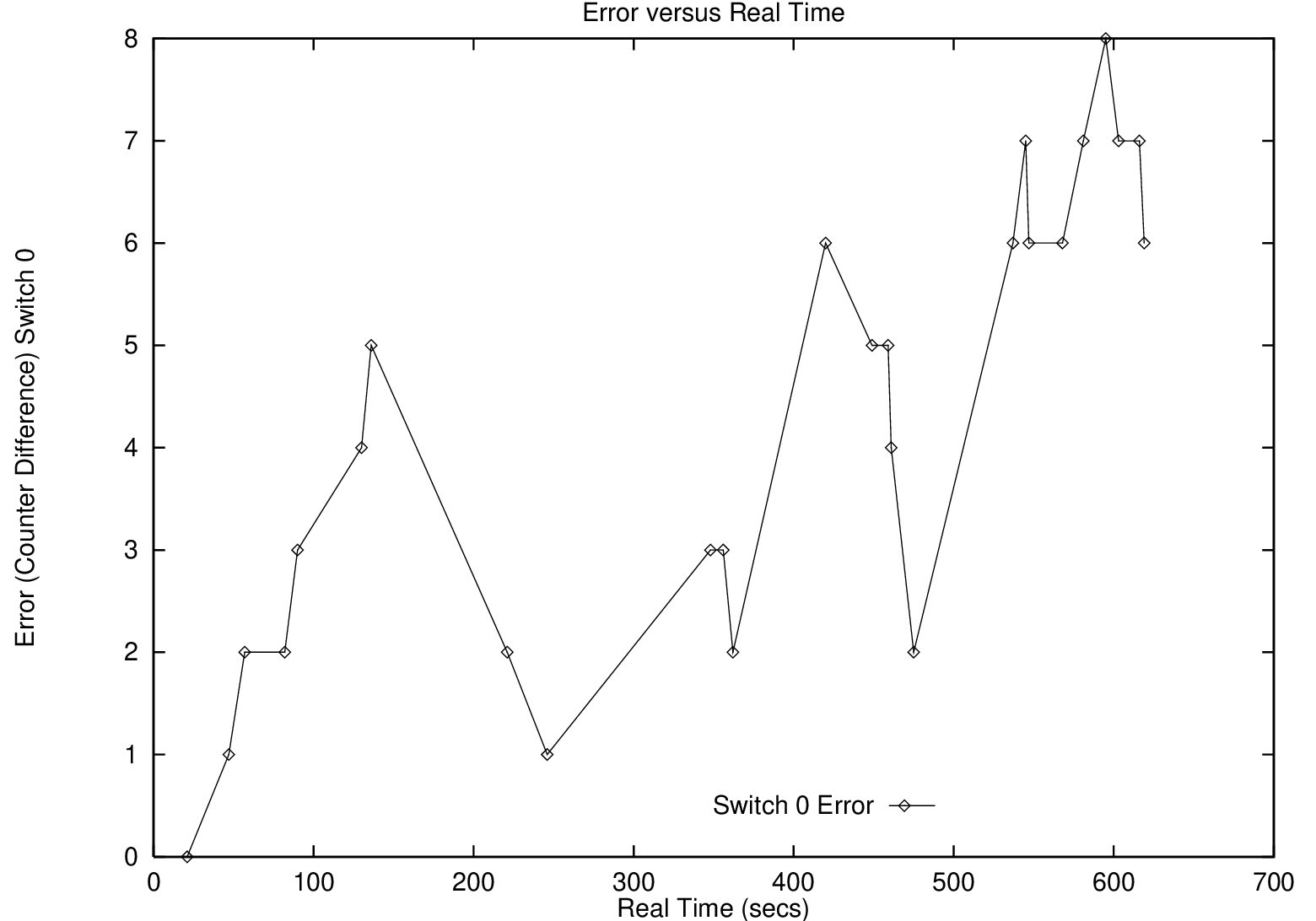,width=6.0in}}
\caption{Amount of Error (5, 10, 1)}
\label{e5101}
\end{figure*}

\begin{figure*}[htbp]
\centerline{\psfig{file=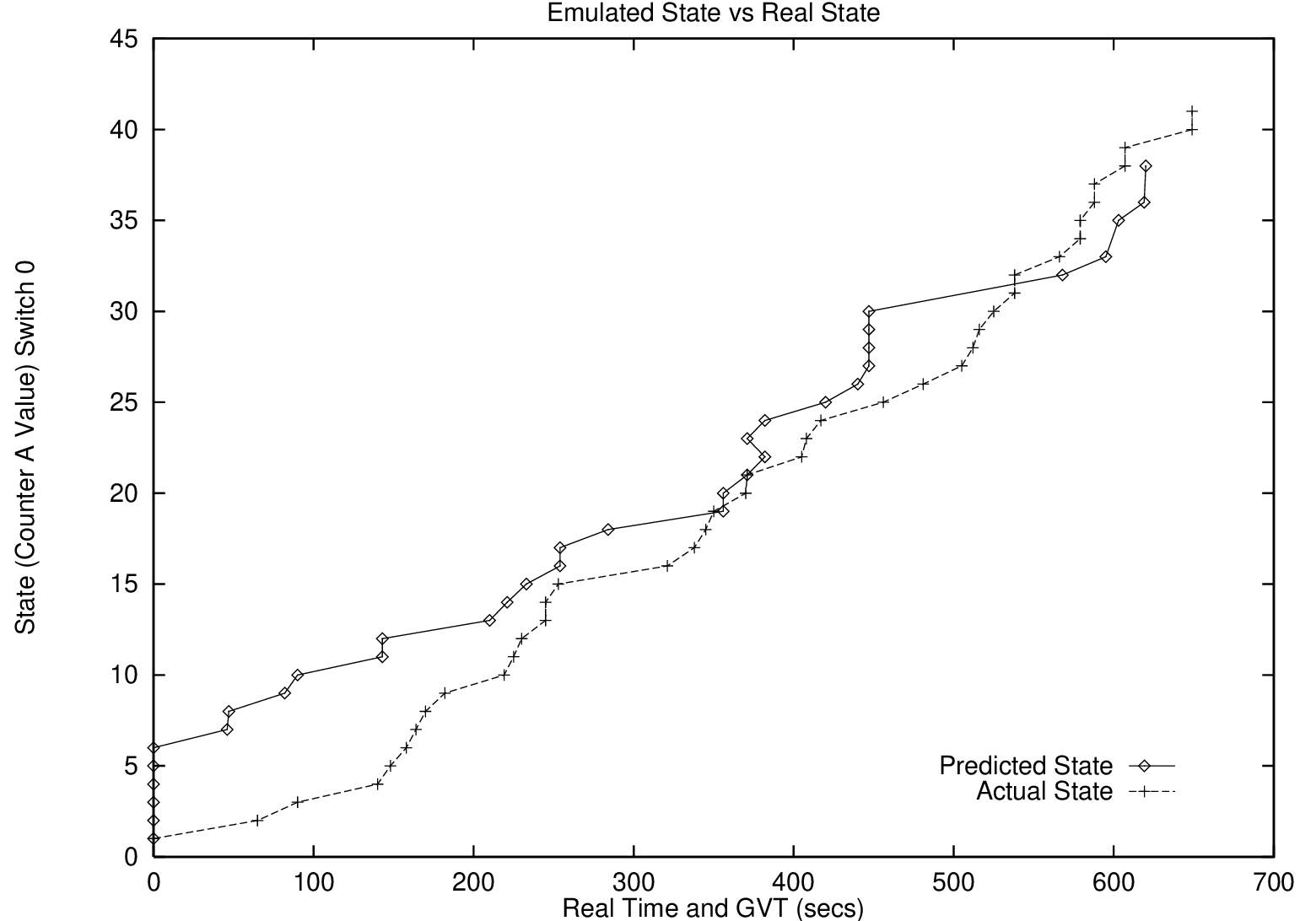,width=6.0in}}
\caption{State (5, 10, 1)}
\label{s5101}
\end{figure*}

The results for the $(5, 10, 1)$ run were similar, except that the 
predictive and actual system comparisons were more frequent because the state 
verification period had been changed from once every 5 seconds to once 
every second. Error was measured as the difference in the predicted \acl{LP} state 
versus the actual system state. 
This run showed errors that were greater than those in the first run, great 
enough to cause state verification\index{Virtual Network Configuration!verification} rollback\index{Rollback}s.
The error levels for both runs are shown in
Figures \ref{e5105} and \ref{e5101}. The state graph for this run is
shown in Figure \ref{s5101}.

\begin{figure*}[htbp]
\centerline{\psfig{file=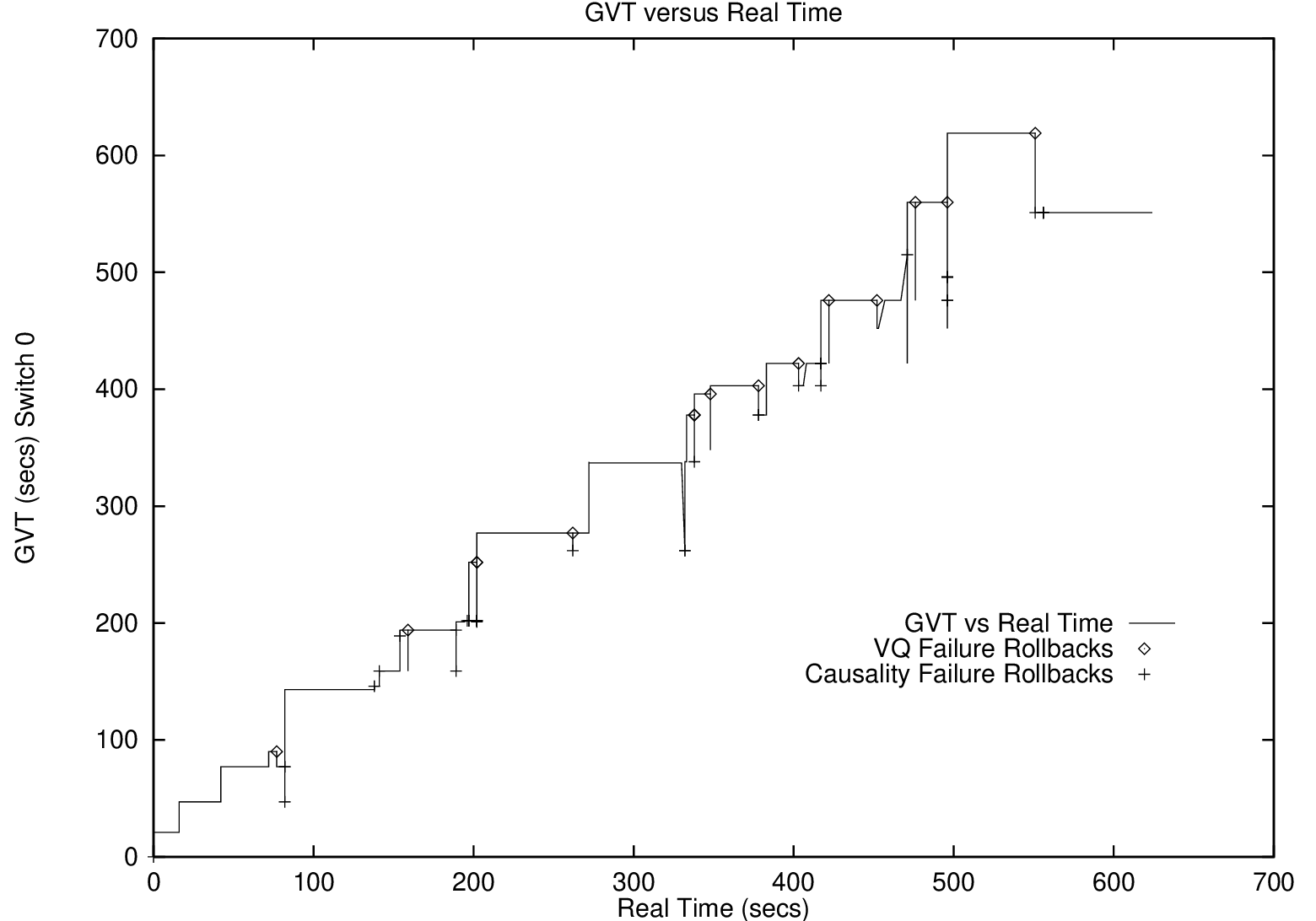,width=6.0in}}
\caption{Rollbacks Due to State\index{State}\index{State} Verification\index{Virtual Network Configuration!verification} Failure (5, 3, 5)}
\label{rb535}
\end{figure*}

\begin{figure*}[htbp]
\centerline{\psfig{file=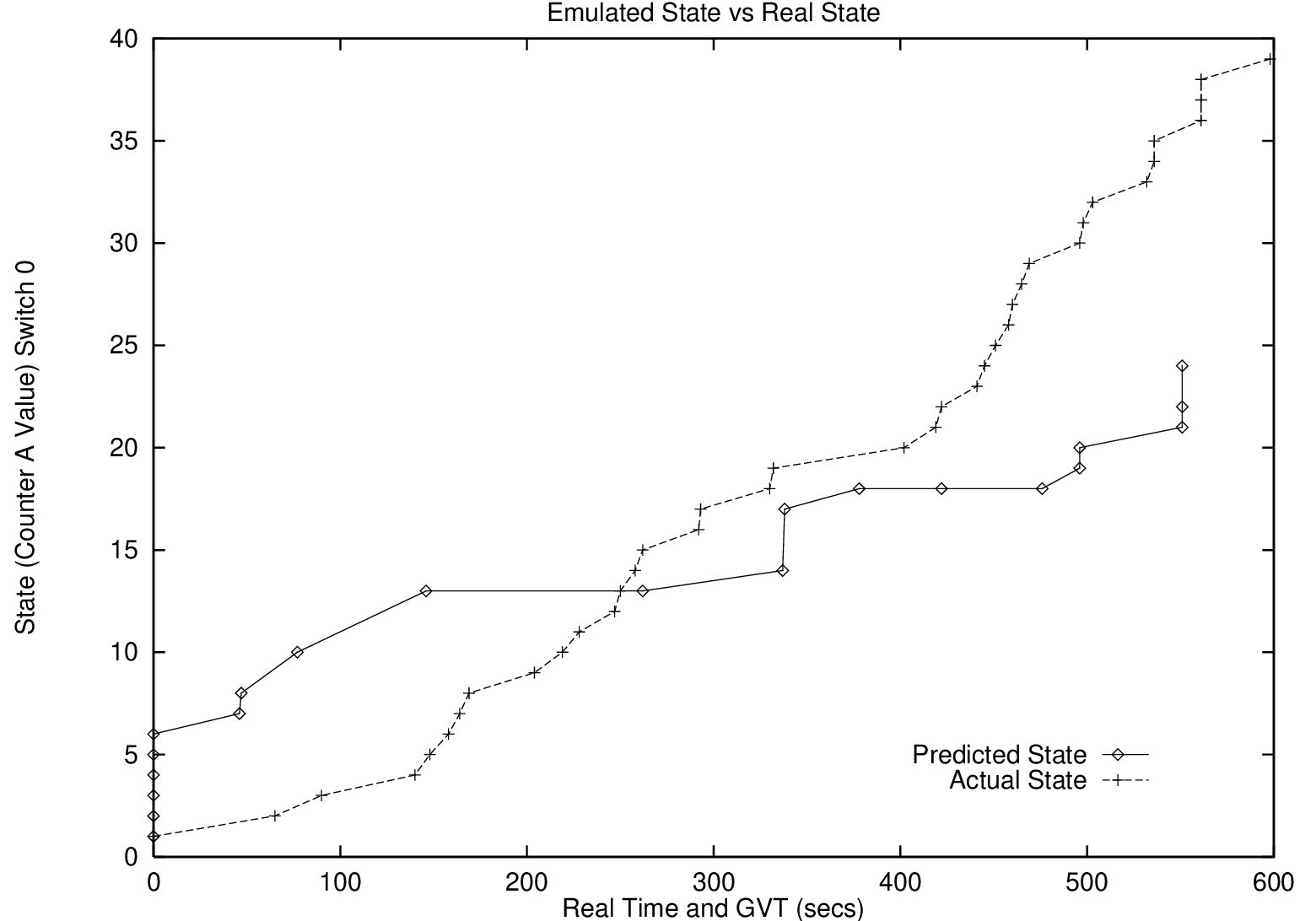,width=6.0in}}
\caption{State (5, 3, 5)}
\label{s535}
\end{figure*}

The next run used $(5, 3, 5)$ parameters. Here we see many more state 
verification failure rollback\index{Rollback}s as shown in Figure \ref{rb535}. This is
expected since the tolerance has been reduced from 10 to 3. The cluster of  
causality rollback\index{Rollback}s near the state verification\index{Virtual Network Configuration!verification} rollback\index{Rollback}s was expected.
These clusters of causality rollback\index{Rollback}s do not appear to significantly           
reduce the feasibility of the system. The real\index{Real}-time 
versus GVT\index{GVT} plot as shown in Figure \ref{rb535} shows much 
larger jumps as the \acl{LP}s were held back due to rollback\index{Rollback}s. The entities 
had a larger variance in their hold times than the $(5, 10, 5)$ run. The
state graph for this run is shown if Figure \ref{s535}. 

\begin{figure*}[htbp]
\centerline{\psfig{file=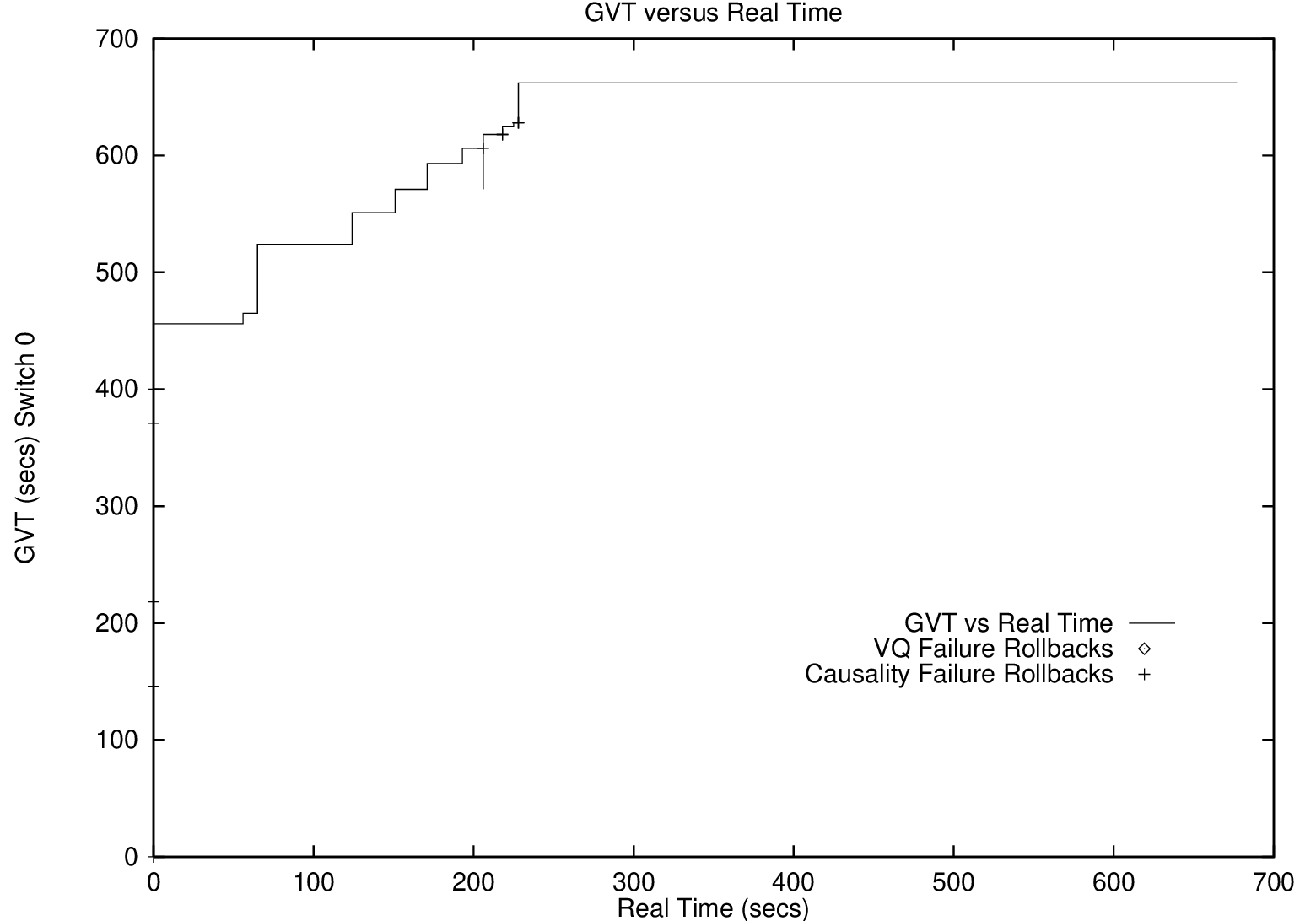,width=6.0in}}
\caption{Rollbacks Due to State\index{State}\index{State} Verification\index{Virtual Network Configuration!verification} Failure (400, 5, 5)}
\label{rb40055}
\end{figure*}

\begin{figure*}[htbp]
\centerline{\psfig{file=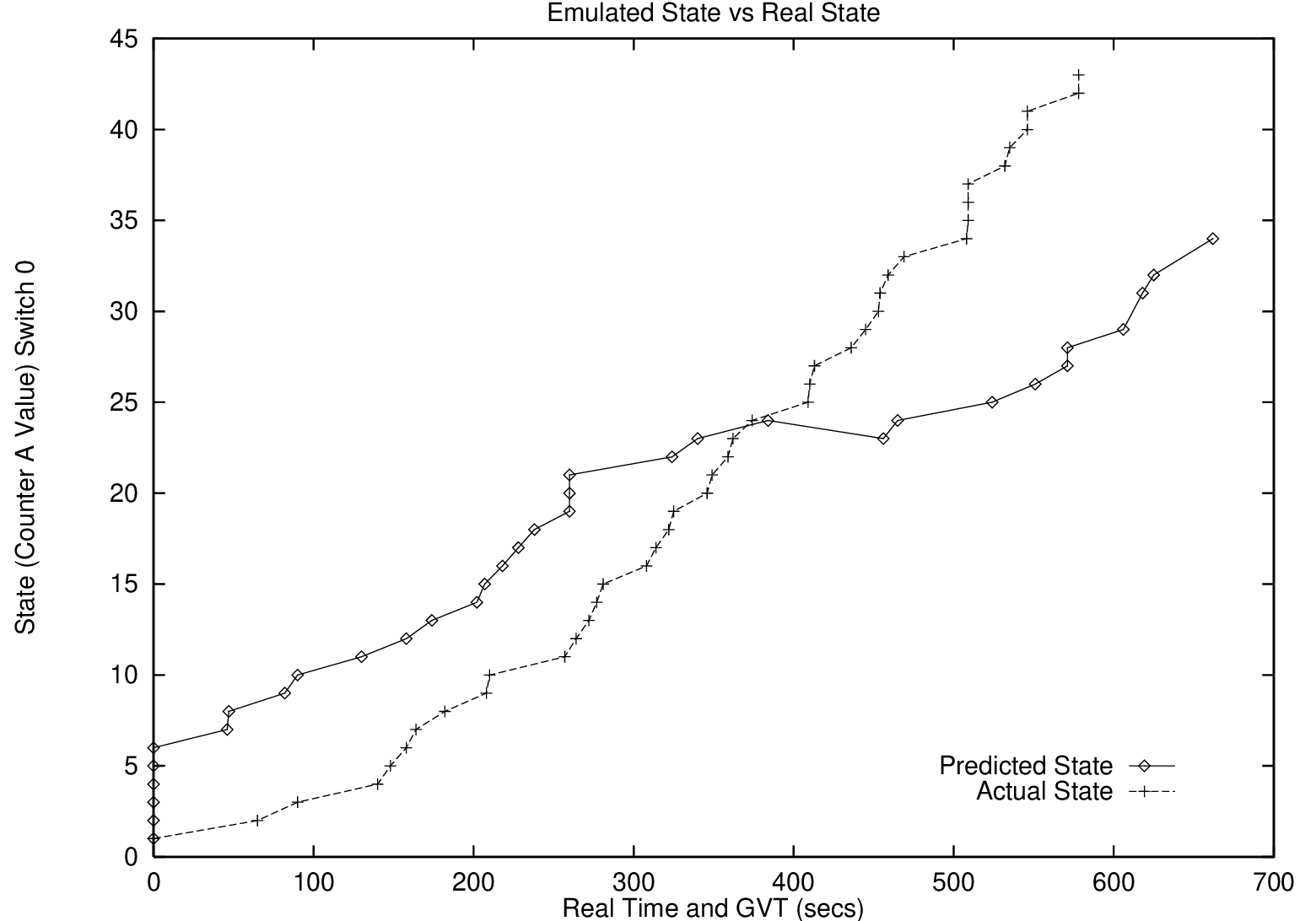,width=6.0in}}
\caption{State (400, 5, 5)}
\label{s40055}
\end{figure*}

A $(400, 5, 5)$ run showed the GVT\index{GVT} jump quickly to 400 and then gradually
increase as the \acl{SLW} maintained a 400 time unit lead
as shown in Figure \ref{rb40055}.
The \acl{LP} hold times were shorter here than an any previous run. The state
graph for this run is shown in Figure \ref{s40055}. 

\subsection{Discussion of Predictive\index{Predictive Network Management!predictive} Management Simulation Results}

Clearly, these results show the system to be
stable with the introduction of state verification\index{Virtual Network Configuration!verification} rollback\index{Rollback}s. The
overhead introduced by these rollback\index{Rollback}s did not greatly impact the
performance, because as previously shown in the GVT\index{GVT} versus time graphs,
Figures \ref{rb5105}, \ref{rb535} and \ref{rb40055}, the system was 
always able to predict up to its lookahead time very quickly.
An unstable system is one in which there exists
exists enough rollback\index{Rollback}s to cause the system to take longer
than real\index{Real}-time to reach the end of the \acl{SLW}.

The inter-\acl{LP} bandwidth overhead\index{Virtual Network Configuration!overhead} is estimated from the
State Figures \ref{s5105}, \ref{s5101}, \ref{s535}, and \ref{s40055}.
The Predicted State\index{State}\index{State} value has been chosen to serve a dual
purpose. It shows the virtual\index{Virtual Network Configuration!verification} message load as well as serving as the
predicted performance metric. Thus, the Predicted State\index{State}\index{State} value is also 
the number of virtual\index{Virtual Network Configuration!verification} messages received by the switch. Since the
rate of change of the state counter is the rate of packets entering
a switch, the slope of the Predicted State\index{State}\index{State} curve in each state graph 
shows the amount of virtual\index{Virtual Network Configuration!verification} message overhead\index{Virtual Network Configuration!overhead}. This does not include 
anti-messages since anti-messages are used only to cancel the effects 
of non-causal messages and are not counted. 
All the simulation runs had less than 4 anti-messages, or less
than 6\% more than the virtual\index{Virtual Network Configuration!verification} message overhead\index{Virtual Network Configuration!overhead}, except
for the (5,3,5) simulation. The (5,3,5) simulation generated a total
of 17 anti-messages which is almost 28\% more than the virtual\index{Virtual Network Configuration!verification}
message overhead\index{Virtual Network Configuration!overhead}. This is expected given the narrow tolerance for
the (5,3,5) run. 

As previously mentioned, the overhead\index{Virtual Network Configuration!overhead} metric of interest for \acl{VNC}
in a predictive\index{Predictive Network Management!predictive} management environment is the amount of polling bandwidth,
since polling status values from managed devices uses the same
bandwidth that is being managed.
In most standards based approaches, network management stations
are sampling counters in managed entities which
simply increment in value until they roll over. A management station
which is simply plotting data will have some fixed polling interval
and record the absolute value of the difference in value of the
counter. Such a graph is not a perfectly accurate representation
of the data, it is merely a statement that sometime within a polling
interval the counter has monotonically increased by some amount.
Spikes in this data, which may be very important to the current state of the
system, may not be noticed if the polling interval
is too long such that a spike followed by low data values expected out
to a normal or low value. Our goal is to determine the minimum
polling interval required to accurately represent the data.

From the information provided by the predictive\index{Predictive Network Management!predictive}
management system, a polling interval which provides the desired
degree of accuracy can be determined and dynamically adjusted; however, the 
cost must be determined.
An upper limit on the number of systems which can be polled
is $N \le {T \over \Delta}$ where $N$ is the number of devices capable of
being polled, $T$ is the polling interval, and $\Delta$ is the time
required for a single poll. Thus although the data accuracy will be
constrained by this upper limit, taking advantage of characteristics
of the data to be monitored can help distribute the polling intervals
efficiently within this constraint.
Assume that $\Delta$ is a calculated and fixed value, as is $N$.
Thus this is a lower bound on the value of $T \ge \Delta N$.

The overhead\index{Virtual Network Configuration!overhead} bandwidth required for use by the management system to 
perform polling is shown in Equation \ref{bw}. The packet size will 
vary depending upon whether it is an \acl{SNMP} or \acl{CMIP} packet 
and the \acl{MIB}
object(s) being polled. The number of packets varies with the amount
of management data requested. Let $P$ be the number of packets,
$S$ be the bits/packet, $N$ be the number of devices polled, and $T$
be the polling period. $Bw$ is the total available bandwidth and
$Bw_{oh}$ is the overhead\index{Virtual Network Configuration!overhead} bandwidth of the management traffic.

\begin{figure*}
\begin{equation}
Bw_{oh}\% = {100.0 P N S Bw \over T}
\label{bw}
\end{equation}
\end{figure*}

It may be desirable to limit the bandwidth used for polling system management
data to be no more than a certain percentage of total bandwidth. Thus
the optimum polling interval will use the least amount of bandwidth
while also maintaining the least amount of variance due to error in the 
data signal. All the required information to maintain the cost versus
accuracy at a desired level is provided by the predictive\index{Predictive Network Management!predictive} network 
management system.

\section{\acl{VNC} Experimental Validation Summary}

This chapter has presented the results of a \acl{VNC} enhanced \acl{RDRN}
orderwire. The experimental validation of \acf{VNC} examined simulations
and implementation\index{Virtual Network Configuration!implementation}s of the \acl{VNC} algorithm in order to determine the 
speedup ($\eta$) and bandwidth overhead\index{Virtual Network Configuration!overhead} ($\beta$) of the \acl{VNC} algorithm. 
The experimental validation also satisfied an existence proof which 
demonstrates the feasibility \acl{VNC}. 

The implementation\index{Virtual Network Configuration!implementation} of the driving\index{Driving Process} process is critical for \acl{VNC}, 
therefore this chapter began with a discussion of an ideal implementation\index{Virtual Network Configuration!implementation} 
for the driving\index{Driving Process} process for \acl{RDRN} position prediction. This
indicated the accuracy that could be expected from a position prediction
process. However, the implementation\index{Virtual Network Configuration!implementation} of the driving\index{Driving Process} process for
the experimental validation in this chapter simulated \acf{GPS} input. 
By simulating the \acl{GPS}\index{GPS}, the amount of error in virtual\index{Virtual Network Configuration!verification} messages was
accurately controlled.

\acf{VNC} was experimental validated in a mobile network
configuration environment and in a simulated predictive\index{Mobile Network!predictive} network 
management environment. In the mobile network environment, a portion
of the \acf{RDRN} \acf{NCP} had been enhanced with \acl{VNC}. Specifically,
the task which processes \textbf{USER\_POS} messages pre-computes
and caches beam tables. The measurements of the \acl{VNC} enhanced
\acf{RDRN} \acf{NCP} and a discussion of the results is in
Section \ref{vncowi}.
In the  simulated predictive\index{Mobile Network!predictive} network
management environment, a standards based management system enhanced
with \acl{VNC} was simulated with Maisie\index{Maisie} \cite{bagrodia}. 

The results of the experimental validation in this section have shown
that the \acl{VNC} algorithm can be stable. In other words, rollback\index{Rollback}
cause by both out-of-tolerance and out-of-order messages do not
cause the system predict events in the past. The results in the
section also indicated that bandwidth overhead\index{Virtual Network Configuration!overhead} was not significantly
more than twice the non-\acl{VNC} bandwidth. Also, this validation
has shown that a system enhanced with \acl{VNC} can be used for
fault prediction rather than to obtain speedup\index{Speedup} as in the \acl{VNC}
\acf{NCP} experimental implementation\index{Virtual Network Configuration!implementation}.
\chapter{Conclusion}
\section{Attributes of Virtual\index{Virtual Network Configuration!verification} Network Configuration Performance}

The focus of this research has been the extension of optimistic
distributed simulation techniques to speedup\index{Speedup} the configuration of
the wireless \acl{ATM} network developed by the \acl{RDRN}
project. This research has resulted in a new algorithm for predictive\index{Mobile Network!predictive} 
real time systems which utilizes results from optimistic distributed\index{Distributed}
simulation and mobile network protocols. The analysis and results
have shown this to be a feasible method for performance enhancement
in a wireless environment where prediction of future location is used 
to speedup\index{Speedup} configuration in the physical\index{Physical} 
and higher protocol layers.
The goal of \acl{VNC} is to provide accurate predictions quickly
enough so that the results are available before they are
needed. Without taking advantage of parallel\index{Parallel}ism,
a less sophisticated algorithm than \acl{VNC} could run ahead of
real-time and cache results for future use. However, simply predicting 
and caching results
ahead of time does not fully utilize inherent parallel\index{Parallel}ism as
long as messages between \acl{LP}s remain strictly synchronized.
Strict synchronization means that processes must wait until
all messages are insured to be processed in order. This research
has shown how the cost of rollback\index{Rollback} overhead\index{Virtual Network Configuration!overhead} relates to the
speedup gained using the \acl{VNC} algorithm.

Examples of physical\index{Physical} and protocol layers which benefit from
\acl{VNC} have been presented. These include beamforming\index{Beamforming},
\acl{ATM} topology\index{Topology} calculation, \acl{ATMARP} service, \acl{NHRP},
IP routing and \acl{PNNI}. A pro-active network management application
has also been explored. It should be noted that the \acl{VNC} algorithm 
has applications\index{Applications} in any real\index{Real}-time system where information about the 
future state of the system can improve performance.

The analysis of \acl{VNC} has focused on three measurements:
speedup, bandwidth overhead\index{Virtual Network Configuration!overhead}, and accuracy. These are a function\index{Function} of 
out-of-order message rate, out-of-tolerance state value
rate, rate of virtual\index{Virtual Network Configuration!verification} messages entering the system,
task execution time, task partitioning into \acl{LP}s, rollback\index{Rollback}
overhead, prediction accuracy as a function\index{Function} of distance into the
future which predictions are attempted, and the effect of
parallelism and optimistic synchronization. All of these factors
have been analyzed. It has been shown both
analytically and through experimental validation that significant
speedup can be gained with \acl{VNC} while
the worst case bandwidth overhead\index{Virtual Network Configuration!overhead} can be kept to slightly more
than two times the bandwidth without \acl{VNC} and even less
with the real\index{Real} message optimization\index{Optimization}. 

The \acl{VNC} algorithm developed in this research
will become increasingly important. As the
physical layer for wireless technology improves, 
bandwidth will become cheaper. While more bandwidth
becomes available, the trend is for smaller geographical
cell sizes for wireless systems, thus increasing the probability
and overhead\index{Virtual Network Configuration!overhead} associated with handoff\index{Handoff}. As shown in this research, 
as the utility for bandwidth goes down relative to the utility 
for speedup\index{Speedup}, the overall utility of \acl{VNC} increases.
\section{Future Work}

The cached information in each \acl{LP} contains useful information
about the accuracy of the driving\index{Driving Process} process. The difference between
the actual state and the cached state can be fed back to the
driving process in order to correct the prediction algorithm.
An interesting technique to accomplish this could utilize 
\acl{PA}\index{Perturbation Analysis} \cite{Ho}.
The technique of \acl{PA} allows a great deal more information to be 
obtained from a simulation execution than explicitly collected statistics. 
It is particularly useful for finding the sensitivity\index{Sensitivity} information of 
simulation parameters from the sample path of a single simulation run.
In a practical sense, sensitivity\index{Sensitivity}
analysis of the \acl{VNC} message or state values qualified with a
tolerance\index{Tolerance} would be useful information in the system 
design. However, this author has not found any work applying Perturbation 
Analysis to optimistic simulation methods in a fundamental manner. Knowledge 
of the sensitivity to error at each \acl{LP} gained through
Perturbation Analysis\index{Perturbation Analysis}
could be used to examine the accumulation of error throughout the
\acl{VNC} system. The tolerances\index{Tolerance} could be automatically adjusted based on
this information similar to back-propagation in a
neural network\index{Neural Network}. The system could be enhanced
with a simultaneous learn phase of operation where the cause of a
rollback is identified and corrected automatically by weighting
input message values or providing feedback\index{Feedback} to the
driving process(es).

It would also be interesting to study the effect of self-similarity\index{Self-similarity}
and emergence\index{Emergence} in the virtual\index{Virtual Network Configuration!virtual message} message values generated by the driving\index{Driving Process}
process on the operation of the \acl{VNC} algorithm and on 
Time Warp algorithms in general. This research assumes that the
driving process will predict future results with greater accuracy
as real\index{Real} time approaches the time of the predicted event. However,
the predictions which appear in the state queue may be chaotic\index{Chaos} or 
self-similar\index{Self-similar} over time. This would certainly be the case if \acl{VNC}
is applied towards predicting LAN or \acl{ATM} switch traffic, both
of which are known to be self-similar\index{Self-similar}.

Active networks\index{Active Networks}, \cite{ANEP, Tenn96} are communications 
networks which view their contents not simply as packets to be transported, 
but also as programs which may be interpreted by intermediate nodes within the 
network \cite{Legedza, Bhattacharjee}.
Thus, active networks become an enabling mechanism for dynamically
loading \acl{VNC} predictive processes into all network nodes including
fixed and wireless networks. In the future, active data which implements
the predictive algorithm in \acl{VNC} may be supplied by network vendors
as commonly as an \acl{SNMP} \acl{MIB} is supplied today.
If we view the \acl{VNC} algorithm in an active network\index{Active Network}
environment, the \acl{VNC} messages may carry programs which are
interpreted by intermediate nodes rather than passive messages. The 
algorithm then becomes much more flexible and powerful. Logical processes may
be created or destroyed as the algorithm executes and the effect of a
rollback becomes much more complex and interesting. The ideas
introduced in this research, particularly self-predicting\index{Self-predicting}
and emergent systems\index{Emergent System}, are clearly a rich area for more 
research.

\appendix                                                
\setcounter{chapter}{0}
\renewcommand{\thechapter}{\Alph{chapter}}


\chapter{VNC Implementation}
\section{Overview of the VNC API}

This section discusses enhancing a \acf{PP} with \acl{VNC}. The 
process descriptions below use the notation for Communicating
Sequential Processes (CSP) \cite{Hoare81}.
In CSP ``X?Y'' indicates process X will wait until a valid message is
received into Y, ``X!Y'' indicates X sends message Y. A guard statement 
is represented by ``X $\rightarrow$ Y'' which indicates that condition
X must be satisfied in order for Y to be executed.
Concurrent operation is indicated by ``X$||$Y'' which means that X
operates in parallel with Y. A ``*'' preceding a statement indicates
that the statement is repeated indefinitely. An alternative command,
represented by ``X$\Box$Y'', indicates
that either X or Y may be executed assuming any guards (conditions)
that they may have are satisfied. If X and Y can both be executed,
then only one is randomly chosen to execute. A familiar example used
to illustrate CSP is shown in Algorithm \ref{examplecsp}. This is the
bounded buffer problem in which a finite size buffer requests
more items from a consumer only when the buffer will not run out
of capacity.

Assume a working \acl{PP} abstracted in Algorithm \ref{normalpp} where
\emph{S} and \emph{D} represent the source and destination of real and
virtual messages. Algorithm \ref{normalvnc} shows the
\acl{PP} converted to a \acl{VNC} \acl{LP} operating with a monotonically 
increasing \acl{LVT}. Note that the actual \acl{VNC} API function names
are used, however, all the function arguments are not shown in order
to simplify the explanation. Each function is described in more
detail later. First the \acl{VNC} API is initialized by vncinit() as 
shown in Algorithm \ref{normalvnc} Line \ref{initvnc}. Then input 
messages are queued in the \acl{QR} by
recvm(). In non-rollback operation (Lines \ref{nrbstart} to
\ref{nrbend}) the function getnextvm() returns the next valid message from
the \acl{QR} to be processed by the \acl{PP}. When the \acl{PP} has a
message to be sent, the message is place in the \acl{QS} by sendvm().
While messages are flowing through the process, the process saves
its state periodically. 
Normal operation of the \acl{VNC} as just described may be interrupted
by a rollback. 
If recvm() returns a non-zero value (Line \ref{nonzero}), then either
an out-of-order or out-of-tolerance message has been received. In
order to perform the rollback, getstate() is called to return the
proper state to which the process must rollback. It is the applications
responsibility to insure that the data returned from getstate() properly
restores the process state. Anti-messages are sent by repeatedly calling
rbq() (Line \ref{getanti}) until rbq() returns a null value. With each call 
of rbq(), an anti-message is returned which is sent to the destination of the
original message.

\begin{algorithm}[htpb]
\begin{algtab}
X::\\
buffer:(0..9) portion; \\
in,out:integer; in := 0; out := 0; \\
$\>$*[in $<$ out + 10; producer?buffer(in mod 10) $\rightarrow$ \\
$\>\>$ in := in + 1; \\
$\>$$\Box$ out $<$ in; consumer?more() $\rightarrow$ \\
$\>\>$ consumer!buffer(out mod 10); out := out + 1;
]
\end{algtab}
\caption{\label{examplecsp}Classical Bounded Buffer in CSP.}
\end{algorithm}

\begin{algorithm}[htpb]
\begin{algtab}
PP::\\
*[S?input; \\
$\>$ output := process(input); \\
$\>\>$ D!output]
\end{algtab}
\caption{\label{normalpp}A Physical Process before VNC Added.}
\end{algorithm}

\begin{algorithm}[htpb]
\begin{algtab}
PP::\\
\alglabel{initvnc} initvnc(); \\
*[S?input; \\
\alglabel{nonzero} $\>$[recvm(input)!=0 $\rightarrow$ getstate(); \\
\alglabel{getanti} $\>$*[rbq()!=NULL $\rightarrow$ D!rbq()] $\Box$ \\
\alglabel{nrbstart} $\>\>$[recvm(input)==0 $\rightarrow$ \\
$\>\>\>$savestate(); \\
$\>\>\>$input := getnextvm(); \\
$\>\>\>$output := process(input); \\
$\>\>\>$sendvm(output); \\
\alglabel{nrbend}$\>\>\>$D!output] \\
$\>$] \\
] \\
\end{algtab}
\caption{\label{normalvnc}A Physical Process with VNC Added.}
\end{algorithm}

\section{VNC API Implementation}

Figure \ref{vncapi} lists the files and their contents for the \acl{VNC} API.
The following functions can be added to a \acl{PP}
in order to implement \acl{VNC}.

\begin{figure}

\[ \left\{ \begin{minipage}{3in}
\begin{description}
\item[vc.h] \acl{VNC} include file.
\item[vcrec.c] Receive a message, determine whether virtual or real, rollback
\item[vcsnd.c] Send a virtual message
\item[vcsendq.c] All queue related functions
\item[vcroll.c] Roll back to given time
\item[vcstate.c] Maintain virtual state
\item[vctime.c] Local virtual time maintenance functions
\item[libvnc.a]  Library of these functions
\end{description}
        \end{minipage}
        \right. \]

\caption{VNC API Files.}
\label{vncapi}
\end{figure}

\subsection{External Library Declarations}

\begin{verbatim}
extern struct vm *vminq; /* receive queue head */
extern struct vm *vmoutq; /* output queue head */
extern struct sq *stateq; /* state queue head */
\end{verbatim}

\subsection{void initvc (struct vcid *id, char *name, int num)}

This function is required for initializing \acl{VNC}. The ``id'' pointer
is used for all other \acl{VNC} functions in order to identify the
\acl{LP}. The ``name'' and ``num'' are input to uniquely identify the 
\acl{LP}. Multiple \acl{LP}s may be run with the same Unix process.

\subsection{time\_t getlvt (struct vcid *id)}

This function Returns the current \acl{LVT} of \acl{LP} ``id''.

\subsection{int setlookahead(struct vcid *id, time\_t t)}

This function sets the maximum amount of time into the future ($\Lambda$) 
that the \acl{LVT} for this process can go.

\subsection{long recvm (struct vcid *id, int anti, long rec, long snd, char *src, char *dst, int ptype, char *m, int msize, int (*comp\_pred) (), int fd, int sg)}

This function should be called each time a message arrives. 
The function comp\_pred() is user defined and described in detail in the
next section. The function recvm() handles both real and virtual messages, 
and will detect if rollback is necessary. It returns time\_t if temporal 
order of messages is violated or a real message is out of tolerance,
otherwise it returns 0. If non-zero, the return value is the time to which 
the process must rollback. The input arguments \emph{anti}, \emph{rec},
\emph{snd}, \emph{src}, \emph{dst} are the message values after a 
message has been depacketize. The input variable \emph{msize} is the
size of the packetized packet and \emph{m} is a pointer to the structure
placed on the \acl{QR}. The input variables \emph{fd} and \emph{sg}
are the \acl{GPS} file descriptor and whether the \acl{GPS} receiver is 
simulated or real.

\subsection{int comp\_pred (struct vcid *vncid, struct vm *s1, struct vm *s2, 
struct sq *stq)}

This function is passed into recvm() and is user defined. It should return
whether the \acl{SQ} differs beyond a tolerable amount ($\Theta$) from the 
predicted value of the \acl{SQ}. If the actual and predicted \acl{SQ} values 
differ beyond $\Theta$, comp\_pred() should return true. The input \emph{stq} 
is a pointer to the \acl{SQ}. The input arguments \emph{s1} and \emph{s2} are 
the current real and predicted virtual messages respectively. These are 
currently not used to determine rollback, however, \emph{s1} and \emph{s2}
may be useful in future modifications to the algorithm.

\subsection{struct vm *getnextm (struct vcid *id)}

This function is called to determine the next message to process. This 
should be the only way the physical process receives input messages. This
insures that all messages are processed first by the \acl{VNC} library
functions. Note that the message space is freed after this call so the 
value must be saved immediately into a local variable.

\subsection{int sendvm (struct vcid *id, int anti, long rec, long snd, char *src, char *dst, int ptype, char *m, int msize)}

This function should be called whenever messages are sent from the process. 
It stores the message in the \acl{QS} and records the \acl{LVT}
at the time the message was sent for rollback purposes.

\subsection{struct vm *rbq (struct vcid *id, struct vm *vncpkt, long t)}

This function should be called repeatedly when recvm() returns a non-zero
value indicating a rollback. The \emph{vncpkt} output pointer is an 
anti-message. rbq() should be called until it returns null. Because the 
messages returned are anti-messages, there is no need to store them in
the \acl{QS}, thus these message should be sent without using sendvm().

\subsection{struct sq *savestate (struct vcid *id, void *m, int msize)}

This function should be called periodically to save the \acl{PP} state for 
rollbacks. The input variable \emph{m} holds the state which is dynamically 
allocated based on the input \emph{msize}. Note that the more often state 
is saved, the less far back rollback has to occur. However, this is a 
tradeoff between storage and efficiency.

\subsection{struct sq *getstate (struct vcid *id, long snd)}

This function will return the state which was saved closest to the input
variable \emph{snd}. This function is used after a rollback to restore a 
valid state. The time is obtained from the return value of recvm().
\section{The VNC Message Receive Function}

There are two small but important sections of \acl{NCP} code which are
illustrate the use of the \acl{VNC} library developed by the author
of this study. The reason for including the actual code in this
appendix is to demonstrate the simplicity and ease of enhancing
a system with \acl{VNC}.
The first section of code receives an \acl{NCP} packet and processes the
packet in \acl{VNC} enhanced mode. At line 28 the packet is received in
non-blocking mode from the orderwire packet radio. The \acl{VNC} related
information is stripped form the packet at line 33. This information
is the send-time, receive-time, and anti-message indicator. The next
line handles the bulk of the \acl{VNC} processing. The function recvm()
checks for causal and out-of-tolerance rollback, updates the local
virtual time, manages the receive queue, and checks for message
annihilation.
Lines 37 through 46 send anti-messages because a rollback has occurred.

\ifisdraft
	\onecolumn
\fi

\Head{}
\File{\1users\1sbush\1rdrn\1ncp\1vncgetpkt.c}{1997}{Mar}{13}{14:51}{1923}
\L{\LB{\K{\#include}_\<\V{stdio}.\V{h}\>}}
\L{\LB{\K{\#include}_\<\V{ctype}.\V{h}\>}}
\L{\LB{\K{\#include}_\S{}\3switch.h\3\SE{}}}
\L{\LB{\K{\#include}_\S{}\3shared.h\3\SE{}}}
\L{\LB{\K{\#include}_\S{}\3vnc.h\3\SE{}}}
\L{\LB{}}
\L{\LB{\K{extern}_\K{struct}_\V{shared\_data}_\*\V{sharedm};}}
\L{\LB{}}
\L{\LB{\K{static}_\K{char}_\*\V{rcs}_=_\S{}\3\$Id: vncgetpkt.tex,v 1.2 1997/05/06 16:05:06 sbush Exp sbush \$\3\SE{};}}
\L{\LB{}}
\L{\LB{\C{}\1\*}}
\L{\LB{_\*_Implement_the_VNC_algorithm_for_incoming_packets.}}
\L{\LB{_\*_struct_sq_\*st_is_a_union_of_all_states_(see_vnc.h).}}
\L{\LB{_\*\1\CE{}}}
\L{\LB{}}
\L{\LB{\K{int}}}
\L{\LB{\Proc{vncgetpkt}\V{vncgetpkt}_(\K{struct}_\V{vcid}_\*\V{vncid},_\K{struct}_\V{sq}_\*\V{st},_\K{int}_(\V{comp\_pred})}}
\L{\LB{_____(\K{struct}_\V{vcid}_\*_\V{vncid},_\K{struct}_\V{vm}_\*_\V{s1},_\K{struct}_\V{vm}_\*_\V{s2},_\K{struct}_\V{sq}_\*_\V{stq}),}}
\L{\LB{}\Tab{8}{___\K{int}_\*\V{gfd},_\K{int}_\*\V{efd},_\K{int}_\*\V{sfd},_\K{struct}_\V{owPacket}_\*\V{op})}}
\L{\LB{\{}}
\L{\LB{__\K{struct}_\V{vm}_\V{vncpkt};}}
\L{\LB{__\V{time\_t}_\V{rt};}}
\L{\LB{__\K{struct}_\V{vm}_\*\V{nxtmsg};}}
\L{\LB{__\K{struct}_\V{gvtS}_\V{GVT};}}
\L{\LB{__\K{int}_\V{pkt\_typ};}}
\L{\LB{__\K{struct}_\V{gps}_\V{pos};}\Tab{32}{\C{}\1\*_Position_packet_and_table_\*\1\CE{}}}
\L{\LB{}}
\L{\LB{__\V{pkt\_typ}_=_\V{getpkt}_(\V{efd},_\V{sfd},_\V{op});}\Tab{40}{\C{}\1\*_read_the_next_message_\*\1\CE{}}}
\L{\LB{__\K{if}_(\V{sharedm}\-\!\>\V{UseVNC}_\&\&_\V{pkt\_typ}_!=_\V{NOPACKET})}}
\L{\LB{____\{}}
\L{\LB{______\C{}\1\*_display_time_and_info_\*\1\CE{}}}
\L{\LB{______\V{displaynow}_(\V{vncid},_\S{}\3VNC_Packet_Received\3\SE{});}}
\L{\LB{______\V{vtpacketize}_(\V{op}\-\!\>\V{m},_\&\V{vncpkt}.\V{anti},_\&\V{vncpkt}.\V{snd\_tm},_\&\V{vncpkt}.\V{rec\_tm});}}
\L{\LB{______\V{rt}_=_\V{recvm}_(\V{vncid},_\V{vncpkt}.\V{anti},_\V{vncpkt}.\V{rec\_tm},}\Tab{56}{\C{}\1\*_check_for_rollback_\*\1\CE{}}}
\L{\LB{}\Tab{16}{__\V{vncpkt}.\V{snd\_tm},_\V{op}\-\!\>\V{m},_\V{strlen}_(\V{op}\-\!\>\V{m}),_(\K{int}_\*)_\V{comp\_pred},}}
\L{\LB{}\Tab{16}{__\*\V{gfd},_(\K{int})_\N{1});}}
\L{\LB{______\K{if}_(\V{rt})}}
\L{\LB{}\Tab{8}{\{}}
\L{\LB{}\Tab{8}{__\V{displaynow}_(\V{vncid},_\S{}\3Rollback\3\SE{});}\Tab{48}{\C{}\1\*_display_time_and_info_\*\1\CE{}}}
\L{\LB{}\Tab{8}{__\V{bcopy}_(\V{getstate}_(\V{vncid},_\V{rt}),_\V{st},_\K{sizeof}_(\K{struct}_\V{sq}));}}
\L{\LB{}\Tab{8}{__\K{while}_((\V{nxtmsg}_=_(\K{struct}_\V{vm}_\*)_\V{getnextm}_(\V{vncid}))_==_\V{NULL})}}
\L{\LB{}\Tab{8}{____\{}}
\L{\LB{}\Tab{8}{______\V{displaynow}_(\V{vncid},_\S{}\3Sending_Anti\-Message\3\SE{});}\Tab{64}{\C{}\1\*_send_anti-messages_\*\1\CE{}}}
\L{\LB{}\Tab{8}{______\V{OrderWireTransmit}_(\V{efd},_\V{sfd},_\V{op}\-\!\>\V{ptype},_\V{op}\-\!\>\V{dst},_\V{op}\-\!\>\V{m},_\V{PPP},_\V{op}\-\!\>\V{src});}}
\L{\LB{}\Tab{8}{____\}}}
\L{\LB{}\Tab{8}{\}}}
\L{\LB{______\V{nxtmsg}_=_\V{getnextm}_(\V{vncid});}\Tab{40}{\C{}\1\*_return_next_message_by_order_of_}}
\L{\LB{}\Tab{40}{_\*_received_time._}}
\L{\LB{}\Tab{40}{_\*\1\CE{}}}
\L{\LB{______\V{parsepacket}_(\V{nxtmsg}\-\!\>\V{m},_\V{op});}\Tab{40}{\C{}\1\*_parse_the_message_\*\1\CE{}}}
\L{\LB{______\V{pkt\_typ}_=_\V{op}\-\!\>\V{ptype};}}
\L{\LB{}}
\L{\LB{______\V{printf}_(\S{}\3(vncgetpkt)_before_deletevminq\2n\3\SE{});}}
\L{\LB{______\V{fflush}_(\V{stdout});}}
\L{\LB{______\V{fflush}_(\V{stderr});}}
\L{\LB{______\V{deletevminq}_(\V{vncid},_\V{nxtmsg});}}
\L{\LB{____\}}}
\L{\LB{__\V{printf}_(\S{}\3(vncgetpkt)_returning_\%s\2n\3\SE{},_\V{displayPktTyp}_(\V{pkt\_typ}));}}
\L{\LB{__\V{fflush}_(\V{stdout});}}
\L{\LB{__\V{fflush}_(\V{stderr});}}
\L{\LB{__\K{return}_\V{pkt\_typ};}}
\L{\LB{\}}}

\ifisdraft
	\twocolumn
\fi
\section{The VNC Message Send Function}

The \acl{VNC} enhanced \acl{NCP} transmit function is shown below. This
function simply sets the \acl{VNC} information appropriately and stores 
the packet in the \acf{QS} by calling sendvc().

\ifisdraft
	\onecolumn
\fi

\Head{}
\File{\1users\1sbush\1rdrn\1ncp\1vnctransmit.c}{1997}{Mar}{ 2}{19:19}{925}
\L{\LB{\K{\#include}_\<\V{stdio}.\V{h}\>}}
\L{\LB{\K{\#include}_\<\V{sys}\1\V{types}.\V{h}\>}}
\L{\LB{\K{\#include}_\S{}\3switch.h\3\SE{}}}
\L{\LB{\K{\#include}_\S{}\3shared.h\3\SE{}}}
\L{\LB{\K{\#include}_\S{}\3vnc.h\3\SE{}}}
\L{\LB{}}
\L{\LB{\K{extern}_\K{struct}_\V{shared\_data}_\*\V{sharedm};}}
\L{\LB{}}
\L{\LB{\K{static}_\K{char}_\*\V{rcs}_=_\S{}\3\$Id: vnctransmit.tex,v 1.2 1997/05/06 16:05:06 sbush Exp sbush \$\3\SE{};}}
\L{\LB{}}
\L{\LB{\C{}\1\*}}
\L{\LB{_\*_Implement_the_VNC_algorithm_for_outgoing_orderwire_packets.}}
\L{\LB{_\*\1\CE{}}}
\L{\LB{}}
\L{\LB{\K{int}}}
\L{\LB{\Proc{vncOrderWireTransmit}\V{vncOrderWireTransmit}_(\K{struct}_\V{vcid}_\*\V{vncid},_\K{int}_\V{gfd},_\K{int}_\V{efd},_\K{int}_\V{sfd},_\K{int}_\V{ptype},}}
\L{\LB{}\Tab{16}{______\K{char}_\*\V{dest},_\K{char}_\*\V{m},_\K{int}_\V{p},_\K{char}_\*\V{callsign})}}
\L{\LB{\{}}
\L{\LB{__\K{int}_\V{res};}}
\L{\LB{__\K{struct}_\V{vm}_\V{vncpkt};}}
\L{\LB{__\V{time\_t}_\V{rt};}}
\L{\LB{__\K{struct}_\V{vm}_\*\V{nxtmsg};}}
\L{\LB{__\K{struct}_\V{gvtS}_\V{GVT};}}
\L{\LB{__\K{struct}_\V{gps}_\V{pos};}\Tab{32}{\C{}\1\*_Position_packet_and_table_\*\1\CE{}}}
\L{\LB{}}
\L{\LB{__\K{if}_(\V{sharedm}\-\!\>\V{UseVNC})}}
\L{\LB{____\{}}
\L{\LB{______\V{vncpkt}.\V{anti}_=_\N{0};}}
\L{\LB{______\V{vncpkt}.\V{snd\_tm}_=_\V{getlvt}_(\V{vncid});}}
\L{\LB{______\V{vncpkt}.\V{rec\_tm}_=_\V{getgpstime}_(\V{gfd},_\N{1});}}
\L{\LB{______\V{vtdepacketize}_(\V{m},_\V{vncpkt}.\V{anti},_\V{vncpkt}.\V{snd\_tm},_\V{vncpkt}.\V{rec\_tm});}}
\L{\LB{______\V{sendvm}_(\V{vncid},_\V{vncpkt}.\V{anti},_\V{vncpkt}.\V{rec\_tm},_\V{vncpkt}.\V{snd\_tm},_\V{m},_\V{strlen}_(\V{m}));}}
\L{\LB{____\}}}
\L{\LB{__\K{return}_\V{OrderWireTransmit}_(\V{efd},_\V{sfd},_\V{ptype},_\V{dest},_\V{m},_\V{p},_\V{callsign});}}
\L{\LB{\}}}

\ifisdraft
	\twocolumn
\fi
\chapter{NCP and VNC MIBS}
\section{The RDRN Simple Network Management MIB}

The following Simple Network Management Protocol (SNMP) Management
Information Base (MIB) is used by the RDRN Network Control Protocol
(NCP) and Link Management Daemon (LMD).

\begin{singlespace}
\footnotesize



\Head{}
\File{\1users\1sbush\1rdrn\1ncp\1ncp.mib}{1997}{Feb}{ 9}{12:48}{35944}
\L{\LB{\V{RDRN}\-\V{NCP}\-\V{MIB}_\K{DEFINITIONS}_::=_\K{BEGIN}}}
\L{\LB{}}
\L{\LB{\K{IMPORTS}}}
\L{\LB{____\K{MODULE}\-\V{IDENTITY},_\K{OBJECT}\-\V{TYPE},_\V{experimental},}}
\L{\LB{____\V{Counter32},_\V{TimeTicks}}}
\L{\LB{____}\Tab{8}{\K{FROM}_\V{SNMPv2}\-\V{SMI}}}
\L{\LB{____\V{DisplayString}}}
\L{\LB{________\K{FROM}_\V{SNMPv2}\-\V{TC};}}
\L{\LB{}}
\L{\LB{\V{rdrnNcpMIB}_\K{MODULE}\-\V{IDENTITY}}}
\L{\LB{_____\K{LAST}\-\V{UPDATED}_\3\N{9701010000Z}\3}}
\L{\LB{_____\V{ORGANIZATION}_\3\V{KU}_\V{TISL}\3}}
\L{\LB{_____\V{CONTACT}\-\V{INFO}}}
\L{\LB{______________\3\V{Steve}_\V{Bush}_\V{sbush}\@\V{tisl}.\V{ukans}.\V{edu}\3}}
\L{\LB{_____\K{DESCRIPTION}}}
\L{\LB{_____________\3\V{Experimental}_\V{MIB}_\V{modules}_\V{for}_\V{the}_\V{Rapidly}_\V{Deployable}}}
\L{\LB{______________\V{Radio}_\V{Network}_(\V{RDRN})_\V{Project}_\V{Network}_\V{Control}_}}
\L{\LB{______________\V{Protocol}_(\V{NCP}).\3}}
\L{\LB{_____::=_\{_\V{experimental}_\V{ncp}(\N{75})_\N{2}_\}}}
\L{\LB{}}
\L{\LB{\-\-}}
\L{\LB{\-\-_\V{NCP}_\V{Initialization}_\V{Group}}}
\L{\LB{\-\-}}
\L{\LB{}}
\L{\LB{\V{initialization}_\K{OBJECT}_\K{IDENTIFIER}_::=_\{_\V{rdrnNcpMIB}_\N{1}_\}}}
\L{\LB{}}
\L{\LB{\V{initContact}_\K{OBJECT}\-\V{TYPE}}}
\L{\LB{}\Tab{8}{\K{SYNTAX}_\V{DisplayString}}}
\L{\LB{}\Tab{8}{\K{MAX}\-\V{ACCESS}_\V{read}\-\V{only}}}
\L{\LB{}\Tab{8}{\K{STATUS}_\V{current}}}
\L{\LB{}\Tab{8}{\K{DESCRIPTION}}}
\L{\LB{}\Tab{16}{\3\V{NCP}_\V{Agent}_\V{Developer},_\V{but}_\V{hopefully}_\V{NOT}_\V{maintainer}.\3}}
\L{\LB{}\Tab{8}{::=_\{_\V{initialization}_\N{1}_\}}}
\L{\LB{}}
\L{\LB{\V{initEntity}_\K{OBJECT}\-\V{TYPE}}}
\L{\LB{________\K{SYNTAX}_\K{INTEGER}_\{}}
\L{\LB{}\Tab{16}{\V{edgeSwitch}(\N{0}),}}
\L{\LB{}\Tab{16}{\V{remoteNode}(\N{1})}}
\L{\LB{}\Tab{8}{\}}}
\L{\LB{________\K{MAX}\-\V{ACCESS}_\V{read}\-\V{only}}}
\L{\LB{________\K{STATUS}_\V{current}}}
\L{\LB{________\K{DESCRIPTION}_______}}
\L{\LB{________________\3\V{Describes}_\V{the}_\V{RDRN}_\V{node}_\V{component}.\3}}
\L{\LB{________::=_\{_\V{initialization}_\N{2}_\}}}
\L{\LB{}}
\L{\LB{\V{initDebug}_\K{OBJECT}\-\V{TYPE}}}
\L{\LB{________\K{SYNTAX}_\K{INTEGER}_(\N{0.\,.100})}}
\L{\LB{________\K{MAX}\-\V{ACCESS}_\V{read}\-\V{write}}}
\L{\LB{________\K{STATUS}_\V{current}}}
\L{\LB{________\K{DESCRIPTION}}}
\L{\LB{________________\3\V{Logging}_\V{level}._\V{Zero}_\V{is}_\V{off};_\V{the}_\V{higher}_\V{the}_\V{number},_\V{the}}}
\L{\LB{_________________\V{more}_\V{logging}_\V{is}_\V{enabled}.\3}}
\L{\LB{________::=_\{_\V{initialization}_\N{3}_\}}}
\L{\LB{}}
\L{\LB{\V{initEncrypt}_\K{OBJECT}\-\V{TYPE}}}
\L{\LB{________\K{SYNTAX}_\K{INTEGER}_\{}}
\L{\LB{}\Tab{16}{\V{clear}(\N{0}),}}
\L{\LB{}\Tab{16}{\V{des}(\N{1})}}
\L{\LB{}\Tab{8}{\}}}
\L{\LB{________\K{MAX}\-\V{ACCESS}_\V{read}\-\V{write}}}
\L{\LB{________\K{STATUS}_\V{current}_____________________}}
\L{\LB{________\K{DESCRIPTION}}}
\L{\LB{}\Tab{16}{\3\V{Indicates}_\V{whether}_\V{data}_\V{encryption}_\V{used}._\V{Both}_\V{endpoints}_\V{of}}}
\L{\LB{_________________\V{a}_\V{packet}_\V{radio}_\V{link}_\V{MUST}_\V{have}_\V{this}_\V{object}_\V{set}_\V{identically}}}
\L{\LB{_________________\V{for}_\V{proper}_\V{communication}.\3}}
\L{\LB{________::=_\{_\V{initialization}_\N{4}_\}}}
\L{\LB{}}
\L{\LB{\V{initVNC}_\K{OBJECT}\-\V{TYPE}}}
\L{\LB{________\K{SYNTAX}_\K{INTEGER}_\{}}
\L{\LB{}\Tab{16}{\V{vncEnabled}(\N{1}),}}
\L{\LB{}\Tab{16}{\V{sequentional}(\N{0})}}
\L{\LB{}\Tab{8}{\}}}
\L{\LB{________\K{MAX}\-\V{ACCESS}_\V{read}\-\V{only}}}
\L{\LB{________\K{STATUS}_\V{current}}}
\L{\LB{________\K{DESCRIPTION}}}
\L{\LB{________________\3\V{Indicates}_\V{whether}_\V{this}_\V{system}_\V{is}_\V{running}_\V{using}}}
\L{\LB{_________________\V{Virtual}_\V{Network}_\V{Configuration}_\V{enhanced}_\V{mode},}}
\L{\LB{_________________\V{or}_\V{sequential}_\V{operation}.\3}}
\L{\LB{________::=_\{_\V{initialization}_\N{5}_\}}}
\L{\LB{}}
\L{\LB{\V{initStacks}_\K{OBJECT}\-\V{TYPE}}}
\L{\LB{________\K{SYNTAX}_\K{INTEGER}_\{}}
\L{\LB{________________\V{noStacks}(\N{1}),}}
\L{\LB{________________\V{useStacks}(\N{0})}}
\L{\LB{________\}}}
\L{\LB{________\K{MAX}\-\V{ACCESS}_\V{read}\-\V{write}}}
\L{\LB{________\K{STATUS}_\V{current}}}
\L{\LB{________\K{DESCRIPTION}}}
\L{\LB{________________\3\V{Indicates}_\V{whether}_\V{link}_\V{level}_\V{stacks}_\V{will}_\V{be}_\V{created}._}}
\L{\LB{_________________\V{Primarily}_\V{used}_\V{for}_\V{testing}_\V{purposes}.\3}}
\L{\LB{________::=_\{_\V{initialization}_\N{6}_\}}}
\L{\LB{}}
\L{\LB{\V{initRadioPort}_\K{OBJECT}\-\V{TYPE}}}
\L{\LB{}\Tab{8}{\K{SYNTAX}_\V{DisplayString}}}
\L{\LB{________\K{MAX}\-\V{ACCESS}_\V{read}\-\V{only}}}
\L{\LB{________\K{STATUS}_\V{current}}}
\L{\LB{________\K{DESCRIPTION}}}
\L{\LB{________________\3\V{Indicates}_\V{the}_\V{packet}_\V{radio}_\V{tty}_\V{port}.\3}}
\L{\LB{________::=_\{_\V{initialization}_\N{7}_\}}}
\L{\LB{}}
\L{\LB{\V{initGpsPort}_\K{OBJECT}\-\V{TYPE}}}
\L{\LB{________\K{SYNTAX}_\V{DisplayString}}}
\L{\LB{________\K{MAX}\-\V{ACCESS}_\V{read}\-\V{only}}}
\L{\LB{________\K{STATUS}_\V{current}}}
\L{\LB{________\K{DESCRIPTION}}}
\L{\LB{________________\3\V{Indicates}_\V{the}_\V{GPS}_\V{tty}_\V{port}.\3}}
\L{\LB{________::=_\{_\V{initialization}_\N{8}_\}}}
\L{\LB{}}
\L{\LB{\V{initSpeed}_\K{OBJECT}\-\V{TYPE}}}
\L{\LB{________\K{SYNTAX}_\V{DisplayString}}}
\L{\LB{________\K{MAX}\-\V{ACCESS}_\V{read}\-\V{only}}}
\L{\LB{________\K{STATUS}_\V{current}}}
\L{\LB{________\K{DESCRIPTION}}}
\L{\LB{________________\3\V{Indicates}_\V{the}_\V{assumed}_\V{initial}_\V{speed}._\V{Primarily}_\V{used}_\V{for}}}
\L{\LB{_________________\V{simulated}_\V{GPS}.\3}}
\L{\LB{________::=_\{_\V{initialization}_\N{9}_\}}}
\L{\LB{}}
\L{\LB{\V{initDirection}_\K{OBJECT}\-\V{TYPE}}}
\L{\LB{________\K{SYNTAX}_\V{DisplayString}}}
\L{\LB{________\K{MAX}\-\V{ACCESS}_\V{read}\-\V{only}}}
\L{\LB{________\K{STATUS}_\V{current}}}
\L{\LB{________\K{DESCRIPTION}}}
\L{\LB{________________\3\V{Indicates}_\V{the}_\V{assumed}_\V{initial}_\V{direction}._\V{Primarily}_\V{used}_\V{for}}}
\L{\LB{_________________\V{simulated}_\V{GPS}.\3}}
\L{\LB{________::=_\{_\V{initialization}_\N{10}_\}}}
\L{\LB{}}
\L{\LB{\V{initXPos}_\K{OBJECT}\-\V{TYPE}}}
\L{\LB{________\K{SYNTAX}_\V{DisplayString}}}
\L{\LB{________\K{MAX}\-\V{ACCESS}_\V{read}\-\V{only}}}
\L{\LB{________\K{STATUS}_\V{current}}}
\L{\LB{________\K{DESCRIPTION}}}
\L{\LB{________________\3\V{Indicates}_\V{the}_\V{assumed}_\V{initial}_\V{X}_\V{axis}_\V{position}._\V{Primarily}_}}
\L{\LB{_________________\V{used}_\V{for}_\V{simulated}_\V{GPS}.\3}}
\L{\LB{________::=_\{_\V{initialization}_\N{11}_\}}}
\L{\LB{}}
\L{\LB{\V{initYPos}_\K{OBJECT}\-\V{TYPE}}}
\L{\LB{________\K{SYNTAX}_\V{DisplayString}}}
\L{\LB{________\K{MAX}\-\V{ACCESS}_\V{read}\-\V{only}}}
\L{\LB{________\K{STATUS}_\V{current}__}}
\L{\LB{________\K{DESCRIPTION}}}
\L{\LB{________________\3\V{Indicates}_\V{the}_\V{assumed}_\V{initial}_\V{Y}_\V{axis}_\V{position}._\V{Primarily}_}}
\L{\LB{_________________\V{used}_\V{for}_\V{simulated}_\V{GPS}.\3}}
\L{\LB{________::=_\{_\V{initialization}_\N{12}_\}}}
\L{\LB{}}
\L{\LB{\V{initSimGps}_\K{OBJECT}\-\V{TYPE}}}
\L{\LB{________\K{SYNTAX}_\K{INTEGER}_\{}}
\L{\LB{}\Tab{16}{\V{gpsReceiver}(\N{0}),}}
\L{\LB{}\Tab{16}{\V{gpsSimulator}(\N{1})}}
\L{\LB{}\Tab{8}{\}}}
\L{\LB{________\K{MAX}\-\V{ACCESS}_\V{read}\-\V{write}}}
\L{\LB{________\K{STATUS}_\V{current}}}
\L{\LB{________\K{DESCRIPTION}}}
\L{\LB{________________\3\V{Indicates}_\V{whether}_\V{the}_\V{GPS}_\V{receiver}_\V{is}_\V{being}_\V{read}_\V{or}}}
\L{\LB{}\Tab{16}{_\V{the}_\V{GPS}_\V{simulation}_\V{code}_\V{is}_\V{being}_\V{used}._\V{This}_\V{value}}}
\L{\LB{_________________\V{can}_\V{be}_\V{set}_\V{while}_\V{running}_\V{without}_\V{harm}._\V{Changing}}}
\L{\LB{_________________\V{from}_\V{gpsReceiver}_\V{to}_\V{gpsSimulator}_\V{will}_\V{cause}_\V{a}_\V{smooth}}}
\L{\LB{_________________\V{transition}_\V{based}_\V{on}_\V{last}_\V{known}_\V{direction}_\V{and}_\V{speed}.}}
\L{\LB{_________________\V{Transitioning}_\V{from}_\V{gps}_\V{Simulator}_\V{to}_\V{gpsReceiver}_\V{may}}}
\L{\LB{_________________\V{cause}_\V{a}_\V{large}_\V{jump}_\V{in}_\V{position}.\3}}
\L{\LB{________::=_\{_\V{initialization}_\N{13}_\}}}
\L{\LB{}}
\L{\LB{\V{initTcpEmulation}_\K{OBJECT}\-\V{TYPE}}}
\L{\LB{________\K{SYNTAX}_\K{INTEGER}_\{}}
\L{\LB{}\Tab{16}{\V{packetRadio}(\N{2}),}}
\L{\LB{}\Tab{16}{\V{tcpEmulation}(\N{1}),}}
\L{\LB{}\Tab{16}{\V{packetRadioAndTcpEmulation}(\N{3})}}
\L{\LB{}\Tab{8}{\}}}
\L{\LB{________\K{MAX}\-\V{ACCESS}_\V{read}\-\V{only}}}
\L{\LB{________\K{STATUS}_\V{current}}}
\L{\LB{________\K{DESCRIPTION}}}
\L{\LB{________________\3\V{Indicates}_\V{whether}_\V{packet}_\V{radios}_\V{or}_\V{TCP}\-\V{IP}_\V{simulation}_\V{of}}}
\L{\LB{_________________\V{packet}_\V{radios}_\V{is}_\V{currently}_\V{in}_\V{use}._\V{This}_\V{value}_\V{cannot}}}
\L{\LB{_________________\V{be}_\V{set}_\V{because}_\V{the}_\V{tcp}_\V{emulation}_\V{requires}_\V{knowledge}}}
\L{\LB{_________________\V{of}_\V{ip}_\V{addresses}_\V{of}_\V{all}_\V{hosts}_\V{in}_\V{the}_\V{simulated}_\V{RDRN}}}
\L{\LB{_________________\V{system}_\V{and}_\V{all}_\V{though}_\V{this}_\V{could}_\V{be}_\V{set}_\V{on}_\V{the}_\V{fly}_}}
\L{\LB{_________________\V{through}_\V{a}_\V{MIB}_\V{table},_\V{I}_\V{don}\4\V{t}_\V{feel}_\V{it}\4\V{s}_\V{worth}_\V{the}_}}
\L{\LB{_________________\V{effort}_\V{to}_\V{develop}_\V{at}_\V{this}_\V{point}.\3}}
\L{\LB{________::=_\{_\V{initialization}_\N{14}_\}}}
\L{\LB{}}
\L{\LB{\-\-}}
\L{\LB{\-\-_\V{NCP}_\V{Configuration}_\V{Group}}}
\L{\LB{\-\-}}
\L{\LB{}}
\L{\LB{\V{configuration}_\K{OBJECT}_\K{IDENTIFIER}_::=_\{_\V{rdrnNcpMIB}_\N{2}_\}}}
\L{\LB{}}
\L{\LB{\V{configTimeout}_\K{OBJECT}\-\V{TYPE}}}
\L{\LB{________\K{SYNTAX}_\K{INTEGER}_(\N{0.\,.2147483647})}}
\L{\LB{________\K{MAX}\-\V{ACCESS}_\V{read}\-\V{write}}}
\L{\LB{________\K{STATUS}_\V{current}______________________________}}
\L{\LB{________\K{DESCRIPTION}_____________}}
\L{\LB{________________\3\V{Indicates}_\V{the}_\V{timeout}_\V{during}_\V{edge}_\V{switch}_\V{reconfiguration}._}}
\L{\LB{_________________\V{Should}_\V{be}_\V{set}_\V{long}_\V{enough}_\V{for}_\V{all}_\V{edge}_\V{switches}_\V{to}_}}
\L{\LB{_________________\V{respond}.\3}}
\L{\LB{________::=_\{_\V{configuration}_\N{1}_\}}}
\L{\LB{}}
\L{\LB{\V{configMaxVel}_\K{OBJECT}\-\V{TYPE}}}
\L{\LB{________\K{SYNTAX}_\V{DisplayString}}}
\L{\LB{________\K{MAX}\-\V{ACCESS}_\V{read}\-\V{write}}}
\L{\LB{________\K{STATUS}_\V{current}}}
\L{\LB{________\K{DESCRIPTION}}}
\L{\LB{________________\3\V{Indicates}_\V{the}_\V{maximum}_\V{of}_\V{a}_\V{uniformly}_\V{distributed}_\V{velocity}}}
\L{\LB{}\Tab{16}{_\V{for}_\V{simulated}_\V{GPS}_\V{operation}.\3}}
\L{\LB{________::=_\{_\V{configuration}_\N{2}_\}}}
\L{\LB{}}
\L{\LB{\V{configSpeed}_\K{OBJECT}\-\V{TYPE}}}
\L{\LB{________\K{SYNTAX}_\V{DisplayString}}}
\L{\LB{________\K{MAX}\-\V{ACCESS}_\V{read}\-\V{write}}}
\L{\LB{________\K{STATUS}_\V{current}}}
\L{\LB{________\K{DESCRIPTION}}}
\L{\LB{________________\3\V{Indicates}_\V{the}_\V{initial}_\V{speed}_\V{for}_\V{simulated}_\V{GPS}_\V{operation}.}}
\L{\LB{}\Tab{16}{_\V{Same}_\V{as}_\V{initSpeed}_\V{for}_\V{historical}_\V{reasons},_\V{e}.\V{g}._\V{this}_\V{one}}}
\L{\LB{_________________\V{set}_\V{through}_\V{a}_\V{file}_\V{while}_\V{the}_\V{other}_\V{is}_\V{set}_\V{via}_\V{arguments}.\3}}
\L{\LB{________::=_\{_\V{configuration}_\N{3}_\}}}
\L{\LB{}}
\L{\LB{\V{configDirection}_\K{OBJECT}\-\V{TYPE}}}
\L{\LB{________\K{SYNTAX}_\V{DisplayString}}}
\L{\LB{________\K{MAX}\-\V{ACCESS}_\V{read}\-\V{write}}}
\L{\LB{________\K{STATUS}_\V{current}}}
\L{\LB{________\K{DESCRIPTION}}}
\L{\LB{________________\3\V{Indicates}_\V{the}_\V{initial}_\V{direction}_\V{in}_\V{degrees}_\V{for}_\V{simulated}_}}
\L{\LB{_________________\V{GPS}_\V{operation}._\V{Same}_\V{as}_\V{initDirection}_\V{for}_\V{historical}}}
\L{\LB{_________________\V{reasons}._\V{This}_\V{one}_\V{is}_\V{read}_\V{from}_\V{a}_\V{file}_\V{while}_\V{the}_\V{other}}}
\L{\LB{_________________\V{is}_\V{set}_\V{as}_\V{a}_\V{program}_\V{argument}_\V{during}_\V{startup}.\3}}
\L{\LB{________::=_\{_\V{configuration}_\N{4}_\}}}
\L{\LB{}}
\L{\LB{\V{configUseRealTopology}_\K{OBJECT}\-\V{TYPE}}}
\L{\LB{________\K{SYNTAX}_\K{INTEGER}_(\N{0.\,.1})}}
\L{\LB{________\K{MAX}\-\V{ACCESS}_\V{read}\-\V{write}}}
\L{\LB{________\K{STATUS}_\V{current}}}
\L{\LB{________\K{DESCRIPTION}__________________}}
\L{\LB{________________\3\V{Indicates}_\V{whether}_\V{edge}_\V{switch}_\V{topology}_\V{code}_\V{is}_\V{called}.}}
\L{\LB{_________________\V{Currently}_\V{this}_\V{code}_\V{uses}_\V{MatLab}_\V{and}_\V{can}_\V{be}_\V{quite}_\V{slow}.\3}}
\L{\LB{________::=_\{_\V{configuration}_\N{5}_\}}}
\L{\LB{}}
\L{\LB{\V{configCallsign}_\K{OBJECT}\-\V{TYPE}}}
\L{\LB{________\K{SYNTAX}_\V{DisplayString}}}
\L{\LB{________\K{MAX}\-\V{ACCESS}_\V{read}\-\V{only}}}
\L{\LB{________\K{STATUS}_\V{current}}}
\L{\LB{________\K{DESCRIPTION}}}
\L{\LB{________________\3\V{Indicates}_\V{this}_\V{nodes}_\V{packet}_\V{radio}_\V{callsign}.\3}}
\L{\LB{________::=_\{_\V{configuration}_\N{6}_\}}}
\L{\LB{}}
\L{\LB{\V{configKey}_\K{OBJECT}\-\V{TYPE}}}
\L{\LB{________\K{SYNTAX}_\V{DisplayString}}}
\L{\LB{________\K{MAX}\-\V{ACCESS}_\V{read}\-\V{only}}}
\L{\LB{________\K{STATUS}_\V{current}}}
\L{\LB{________\K{DESCRIPTION}}}
\L{\LB{________________\3\V{Indicates}_\V{the}_\V{current}_\V{DES}_\V{key}._\V{Probably}_\V{not}_\V{secure}_\V{to}}}
\L{\LB{_________________\V{put}_\V{this}_\V{here},_\V{but}_\V{useful}_\V{for}_\V{debugging}.\3}}
\L{\LB{________::=_\{_\V{configuration}_\N{7}_\}}}
\L{\LB{}}
\L{\LB{\V{configUserSwitch}_\K{OBJECT}\-\V{TYPE}}}
\L{\LB{________\K{SYNTAX}_\V{DisplayString}}}
\L{\LB{________\K{MAX}\-\V{ACCESS}_\V{read}\-\V{only}}}
\L{\LB{________\K{STATUS}_\V{current}}}
\L{\LB{________\K{DESCRIPTION}}}
\L{\LB{________________\3\V{If}_\V{this}_\V{is}_\V{a}_\V{remote}_\V{node},_\V{this}_\V{is}_\V{the}_\V{associated}_\V{edge}}}
\L{\LB{_________________\V{switch}.\3}}
\L{\LB{________::=_\{_\V{configuration}_\N{8}_\}}}
\L{\LB{}}
\L{\LB{\V{configUpdatePeriod}_\K{OBJECT}\-\V{TYPE}}}
\L{\LB{________\K{SYNTAX}_\K{INTEGER}_(\N{0.\,.2147483647})}}
\L{\LB{________\K{MAX}\-\V{ACCESS}_\V{read}\-\V{write}}}
\L{\LB{________\K{STATUS}_\V{current}}}
\L{\LB{________\K{DESCRIPTION}}}
\L{\LB{________________\3\V{Indicates}_\V{the}_\V{remote}_\V{node}\4\V{s}_\V{user}_\V{position}_\V{update}_\V{rate}.\3}}
\L{\LB{________::=_\{_\V{configuration}_\N{9}_\}}}
\L{\LB{}}
\L{\LB{\V{configDistanceTolerance}_\K{OBJECT}\-\V{TYPE}}}
\L{\LB{________\K{SYNTAX}_\K{INTEGER}_(\N{0.\,.2147483647})}}
\L{\LB{________\K{MAX}\-\V{ACCESS}_\V{read}\-\V{write}}}
\L{\LB{________\K{STATUS}_\V{current}}}
\L{\LB{________\K{DESCRIPTION}}}
\L{\LB{________________\3\V{Indicates}_\V{the}_\V{amount}_\V{of}_\V{movement}_\V{allowed}_\V{before}}}
\L{\LB{_________________\V{beam}_\V{angles}_\V{must}_\V{be}_\V{recomputed}.\3}}
\L{\LB{________::=_\{_\V{configuration}_\N{10}_\}}}
\L{\LB{}}
\L{\LB{\V{configAntennaAngle}_\K{OBJECT}\-\V{TYPE}}}
\L{\LB{________\K{SYNTAX}_\K{INTEGER}_(\N{0.\,.2147483647})}}
\L{\LB{________\K{MAX}\-\V{ACCESS}_\V{read}\-\V{write}}}
\L{\LB{________\K{STATUS}_\V{current}}}
\L{\LB{________\K{DESCRIPTION}}}
\L{\LB{________________\3\V{Indicates}_\V{the}_\V{direction}_\V{of}_\V{the}_\V{physical}_\V{position}}}
\L{\LB{_________________\V{of}_\V{the}_\V{linear}_\V{array}_\V{from}_\V{pointing}_\V{North}.\3}}
\L{\LB{________::=_\{_\V{configuration}_\N{11}_\}}}
\L{\LB{}}
\L{\LB{\-\-}}
\L{\LB{\-\-_\V{Current}_\V{NCP}_\V{AX}.\N{25}_\V{Connection}_\V{Group}}}
\L{\LB{\-\-}}
\L{\LB{}}
\L{\LB{\V{radioStreams}_\K{OBJECT}_\K{IDENTIFIER}_::=_\{_\V{rdrnNcpMIB}_\N{3}_\}}}
\L{\LB{}}
\L{\LB{\V{radioStreamsTable}_\K{OBJECT}\-\V{TYPE}}}
\L{\LB{}\Tab{8}{\K{SYNTAX}_\K{SEQUENCE}_\V{OF}_\V{RadioStreamsEntry}}}
\L{\LB{}\Tab{8}{\K{MAX}\-\V{ACCESS}_\V{not}\-\V{accessible}}}
\L{\LB{}\Tab{8}{\K{STATUS}_\V{current}}}
\L{\LB{}\Tab{8}{\K{DESCRIPTION}}}
\L{\LB{}\Tab{16}{\3\V{Table}_\V{of}_\V{packet}_\V{currently}_\V{established}_\V{packet}_\V{radio}_\V{streams}}}
\L{\LB{_________________\V{and}_\V{callsigns}.\3}}
\L{\LB{}\Tab{8}{::=_\{_\V{radioStreams}_\N{1}_\}}}
\L{\LB{}}
\L{\LB{\V{radioStreamsEntry}_\K{OBJECT}\-\V{TYPE}}}
\L{\LB{}\Tab{8}{\K{SYNTAX}_\V{RadioStreamsEntry}}}
\L{\LB{}\Tab{8}{\K{MAX}\-\V{ACCESS}_\V{not}\-\V{accessible}}}
\L{\LB{}\Tab{8}{\K{STATUS}_\V{current}}}
\L{\LB{________\K{DESCRIPTION}}}
\L{\LB{________________\3\V{Table}_\V{of}_\V{packet}_\V{currently}_\V{established}_\V{packet}_\V{radio}_\V{streams}}}
\L{\LB{_________________\V{and}_\V{callsigns}.\3}}
\L{\LB{}\Tab{8}{\K{INDEX}_\{_\V{radioStreamsID}_\}}}
\L{\LB{}\Tab{8}{::=_\{_\V{radioStreamsTable}_\N{1}_\}}}
\L{\LB{}}
\L{\LB{\V{RadioStreamsEntry}_::=_\K{SEQUENCE}_\{}}
\L{\LB{}\Tab{8}{\V{radioStreamsID}_}}
\L{\LB{}\Tab{16}{\V{DisplayString},}}
\L{\LB{}\Tab{8}{\V{radioStreamsCallsign}}}
\L{\LB{}\Tab{16}{\V{DisplayString}}}
\L{\LB{}\Tab{8}{\}}}
\L{\LB{}}
\L{\LB{\V{radioStreamsID}_\K{OBJECT}\-\V{TYPE}}}
\L{\LB{}\Tab{8}{\K{SYNTAX}_\V{DisplayString}}}
\L{\LB{}\Tab{8}{\K{MAX}\-\V{ACCESS}_\V{read}\-\V{only}}}
\L{\LB{}\Tab{8}{\K{STATUS}_\V{current}}}
\L{\LB{}\Tab{8}{\K{DESCRIPTION}}}
\L{\LB{}\Tab{16}{\3\V{The}_\V{Packet}_\V{Radio}_\V{stream}_\V{identifier}_\V{for}_\V{this}}}
\L{\LB{_________________\V{ARQ}_\V{AX}.\N{25}_\V{connection}.\3}}
\L{\LB{}\Tab{8}{::=_\{_\V{radioStreamsEntry}_\N{1}_\}}}
\L{\LB{}}
\L{\LB{\V{radioStreamsCallsign}_\K{OBJECT}\-\V{TYPE}}}
\L{\LB{}\Tab{8}{\K{SYNTAX}_\V{DisplayString}}}
\L{\LB{}\Tab{8}{\K{MAX}\-\V{ACCESS}_\V{read}\-\V{only}}}
\L{\LB{}\Tab{8}{\K{STATUS}_\V{current}}}
\L{\LB{}\Tab{8}{\K{DESCRIPTION}}}
\L{\LB{}\Tab{16}{\3\V{This}_\V{is}_\V{the}_\V{callsign}_\V{of}_\V{the}_\V{packet}_\V{radio}_\V{with}}}
\L{\LB{_________________\V{the}_\V{associated}_\V{stream}_\V{identifier}.\3}}
\L{\LB{}\Tab{8}{::=_\{_\V{radioStreamsEntry}_\N{2}_\}}}
\L{\LB{}}
\L{\LB{\-\-}}
\L{\LB{\-\-_\V{Known}_\V{Edge}_\V{Switch}_\V{Group}}}
\L{\LB{\-\-}}
\L{\LB{}}
\L{\LB{\V{edgeSwitchNetwork}_\K{OBJECT}_\K{IDENTIFIER}_::=_\{_\V{rdrnNcpMIB}_\N{4}_\}}}
\L{\LB{}}
\L{\LB{\V{edgeSwitchTable}_\K{OBJECT}\-\V{TYPE}}}
\L{\LB{________\K{SYNTAX}_\K{SEQUENCE}_\V{OF}_\V{EdgeSwitchEntry}}}
\L{\LB{________\K{MAX}\-\V{ACCESS}_\V{not}\-\V{accessible}}}
\L{\LB{________\K{STATUS}_\V{current}}}
\L{\LB{________\K{DESCRIPTION}}}
\L{\LB{________________\3\V{Table}_\V{of}_\V{RDRN}_\V{edge}_\V{switches}_\V{known}_\V{by}_\V{this}_\V{edge}_\V{switch}.\3}}
\L{\LB{________::=_\{_\V{edgeSwitchNetwork}_\N{1}_\}}}
\L{\LB{}}
\L{\LB{\V{edgeSwitchEntry}_\K{OBJECT}\-\V{TYPE}}}
\L{\LB{________\K{SYNTAX}_\V{EdgeSwitchEntry}}}
\L{\LB{________\K{MAX}\-\V{ACCESS}_\V{not}\-\V{accessible}}}
\L{\LB{________\K{STATUS}_\V{current}}}
\L{\LB{________\K{DESCRIPTION}}}
\L{\LB{________________\3\V{Table}_\V{of}_\V{RDRN}_\V{edge}_\V{switches}_\V{known}_\V{by}_\V{this}_\V{edge}_\V{switch}.\3}}
\L{\LB{}\Tab{8}{\K{INDEX}_\{_\V{edgeSwitchCallsign}_\}}}
\L{\LB{________::=_\{_\V{edgeSwitchTable}_\N{1}_\}}}
\L{\LB{}}
\L{\LB{\V{EdgeSwitchEntry}_::=_\K{SEQUENCE}_\{}}
\L{\LB{}\Tab{8}{\V{edgeSwitchCallsign}}}
\L{\LB{}\Tab{16}{\V{DisplayString},}}
\L{\LB{}\Tab{8}{\V{edgeSwitchHostname}}}
\L{\LB{}\Tab{16}{\V{DisplayString},}}
\L{\LB{}\Tab{8}{\V{edgeSwitchATMAddr}}}
\L{\LB{}\Tab{16}{\V{DisplayString},}}
\L{\LB{}\Tab{8}{\V{edgeSwitchLatDeg}}}
\L{\LB{}\Tab{16}{\V{DisplayString},}}
\L{\LB{}\Tab{8}{\V{edgeSwitchLatMin}}}
\L{\LB{}\Tab{16}{\V{DisplayString},}}
\L{\LB{}\Tab{8}{\V{edgeSwitchLatDir}}}
\L{\LB{}\Tab{16}{\V{DisplayString},}}
\L{\LB{________\V{edgeSwitchLonDeg}}}
\L{\LB{________________\V{DisplayString},}}
\L{\LB{________\V{edgeSwitchLonMin}}}
\L{\LB{________________\V{DisplayString},}}
\L{\LB{________\V{edgeSwitchLonDir}}}
\L{\LB{________________\V{DisplayString},}}
\L{\LB{}\Tab{8}{\V{edgeSwitchXPos}}}
\L{\LB{}\Tab{16}{\V{DisplayString},}}
\L{\LB{}\Tab{8}{\V{edgeSwitchYPos}}}
\L{\LB{}\Tab{16}{\V{DisplayString},}}
\L{\LB{}\Tab{8}{\V{edgeSwitchSpeed}}}
\L{\LB{}\Tab{16}{\V{DisplayString},}}
\L{\LB{}\Tab{8}{\V{edgeSwitchDir}}}
\L{\LB{}\Tab{16}{\V{DisplayString},}}
\L{\LB{}\Tab{8}{\V{edgeSwitchAvaliability}}}
\L{\LB{}\Tab{16}{\K{INTEGER},}}
\L{\LB{}\Tab{8}{\V{edgeSwitchNSat}}}
\L{\LB{}\Tab{16}{\K{INTEGER},}}
\L{\LB{}\Tab{8}{\V{edgeSwitchHDop}}}
\L{\LB{}\Tab{16}{\K{INTEGER},}}
\L{\LB{}\Tab{8}{\V{edgeSwitchElevation}}}
\L{\LB{}\Tab{16}{\V{DisplayString},}}
\L{\LB{}\Tab{8}{\V{edgeSwitchHeight}}}
\L{\LB{}\Tab{16}{\V{DisplayString},}}
\L{\LB{}\Tab{8}{\V{edgeSwitchIndex}}}
\L{\LB{}\Tab{16}{\V{DisplayString},}}
\L{\LB{}\Tab{8}{\V{edgeSwitchTime}}}
\L{\LB{}\Tab{16}{\V{TimeTicks}}}
\L{\LB{\}}}
\L{\LB{}}
\L{\LB{\V{edgeSwitchCallsign}_\K{OBJECT}\-\V{TYPE}}}
\L{\LB{________________\K{SYNTAX}_\V{DisplayString}}}
\L{\LB{}\Tab{16}{\K{MAX}\-\V{ACCESS}_\V{read}\-\V{only}}}
\L{\LB{}\Tab{16}{\K{STATUS}_\V{current}}}
\L{\LB{}\Tab{16}{\K{DESCRIPTION}}}
\L{\LB{}\Tab{24}{\3\V{Edge}_\V{Switch}_\V{Callsign}.\3}}
\L{\LB{}\Tab{16}{::=_\{_\V{edgeSwitchEntry}_\N{1}_\}}}
\L{\LB{}}
\L{\LB{\V{edgeSwitchHostname}_\K{OBJECT}\-\V{TYPE}}}
\L{\LB{________________\K{SYNTAX}_\V{DisplayString}}}
\L{\LB{________________\K{MAX}\-\V{ACCESS}_\V{read}\-\V{only}}}
\L{\LB{________________\K{STATUS}_\V{current}}}
\L{\LB{________________\K{DESCRIPTION}}}
\L{\LB{________________________\3\V{Edge}_\V{Switch}_\V{IP}_\V{Hostname}.\3}}
\L{\LB{________________::=_\{_\V{edgeSwitchEntry}_\N{2}_\}}}
\L{\LB{}}
\L{\LB{\V{edgeSwitchATMAddr}_\K{OBJECT}\-\V{TYPE}}}
\L{\LB{________________\K{SYNTAX}_\V{DisplayString}}}
\L{\LB{________________\K{MAX}\-\V{ACCESS}_\V{read}\-\V{only}}}
\L{\LB{________________\K{STATUS}_\V{current}}}
\L{\LB{________________\K{DESCRIPTION}}}
\L{\LB{________________________\3\V{Edge}_\V{Switch}_\V{ATM}_\V{Address}.\3}}
\L{\LB{________________::=_\{_\V{edgeSwitchEntry}_\N{3}_\}}}
\L{\LB{}}
\L{\LB{\V{edgeSwitchLatDeg}_\K{OBJECT}\-\V{TYPE}}}
\L{\LB{________________\K{SYNTAX}_\V{DisplayString}}}
\L{\LB{________________\K{MAX}\-\V{ACCESS}_\V{read}\-\V{only}}}
\L{\LB{________________\K{STATUS}_\V{current}}}
\L{\LB{________________\K{DESCRIPTION}}}
\L{\LB{________________________\3\V{Edge}_\V{Switch}_\V{Latitude}_\-_\V{degrees}.\3}}
\L{\LB{________________::=_\{_\V{edgeSwitchEntry}_\N{4}_\}}}
\L{\LB{}}
\L{\LB{\V{edgeSwitchLatMin}_\K{OBJECT}\-\V{TYPE}}}
\L{\LB{________________\K{SYNTAX}_\V{DisplayString}}}
\L{\LB{________________\K{MAX}\-\V{ACCESS}_\V{read}\-\V{only}}}
\L{\LB{________________\K{STATUS}_\V{current}}}
\L{\LB{________________\K{DESCRIPTION}}}
\L{\LB{________________________\3\V{Edge}_\V{Switch}_\V{Latitude}_\-_\V{minutes}.\3}}
\L{\LB{________________::=_\{_\V{edgeSwitchEntry}_\N{5}_\}}}
\L{\LB{}}
\L{\LB{\V{edgeSwitchLatDir}_\K{OBJECT}\-\V{TYPE}}}
\L{\LB{________________\K{SYNTAX}_\V{DisplayString}}}
\L{\LB{________________\K{MAX}\-\V{ACCESS}_\V{read}\-\V{only}}}
\L{\LB{________________\K{STATUS}_\V{current}}}
\L{\LB{________________\K{DESCRIPTION}}}
\L{\LB{________________________\3\V{Edge}_\V{Switch}_\V{Latitude}_\V{Hemisphere}.\3}}
\L{\LB{________________::=_\{_\V{edgeSwitchEntry}_\N{6}_\}}}
\L{\LB{}}
\L{\LB{\V{edgeSwitchLonDeg}_\K{OBJECT}\-\V{TYPE}}}
\L{\LB{________________\K{SYNTAX}_\V{DisplayString}}}
\L{\LB{________________\K{MAX}\-\V{ACCESS}_\V{read}\-\V{only}}}
\L{\LB{________________\K{STATUS}_\V{current}}}
\L{\LB{________________\K{DESCRIPTION}}}
\L{\LB{________________________\3\V{Edge}_\V{Switch}_\V{Longitude}_\-_\V{degrees}.\3}}
\L{\LB{________________::=_\{_\V{edgeSwitchEntry}_\N{7}_\}}}
\L{\LB{}}
\L{\LB{\V{edgeSwitchLonMin}_\K{OBJECT}\-\V{TYPE}}}
\L{\LB{________________\K{SYNTAX}_\V{DisplayString}}}
\L{\LB{________________\K{MAX}\-\V{ACCESS}_\V{read}\-\V{only}}}
\L{\LB{________________\K{STATUS}_\V{current}}}
\L{\LB{________________\K{DESCRIPTION}}}
\L{\LB{________________________\3\V{Edge}_\V{Switch}_\V{Longitude}_\-_\V{minutes}.\3}}
\L{\LB{________________::=_\{_\V{edgeSwitchEntry}_\N{8}_\}}}
\L{\LB{}}
\L{\LB{\V{edgeSwitchLonDir}_\K{OBJECT}\-\V{TYPE}}}
\L{\LB{________________\K{SYNTAX}_\V{DisplayString}}}
\L{\LB{________________\K{MAX}\-\V{ACCESS}_\V{read}\-\V{only}}}
\L{\LB{________________\K{STATUS}_\V{current}}}
\L{\LB{________________\K{DESCRIPTION}}}
\L{\LB{________________________\3\V{Edge}_\V{Switch}_\V{Longitude}_\-_\V{Hemisphere}.\3}}
\L{\LB{________________::=_\{_\V{edgeSwitchEntry}_\N{9}_\}}}
\L{\LB{}}
\L{\LB{\V{edgeSwitchXPos}_\K{OBJECT}\-\V{TYPE}}}
\L{\LB{________________\K{SYNTAX}_\V{DisplayString}}}
\L{\LB{________________\K{MAX}\-\V{ACCESS}_\V{read}\-\V{only}}}
\L{\LB{________________\K{STATUS}_\V{current}}}
\L{\LB{________________\K{DESCRIPTION}}}
\L{\LB{________________________\3\V{Edge}_\V{Switch}_\V{Cartesian}_\V{X}_\V{Coordinate}._\V{Units}_\V{are}}}
\L{\LB{_________________________\V{meters}.\3}}
\L{\LB{________________::=_\{_\V{edgeSwitchEntry}_\N{10}_\}}}
\L{\LB{}}
\L{\LB{\V{edgeSwitchYPos}_\K{OBJECT}\-\V{TYPE}}}
\L{\LB{________________\K{SYNTAX}_\V{DisplayString}}}
\L{\LB{________________\K{MAX}\-\V{ACCESS}_\V{read}\-\V{only}}}
\L{\LB{________________\K{STATUS}_\V{current}}}
\L{\LB{}\Tab{16}{\K{DESCRIPTION}}}
\L{\LB{________________________\3\V{Edge}_\V{Switch}_\V{Cartesian}_\V{Y}_\V{Coordinate}._\V{Units}_\V{are}}}
\L{\LB{_________________________\V{meters}.\3}}
\L{\LB{________________::=_\{_\V{edgeSwitchEntry}_\N{11}_\}}}
\L{\LB{}}
\L{\LB{\V{edgeSwitchSpeed}_\K{OBJECT}\-\V{TYPE}}}
\L{\LB{________________\K{SYNTAX}_\V{DisplayString}}}
\L{\LB{________________\K{MAX}\-\V{ACCESS}_\V{read}\-\V{only}}}
\L{\LB{________________\K{STATUS}_\V{current}}}
\L{\LB{}\Tab{16}{\K{DESCRIPTION}}}
\L{\LB{________________________\3\V{Edge}_\V{Switch}_\V{current}_\V{speed}.\3}}
\L{\LB{________________::=_\{_\V{edgeSwitchEntry}_\N{12}_\}}}
\L{\LB{}}
\L{\LB{\V{edgeSwitchDir}_\K{OBJECT}\-\V{TYPE}}}
\L{\LB{________________\K{SYNTAX}_\V{DisplayString}}}
\L{\LB{________________\K{MAX}\-\V{ACCESS}_\V{read}\-\V{only}}}
\L{\LB{________________\K{STATUS}_\V{current}}}
\L{\LB{}\Tab{16}{\K{DESCRIPTION}}}
\L{\LB{________________________\3\V{Edge}_\V{Switch}_\V{current}_\V{direction}_\V{in}_\V{degrees}}}
\L{\LB{_________________________\N{0}_\V{degrees}_\V{is}_\V{North}.\3}}
\L{\LB{________________::=_\{_\V{edgeSwitchEntry}_\N{13}_\}}}
\L{\LB{}}
\L{\LB{\V{edgeSwitchAvaliability}_\K{OBJECT}\-\V{TYPE}}}
\L{\LB{________________\K{SYNTAX}_\K{INTEGER}_(\N{0.\,.2})}}
\L{\LB{________________\K{MAX}\-\V{ACCESS}_\V{read}\-\V{only}}}
\L{\LB{________________\K{STATUS}_\V{current}}}
\L{\LB{________________\K{DESCRIPTION}}}
\L{\LB{________________________\3\V{Edge}_\V{Switch}_\V{GPS}_\V{quality}_\V{information}.}}
\L{\LB{}\Tab{24}{_\N{0}_\-_\V{fix}_\V{not}_\V{available}}}
\L{\LB{}\Tab{24}{_\N{1}_\-_\V{Non}_\V{differential}_\V{GPS}_\V{fix}_\V{available}}}
\L{\LB{}\Tab{24}{_\N{2}_\-_\V{Differential}_\V{GPS}_\V{fix}_\V{available}.\3}}
\L{\LB{________________::=_\{_\V{edgeSwitchEntry}_\N{14}_\}}}
\L{\LB{}}
\L{\LB{\V{edgeSwitchNSat}_\K{OBJECT}\-\V{TYPE}}}
\L{\LB{________________\K{SYNTAX}_\K{INTEGER}_(\N{0.\,.8})}}
\L{\LB{________________\K{MAX}\-\V{ACCESS}_\V{read}\-\V{only}}}
\L{\LB{________________\K{STATUS}_\V{current}}}
\L{\LB{________________\K{DESCRIPTION}}}
\L{\LB{________________________\3\V{Edge}_\V{Switch}_\V{GPS}_\V{number}_\V{of}_\V{satellites}_\V{in}_\V{use}.}}
\L{\LB{_________________________\V{From}_\N{0}_\V{to}_\N{8}_\V{satellites}.\3}}
\L{\LB{________________::=_\{_\V{edgeSwitchEntry}_\N{15}_\}}}
\L{\LB{}}
\L{\LB{\V{edgeSwitchHDop}_\K{OBJECT}\-\V{TYPE}}}
\L{\LB{________________\K{SYNTAX}_\K{INTEGER}_(\N{0.\,.999})}}
\L{\LB{________________\K{MAX}\-\V{ACCESS}_\V{read}\-\V{only}}}
\L{\LB{________________\K{STATUS}_\V{current}}}
\L{\LB{________________\K{DESCRIPTION}}}
\L{\LB{________________________\3\V{Edge}_\V{Switch}_\V{GPS}_\V{horizontal}_\V{dilution}_\V{of}_\V{position}.}}
\L{\LB{_________________________\V{Varies}_\V{from}_\N{1.0}_\V{to}_\N{99.9.}\3}}
\L{\LB{________________::=_\{_\V{edgeSwitchEntry}_\N{16}_\}}}
\L{\LB{}}
\L{\LB{\V{edgeSwitchElevation}_\K{OBJECT}\-\V{TYPE}}}
\L{\LB{________________\K{SYNTAX}_\V{DisplayString}}}
\L{\LB{________________\K{MAX}\-\V{ACCESS}_\V{read}\-\V{only}}}
\L{\LB{________________\K{STATUS}_\V{current}}}
\L{\LB{________________\K{DESCRIPTION}}}
\L{\LB{________________________\3\V{Edge}_\V{Switch}_\V{current}_\V{height}_\V{with}_\V{respect}_\V{to}_\V{sea}}}
\L{\LB{_________________________\V{level}_\V{in}_\V{meters}.\3}}
\L{\LB{________________::=_\{_\V{edgeSwitchEntry}_\N{17}_\}}}
\L{\LB{}}
\L{\LB{\V{edgeSwitchHeight}_\K{OBJECT}\-\V{TYPE}}}
\L{\LB{________________\K{SYNTAX}_\V{DisplayString}}}
\L{\LB{________________\K{MAX}\-\V{ACCESS}_\V{read}\-\V{only}}}
\L{\LB{________________\K{STATUS}_\V{current}}}
\L{\LB{________________\K{DESCRIPTION}}}
\L{\LB{________________________\3\V{Edge}_\V{Switch}_\V{current}_\V{geoidal}_\V{height}_\V{in}_\V{meters}.\3}}
\L{\LB{________________::=_\{_\V{edgeSwitchEntry}_\N{18}_\}}}
\L{\LB{}}
\L{\LB{\V{edgeSwitchIndex}_\K{OBJECT}\-\V{TYPE}}}
\L{\LB{________________\K{SYNTAX}_\V{DisplayString}}}
\L{\LB{________________\K{MAX}\-\V{ACCESS}_\V{read}\-\V{only}}}
\L{\LB{________________\K{STATUS}_\V{current}}}
\L{\LB{________________\K{DESCRIPTION}}}
\L{\LB{________________________\3\V{Internal}_\V{index}_\V{for}_\V{use}_\V{in}_\V{ES}\1\V{RN}_\V{associations}.\3}}
\L{\LB{________________::=_\{_\V{edgeSwitchEntry}_\N{19}_\}}}
\L{\LB{}}
\L{\LB{\V{edgeSwitchTime}_\K{OBJECT}\-\V{TYPE}}}
\L{\LB{________________\K{SYNTAX}_\V{TimeTicks}}}
\L{\LB{________________\K{MAX}\-\V{ACCESS}_\V{read}\-\V{only}}}
\L{\LB{________________\K{STATUS}_\V{current}}}
\L{\LB{________________\K{DESCRIPTION}}}
\L{\LB{________________________\3\V{GPS}_\V{time}_\V{at}_\V{which}_\V{this}_\V{information}_\V{was}_\V{valid}.\3}}
\L{\LB{________________::=_\{_\V{edgeSwitchEntry}_\N{20}_\}}}
\L{\LB{}}
\L{\LB{\-\-}}
\L{\LB{\-\-_\V{Associated}_\V{Remote}_\V{Nodes}_\V{Group}}}
\L{\LB{\-\-}}
\L{\LB{}}
\L{\LB{\V{remoteNodeNetwork}_\K{OBJECT}_\K{IDENTIFIER}_::=_\{_\V{rdrnNcpMIB}_\N{5}_\}}}
\L{\LB{}}
\L{\LB{\V{remoteNodeTable}_\K{OBJECT}\-\V{TYPE}}}
\L{\LB{________\K{SYNTAX}_\K{SEQUENCE}_\V{OF}_\V{RemoteNodeEntry}}}
\L{\LB{________\K{MAX}\-\V{ACCESS}_\V{not}\-\V{accessible}}}
\L{\LB{________\K{STATUS}_\V{current}}}
\L{\LB{________\K{DESCRIPTION}_________________________________}}
\L{\LB{________________\3\V{Table}_\V{of}_\V{RDRN}_\V{remote}_\V{nodes}_\V{known}_\V{by}_\V{this}_\V{edge}_\V{switch}.\3}}
\L{\LB{________::=_\{_\V{remoteNodeNetwork}_\N{1}_\}}}
\L{\LB{}}
\L{\LB{\V{remoteNodeEntry}_\K{OBJECT}\-\V{TYPE}}}
\L{\LB{________\K{SYNTAX}_\V{RemoteNodeEntry}}}
\L{\LB{________\K{MAX}\-\V{ACCESS}_\V{not}\-\V{accessible}}}
\L{\LB{________\K{STATUS}_\V{current}}}
\L{\LB{________\K{DESCRIPTION}}}
\L{\LB{________________\3\V{Table}_\V{of}_\V{RDRN}_\V{remote}_\V{nodes}_\V{known}_\V{by}_\V{this}_\V{edge}_\V{switch}.\3}}
\L{\LB{}\Tab{8}{\K{INDEX}_\{_\V{remoteNodeCallsign}_\}}}
\L{\LB{________::=_\{_\V{remoteNodeTable}_\N{1}_\}}}
\L{\LB{}}
\L{\LB{\V{RemoteNodeEntry}_::=_\K{SEQUENCE}_\{}}
\L{\LB{________\V{remoteNodeCallsign}}}
\L{\LB{________________\V{DisplayString},}}
\L{\LB{________\V{remoteNodeHostname}}}
\L{\LB{________________\V{DisplayString},}}
\L{\LB{________\V{remoteNodeATMAddr}}}
\L{\LB{________________\V{DisplayString},}}
\L{\LB{________\V{remoteNodeLatDeg}}}
\L{\LB{________________\V{DisplayString},}}
\L{\LB{________\V{remoteNodeLatMin}}}
\L{\LB{________________\V{DisplayString},}}
\L{\LB{________\V{remoteNodeLatDir}}}
\L{\LB{________________\V{DisplayString},}}
\L{\LB{________\V{remoteNodeLonDeg}}}
\L{\LB{________________\V{DisplayString},}}
\L{\LB{________\V{remoteNodeLonMin}}}
\L{\LB{________________\V{DisplayString},}}
\L{\LB{________\V{remoteNodeLonDir}}}
\L{\LB{________________\V{DisplayString},}}
\L{\LB{________\V{remoteNodeXPos}}}
\L{\LB{________________\V{DisplayString},}}
\L{\LB{________\V{remoteNodeYPos}}}
\L{\LB{________________\V{DisplayString},}}
\L{\LB{________\V{remoteNodeSpeed}}}
\L{\LB{________________\V{DisplayString},}}
\L{\LB{________\V{remoteNodeDir}}}
\L{\LB{________________\V{DisplayString},}}
\L{\LB{________\V{remoteNodeAvaliability}}}
\L{\LB{________________\K{INTEGER},}}
\L{\LB{________\V{remoteNodeNSat}}}
\L{\LB{________________\K{INTEGER},}}
\L{\LB{________\V{remoteNodeHDop}}}
\L{\LB{________________\K{INTEGER},}}
\L{\LB{________\V{remoteNodeElevation}}}
\L{\LB{________________\V{DisplayString},}}
\L{\LB{________\V{remoteNodeHeight}}}
\L{\LB{________________\V{DisplayString},}}
\L{\LB{________\V{remoteNodeIndex}}}
\L{\LB{________________\V{DisplayString},}}
\L{\LB{}\Tab{8}{\V{remoteNodeTime}}}
\L{\LB{}\Tab{16}{\V{TimeTicks},}}
\L{\LB{________\V{remoteNodeDistance}}}
\L{\LB{}\Tab{16}{\V{DisplayString}}}
\L{\LB{\}}}
\L{\LB{}}
\L{\LB{\V{remoteNodeCallsign}_\K{OBJECT}\-\V{TYPE}}}
\L{\LB{________________\K{SYNTAX}_\V{DisplayString}}}
\L{\LB{________________\K{MAX}\-\V{ACCESS}_\V{read}\-\V{only}}}
\L{\LB{________________\K{STATUS}_\V{current}}}
\L{\LB{________________\K{DESCRIPTION}}}
\L{\LB{________________________\3\V{Remote}_\V{Node}_\V{Callsign}.\3}}
\L{\LB{________________::=_\{_\V{remoteNodeEntry}_\N{1}_\}}}
\L{\LB{}}
\L{\LB{\V{remoteNodeHostname}_\K{OBJECT}\-\V{TYPE}}}
\L{\LB{________________\K{SYNTAX}_\V{DisplayString}}}
\L{\LB{________________\K{MAX}\-\V{ACCESS}_\V{read}\-\V{only}}}
\L{\LB{________________\K{STATUS}_\V{current}}}
\L{\LB{________________\K{DESCRIPTION}}}
\L{\LB{________________________\3\V{Remote}_\V{Node}_\V{IP}_\V{hostname}.\3}}
\L{\LB{________________::=_\{_\V{remoteNodeEntry}_\N{2}_\}}}
\L{\LB{}}
\L{\LB{\V{remoteNodeATMAddr}_\K{OBJECT}\-\V{TYPE}}}
\L{\LB{________________\K{SYNTAX}_\V{DisplayString}}}
\L{\LB{________________\K{MAX}\-\V{ACCESS}_\V{read}\-\V{only}}}
\L{\LB{________________\K{STATUS}_\V{current}}}
\L{\LB{________________\K{DESCRIPTION}}}
\L{\LB{________________________\3\V{Remote}_\V{Node}_\V{ATM}_\V{Address}.\3}}
\L{\LB{________________::=_\{_\V{remoteNodeEntry}_\N{3}_\}}}
\L{\LB{}}
\L{\LB{\V{remoteNodeLatDeg}_\K{OBJECT}\-\V{TYPE}}}
\L{\LB{________________\K{SYNTAX}_\V{DisplayString}}}
\L{\LB{________________\K{MAX}\-\V{ACCESS}_\V{read}\-\V{only}}}
\L{\LB{________________\K{STATUS}_\V{current}}}
\L{\LB{________________\K{DESCRIPTION}}}
\L{\LB{________________________\3\V{Remote}_\V{Node}_\V{Latitude}_\-_\V{degrees}.\3}}
\L{\LB{________________::=_\{_\V{remoteNodeEntry}_\N{4}_\}}}
\L{\LB{}}
\L{\LB{\V{remoteNodeLatMin}_\K{OBJECT}\-\V{TYPE}}}
\L{\LB{________________\K{SYNTAX}_\V{DisplayString}}}
\L{\LB{________________\K{MAX}\-\V{ACCESS}_\V{read}\-\V{only}}}
\L{\LB{________________\K{STATUS}_\V{current}}}
\L{\LB{________________\K{DESCRIPTION}_____}}
\L{\LB{________________________\3\V{Remote}_\V{Node}_\V{Latitude}_\-_\V{minutes}.\3}}
\L{\LB{________________::=_\{_\V{remoteNodeEntry}_\N{5}_\}}}
\L{\LB{}}
\L{\LB{\V{remoteNodeLatDir}_\K{OBJECT}\-\V{TYPE}}}
\L{\LB{________________\K{SYNTAX}_\V{DisplayString}}}
\L{\LB{________________\K{MAX}\-\V{ACCESS}_\V{read}\-\V{only}}}
\L{\LB{________________\K{STATUS}_\V{current}}}
\L{\LB{________________\K{DESCRIPTION}_____}}
\L{\LB{________________________\3\V{Remote}_\V{Node}_\V{Latitude}_\V{Hemisphere}.\3}}
\L{\LB{________________::=_\{_\V{remoteNodeEntry}_\N{6}_\}}}
\L{\LB{}}
\L{\LB{\V{remoteNodeLonDeg}_\K{OBJECT}\-\V{TYPE}}}
\L{\LB{________________\K{SYNTAX}_\V{DisplayString}}}
\L{\LB{________________\K{MAX}\-\V{ACCESS}_\V{read}\-\V{only}}}
\L{\LB{________________\K{STATUS}_\V{current}}}
\L{\LB{________________\K{DESCRIPTION}_____}}
\L{\LB{________________________\3\V{Remote}_\V{Node}_\V{Longitude}_\-_\V{degrees}.\3}}
\L{\LB{________________::=_\{_\V{remoteNodeEntry}_\N{7}_\}}}
\L{\LB{}}
\L{\LB{\V{remoteNodeLonMin}_\K{OBJECT}\-\V{TYPE}}}
\L{\LB{________________\K{SYNTAX}_\V{DisplayString}}}
\L{\LB{________________\K{MAX}\-\V{ACCESS}_\V{read}\-\V{only}}}
\L{\LB{________________\K{STATUS}_\V{current}}}
\L{\LB{________________\K{DESCRIPTION}}}
\L{\LB{________________________\3\V{Remote}_\V{Node}_\V{Longitude}_\-_\V{minutes}.\3}}
\L{\LB{________________::=_\{_\V{remoteNodeEntry}_\N{8}_\}}}
\L{\LB{}}
\L{\LB{\V{remoteNodeLonDir}_\K{OBJECT}\-\V{TYPE}}}
\L{\LB{________________\K{SYNTAX}_\V{DisplayString}}}
\L{\LB{________________\K{MAX}\-\V{ACCESS}_\V{read}\-\V{only}}}
\L{\LB{________________\K{STATUS}_\V{current}}}
\L{\LB{________________\K{DESCRIPTION}}}
\L{\LB{________________________\3\V{Remote}_\V{Node}_\V{Longitude}_\-_\V{Hemisphere}.\3}}
\L{\LB{________________::=_\{_\V{remoteNodeEntry}_\N{9}_\}}}
\L{\LB{}}
\L{\LB{\V{remoteNodeXPos}_\K{OBJECT}\-\V{TYPE}}}
\L{\LB{________________\K{SYNTAX}_\V{DisplayString}}}
\L{\LB{________________\K{MAX}\-\V{ACCESS}_\V{read}\-\V{only}}}
\L{\LB{________________\K{STATUS}_\V{current}}}
\L{\LB{________________\K{DESCRIPTION}}}
\L{\LB{________________________\3\V{Remote}_\V{Node}_\V{Cartesian}_\V{X}_\V{Coordinate}._\V{Units}_\V{are}}}
\L{\LB{_________________________\V{meters}.\3}}
\L{\LB{________________::=_\{_\V{remoteNodeEntry}_\N{10}_\}}}
\L{\LB{}}
\L{\LB{\V{remoteNodeYPos}_\K{OBJECT}\-\V{TYPE}}}
\L{\LB{________________\K{SYNTAX}_\V{DisplayString}}}
\L{\LB{________________\K{MAX}\-\V{ACCESS}_\V{read}\-\V{only}}}
\L{\LB{________________\K{STATUS}_\V{current}}}
\L{\LB{________________\K{DESCRIPTION}}}
\L{\LB{________________________\3\V{Remote}_\V{Node}_\V{Cartesian}_\V{Y}_\V{Coordinate}._\V{Units}_\V{are}}}
\L{\LB{_________________________\V{meters}.\3}}
\L{\LB{________________::=_\{_\V{remoteNodeEntry}_\N{11}_\}}}
\L{\LB{}}
\L{\LB{\V{remoteNodeSpeed}_\K{OBJECT}\-\V{TYPE}}}
\L{\LB{________________\K{SYNTAX}_\V{DisplayString}}}
\L{\LB{________________\K{MAX}\-\V{ACCESS}_\V{read}\-\V{only}}}
\L{\LB{________________\K{STATUS}_\V{current}}}
\L{\LB{________________\K{DESCRIPTION}__________________________________}}
\L{\LB{________________________\3\V{Remote}_\V{Node}_\V{current}_\V{speed}.\3}}
\L{\LB{________________::=_\{_\V{remoteNodeEntry}_\N{12}_\}}}
\L{\LB{}}
\L{\LB{\V{remoteNodeDir}_\K{OBJECT}\-\V{TYPE}}}
\L{\LB{________________\K{SYNTAX}_\V{DisplayString}}}
\L{\LB{________________\K{MAX}\-\V{ACCESS}_\V{read}\-\V{only}}}
\L{\LB{________________\K{STATUS}_\V{current}}}
\L{\LB{________________\K{DESCRIPTION}___________________________________________}}
\L{\LB{________________________\3\V{Remote}_\V{Node}_\V{current}_\V{direction}_\V{in}_\V{degrees}}}
\L{\LB{_________________________\N{0}_\V{degrees}_\V{is}_\V{North}.\3}}
\L{\LB{________________::=_\{_\V{remoteNodeEntry}_\N{13}_\}}}
\L{\LB{}}
\L{\LB{\V{remoteNodeAvaliability}_\K{OBJECT}\-\V{TYPE}}}
\L{\LB{________________\K{SYNTAX}_\K{INTEGER}_(\N{0.\,.2})}}
\L{\LB{________________\K{MAX}\-\V{ACCESS}_\V{read}\-\V{only}}}
\L{\LB{________________\K{STATUS}_\V{current}}}
\L{\LB{________________\K{DESCRIPTION}}}
\L{\LB{________________________\3\V{Remote}_\V{Node}_\V{GPS}_\V{quality}_\V{information}.}}
\L{\LB{_________________________\N{0}_\-_\V{fix}_\V{not}_\V{available}}}
\L{\LB{_________________________\N{1}_\-_\V{Non}_\V{differential}_\V{GPS}_\V{fix}_\V{available}}}
\L{\LB{_________________________\N{2}_\-_\V{Differential}_\V{GPS}_\V{fix}_\V{available}.\3}}
\L{\LB{________________::=_\{_\V{remoteNodeEntry}_\N{14}_\}}}
\L{\LB{}}
\L{\LB{\V{remoteNodeNSat}_\K{OBJECT}\-\V{TYPE}}}
\L{\LB{________________\K{SYNTAX}_\K{INTEGER}_(\N{0.\,.8})}}
\L{\LB{________________\K{MAX}\-\V{ACCESS}_\V{read}\-\V{only}}}
\L{\LB{________________\K{STATUS}_\V{current}}}
\L{\LB{________________\K{DESCRIPTION}}}
\L{\LB{________________________\3\V{Remote}_\V{Node}_\V{GPS}_\V{number}_\V{of}_\V{satellites}_\V{in}_\V{use}.}}
\L{\LB{_________________________\V{From}_\N{0}_\V{to}_\N{8}_\V{satellites}.\3}}
\L{\LB{________________::=_\{_\V{remoteNodeEntry}_\N{15}_\}}}
\L{\LB{}}
\L{\LB{\V{remoteNodeHDop}_\K{OBJECT}\-\V{TYPE}}}
\L{\LB{________________\K{SYNTAX}_\K{INTEGER}_(\N{0.\,.999})}}
\L{\LB{________________\K{MAX}\-\V{ACCESS}_\V{read}\-\V{only}}}
\L{\LB{________________\K{STATUS}_\V{current}}}
\L{\LB{________________\K{DESCRIPTION}}}
\L{\LB{________________________\3\V{Remote}_\V{Node}_\V{GPS}_\V{horizontal}_\V{dilution}_\V{of}_\V{position}.}}
\L{\LB{_________________________\V{Varies}_\V{from}_\N{1.0}_\V{to}_\N{99.9.}\3}}
\L{\LB{________________::=_\{_\V{remoteNodeEntry}_\N{16}_\}}}
\L{\LB{}}
\L{\LB{\V{remoteNodeElevation}_\K{OBJECT}\-\V{TYPE}}}
\L{\LB{________________\K{SYNTAX}_\V{DisplayString}}}
\L{\LB{________________\K{MAX}\-\V{ACCESS}_\V{read}\-\V{only}}}
\L{\LB{________________\K{STATUS}_\V{current}}}
\L{\LB{________________\K{DESCRIPTION}}}
\L{\LB{________________________\3\V{Remote}_\V{Node}_\V{current}_\V{height}_\V{with}_\V{respect}_\V{to}_\V{sea}}}
\L{\LB{_________________________\V{level}_\V{in}_\V{meters}.\3}}
\L{\LB{________________::=_\{_\V{remoteNodeEntry}_\N{17}_\}}}
\L{\LB{}}
\L{\LB{\V{remoteNodeHeight}_\K{OBJECT}\-\V{TYPE}}}
\L{\LB{________________\K{SYNTAX}_\V{DisplayString}}}
\L{\LB{________________\K{MAX}\-\V{ACCESS}_\V{read}\-\V{only}}}
\L{\LB{________________\K{STATUS}_\V{current}}}
\L{\LB{________________\K{DESCRIPTION}}}
\L{\LB{________________________\3\V{Remote}_\V{Node}_\V{current}_\V{geoidal}_\V{height}_\V{in}_\V{meters}.\3}}
\L{\LB{________________::=_\{_\V{remoteNodeEntry}_\N{18}_\}}}
\L{\LB{}}
\L{\LB{\V{remoteNodeIndex}_\K{OBJECT}\-\V{TYPE}}}
\L{\LB{________________\K{SYNTAX}_\V{DisplayString}}}
\L{\LB{________________\K{MAX}\-\V{ACCESS}_\V{read}\-\V{only}}}
\L{\LB{________________\K{STATUS}_\V{current}}}
\L{\LB{________________\K{DESCRIPTION}______________________________________________}}
\L{\LB{________________________\3\V{Internal}_\V{index}_\V{for}_\V{use}_\V{in}_\V{ES}\1\V{RN}_\V{associations}.\3}}
\L{\LB{________________::=_\{_\V{remoteNodeEntry}_\N{19}_\}}}
\L{\LB{}}
\L{\LB{\V{remoteNodeTime}_\K{OBJECT}\-\V{TYPE}}}
\L{\LB{________________\K{SYNTAX}_\V{TimeTicks}}}
\L{\LB{________________\K{MAX}\-\V{ACCESS}_\V{read}\-\V{only}}}
\L{\LB{________________\K{STATUS}_\V{current}}}
\L{\LB{________________\K{DESCRIPTION}}}
\L{\LB{________________________\3\V{GPS}_\V{time}_\V{at}_\V{which}_\V{this}_\V{information}_\V{was}_\V{valid}.\3}}
\L{\LB{________________::=_\{_\V{remoteNodeEntry}_\N{20}_\}}}
\L{\LB{}}
\L{\LB{\V{remoteNodeDistance}_\K{OBJECT}\-\V{TYPE}}}
\L{\LB{________________\K{SYNTAX}_\V{DisplayString}}}
\L{\LB{________________\K{MAX}\-\V{ACCESS}_\V{read}\-\V{only}}}
\L{\LB{________________\K{STATUS}_\V{current}}}
\L{\LB{________________\K{DESCRIPTION}}}
\L{\LB{________________________\3\V{Distance}_\V{of}_\V{this}_\V{remote}_\V{node}_\V{from}_\V{its}_\V{associated}_}}
\L{\LB{_________________________\V{base}_\V{station}._\V{A}_\V{value}_\V{of}_\N{0.0}_\V{means}_\V{that}_\V{no}_\V{base}_}}
\L{\LB{_________________________\V{station}_\V{has}_\V{been}_\V{associated}_\V{yet}.\3}}
\L{\LB{________________::=_\{_\V{remoteNodeEntry}_\N{21}_\}}}
\L{\LB{}}
\L{\LB{\-\-}}
\L{\LB{\-\-_\V{GPS}_\V{Specific}_\V{Stats}_\V{Group}}}
\L{\LB{\-\-}}
\L{\LB{}}
\L{\LB{\V{gps}_\K{OBJECT}_\K{IDENTIFIER}_::=_\{_\V{rdrnNcpMIB}_\N{6}_\}}}
\L{\LB{}}
\L{\LB{\V{gpsPosRead}_\K{OBJECT}\-\V{TYPE}}}
\L{\LB{}\Tab{8}{\K{SYNTAX}_\V{Counter32}}}
\L{\LB{}\Tab{8}{\K{MAX}\-\V{ACCESS}_\V{read}\-\V{only}}}
\L{\LB{}\Tab{8}{\K{STATUS}_\V{current}}}
\L{\LB{}\Tab{8}{\K{DESCRIPTION}}}
\L{\LB{}\Tab{16}{\3\V{The}_\V{total}_\V{number}_\V{of}_\V{GPS}_\V{position}_\V{records}_\V{that}_\V{have}}}
\L{\LB{_________________\V{been}_\V{read}_\V{since}_\V{initialization}.\3}}
\L{\LB{}\Tab{8}{::=_\{_\V{gps}_\N{1}_\}}}
\L{\LB{}}
\L{\LB{\-\-______________________}}
\L{\LB{\-\-_\V{Packet}_\V{Radio}_\V{Specific}_\V{Stats}_\V{Group}}}
\L{\LB{\-\-___}}
\L{\LB{}}
\L{\LB{\V{packetRadio}_\K{OBJECT}_\K{IDENTIFIER}_::=_\{_\V{rdrnNcpMIB}_\N{7}_\}}}
\L{\LB{}}
\L{\LB{\V{packetsIn}_\K{OBJECT}\-\V{TYPE}}}
\L{\LB{________\K{SYNTAX}_\V{Counter32}_}}
\L{\LB{________\K{MAX}\-\V{ACCESS}_\V{read}\-\V{only}}}
\L{\LB{________\K{STATUS}_\V{current}}}
\L{\LB{________\K{DESCRIPTION}}}
\L{\LB{________________\3\V{The}_\V{total}_\V{number}_\V{of}_\V{NCP}_\V{packets}_\V{which}_\V{have}}}
\L{\LB{_________________\V{been}_\V{read}_\V{since}_\V{initialization}.\3}}
\L{\LB{________::=_\{_\V{packetRadio}_\N{1}_\}}}
\L{\LB{}}
\L{\LB{\V{packetsOut}_\K{OBJECT}\-\V{TYPE}}}
\L{\LB{________\K{SYNTAX}_\V{Counter32}}}
\L{\LB{________\K{MAX}\-\V{ACCESS}_\V{read}\-\V{only}}}
\L{\LB{________\K{STATUS}_\V{current}}}
\L{\LB{________\K{DESCRIPTION}}}
\L{\LB{________________\3\V{The}_\V{total}_\V{number}_\V{of}_\V{NCP}_\V{packets}_\V{which}_\V{have}}}
\L{\LB{_________________\V{been}_\V{written}_\V{to}_\V{the}_\V{packet}_\V{radio}_\V{since}_\V{initialization}.\3}}
\L{\LB{________::=_\{_\V{packetRadio}_\N{2}_\}}}
\L{\LB{}}
\L{\LB{\V{packetsBad}_\K{OBJECT}\-\V{TYPE}}}
\L{\LB{________\K{SYNTAX}_\V{Counter32}}}
\L{\LB{________\K{MAX}\-\V{ACCESS}_\V{read}\-\V{only}}}
\L{\LB{________\K{STATUS}_\V{current}}}
\L{\LB{________\K{DESCRIPTION}}}
\L{\LB{________________\3\V{The}_\V{total}_\V{number}_\V{of}_\V{NCP}_\V{packets}_\V{which}_\V{have}}}
\L{\LB{_________________\V{been}_\V{read}_\V{and}_\V{discarded}_\V{as}_\V{unreadable}_\V{or}_\V{incomplete}}}
\L{\LB{_________________\V{since}_\V{initialization}.\3}}
\L{\LB{________::=_\{_\V{packetRadio}_\N{3}_\}}}
\L{\LB{}}
\L{\LB{\-\-______________________}}
\L{\LB{\-\-_\V{NCP}_\V{Level}_\V{Specific}_\V{Stats}_\V{Group}}}
\L{\LB{\-\-___}}
\L{\LB{}}
\L{\LB{\V{ncpLevel}_\K{OBJECT}_\K{IDENTIFIER}_::=_\{_\V{rdrnNcpMIB}_\N{8}_\}}}
\L{\LB{}}
\L{\LB{\V{ncpHandoffs}_\K{OBJECT}\-\V{TYPE}}}
\L{\LB{________\K{SYNTAX}_\V{Counter32}}}
\L{\LB{________\K{MAX}\-\V{ACCESS}_\V{read}\-\V{only}}}
\L{\LB{________\K{STATUS}_\V{current}}}
\L{\LB{________\K{DESCRIPTION}}}
\L{\LB{________________\3\V{The}_\V{total}_\V{number}_\V{handoffs}_\V{since}_\V{since}_\V{initialization}.\3}}
\L{\LB{________::=_\{_\V{ncpLevel}_\N{1}_\}}}
\L{\LB{}}
\L{\LB{\-\-}}
\L{\LB{\-\-_\V{Existing}_\V{ATM}_\V{Stacks}_\V{Group}}}
\L{\LB{\-\-}}
\L{\LB{}}
\L{\LB{\V{atmStacks}_\K{OBJECT}_\K{IDENTIFIER}_::=_\{_\V{rdrnNcpMIB}_\N{9}_\}}}
\L{\LB{}}
\L{\LB{\V{atmStacksTable}_\K{OBJECT}\-\V{TYPE}}}
\L{\LB{________\K{SYNTAX}_\K{SEQUENCE}_\V{OF}_\V{AtmStacksEntry}}}
\L{\LB{________\K{MAX}\-\V{ACCESS}_\V{not}\-\V{accessible}}}
\L{\LB{________\K{STATUS}_\V{current}}}
\L{\LB{________\K{DESCRIPTION}}}
\L{\LB{________________\3\V{Table}_\V{of}_\V{currently}_\V{active}_\V{AHDLC}_\V{stacks}.\3}}
\L{\LB{________::=_\{_\V{atmStacks}_\N{1}_\}}}
\L{\LB{}}
\L{\LB{\V{atmStacksEntry}_\K{OBJECT}\-\V{TYPE}}}
\L{\LB{________\K{SYNTAX}_\V{AtmStacksEntry}}}
\L{\LB{________\K{MAX}\-\V{ACCESS}_\V{not}\-\V{accessible}}}
\L{\LB{________\K{STATUS}_\V{current}}}
\L{\LB{________\K{DESCRIPTION}}}
\L{\LB{________________\3\V{Table}_\V{of}_\V{currently}_\V{active}_\V{AHDLC}_\V{stacks}.\3}}
\L{\LB{}\Tab{8}{\K{INDEX}_\{_\V{atmStacksATMaddr}_\}}}
\L{\LB{________::=_\{_\V{atmStacksTable}_\N{1}_\}}}
\L{\LB{}}
\L{\LB{\V{AtmStacksEntry}_::=_\K{SEQUENCE}_\{}}
\L{\LB{________\V{atmStacksBeam}}}
\L{\LB{________________\K{INTEGER},}}
\L{\LB{________\V{atmStacksSlot}}}
\L{\LB{________________\K{INTEGER},}}
\L{\LB{}\Tab{8}{\V{atmStacksATMaddr}}}
\L{\LB{}\Tab{16}{\V{DisplayString},}}
\L{\LB{}\Tab{8}{\V{atmStacksAhdlcState}}}
\L{\LB{}\Tab{16}{\K{INTEGER}}}
\L{\LB{________\}}}
\L{\LB{}}
\L{\LB{\V{atmStacksBeam}_\K{OBJECT}\-\V{TYPE}}}
\L{\LB{________\K{SYNTAX}_\K{INTEGER}_(\N{0.\,.8})}}
\L{\LB{________\K{MAX}\-\V{ACCESS}_\V{read}\-\V{only}}}
\L{\LB{________\K{STATUS}_\V{current}}}
\L{\LB{________\K{DESCRIPTION}}}
\L{\LB{________________\3\V{The}_\V{beam}_\V{number}_\V{for}_\V{this}_\V{AHDLC}_\V{stack}.\3}}
\L{\LB{________::=_\{_\V{atmStacksEntry}_\N{1}_\}}}
\L{\LB{}}
\L{\LB{\V{atmStacksSlot}_\K{OBJECT}\-\V{TYPE}}}
\L{\LB{________\K{SYNTAX}_\K{INTEGER}_(\N{0.\,.256})}}
\L{\LB{________\K{MAX}\-\V{ACCESS}_\V{read}\-\V{only}}}
\L{\LB{________\K{STATUS}_\V{current}}}
\L{\LB{________\K{DESCRIPTION}}}
\L{\LB{________________\3\V{The}_\V{slot}_\V{number}_\V{for}_\V{this}_\V{AHDLC}_\V{stack}.\3}}
\L{\LB{________::=_\{_\V{atmStacksEntry}_\N{2}_\}}}
\L{\LB{}}
\L{\LB{\V{atmStacksATMaddr}_\K{OBJECT}\-\V{TYPE}}}
\L{\LB{________\K{SYNTAX}_\V{DisplayString}}}
\L{\LB{________\K{MAX}\-\V{ACCESS}_\V{read}\-\V{only}}}
\L{\LB{________\K{STATUS}_\V{current}}}
\L{\LB{________\K{DESCRIPTION}}}
\L{\LB{________________\3\V{The}_\V{ATM}_\V{address}_\V{of}_\V{the}_\V{current}_\V{AHDLC}_\V{stack}.\3}}
\L{\LB{________::=_\{_\V{atmStacksEntry}_\N{3}_\}}}
\L{\LB{}}
\L{\LB{\V{atmStacksAhdlcState}_\K{OBJECT}\-\V{TYPE}}}
\L{\LB{________\K{SYNTAX}_\K{INTEGER}_\{__}}
\L{\LB{________________\V{down}(\N{0}),___}}
\L{\LB{________________\V{up}(\N{1})}}
\L{\LB{________\}}}
\L{\LB{________\K{MAX}\-\V{ACCESS}_\V{read}\-\V{only}_________}}
\L{\LB{________\K{STATUS}_\V{current}}}
\L{\LB{________\K{DESCRIPTION}}}
\L{\LB{________________\3\V{This}_\V{is}_\V{the}_\V{state}_\V{of}_\V{the}_\V{AHDLC}_\V{layer}.\3_________}}
\L{\LB{________::=_\{_\V{atmStacksEntry}_\N{4}_\}____}}
\L{\LB{}}
\L{\LB{\-\-}}
\L{\LB{\-\-_\V{The}_\V{Beam}_\V{Coverage}_\V{Group}}}
\L{\LB{\-\-}}
\L{\LB{}}
\L{\LB{\V{beamCoverage}_\K{OBJECT}_\K{IDENTIFIER}_::=_\{_\V{rdrnNcpMIB}_\N{10}_\}}}
\L{\LB{}}
\L{\LB{\V{beamCovTable}_\K{OBJECT}\-\V{TYPE}}}
\L{\LB{________\K{SYNTAX}_\K{SEQUENCE}_\V{OF}_\V{BeamCovEntry}}}
\L{\LB{________\K{MAX}\-\V{ACCESS}_\V{not}\-\V{accessible}}}
\L{\LB{________\K{STATUS}_\V{current}}}
\L{\LB{________\K{DESCRIPTION}}}
\L{\LB{________________\3\V{Table}_\V{of}_\V{beam}_\V{coverage}_\V{information}.\3}}
\L{\LB{________::=_\{_\V{beamCoverage}_\N{1}_\}}}
\L{\LB{}}
\L{\LB{\V{beamCovEntry}_\K{OBJECT}\-\V{TYPE}}}
\L{\LB{________\K{SYNTAX}_\V{BeamCovEntry}}}
\L{\LB{________\K{MAX}\-\V{ACCESS}_\V{not}\-\V{accessible}}}
\L{\LB{________\K{STATUS}_\V{current}}}
\L{\LB{________\K{DESCRIPTION}}}
\L{\LB{________________\3\V{Table}_\V{of}_\V{beam}_\V{coverage}_\V{information}.\3}}
\L{\LB{}\Tab{8}{\K{INDEX}_\{_\V{beamCovAngle}_\}}}
\L{\LB{________::=_\{_\V{beamCovTable}_\N{1}_\}}}
\L{\LB{}}
\L{\LB{\V{BeamCovEntry}_::=_\K{SEQUENCE}_\{}}
\L{\LB{________\V{beamCovAngle}}}
\L{\LB{________________\V{DisplayString},}}
\L{\LB{________\V{beamCovPower}}}
\L{\LB{________________\V{DisplayString},}}
\L{\LB{________\V{beamCovSIR}}}
\L{\LB{________________\V{DisplayString}}}
\L{\LB{________\}}}
\L{\LB{}}
\L{\LB{\V{beamCovAngle}_\K{OBJECT}\-\V{TYPE}}}
\L{\LB{________\K{SYNTAX}_\V{DisplayString}}}
\L{\LB{________\K{MAX}\-\V{ACCESS}_\V{read}\-\V{only}}}
\L{\LB{________\K{STATUS}_\V{current}}}
\L{\LB{________\K{DESCRIPTION}}}
\L{\LB{________________\3\V{The}_\V{beam}_\V{angle}_\V{of}_\V{this}_\V{beam}.\3}}
\L{\LB{________::=_\{_\V{beamCovEntry}_\N{1}_\}}}
\L{\LB{}}
\L{\LB{\V{beamCovPower}_\K{OBJECT}\-\V{TYPE}}}
\L{\LB{________\K{SYNTAX}_\V{DisplayString}}}
\L{\LB{________\K{MAX}\-\V{ACCESS}_\V{read}\-\V{only}}}
\L{\LB{________\K{STATUS}_\V{current}}}
\L{\LB{________\K{DESCRIPTION}}}
\L{\LB{________________\3\V{The}_\V{power}_\V{level}_\V{of}_\V{this}_\V{beam}.\3}}
\L{\LB{________::=_\{_\V{beamCovEntry}_\N{2}_\}}}
\L{\LB{}}
\L{\LB{\V{beamCovSIR}_\K{OBJECT}\-\V{TYPE}}}
\L{\LB{________\K{SYNTAX}_\V{DisplayString}}}
\L{\LB{________\K{MAX}\-\V{ACCESS}_\V{read}\-\V{only}}}
\L{\LB{________\K{STATUS}_\V{current}}}
\L{\LB{________\K{DESCRIPTION}}}
\L{\LB{________________\3\V{The}_\V{signal}_\V{to}_\V{interference}_\V{ratio}_\V{for}_\V{this}_\V{beam}.\3}}
\L{\LB{________::=_\{_\V{beamCovEntry}_\N{3}_\}}}
\L{\LB{}}
\L{\LB{\-\-}}
\L{\LB{\-\-_\V{The}_\V{NCP}_\V{Status}_\V{Group}}}
\L{\LB{\-\-}}
\L{\LB{}}
\L{\LB{\V{status}_\K{OBJECT}_\K{IDENTIFIER}_::=_\{_\V{rdrnNcpMIB}_\N{11}_\}}}
\L{\LB{}}
\L{\LB{\V{statState}_\K{OBJECT}\-\V{TYPE}___}}
\L{\LB{________\K{SYNTAX}_\K{INTEGER}_\{___}}
\L{\LB{________________\V{down}(\N{0}),}}
\L{\LB{________________\V{configuring}(\N{1}),__}}
\L{\LB{________________\V{associating}(\N{2}),______}}
\L{\LB{________________\V{active}(\N{3}),}}
\L{\LB{________________\V{delay}(\N{4})}}
\L{\LB{________\}}}
\L{\LB{________\K{MAX}\-\V{ACCESS}_\V{read}\-\V{only}_______________________________}}
\L{\LB{________\K{STATUS}_\V{current}______}}
\L{\LB{________\K{DESCRIPTION}_________________________}}
\L{\LB{________________\3\V{Describes}_\V{the}_\V{NCP}_\V{state}_\V{of}_\V{operation}:}}
\L{\LB{________________________\V{down}________\-_\V{node}_\V{administratively}_\V{idle}.}}
\L{\LB{________________________\V{configuring}_\-_\V{es}\-\V{es}_\V{topology}_\V{configuration}_\V{or}}}
\L{\LB{______________________________________\V{initial}_\V{startup}.}}
\L{\LB{________________________\V{associating}_\-_\V{handoff}_\V{in}_\V{process}.}}
\L{\LB{________________________\V{active}______\-_\V{other}_\V{normal}_\V{operation}.}}
\L{\LB{________________________\V{delay}_______\-_\V{node}_\V{is}_\V{in}_\V{a}_\V{delay}_\V{mode},_\V{e}.\V{g}.}}
\L{\LB{______________________________________\V{remote}_\V{node}_\V{waiting}_\V{between}_}}
\L{\LB{_________________\V{updates}.\3}}
\L{\LB{________::=_\{_\V{status}_\N{1}_\}}}
\L{\LB{}}
\L{\LB{\V{statAdminState}_\K{OBJECT}\-\V{TYPE}}}
\L{\LB{________\K{SYNTAX}_\K{INTEGER}_\{}}
\L{\LB{________________\V{idle}(\N{0}),}}
\L{\LB{________________\V{active}(\N{1})}}
\L{\LB{________\}}}
\L{\LB{________\K{MAX}\-\V{ACCESS}_\V{read}\-\V{write}}}
\L{\LB{________\K{STATUS}_\V{current}}}
\L{\LB{________\K{DESCRIPTION}}}
\L{\LB{________________\3\V{Indicates}_\V{whether}_\V{the}_\V{process}_\V{is}_\V{in}_\V{an}_\V{idle}_\V{mode}}}
\L{\LB{_________________\V{or}_\V{active}._\V{Setting}_\V{this}_\V{object}_\V{to}_\V{idle}_\V{causes}_\V{the}}}
\L{\LB{_________________\V{NCP}_\V{process}_\V{to}_\V{remain}_\V{alive},_\V{but}_\V{stop}_\V{processing}}}
\L{\LB{_________________\V{packets}._\V{A}_\V{transition}_\V{setting}_\V{from}_\V{idle}_\V{to}_\V{active}}}
\L{\LB{_________________\V{causes}_\V{the}_\V{NCP}_\V{process}_\V{to}_\V{re}\-\V{initialize}_\V{as}_\V{though}}}
\L{\LB{_________________\V{just}_\V{started}.\3}}
\L{\LB{________::=_\{_\V{status}_\N{2}_\}}}
\L{\LB{}}
\L{\LB{\V{statTopologyTime}_\K{OBJECT}\-\V{TYPE}}}
\L{\LB{________\K{SYNTAX}_\K{INTEGER}_(\N{0.\,.2147483647})}}
\L{\LB{________\K{MAX}\-\V{ACCESS}_\V{read}\-\V{only}}}
\L{\LB{________\K{STATUS}_\V{current}}}
\L{\LB{________\K{DESCRIPTION}}}
\L{\LB{________________\3\V{The}_\V{time}_\V{in}_\V{micro}\-\V{seconds}_\V{to}_\V{do}_\V{the}_\V{last}_\V{optimal}_}}
\L{\LB{_________________\V{topology}_\V{calculation}._\V{A}_\V{zero}_\V{value}_\V{means}_\V{that}}}
\L{\LB{_________________\V{the}_\V{topology}_\V{calculation}_\V{has}_\V{not}_\V{been}_\V{performed}}}
\L{\LB{_________________\V{yet}.\3}}
\L{\LB{________::=_\{_\V{status}_\N{3}_\}}}
\L{\LB{}}
\L{\LB{\V{statBeamformTime}_\K{OBJECT}\-\V{TYPE}}}
\L{\LB{________\K{SYNTAX}_\K{INTEGER}_(\N{0.\,.2147483647})}}
\L{\LB{________\K{MAX}\-\V{ACCESS}_\V{read}\-\V{only}___}}
\L{\LB{________\K{STATUS}_\V{current}}}
\L{\LB{________\K{DESCRIPTION}_________}}
\L{\LB{________________\3\V{The}_\V{time}_\V{in}_\V{micro}\-\V{seconds}_\V{to}_\V{do}_\V{the}_\V{last}_\V{beamform}.}}
\L{\LB{_________________\V{A}_\V{zero}_\V{values}_\V{indicates}_\V{that}_\V{beamforming}_\V{has}_\V{not}}}
\L{\LB{_________________\V{been}_\V{performed}_\V{yet}.\3____________________________}}
\L{\LB{________::=_\{_\V{status}_\N{4}_\}______}}
\L{\LB{}}
\L{\LB{\V{statNcpTransferTime}_\K{OBJECT}\-\V{TYPE}}}
\L{\LB{________\K{SYNTAX}_\K{INTEGER}_(\N{0.\,.2147483647})}}
\L{\LB{________\K{MAX}\-\V{ACCESS}_\V{read}\-\V{only}}}
\L{\LB{________\K{STATUS}_\V{current}}}
\L{\LB{________\K{DESCRIPTION}}}
\L{\LB{________________\3\V{The}_\V{time}_\V{in}_\V{micro}\-\V{seconds}_\V{of}_\V{the}_\V{last}_\V{NCP}_\V{packet}}}
\L{\LB{_________________\V{transmission}._\V{A}_\V{zero}_\V{value}_\V{means}_\V{that}_\V{no}_\V{packet}}}
\L{\LB{_________________\V{measurements}_\V{are}_\V{being}_\V{taken}.\3}}
\L{\LB{________::=_\{_\V{status}_\N{5}_\}}}
\L{\LB{}}
\L{\LB{\-\-___________________________________________________________________}}
\L{\LB{\-\-_\V{The}_\V{VNC}\-\V{NCP}_\V{Group}}}
\L{\LB{\-\-_____________________________}}
\L{\LB{}}
\L{\LB{\V{vncNcp}_\K{OBJECT}_\K{IDENTIFIER}_::=_\{_\V{rdrnNcpMIB}_\N{12}_\}}}
\L{\LB{}}
\L{\LB{\V{vncNcpHandoff}_\K{OBJECT}_\K{IDENTIFIER}_::=_\{_\V{vncNcp}_\N{1}_\}}}
\L{\LB{}}
\L{\LB{\V{vncNcpHandoffTable}_\K{OBJECT}\-\V{TYPE}}}
\L{\LB{________\K{SYNTAX}_\K{SEQUENCE}_\V{OF}_\V{VncNcpHandoffEntry}}}
\L{\LB{________\K{MAX}\-\V{ACCESS}_\V{not}\-\V{accessible}}}
\L{\LB{________\K{STATUS}_\V{current}}}
\L{\LB{________\K{DESCRIPTION}}}
\L{\LB{________________\3\V{VNC}_\V{Handoff}_\V{Prediction}_\V{Table}.\3}}
\L{\LB{________::=_\{_\V{vncNcpHandoff}_\N{1}_\}}}
\L{\LB{}}
\L{\LB{\V{vncNcpHandoffEntry}_\K{OBJECT}\-\V{TYPE}}}
\L{\LB{________\K{SYNTAX}_\V{VncNcpHandoffEntry}}}
\L{\LB{________\K{MAX}\-\V{ACCESS}_\V{not}\-\V{accessible}}}
\L{\LB{________\K{STATUS}_\V{current}}}
\L{\LB{________\K{DESCRIPTION}}}
\L{\LB{________________\3\V{VNC}_\V{Handoff}_\V{Prediction}_\V{Table}.\3}}
\L{\LB{}\Tab{8}{\K{INDEX}_\{_\V{vncNcpHandoffTime}_\}}}
\L{\LB{________::=_\{_\V{vncNcpHandoffTable}_\N{1}_\}}}
\L{\LB{}}
\L{\LB{\V{VncNcpHandoffEntry}_::=_\K{SEQUENCE}_\{}}
\L{\LB{}\Tab{8}{\V{vncNcpHandoffProb}}}
\L{\LB{________________\K{INTEGER},}}
\L{\LB{}\Tab{8}{\V{vncNcpHandoffTime}}}
\L{\LB{________________\V{TimeTicks},}}
\L{\LB{}\Tab{8}{\V{vncNcpHandoffName}}}
\L{\LB{}\Tab{16}{\V{DisplayString}}}
\L{\LB{}\Tab{8}{\}}}
\L{\LB{}}
\L{\LB{\V{vncNcpHandoffProb}_\K{OBJECT}\-\V{TYPE}}}
\L{\LB{________\K{SYNTAX}_\K{INTEGER}_(\N{0.\,.2147483647})}}
\L{\LB{________\K{MAX}\-\V{ACCESS}_\V{read}\-\V{only}}}
\L{\LB{________\K{STATUS}_\V{current}}}
\L{\LB{________\K{DESCRIPTION}}}
\L{\LB{________________\3\V{This}_\V{is}_\V{the}_\V{probability}_\V{that}_\V{this}_\V{new}_\V{ES}\-\V{RN}_\V{association}_}}
\L{\LB{_________________\V{will}_\V{occur}.\3}}
\L{\LB{________::=_\{_\V{vncNcpHandoffEntry}_\N{1}_\}}}
\L{\LB{}}
\L{\LB{\V{vncNcpHandoffTime}_\K{OBJECT}\-\V{TYPE}}}
\L{\LB{________\K{SYNTAX}_\V{TimeTicks}}}
\L{\LB{________\K{MAX}\-\V{ACCESS}_\V{read}\-\V{only}}}
\L{\LB{________\K{STATUS}_\V{current}}}
\L{\LB{________\K{DESCRIPTION}}}
\L{\LB{________________\3\V{This}_\V{is}_\V{the}_\V{predicted}_\V{time}_\V{of}_\V{the}_\V{new}_\V{ES}\-\V{RN}_\V{association}.\3}}
\L{\LB{________::=_\{_\V{vncNcpHandoffEntry}_\N{2}_\}}}
\L{\LB{}}
\L{\LB{\V{vncNcpHandoffName}_\K{OBJECT}\-\V{TYPE}}}
\L{\LB{________\K{SYNTAX}_\V{DisplayString}}}
\L{\LB{________\K{MAX}\-\V{ACCESS}_\V{read}\-\V{only}}}
\L{\LB{________\K{STATUS}_\V{current}}}
\L{\LB{________\K{DESCRIPTION}}}
\L{\LB{________________\3\V{This}_\V{is}_\V{the}_\V{name}_\V{of}_\V{the}_\V{new}_\V{node}_\V{predicted}_\V{to}_\V{be}_\V{associated}}}
\L{\LB{_________________\V{with}_\V{the}_\V{current}_\V{node}._\V{This}_\V{should}_\V{point}_\V{to}_\V{the}_\V{remoteNode}}}
\L{\LB{_________________\V{table}_\V{already}_\V{defined}_\V{but}_\V{with}_\V{predicted}_\V{information}.\3}}
\L{\LB{________::=_\{_\V{vncNcpHandoffEntry}_\N{3}_\}}}
\L{\LB{}}
\L{\LB{\V{vncNcpBeamform}_\K{OBJECT}_\K{IDENTIFIER}_::=_\{_\V{vncNcp}_\N{2}_\}}}
\L{\LB{}}
\L{\LB{\V{vncNcpBeamformTable}_\K{OBJECT}\-\V{TYPE}}}
\L{\LB{________\K{SYNTAX}_\K{SEQUENCE}_\V{OF}_\V{VncNcpBeamformEntry}}}
\L{\LB{________\K{MAX}\-\V{ACCESS}_\V{not}\-\V{accessible}}}
\L{\LB{________\K{STATUS}_\V{current}}}
\L{\LB{________\K{DESCRIPTION}}}
\L{\LB{________________\3\V{VNC}_\V{Beamform}_\V{prediction}_\V{table}.\3}}
\L{\LB{________::=_\{_\V{vncNcpBeamform}_\N{1}_\}}}
\L{\LB{}}
\L{\LB{\V{vncNcpBeamformEntry}_\K{OBJECT}\-\V{TYPE}}}
\L{\LB{________\K{SYNTAX}_\V{VncNcpBeamformEntry}}}
\L{\LB{________\K{MAX}\-\V{ACCESS}_\V{not}\-\V{accessible}}}
\L{\LB{________\K{STATUS}_\V{current}}}
\L{\LB{________\K{DESCRIPTION}}}
\L{\LB{________________\3\V{VNC}_\V{Beamform}_\V{prediction}_\V{table}.\3}}
\L{\LB{}\Tab{8}{\K{INDEX}_\{_\V{vncNcpBeamformTime}_\}}}
\L{\LB{________::=_\{_\V{vncNcpBeamformTable}_\N{1}_\}}}
\L{\LB{}}
\L{\LB{\V{VncNcpBeamformEntry}_::=_\K{SEQUENCE}_\{}}
\L{\LB{}\Tab{8}{\V{vncNcpBeamformProb}}}
\L{\LB{________________\K{INTEGER},}}
\L{\LB{}\Tab{8}{\V{vncNcpBeamformTime}}}
\L{\LB{________________\V{TimeTicks},}}
\L{\LB{}\Tab{8}{\V{vncNcpBeamformAngle}}}
\L{\LB{}\Tab{16}{\V{DisplayString}}}
\L{\LB{}\Tab{8}{\}}}
\L{\LB{}}
\L{\LB{\V{vncNcpBeamformProb}_\K{OBJECT}\-\V{TYPE}}}
\L{\LB{________\K{SYNTAX}_\K{INTEGER}_(\N{0.\,.2147483647})}}
\L{\LB{________\K{MAX}\-\V{ACCESS}_\V{read}\-\V{only}}}
\L{\LB{________\K{STATUS}_\V{current}}}
\L{\LB{________\K{DESCRIPTION}}}
\L{\LB{________________\3\V{This}_\V{is}_\V{the}_\V{probability}_\V{that}_\V{this}_\V{beam}_\V{information}}}
\L{\LB{_________________\V{will}_\V{be}_\V{valid}.\3}}
\L{\LB{________::=_\{_\V{vncNcpBeamformEntry}_\N{1}_\}}}
\L{\LB{}}
\L{\LB{\V{vncNcpBeamformTime}_\K{OBJECT}\-\V{TYPE}}}
\L{\LB{________\K{SYNTAX}_\V{TimeTicks}}}
\L{\LB{________\K{MAX}\-\V{ACCESS}_\V{read}\-\V{only}}}
\L{\LB{________\K{STATUS}_\V{current}}}
\L{\LB{________\K{DESCRIPTION}}}
\L{\LB{________________\3\V{This}_\V{is}_\V{the}_\V{predicted}_\V{time}_\V{of}_\V{this}_\V{new}_\V{beam}_}}
\L{\LB{_________________\V{configuration}_\V{will}_\V{be}_\V{valid}.\3}}
\L{\LB{________::=_\{_\V{vncNcpBeamformEntry}_\N{2}_\}}}
\L{\LB{}}
\L{\LB{\V{vncNcpBeamformAngle}_\K{OBJECT}\-\V{TYPE}}}
\L{\LB{________\K{SYNTAX}_\V{DisplayString}}}
\L{\LB{________\K{MAX}\-\V{ACCESS}_\V{read}\-\V{only}}}
\L{\LB{________\K{STATUS}_\V{current}}}
\L{\LB{________\K{DESCRIPTION}}}
\L{\LB{________________\3\V{This}_\V{is}_\V{the}_\V{predicted}_\V{angle}_\V{of}_\V{the}_\V{beam}._\V{Should}_\V{point}}}
\L{\LB{_________________\V{to}_\V{beamtable}_\V{already}_\V{defined}_\V{but}_\V{with}_\V{predicted}_\V{values}.\3_}}
\L{\LB{________::=_\{_\V{vncNcpBeamformEntry}_\N{3}_\}}}
\L{\LB{\K{END}}}

\end{singlespace}
\normalsize

The following SNMP MIB is part of the RDRN NCP and LMD. This MIB
contains the VNC specific portion of a process and is intended to
be generic enough for use by any process using the VNC algorithm.
It is a table of the most important parameters of the VNC algorithm. 
Currently, the remote node GPS, remote node NCP/LMD, and the edge 
switch have been implemented as LPs with this MIB. When the RN and ES
LMDs are running, live data from this table can be viewed from
http://www.tisl.ukans.edu/\~{}sbush/rdrn/ncp.html.

\footnotesize
\begin{singlespace}



\Head{}
\File{\1users\1sbush\1rdrn\1ncp\1vnc.mib}{1997}{Feb}{ 9}{12:48}{4256}
\L{\LB{\V{RDRN}\-\V{VNC}\-\V{MIB}_\K{DEFINITIONS}_::=_\K{BEGIN}}}
\L{\LB{}}
\L{\LB{\K{IMPORTS}}}
\L{\LB{____\K{MODULE}\-\V{IDENTITY},_\K{OBJECT}\-\V{TYPE},_\V{experimental},}}
\L{\LB{____\V{Counter32},_\V{TimeTicks}}}
\L{\LB{____}\Tab{8}{\K{FROM}_\V{SNMPv2}\-\V{SMI}}}
\L{\LB{____\V{DisplayString}}}
\L{\LB{________\K{FROM}_\V{SNMPv2}\-\V{TC};}}
\L{\LB{}}
\L{\LB{\V{rdrnVncMIB}_\K{MODULE}\-\V{IDENTITY}}}
\L{\LB{_____\K{LAST}\-\V{UPDATED}_\3\N{9701010000Z}\3}}
\L{\LB{_____\V{ORGANIZATION}_\3\V{KU}_\V{TISL}\3}}
\L{\LB{_____\V{CONTACT}\-\V{INFO}}}
\L{\LB{______________\3\V{Steve}_\V{Bush}_\V{sbush}\@\V{tisl}.\V{ukans}.\V{edu}\3}}
\L{\LB{_____\K{DESCRIPTION}}}
\L{\LB{_____________\3\V{Experimental}_\V{MIB}_\V{modules}_\V{for}_\V{the}_\V{Rapidly}_\V{Deployable}}}
\L{\LB{______________\V{Radio}_\V{Network}_(\V{RDRN})_\V{Project}_\V{Network}_\V{Control}_}}
\L{\LB{______________\V{Protocol}_(\V{NCP})_\V{enhanced}_\V{with}_\V{the}_\V{Virtual}_\V{Network}}}
\L{\LB{______________\V{Configuration}.\3}}
\L{\LB{_____::=_\{_\V{experimental}_\V{ncp}(\N{75})_\N{4}_\}}}
\L{\LB{}}
\L{\LB{\-\-}}
\L{\LB{\-\-_\V{Logical}_\V{Process}_\V{Table}}}
\L{\LB{\-\-}}
\L{\LB{}}
\L{\LB{\V{logicalProcess}_\K{OBJECT}_\K{IDENTIFIER}_::=_\{_\V{rdrnVncMIB}(\N{4})_\N{1}_\}}}
\L{\LB{}}
\L{\LB{\V{logicalProcessTable}_\K{OBJECT}\-\V{TYPE}}}
\L{\LB{}\Tab{8}{\K{SYNTAX}_\K{SEQUENCE}_\V{OF}_\V{LogicalProcessEntry}}}
\L{\LB{}\Tab{8}{\K{MAX}\-\V{ACCESS}_\V{not}\-\V{accessible}}}
\L{\LB{}\Tab{8}{\K{STATUS}_\V{current}}}
\L{\LB{}\Tab{8}{\K{DESCRIPTION}}}
\L{\LB{}\Tab{16}{\3\V{Table}_\V{of}_\V{VNC}_\V{LP}_\V{information}.\3}}
\L{\LB{}\Tab{8}{::=_\{_\V{logicalProcess}_\N{1}_\}}}
\L{\LB{}}
\L{\LB{\V{logicalProcessEntry}_\K{OBJECT}\-\V{TYPE}}}
\L{\LB{}\Tab{8}{\K{SYNTAX}_\V{LogicalProcessEntry}}}
\L{\LB{}\Tab{8}{\K{MAX}\-\V{ACCESS}_\V{not}\-\V{accessible}}}
\L{\LB{}\Tab{8}{\K{STATUS}_\V{current}}}
\L{\LB{________\K{DESCRIPTION}}}
\L{\LB{________________\3\V{Table}_\V{of}_\V{VNC}_\V{LP}_\V{information}.\3}}
\L{\LB{}\Tab{8}{\K{INDEX}_\{_\V{logicalProcessID}_\}}}
\L{\LB{}\Tab{8}{::=_\{_\V{logicalProcessTable}_\N{1}_\}}}
\L{\LB{}}
\L{\LB{\V{LogicalProcessEntry}_::=_\K{SEQUENCE}_\{}}
\L{\LB{}\Tab{8}{\V{logicalProcessID}_}}
\L{\LB{}\Tab{16}{\V{DisplayString},}}
\L{\LB{}\Tab{8}{\V{logicalProcessLVT}}}
\L{\LB{}\Tab{16}{\K{INTEGER},}}
\L{\LB{________\V{logicalProcessQRSize}}}
\L{\LB{}\Tab{16}{\K{INTEGER},}}
\L{\LB{________\V{logicalProcessQSSize}}}
\L{\LB{}\Tab{16}{\K{INTEGER},}}
\L{\LB{________\V{logicalProcessRollbacks}}}
\L{\LB{}\Tab{16}{\K{INTEGER},}}
\L{\LB{________\V{logicalProcessSQSize}}}
\L{\LB{}\Tab{16}{\K{INTEGER},}}
\L{\LB{________\V{logicalProcessTolerance}}}
\L{\LB{}\Tab{16}{\K{INTEGER},}}
\L{\LB{________\V{logicalProcessGVT}}}
\L{\LB{}\Tab{16}{\K{INTEGER},}}
\L{\LB{________\V{logicalProcessLookAhead}}}
\L{\LB{}\Tab{16}{\K{INTEGER},}}
\L{\LB{________\V{logicalProcessGvtUpdate}}}
\L{\LB{}\Tab{16}{\K{INTEGER},}}
\L{\LB{________\V{logicalProcessStepSize}}}
\L{\LB{}\Tab{16}{\K{INTEGER}}}
\L{\LB{}\Tab{8}{\}}}
\L{\LB{}}
\L{\LB{\V{logicalProcessID}_\K{OBJECT}\-\V{TYPE}}}
\L{\LB{}\Tab{8}{\K{SYNTAX}_\V{DisplayString}}}
\L{\LB{}\Tab{8}{\K{MAX}\-\V{ACCESS}_\V{read}\-\V{only}}}
\L{\LB{}\Tab{8}{\K{STATUS}_\V{current}}}
\L{\LB{}\Tab{8}{\K{DESCRIPTION}}}
\L{\LB{}\Tab{16}{\3\V{The}_\V{LP}_\V{identifier}.\3}}
\L{\LB{}\Tab{8}{::=_\{_\V{logicalProcessEntry}_\N{1}_\}}}
\L{\LB{}}
\L{\LB{\V{logicalProcessLVT}_\K{OBJECT}\-\V{TYPE}}}
\L{\LB{}\Tab{8}{\K{SYNTAX}_\K{INTEGER}_(\N{0.\,.2147483647})}}
\L{\LB{}\Tab{8}{\K{MAX}\-\V{ACCESS}_\V{read}\-\V{only}}}
\L{\LB{}\Tab{8}{\K{STATUS}_\V{current}}}
\L{\LB{}\Tab{8}{\K{DESCRIPTION}}}
\L{\LB{}\Tab{16}{\3\V{This}_\V{is}_\V{the}_\V{LP}_\V{Local}_\V{Virtual}_\V{Time}.\3}}
\L{\LB{}\Tab{8}{::=_\{_\V{logicalProcessEntry}_\N{2}_\}}}
\L{\LB{}}
\L{\LB{\V{logicalProcessQRSize}_\K{OBJECT}\-\V{TYPE}}}
\L{\LB{________\K{SYNTAX}_\K{INTEGER}_(\N{0.\,.2147483647})}}
\L{\LB{________\K{MAX}\-\V{ACCESS}_\V{read}\-\V{only}}}
\L{\LB{________\K{STATUS}_\V{current}}}
\L{\LB{________\K{DESCRIPTION}}}
\L{\LB{________________\3\V{This}_\V{is}_\V{the}_\V{LP}_\V{Receive}_\V{Queue}_\V{Size}.\3}}
\L{\LB{________::=_\{_\V{logicalProcessEntry}_\N{3}_\}}}
\L{\LB{}}
\L{\LB{\V{logicalProcessQSSize}_\K{OBJECT}\-\V{TYPE}}}
\L{\LB{________\K{SYNTAX}_\K{INTEGER}_(\N{0.\,.2147483647})}}
\L{\LB{________\K{MAX}\-\V{ACCESS}_\V{read}\-\V{only}}}
\L{\LB{________\K{STATUS}_\V{current}}}
\L{\LB{________\K{DESCRIPTION}}}
\L{\LB{________________\3\V{This}_\V{is}_\V{the}_\V{LP}_\V{send}_\V{queue}_\V{size}.\3}}
\L{\LB{________::=_\{_\V{logicalProcessEntry}_\N{4}_\}}}
\L{\LB{}}
\L{\LB{\V{logicalProcessRollbacks}_\K{OBJECT}\-\V{TYPE}}}
\L{\LB{________\K{SYNTAX}_\K{INTEGER}_(\N{0.\,.2147483647})}}
\L{\LB{________\K{MAX}\-\V{ACCESS}_\V{read}\-\V{only}}}
\L{\LB{________\K{STATUS}_\V{current}}}
\L{\LB{________\K{DESCRIPTION}}}
\L{\LB{________________\3\V{This}_\V{is}_\V{the}_\V{number}_\V{of}_\V{rollbacks}_\V{this}_\V{LP}_\V{has}_\V{suffered}.\3}}
\L{\LB{________::=_\{_\V{logicalProcessEntry}_\N{5}_\}}}
\L{\LB{}}
\L{\LB{\V{logicalProcessSQSize}_\K{OBJECT}\-\V{TYPE}}}
\L{\LB{________\K{SYNTAX}_\K{INTEGER}_(\N{0.\,.2147483647})}}
\L{\LB{________\K{MAX}\-\V{ACCESS}_\V{read}\-\V{only}}}
\L{\LB{________\K{STATUS}_\V{current}}}
\L{\LB{________\K{DESCRIPTION}}}
\L{\LB{________________\3\V{This}_\V{is}_\V{the}_\V{LP}_\V{state}_\V{queue}_\V{size}.\3}}
\L{\LB{________::=_\{_\V{logicalProcessEntry}_\N{6}_\}}}
\L{\LB{}}
\L{\LB{\V{logicalProcessTolerance}_\K{OBJECT}\-\V{TYPE}}}
\L{\LB{________\K{SYNTAX}_\K{INTEGER}_(\N{0.\,.2147483647})}}
\L{\LB{________\K{MAX}\-\V{ACCESS}_\V{read}\-\V{only}}}
\L{\LB{________\K{STATUS}_\V{current}}}
\L{\LB{________\K{DESCRIPTION}}}
\L{\LB{________________\3\V{This}_\V{is}_\V{the}_\V{allowable}_\V{deviation}_\V{between}_\V{process}\4\V{s}_}}
\L{\LB{_________________\V{predicted}_\V{state}_\V{and}_\V{the}_\V{actual}_\V{state}.\3}}
\L{\LB{________::=_\{_\V{logicalProcessEntry}_\N{7}_\}}}
\L{\LB{}}
\L{\LB{\V{logicalProcessGVT}__\K{OBJECT}\-\V{TYPE}}}
\L{\LB{________\K{SYNTAX}_\K{INTEGER}_(\N{0.\,.2147483647})}}
\L{\LB{________\K{MAX}\-\V{ACCESS}_\V{read}\-\V{only}______________________}}
\L{\LB{________\K{STATUS}_\V{current}________}}
\L{\LB{________\K{DESCRIPTION}}}
\L{\LB{________________\3\V{This}_\V{is}_\V{this}_\V{system}\4\V{s}_\V{notion}_\V{of}_\V{Global}_\V{Virtual}_\V{Time}.\3}}
\L{\LB{________::=_\{_\V{logicalProcessEntry}_\N{8}_\}}}
\L{\LB{}}
\L{\LB{\V{logicalProcessLookAhead}__\K{OBJECT}\-\V{TYPE}}}
\L{\LB{________\K{SYNTAX}_\K{INTEGER}_(\N{0.\,.2147483647})}}
\L{\LB{________\K{MAX}\-\V{ACCESS}_\V{read}\-\V{only}}}
\L{\LB{________\K{STATUS}_\V{current}}}
\L{\LB{________\K{DESCRIPTION}}}
\L{\LB{________________\3\V{This}_\V{is}_\V{this}_\V{system}\4\V{s}_\V{maximum}_\V{time}_\V{into}_\V{which}_\V{it}_\V{can}}}
\L{\LB{_________________\V{predict}.\3}}
\L{\LB{________::=_\{_\V{logicalProcessEntry}_\N{9}_\}}}
\L{\LB{}}
\L{\LB{\V{logicalProcessGvtUpdate}_\K{OBJECT}\-\V{TYPE}}}
\L{\LB{________\K{SYNTAX}_\K{INTEGER}_(\N{0.\,.2147483647})}}
\L{\LB{________\K{MAX}\-\V{ACCESS}_\V{read}\-\V{only}}}
\L{\LB{________\K{STATUS}_\V{current}}}
\L{\LB{________\K{DESCRIPTION}}}
\L{\LB{________________\3\V{This}_\V{is}_\V{the}_\V{GVT}_\V{update}_\V{rate}.\3}}
\L{\LB{________::=_\{_\V{logicalProcessEntry}_\N{10}_\}}}
\L{\LB{}}
\L{\LB{\V{logicalProcessStepSize}_\K{OBJECT}\-\V{TYPE}}}
\L{\LB{________\K{SYNTAX}_\K{INTEGER}_(\N{0.\,.2147483647})}}
\L{\LB{________\K{MAX}\-\V{ACCESS}_\V{read}\-\V{only}}}
\L{\LB{________\K{STATUS}_\V{current}}}
\L{\LB{________\K{DESCRIPTION}}}
\L{\LB{________________\3\V{This}_\V{is}_\V{the}_\V{time}_\V{until}_\V{next}_\V{virtual}_\V{message}.\3}}
\L{\LB{________::=_\{_\V{logicalProcessEntry}_\N{11}_\}}}
\L{\LB{\K{END}}}
\L{\LB{}}

\end{singlespace}
\normalsize
\chapter{Acronyms}
\section{Acronyms}

\begin{singlespace}

\begin{acronym}
 \acro{ABR}{Available Bit Rate}\index{Available Bit Rate}.
ABR is an ATM layer service category for
which the limiting ATM layer transfer characteristics provided
by the network may change subsequent to connection
establishment. A flow control mechanism is specified which
supports several types of feedback to control the source rate
in response to changing ATM layer transfer characteristics. It
is expected that an end-system that adapts its traffic in
accordance with the feedback will experience a low cell loss
ratio and obtain a fair share of the available bandwidth
according to a network specific allocation policy. Cell delay
variation is not controlled in this service, although admitted
cells are not delayed unnecessarily.
 \acro{AHDLC}{Adaptive High Level Data Link Protocol}\index{Adaptive High Level Data Link Protocol}.
An Adaptive HDLC-like link layer protocol used in the \acf{RDRN}.
The frame size and retry mechanism are adaptive.
 \acro{AMPS}{Advanced Mobile Phone System}\index{Advanced Mobile Phone System}.
\acl{AMPS} \cite{AMPS} is the North American analog cellular phone standard. 
\acl{AMPS} operates in the 800 MHz and 900 MHz bands. About 85\%
of \acl{AMPS} subscribers are in the U.S. 
 \acro{ANO}{Attach New to Old}\index{Attach New to Old}.
The node where old and new connections meet after a \acf{HO} is called 
the \acl{ANO} point. A term defined in \cite{Hauwermeiren} for mobile 
wireless \acf{ATM} architecture along with \acf{MT}, \acf{BTS}, \acf{HO}, 
and \acf{CSS}. These elements are shown in Figure \ref{mobreq}.
 \acro{ARP}{Address Resolution Protocol}\index{Address Resolution Protocol}.
A TCP/IP Protocol that dynamically binds a Network Layer IP
address to a Data Link Layer physical hardware address.
 \acro{ATMARP}{Asynchronous Transfer Mode Address Resolution Protocol}\index{Asynchronous Transfer Mode Address Resolution Protocol}.
A TCP/IP Protocol for dynamically binding a Network Layer IP
address to an ATM address. Described in \cite{RFC1577}.
 \acro{ATM}{Asynchronous Transfer Mode}\index{Asynchronous Transfer Mode}.
A transfer mode in which the information is organized into cells. 
It is asynchronous in the sense that the recurrence of cells containing 
information from an individual user is not necessarily periodic.
 \acro{BTS}{Base Transceiver Station}\index{Base Transceiver Station}.
A non-mobile base station which has both wired and wireless \acf{ATM}
ports. A term defined in \cite{Hauwermeiren} for mobile wireless \acf{ATM}
architecture along with \acf{MT}, \acf{ANO}, \acf{HO}, and \acf{CSS}.
These elements are shown in Figure \ref{mobreq}.
 \acro{C/E}{Condition Event Network}\index{Condition Event Network}.
A \acl{C/E} network consists of state and transition elements which
contain tokens. Tokens reside in state elements. When all state elements
leading to a transition element contain a token, several changes take
place in the \acl{C/E} network. First,
the tokens are removed from the conditions which triggered the event, the
event occurs, and finally tokens are placed in all state outputs from the
transition which was triggered. Multiple tokens in a condition
and the uniqueness of the tokens is irrelevant in a \acl{C/E} Net.
 \acro{CDMA}{Code Division Multiple Access}\index{Code Division Multiple Access}.
A coding scheme, used as a modulation     
technique, in which multiple channels are independently coded for
transmission over a single wide-band channel. Note 1: In some              
communication systems, CDMA is used as an access method that permits            
carriers from different stations to use the same transmission equipment by      
using a wider bandwidth than the individual carriers. On reception, each        
carrier can be distinguished from the others by means of a specific             
modulation code, thereby allowing for the reception of signals that were        
originally overlapping in frequency and time. Thus, several transmissions       
can occur simultaneously within the same bandwidth, with the mutual             
interference reduced by the degree of orthogonality of the unique codes used    
in each transmission. Note 2: CDMA permits a more uniform distribution of       
energy in the emitted bandwidth.
 \acro{CDPD}{Cellular Digital Packet Data}\index{Cellular Digital Packet Data}.
\acl{CDPD} is described in \cite{CDPD} and \cite{Kunzinger} protocol. 
The \acl{CDPD} protocol operates much like the Mobile IP protocol for 
mobile cellular voice networks and is designed to co-exist with the 
\acl{AMPS}\cite{AMPS}. \acl{CDPD} uses registration and encapsulation to 
forward packets to the current location of the mobile host just as in 
Mobile-IP. \acl{CDPD} was developed independently of Mobile-IP.
 \acro{CE}{Clustered Environment}\index{Clustered Environment}.
One of the contributions of \cite{Avril95} in \acl{CTW} is an attempt 
to efficiently control a cluster of \acl{LP}s on a processor by means of the     
\acl{CE}. The \acl{CE} allows multiple \acf{LP} to behave as individual \acl{LP}s
as in the basic time warp algorithm or as a single collective \acl{LP}.
 \acro{CMB}{Chandy-Misra-Bryant}\index{Chandy-Misra-Bryant}.
A Conservative distributed simulation synchronization technique.
 \acro{CMIP}{Common Management Information Protocol}\index{Common Management Information Protocol}.
A protocol used by an application process to exchange information and 
commands for the purpose of managing remote computer and communications 
resources. Described in \cite{CMIP}.
 \acro{CRC}{Cyclic Redundancy Checksum}\index{Cyclic Redundancy Checksum}.
An error-detection scheme that (a) uses
parity bits generated by polynomial encoding of digital signals, (b) appends
those parity bits to the digital signal, and (c) uses decoding algorithms
that detect errors in the received digital signal. Note: Error correction,
if required, may be accomplished through the use of an automatic            
repeat-request (ARQ) system.
 \acro{CSS}{Cell Site Switch}\index{Cell Site Switch}.
A wired \acf{ATM} switch near the \acf{BTS}.
A term defined in \cite{Hauwermeiren} for mobile wireless \acf{ATM}
architecture along with \acf{MT}, \acf{ANO}, \acf{HO}, and \acf{BTS}.
These elements are shown in Figure \ref{mobreq}.
 \acro{CS}{Current State}\index{Current State}.
The current value of all information concerning a \acf{PP} encapsulated
by a \acf{LP} and all the structures associated with the \acl{LP}.
 \acro{CTW}{Clustered Time Warp}\index{Clustered Time Warp}.
\acl{CTW} is an optimistic distributed simulation mechanism described 
in \cite{Avril95}.
 \acro{DES}{Data Encryption Standard}\index{Data Encryption Standard}.
A cryptographic algorithm for the         
protection of unclassified computer data and published by the National
Institute of Standards and Technology in Federal Information Processing
Standard Publication 46-1. Note: DES is not approved for protection of
national security classified information.
 \acro{DHCP}{Dynamic Host Configuration Protocol}\index{Dynamic Host Configuration Protocol}.
The \acl{DHCP}\index{Dynamic Host Configuration Protocol} \cite{RFC1541} 
provides for automatic configuration of a host from data stored on the 
network.
 \acro{EN}{Edge Node}\index{Edge Node}.
A \acf{RDRN} \cite{BushWINET97} node which provides connectivity between 
a wired and wireless \acf{ATM} network.
 \acro{ES}{Edge Switch}\index{Edge Switch}.
A component of the \acf{RDRN} \cite{BushWINET97} \acf{EN} which contains 
internal \acf{ATM} switching capability for wireless input ports.
 \acro{ESN}{Electronic Serial Number}\index{Electronic Serial Number}\.
The \acl{ESN} is stored along with the \acf{MIN} in the \acf{HLR}
in the \acf{AMPS} \cite{AMPS}.
 \acro{FH}{Fixed Host}\index{Fixed Host}.
A host directly connected to a wired network.
 \acro{FSM}{Finite State Machine}\index{Finite State Machine}.
A five-tuple consisting of a set of states, an input alphabet, an
output alphabet, a next-state transition function, and an output
function. Used to formally describe the operation of a protocol.
 \acro{GFC}{Generic Flow Control}\index{Generic Flow Control}.
\acl{GFC} is a field in the ATM header which can be used to provide local 
functions (e.g., flow control). It has local significance only and the 
value encoded in the field is not carried end-to-end.
 \acro{GPS}{Global Positioning System}\index{Global Positioning System}.
A satellite-based positioning service developed
and operated by the Department of Defense.
 \acro{GSM}{Global System for Mobile Communication}\index{Global System for Mobile Communication}.
A pan-European standard for digital mobile telephony which provides a much
higher capacity than traditional analog telephones as well as diversified
services (voice, data) and a greater transmission security through information
encoding for users across Europe. \acl{GSM} is described in \cite{GSM}.
 \acro{GSV}{Global Synchronic Distance}\index{Global Synchronic Distance}.
The maximum Synchronic Distance in a Petri-Net model of a system.
 \acro{GVT}{Global Virtual Time}\index{Global Virtual Time}.
The largest time beyond which a rollback based system will 
never rollback.
 \acro{HDLC}{High Level Data Link Protocol}\index{High Level Data Link Protocol}.
A link-level protocol used to facilitate reliable 
point to point transmission of a data packet. A
subset of HDLC, known as LAP-B is the Layer two protocol
for CCITT Recommendation X.25.
 \acro{HLR}{Home Location Register}\index{Home Location Register}.
The \acl{HLR} and \acl{VLR} are used in mobile cellular voice networks
\cite{AMPS} and \cite{GSM}. These agents maintain the information
concerning whether a mobile node is currently associated with its home
network or is a visiting mobile node, that is, not currently in its home
network.
 \acro{HO}{Handoff}\index{Handoff}.
A change in association of a mobile node from one base station to
another. A term defined in \cite{Hauwermeiren} for mobile wireless 
\acf{ATM} architecture along with \acf{MT}, \acf{ANO}, \acf{HO}, 
and \acf{BTS}. These elements are shown in Figure \ref{mobreq}.
 \acro{IETF}{Internet Engineering Task Force}\index{Internet Engineering Task Force}.
The main standards organization for the Internet. The IETF is
a large open international community of network
designers, operators, vendors, and researchers
concerned with the evolution of the Internet
architecture and the smooth operation of the
Internet. It is open to any interested individual.
 \acro{IPC}{Inter-Processor Communication}\index{Inter-Processor Communication}.
Communication among Unix processes. This may take place via
sockets, shared memory, or semaphores.
 \acro{LIS}{Logical IP Subnet}\index{Logical IP Subnet}.
In the LIS scenario described in \cite{RFC1577}, each separate 
administrative entity configures its hosts and routers within a 
closed logical IP subnetwork. Each LIS operates and communicates 
independently of other ac{LIS}s on the same ATM network. Hosts 
connected to ATM communicate directly to other hosts within the 
same LIS. Communication to hosts outside of the local LIS is provided 
via an IP router. This router is an ATM Endpoint attached to the ATM 
network that is configured as a member of one or more LISs.  This 
configuration may result in a number of disjoint LISs operating over 
the same ATM network. Hosts of differing IP subnets MUST communicate 
via an intermediate IP router even though it may be possible to open 
a direct VC between the two IP members over the ATM network.
 \acro{LN}{Logical Node}\index{Logical Node}.
A logical node that represents a lower level peer group as a
single point for purposes of operating at one level of the
PNNI routing hierarchy.
 \acro{LP}{Logical Process}\index{Logical Process}.
An \acl{LP} consists of the \acf{PP} and additional
data structures and instructions which maintain message
order and correct operation as a system executes ahead of real
time.
 \acro{LVT}{Local Virtual Time}\index{Local Virtual Time}.
The \acf{LP} contains its notion of time known as \acf{LVT}.
 \acro{MAC}{Media Access Control}\index{Media Access Control}.
The part of the data link layer that supports topology-dependent 
functions and uses  the services of the physical layer to provide 
services to the logical link control sub-layer.
 \acro{MH}{Mobile Host}\index{Mobile Host}.
A term used to describe a mobile end system in \cite{Seshan}.
 \acro{MIB}{Management Information Base}\index{Management Information Base}.
A collection of objects which can be accessed by a network management 
protocol.
 \acro{MIN}{Mobile Identification Number}\index{Mobile Identification Number}.
The \acl{MIN} is stored along with the \acf{ESN} in the \acf{HLR}
in the \acf{AMPS} \cite{AMPS}.
 \acro{MSC}{Mobile Switching Center}\index{Mobile Switching Center}.
The \acl{HLR} and \acl{VLR} reside on the \acl{MSC} which is connected    
to each mobile base station in the \acf{AMPS}. The \acl{MSC} controls
call setup, call transfer, billing, interaction with the PSTN, and
other functions using \acf{SS7}.
 \acro{MSR}{Mobile Support Router}\index{Mobile Support Router}.
The mobile TCP/IP known as I-TCP requires \acl{MSR} which have the 
ability to transfer images of sockets involved in an established 
TCP connection from one MSR to another.
 \acro{MTW}{Moving Time Windows}\index{Moving Time Windows}.
\acl{MTW} is a distributed simulation algorithm that controls the amount 
of aggressiveness in the system by means of a moving time window \acl{MTW}.
The trade-off in having no roll-backs in this algorithm is loss of
fidelity in the simulation results.
 \acro{MT}{Mobile Terminal}\index{Mobile Terminal}.
A mobile host end system which receives wireless \acf{ATM} cells.
A term defined in \cite{Hauwermeiren} for mobile wireless \acf{ATM}           
architecture along with \acf{CSS}, \acf{ANO}, \acf{HO}, and \acf{BTS}.
These elements are shown in Figure \ref{mobreq}.
 \acro{NCP}{Network Control Protocol}\index{Network Control Protocol}.
The \acf{RDRN} protocol which supports wireless \acf{ATM} 
communications.
 \acro{NFT}{No False Time-stamps}\index{No False Time-stamps}.
\acl{NFT} Time Warp assumes that if an incorrect computation produces and
incorrect event ($E_{i,T}$), then it must be the case that the correct
computation also produces an event ($E_{i,T}$) with the same timestamp
This simplification makes the analysis in \cite{Ghosh93}
tractable.
 \acro{NHRP}{Next Hop Resolution Protocol}\index{Next Hop Resolution Protocol}.
The \acl{NHRP} \cite{NBMA} allows \acl{ATMARP} to take place across 
\acf{LIS}.
 \acro{NPSI}{Near Perfect State Information}\index{Near Perfect State Information}.
The \acl{NPSI} Adaptive Synchronization Algorithms for PDES are
discussed in \cite{Srini44} and \cite{Srini20}. The adaptive
algorithms use feedback from the simulation itself in order to adapt.
The NPSI system requires an overlay system to return feedback information 
to the \acl{LP}s. The \acl{NPSI} Adaptive Synchronization Algorithm examines 
the system state (or an approximation of the state) calculates an error 
potential for future error, then translates the error potential into a 
value which controls the amount of optimism.
 \acro{NTP}{Network Time Protocol}\index{Network Time Protocol}.
A TCP/IP time synchronization mechanism. \acl{NTP} \cite{RFC958} 
is not required in \acf{VNC} on the \acf{RDRN} because
each host in the \acl{RDRN} network has its own \acl{GPS}
receiver.
 \acro{PA}{Perturbation Analysis}\index{Perturbation Analysis}.
The technique of \acl{PA} allows a great deal more information to be
obtained from a single simulation execution than explicitly collected 
statistics. It is particularly useful for finding the sensitivity 
information of simulation parameters from the sample path of a single 
simulation run. May be an ideal way for VNC to automatically adjust 
tolerances and provide feedback to driving process(es).
Briefly, assume a sample path, $(\Theta,\xi)$ from a simulation.
$\Theta$ is vector of all parameters $\xi$ is vector of all random 
occurrences. $L(\Theta, \xi)$ is the sample performance.
$J(\Theta)$ is the average performance, $E[L(\Theta,\xi)].$
Parameter changes cause perturbations in event timing.
Perturbations in event timing propagate to other events.
This induces perturbations in $L$.
If perturbations into $(\Theta, \xi)$ are small, assume event
trace $(\Theta+d\Theta, \xi)$ remains unchanged.
Then ${{dL(\Theta,\xi)} \over {d\Theta}}$ can be calculated.
From this, the gradient of $J(\Theta)$ can be obtained which
provides the sensitivity of performance to parameter changes.
PA can be used to adjust tolerances while VNC is executing 
because event times are readily available in the \acf{SQ}.
 \acro{PBS}{Portable Base Station}\index{Portable Base Station}.
\acl{PBS} nodes are connected via Virtual Path Trees which are
part of the wireless \acl{ATM} BAHAMA \cite{BAHAMA, Veer, Veer97} 
network. 
 \acro{PCN}{Personal Communications Network}\index{Personal Communications Network}.
A set of capabilities
that allows some combination of terminal mobility, personal mobility, and
service profile management. Note 1: The flexibility offered by PCS can    
supplement existing telecommunications services, such as cellular radio,    
used for NS/EP missions. Note 2: PCS and UPT are sometimes mistakenly
assumed to be the same service concept. UPT allows complete personal        
mobility across multiple networks and service providers. PCS may use UPT    
concepts to improve subscriber mobility in allowing roaming to different    
service providers, but UPT and PCS are not the same service concept.        
Contrast with Universal Personal Telecommunications service.
 \acro{PDES}{Parallel Discrete Event Simulation}\index{Parallel Discrete Event Simulation}.
\acl{PDES} is a class of simulation algorithms which partition a
simulation into individual events and synchronizes the time
the events are executed on multiple processors such that the real
time to execute the simulation is as fast as possible.
 \acro{PDU}{Protocol Data Unit}\index{Protocol Data Unit}.
1. Information that is delivered as a unit among  
peer entities of a network and that may contain control information, address
information, or data. 2. In layered systems, a unit of data that is      
specified in a protocol of a given layer and that consists of  
protocol-control information of the given layer and possibly user data of   
that layer.
 \acro{PGL}{Peer Group Leader}\index{Peer Group Leader}.
A node which has been elected to perform some of the functions
associated with a logical group node.
 \acro{PG}{Peer Group}\index{Peer Group}.
A set of logical nodes which are grouped for purposes of
creating a routing hierarchy. PTSEs are exchanged among all
members of the group.
 \acro{P/T}{Place Transition Net}\index{Place Transition Net}.
A P/T Network is exactly like a \acf{C/E} Net except that a P/T Net 
allows multiple tokens in a place and multiple tokens may be required 
to cause a transition to fire.
 \acro{PIPS}{Partially Implemented Performance Specification}\index{Partially Implemented Performance Specification}.
\acl{PIPS} is a hybrid simulation and real-time system which is described 
in \cite{Bagrodia91}. Components of a
performance specification for a distributed system are implemented
while the remainder of the system is simulated. More components are
implemented and tested with the simulated system in an iterative
manner until the entire distributed system is implemented.
 \acro{PNNI}{Private Network-Network Interface}\index{Private Network-Network Interface}.
A routing information protocol that enables extremely scalable, 
full function, dynamic multi-vendor ATM switches to be integrated in 
the same network.
 \acro{PP}{Physical Process}\index{Physical Process}.
A \acl{PP} is nothing more than an executing task defined by
program code. An example of a \acl{PP} is the \acf{RDRN} beam table creation
task. The beam table creation task generates a table of complex
weights which controls the angle of the radio beams based
on position input.
 \acro{Q.2931}{Q.2931}\index{Q.2931}.
Q.2931 is the ITU standard for \acf{ATM} signaling described in 
\cite{Q2931}.
 \acro{QR}{Receive Queue}\index{Receive Queue}.
A queue used in the \acf{VNC} algorithm to hold incoming messages
to a \acf{LP}. The messages are stored in the queue in order
by receive time.
 \acro{QS}{Send Queue}\index{Send Queue}.
A queue used in the \acf{VNC} algorithm to hold copies of messages
which have been sent by a \acf{LP}. The messages in the \acf{QS}
may be sent as anti-messages if a rollback occurs.
 \acro{QoS}{Quality of Service}\index{Quality of Service}.
Quality of Service is defined on an end-to-end basis in terms of the 
following attributes of the end-to-end ATM connection: Cell Loss Ratio, 
Cell Transfer Delay, Cell Delay Variation.
 \acro{RDRN}{Rapidly Deployable Radio Network}\index{Rapidly Deployable Radio Network}.
This project, funded by the Information Technology Office (ITO) of the 
Advanced Research Projects Agency, involves the design and implementation 
of a reconfigurable ATM wireless network which uses antenna beamforming 
for improved spatial reuse of frequencies. The RDRN project is associated 
with the ARPA Global Mobile Information Systems initiative. 
 \acro{RN}{Remote Node}\index{Remote Node}.
A host station end system in the \acf{RDRN}.
 \acro{RT}{Real Time}\index{Real Time}.
The current wall clock time.
 \acro{SID}{Station Identification}\index{Station Identification}.
The \acl{SID} is used in the \acf{AMPS} \cite{AMPS} to identify a base station.
 \acro{SLP}{Single Processor Logical Process}\index{Single Processor Logical Process}.
Multiple \acf{LP} executing on a single processor.
 \acro{SLW}{Sliding Lookahead Window}\index{Sliding Lookahead Window}.
The \acl{SLW} is used in \acf{VNC} to limit or throttle the prediction
rate of the \acl{VNC} system. The \acl{SLW} is defined as the maximum
time into the future for which the ac{VNC} system may predict events.
 \acro{SNMP}{Simple Network Management Protocol}\index{Simple Network Management Protocol}.
The Transmission Control         
Protocol/Internet Protocol (TCP/IP) standard protocol that (a) is used to   
manage and control IP gateways and the networks to which they are attached,
(b) uses IP directly, bypassing the masking effects of TCP error correction,
(c) has direct access to IP datagrams on a network that may be operating 
abnormally, thus requiring management, (d) defines a set of variables that  
the gateway must store, and (e) specifies that all control operations on the
gateway are a side-effect of fetching or storing those data variables, i.e.,
operations that are analogous to writing commands and reading status.
\acl{SNMP} is described in \cite{SNMP}.
 \acro{SQ}{State Queue}\index{State Queue}.
The \acl{SQ} is used in \acf{VNC} as a \acf{LP} structure to hold 
saved state information for use in case of a rollback.
The \acl{SQ} is the cache into which pre-computed results are stored.
 \acro{SS7}{Signaling System 7}\index{Signaling System 7}\.
A common-channel signaling system defined by  
the CCITT in the 1988 Blue Book, in Recommendations Q.771 through Q.774.    
Note: SS7 is a prerequisite for implementation of an Integrated Services   
Digital Network (ISDN).
 \acro{TDMA}{Time Division Multiple Access}\index{Time Division Multiple Access}.
A communications technique that uses a
common channel (multi-point or broadcast) for communications among multiple
users by allocating unique time slots to different users. Note: TDMA
is used extensively in satellite systems, local area networks, physical
security systems, and combat-net radio systems.
 \acro{TDN}{Temporary Directory Number}\index{Temporary Directory Number}.
If the visiting mobile node can be reached in the \acf{AMPS}, the visiting
mobile's \acf{MSC} will respond with a \acl{TDN}.
 \acro{TNC}{Terminal Node Controller}\index{Terminal Node Controller}.
The \acl{TNC} automatically divides the message into packets, keys the
transmitter and sends the packets. While receiving packets, the
TNC automatically decodes, checks for errors, and displays the
received messages. In addition, any packet TNC can be used a
packet relay station, sometimes called a digipeater. This allows
for greater range by stringing several packet stations together.
 \acro{TOE}{Time of Expiry}\index{Time of Expiry}.
A priority queuing mechanism proposed in \cite{Raychaudhuri} for 
queuing wireless \acf{ATM} cells. Call request packets indicate
an expiry time for data to be transmitted. Cells are then served
in order of expiry with the smallest expiry time served first.
The \acl{TOE} mechanism is proposed to reduce packet loss for
real-time data.
 \acro{TR}{Receive Time}\index{Receive Time}.
The time a \acf{VNC} message value is predicted to be valid.
 \acro{TS}{Send Time}\index{Send Time}.
The \acf{LVT} that a virtual message has been sent. This value
is carried within the header of the message. The \acl{TS} is
used for canceling the effects of false messages.
 \acro{VC}{Virtual Circuit}\index{Virtual Circuit}.
A communications channel that provides for the sequential
unidirectional transport of ATM cells.
 \acro{VCI}{Virtual Circuit Identifier}\index{Virtual Circuit Identifier}.
Virtual Channel Identifier: A unique numerical tag as defined
by a 16 bit field in the ATM cell header that identifies a
virtual channel, over which the cell is to travel.
 \acro{VLR}{Visiting Location Register}\index{Visiting Location Register}.
The \acl{VLR} and \acl{HLR} are used in mobile cellular voice networks
\cite{AMPS} and \cite{GSM}. These agents maintain the information
concerning whether a mobile node is currently associated with its home
network or is a visiting mobile node, that is, not currently in its home
network.
 \acro{VNC}{Virtual Network Configuration}\index{Virtual Network Configuration}.
\acl{VNC} allows future states of a system to be predicted and used
efficiently via optimistic distributed simulation technique 
applied to a real time system.
 \acro{VP}{Virtual Path}\index{Virtual Path}.
A unidirectional logical association or bundle of \acf{VC}s.
 \acro{VPI}{Virtual Path Identifier}\index{Virtual Path Identifier}.
An eight bit field in the ATM cell header which indicates the 
\acf{VP} over which the cell should be routed.
 \acro{VTRP}{Virtual Trees Routing Protocol}\index{Virtual Trees Routing Protocol}.
A wireless \acl{ATM} routing protocol used in the BAHAMA 
\cite{BAHAMA, Veer, Veer97} network. \acf{PBS} nodes are
connected via Virtual Path Trees which are
preconfigured \acf{ATM} \acf{VP}s. These trees can change based on
the topology as described in the {\em Virtual Trees Routing Protocol}
\cite{VTRP}.
\end{acronym}

\end{singlespace}
\chapter{Network Management Simulation Architecture}
The simulation has been implemented with Maisie \cite{bagrodia}.
Its suitability for this has been demonstrated in the \acl{RDRN}\index{RDRN}
network management and control design and development and in
\cite{Short} to develop a mobile wireless network
parallel simulation\index{Parallel Simulation} environment. The 
parallel simulation\index{Parallel Simulation} 
environment shows a speedup over the currently used commercial 
sequential simulation packages.  The environment 
and a set of modules which have been developed for mobile network
simulation are described in \cite{Short}.
Maisie uses a language which has been influenced by a classic work describing 
the characteristics of a parallel programming\index{Parallel Programming}
language structure \cite{Hoare81}. The programming features
developed here are used in many parallel programming 
languages\index{Parallel Programming!language} besides Maisie\index{Maisie}.
Since every Maisie 
entity has a built-in input queue, each \acl{LP} is comprised of three
additional Maisie entities:
\begin{itemize}
\item An entity which represents the \acl{PP}\index{PP}
\item An entity for the \acl{LP} state queue\index{State Queue}
\item An entity for the \acl{LP} output message queue\index{Send Queue}
\end{itemize}

There is also a gvt entity for the calculation of Global Virtual 
Time (\acl{GVT}\index{Global Virtual Time}). All three of the above 
entities work together to implement Virtual Time as described 
in \cite{Jefferson82}. The first entity above, representing the \acl{PP}, 
contains a delay mechanism in order to implement the sliding 
lookahead window\index{Sliding Lookahead Window|see{$\Lambda$}}.
The gvt process should notify all processes to cease forward simulation 
when \acl{GVT} reaches the end of the window. In this version of \acl{VNC}, each \acl{LP}
simply compares its LVT to the current time and holds processing
until current time is back within the lookahead sliding window.

Determination of Global Virtual Time (\acl{GVT}) should be done as defined by
\cite{Lazowaska90}.
This algorithm allows \acl{GVT} to be determined in a message-passing environment
as opposed to the easier case of a shared memory environment. It also
allows normal processing to continue during the \acl{GVT} determination phase.
However, in this implementation each output message is sent to the gvt 
entity as well as to its proper destination. In addition, the gvt entity 
checks all \acl{LP}s for their current LVT and chooses the minimum message send time
and LVT as the current \acl{GVT}. The gvt entity is allowed to execute in parallel 
with the other entities in this simulation, it does not stop the other
entities while performing its computation and thus may not always be 
perfectly accurate. However, the results were close enough for the purpose
of these experiments.

State adjustment\index{State Adjustment|see{VNC}} rollbacks are the 
most critical part of \acl{VNC}.
They are handled in a slightly different fashion from causality\index{VNC!causality}
failure rollbacks. A state verification failure
causes the \acl{LP} state to be corrected at the time of the state 
verification which failed. 
The state, $S_{v}$, has been obtained from the actual device from 
the state adjustment query at time $t_{v}$.
The \acl{LP} rolls back to exactly $t_{v}$ with state, $S_{v}$.
States greater than $t_{v}$ are removed from the state 
queue\index{State Queue}. 
Anti-messages\index{Anti-Message} are sent from the output message queue 
for all messages
greater than $t_{v}$. The \acl{LP} continues forward execution from this 
point.
Note that this implies that the message and state queues cannot
be purged of elements which are older than the \acl{GVT}. Only elements which
are older than real time can be purged.
\chapter{Complexity in Self-Predictive Systems}
\label{philo}

\begin{figure*}[htbp]                                              
        \centerline{\psfig{file=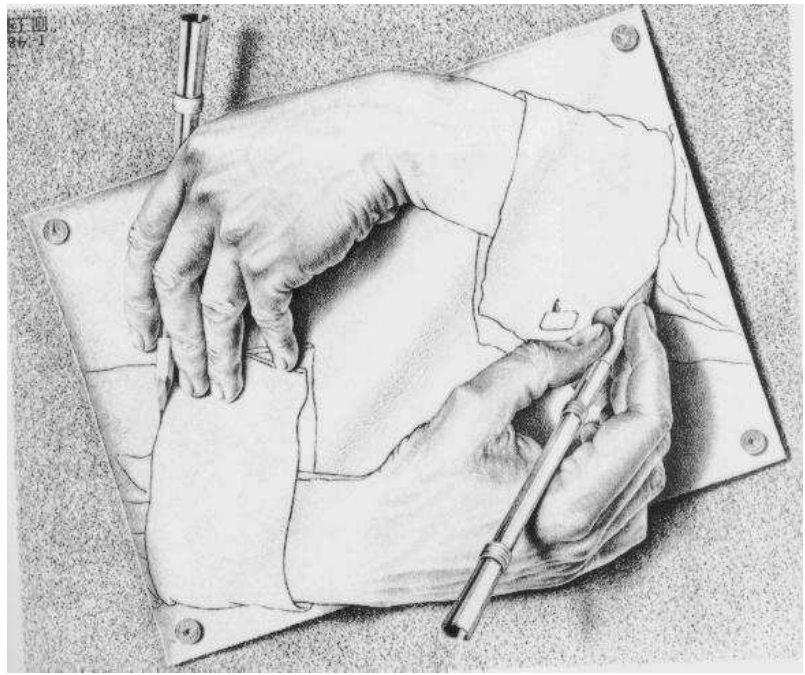,height=3.5in}}
        \caption{A Tangled Hierarchy. \emph{Drawing by M. C. Escher}.}
        \label{th}
\end{figure*}

A fascinating perspective on the topic of self-predictive systems
is found in \emph{G\"odel, Escher, and Bach: An Eternal Golden Braid} 
which is a wonderful look at the nature of Human and Artificial 
Intelligence\index{Artificial Intelligence}. A central point in 
\cite{Hofstadter} is that intelligence is a Tangled 
Hierarchy\index{Tangled Hierarchy}, illustrated in Figure \ref{th}. A 
hand performing the act of drawing is expected to be in a level above 
the hand being drawn. When the two levels are folded together a Tangled 
Hierarchy\index{Tangled Hierarchy} results, which is expressed much more 
elegantly in \cite{Hofstadter}. \acl{VNC} as presented in this work is a 
Tangled Hierarchy\index{Tangled Hierarchy} on several levels: 
simulation-reality and also present-future time. One of the hands in 
Figure \ref{th} represents prediction based on simulation and the other
represents reality, each modifying the other in the \acl{VNC} algorithm.
However, there is a much deeper mathematical relationship present in this 
algorithm which relates to G\"odel's Theorem\index{G\"odel's Theorem}. In 
a nutshell, G\"odel's theorem states that no formal system can describe 
itself with complete fidelity. This places a fundamental limitation on the 
ability of mathematics to describe itself. The implication to artificial 
intelligence\index{Artificial Intelligence} is that 
the human mind can never fully understand its own operation, or possibly
that if one could fully understand how one thinks while one is thinking, 
then one would cease to ``be''. In the much more mundane
\acl{VNC} algorithm, a system is in some sense attempting to use itself to 
predict its own future state with the goal of perfect fidelity. If
G\"odel's Theorem\index{G\"odel's Theorem} applies, then perfect 
fidelity is an impossible goal. However, by allowing for a given tolerance
in the amount of error and assuming accuracy in prediction which increases
as real time approaches the actual time of an event, this study assumes 
that a useful self-predictive system can be implemented. 

\bibliography {/home/bushsf/ref/mob,/home/bushsf/ref/twe,/home/bushsf/ref/psim,/home/bushsf/ref/atmmob,/home/bushsf/ref/standards,/home/bushsf/ref/vci,/home/bushsf/ref/handoff,/home/bushsf/ref/theorem_proving,/home/bushsf/ref/topo,/home/bushsf/ref/signaling,/home/bushsf/ref/an}

\end {document}